\documentclass[prd,nofootinbib,twocolumn,superscriptaddress,balancelastpage]{revtex4-2}
\pdfoutput=1
\usepackage[utf8]{inputenc}
\usepackage{amsmath,amssymb,bm,slashed,braket}
\usepackage{dsfont}
\usepackage{amsfonts}
\usepackage{graphicx}
\usepackage{slashed}
\usepackage[dvipsnames,table]{xcolor}
\usepackage[normalem]{ulem}
\usepackage{hyperref}
\hypersetup{colorlinks,citecolor= blue,linkcolor= blue, urlcolor=blue}
\usepackage{array,boldline,makecell,booktabs}
\usepackage{verbatim}
\usepackage{epsfig}
\usepackage{multirow}
\usepackage{hhline}
\usepackage{bbold}
\usepackage{float}
\usepackage[capitalise]{cleveref}
\usepackage{orcidlink}
\usepackage{graphicx}
\usepackage{placeins} 
\allowdisplaybreaks[4]

\newcommand{\calO}{\mathcal{O}}
\newcommand{\tO}{\tilde{\mathcal{O}}}
\newcommand{\calP}{\mathcal{P}}
\newcommand{\calN}{\mathcal{N}}
\newcommand{\calC}{\mathcal{C}}
\newcommand{\calK}{\mathcal{K}}
\newcommand{\C}{\textsc{c}}
\newcommand{\tN}{\textsc{n}}
\newcommand{\tL}{\textsc{l}}
\newcommand{\tR}{\textsc{r}}
\newcommand{\tS}{\textsc{s}}
\newcommand{\tV}{\textsc{v}}
\newcommand{\tT}{\textsc{t}}
\newcommand{\nn}{{\nonumber}}
\newcommand{\tf}{\texorpdfstring}

\begin{document} 

\title{Nucleon decays into three leptons plus a kaon}

\author{Jin-Han Liang\,\orcidlink{0000-0002-6141-216X}}
\email{jinhanliang@wyu.edu.cn}
\affiliation{School of Applied Physics and Materials, Wuyi University,
Jiangmen 529020, China}
\author{Xiao-Dong Ma\,\orcidlink{0000-0001-7207-7793}}
\email{maxid@scnu.edu.cn}
\affiliation{State Key Laboratory of Nuclear Physics and Technology,
Institute of Quantum Matter, South China Normal University,
Guangzhou 510006, China}
\affiliation{Guangdong Basic Research Center of Excellence for
Structure and Fundamental Interactions of Matter, Guangdong
Provincial Key Laboratory of Nuclear Science,
Guangzhou 510006, China}
\author{Xiang Zhao\,\orcidlink{0009-0008-6024-7722}}
\email{zhaox@m.scnu.edu.cn}
\affiliation{State Key Laboratory of Nuclear Physics and Technology,
Institute of Quantum Matter, South China Normal University,
Guangzhou 510006, China}
\affiliation{Guangdong Basic Research Center of Excellence for
Structure and Fundamental Interactions of Matter, Guangdong
Provincial Key Laboratory of Nuclear Science,
Guangzhou 510006, China}

\begin{abstract}
We systematically investigate baryon-number-violating (BNV) nucleon decays into three leptons and a kaon within the framework of low-energy effective field theory (LEFT). We identify all kinematically allowed four-body decay modes involving a $K^+$ or $K^0$ meson and derive their decay widths induced by local dimension-9 operators in terms of the LEFT Wilson coefficients. 
These results are obtained by classifying the relevant operators, which contain three lepton and three quark fields, including one strange quark, according to their transformation properties under the chiral symmetry group, matching them onto chiral perturbation theory. 
By reanalyzing existing experimental data, we establish stringent bounds on the partial lifetimes of these four-body decay modes and derive corresponding constraints on the dimension-9 Wilson coefficients. 
Our results provide a convenient EFT framework 
for connecting the underlying BNV interactions to experimental searches for these exotic nucleon decays. 
\end{abstract}

\maketitle
\section{Introduction}

Exotic nucleon decays into three leptons offer a unique probe of baryon-number-violating (BNV) physics with nontrivial lepton flavor structures~\cite{McGrew:1999nd,Super-Kamiokande:2014pqx,Super-Kamiokande:2020tor,Hambye:2017qix,Dorsner:2022twk,Liao:2025vlj,Bhoonah:2025qzd}.
In recent studies \cite{Chen:2025mjt,Liao:2025wxk}, we conducted a systematic investigation of proton and neutron decays into three leptons, $\tN \to l_x l_y l _z$ (with $\tN$ being either a proton $p$ or a neutron $n$, and $l_{x,y,z}=e^\pm,\mu^\pm,\nu_{e,\mu,\tau},\bar\nu_{e,\mu,\tau}$ representing lepton states), mediated by both dimension-6 (dim-6) and dim-9 BNV operators in low-energy effective field theory (LEFT).  
Classifying these decays using a generalized lepton flavor number $\Delta F_L \equiv |\Delta F_e|+|\Delta F_\mu|+|\Delta F_\tau|$~\cite{Chen:2025mjt}, where $\Delta F_i$ denotes the net change in flavor $i=e,\mu,\tau$, 
all triple-lepton modes fall into either $\Delta F_L=1$ or $\Delta F_L=3$.
Our analysis revealed that dim-6 BNV operators contribute exclusively to $\Delta F_L=1$ decays via noncontact diagrams, but their effects are generally negligible due to stringent constraints from conventional two-body nucleon decay modes (into a lepton and a meson).
Consequently, dim-9 operators could dominate these processes.
niIn contrast, $\Delta F_L=3$ decays are exclusively induced by dim-9 BNV operators at leading order.

However, within the EFT framework, three-body nucleon purely-triple-lepton decays are only sensitive to operators involving up and down quarks, excluding those with strange quarks.
In this work, we extend our analysis to include dim-9 BNV operators with a strange quark, which can mediate four-body nucleon decay channels featuring three leptons and an additional kaon.
These rare processes provide a novel pathway to probe BNV physics involving strange quarks, complementing existing studies. Notably, certain UV models or effective operators can induce such decays at leading order~\cite{Pati:1983jk,Faroughy:2014tfa,Hambye:2017qix}. 
Building upon our previous works in~\cite{Liao:2025vlj,Liao:2025wxk}, we present a comprehensive analysis of four-body nucleon decays into three leptons plus either a $K^+$ or $K^0$ meson.
First, we classify all dim-9 operators with the field configuration $l^3qqs$ (where $l$ is a generic lepton and $q=u,d$) according to the four general chiral structures and their chirality-flipped counterparts, as identified in~\cite{Liao:2025vlj}.
Second, since nucleon decays occur at a sub-GeV energy scale, we match these LEFT operators to chiral perturbation theory (ChPT) and derive their chiral realizations to systematically evaluate their hadronic matrix elements.
Within the ChPT framework, we compute the decay widths for all relevant four-body decay modes induced by these dim-9 operators and express them in terms of the LEFT Wilson coefficients (WCs). These results provide a general, model-independent parametrization for such decays, while also establishing a direct connection between experimental searches and ultraviolet (UV)-complete scenarios upon integrating out heavy new particles. 

\begin{table*}[!t]
\center
\resizebox{\linewidth}{!}{
\renewcommand{\arraystretch}{1.2}
\begin{tabular}{ l | l l | l l | l l}
\hline
Class
& \multicolumn{2}{c|}{$l^3uus$ configuration}
& \multicolumn{2}{c|}{$l^3dds$ configuration}
& \multicolumn{2}{c}{$l^3uds$ configuration}
\\\hline
$\Delta L = +1$
&$p \to \ell_x^+ \ell_y^+ \ell_z^-K^0$ 
&$p \to \nu_z \bar \nu_y \ell_x^+K^0$
& 
&$n \to \bar\nu_x \bar \nu_y \ell^-_z K^+$ (a)
&\makecell[l]{
$p \to \ell_z^- \ell_y^+ \bar\nu_x K^+$\\
$n \to \ell_z^- \ell_y^+ \bar\nu_x K^0$} (b)
&\makecell[l]{
$p \to \bar\nu_x \bar \nu_y \nu_z K^+$\\
$n \to \bar\nu_x \bar \nu_y \nu_z K^0$} (c)
\\\hline
$\Delta L = -1$
& 
&$p \to \nu_x \nu_y \ell_z^+K^0$ (d)
&$n \to  \ell^-_x \ell^-_y \ell^+_z K^+$ 
&$n\to \nu_x \bar\nu_z \ell_y^-K^+$
&\makecell[l]{
$p \to \ell_x^- \ell_z^+\nu_y K^+$\\
$n \to \ell_x^-\ell_z^+ \nu_y K^0$} (b$'$)
&\makecell[l]{
$p \to \nu_x \nu_y \bar\nu_z K^+$\\
$n \to \nu_x \nu_y \bar\nu_z K^0$} (c$'$)
\\\hline
$\Delta L = +3$
&
&$p \to \bar\nu_x \bar\nu_y \ell_z^+ K^0$ (d$'$)
&
& 
&
&\makecell[l]{
$p \to \bar\nu_x \bar\nu_y \bar\nu_z K^+$\\
$n \to \bar\nu_x \bar\nu_y \bar\nu_z K^0$} (e)
\\\hline
$\Delta L = -3$
& & &
&$n \to \nu_x \nu_y \ell^-_z K^+$ (a$'$)
& 
&\makecell[l]{
$p \to \nu_x \nu_y \nu_z K^+$\\
$n \to \nu_x \nu_y \nu_z K^0$ } (e$'$)
\\\hline
\end{tabular}}
\caption{Classification of all possible four-body nucleon decays into three leptons and a $K^+$ or $K^0$ meson that are induced by dim-9 BNV operators involving a single strange quark. 
Pairs of processes indicated by (x) and (x') are 
related by the interchange of neutrinos with antineutrinos.}
\label{tab:n2lllK}
\end{table*}

Currently, no dedicated experimental searches have been conducted for these decay modes. However, because their experimental signatures in water-Cherenkov detector are similar to those of conventional two-body decays, including $p\to (e^+,\mu^+)K^0$,
$n\to (e^+,\mu^+)\pi^-$,
$p\to \nu K^+$, 
$n\to \nu K^0$, and the exotic three-body decays $p\to(e^+,\mu^+)\pi^0\pi^0$, 
existing data from searches for the latter can be reanalyzed to constrain the former.
Following the methodology of \cite{Liang:2026vqk}, we reinterpret existing Super-Kamiokande (Super-K) data~\cite{Super-Kamiokande:2005lev,Super-Kamiokande:2022egr,Super-Kamiokande:2017gev,Super-Kamiokande:2014otb,Kamiokande:2026zwu} to derive conservative lower limits on the partial lifetimes (i.e., the inverse decay width) of 26 distinct four-body decay modes involving a kaon. The resulting recast limits range from $10^{31}$ to $10^{34}~\text{yr}$, providing the first estimates of the experimental constraints on these modes.  
We further use these lifetime limits to constrain
the WCs of the relevant dim-9 LEFT operators, probing effective scales of $40\text{--}200~\text{TeV}$, depending on specific operators. 

The remainder of this paper is organized as follows. 
In \cref{sec:dim9basis}, we classify the four-body nucleon decay modes and the dim-9 BNV LEFT operators involving a strange quark. In \cref{sec:chpt}, we match these operators to the ChPT and derive hadron-level interaction vertices. 
In \cref{sec:decay_width}, we detail the calculation of decay widths and four-body phase space integration.  
The details of the reinterpretation of existing experimental data to derive lifetime limits on all relevant four-body decay modes, along with the resulting constraints on the dim-9 LEFT WCs, are presented in \cref{sec:constraints}.
We conclude in \cref{sec:summary}.
Detailed results, including SMEFT examples, 
four-body phase space integration, matrix elements for all relevant decay modes, and the decay widths expressed in terms of dim-9 LEFT WCs, are compiled in Appendices \ref{app:smeft_case}, \ref{app:4body_PS}, \ref{app:decay_width_amp}, and \ref{app:decay_width_WCs}, respectively. 

\section{Dimension-9 BNV LEFT operator basis involving a strange quark}
\label{sec:dim9basis}

For the dim-9 BNV operators with the field configuration $l^3qqs$, where $l$ is a generic lepton field and $q$ represents either an up or down quark, the induced nucleon decays involve three leptons accompanied only by a $K^+$ or $K^0$. 
Processes involving a $K^-$ or $\bar K^0$ carry the opposite strangeness and therefore cannot be induced by these operators at leading order. Such processes can arise only through additional weak interactions or from higher-dimensional operators containing five quark fields, such as $l^3\overline{s}qqqq$. 

Based on the quark-field configurations and the lepton number change $\Delta L$, defined as the lepton number of the initial state minus that of the final state ($\Delta L = L_{\text{initial}} - L_{\text{final}}$), 
all relevant processes are classified in \cref{tab:n2lllK}. In the table, the indices $x,y,z=e,\mu,\tau$ label the three lepton flavors, with the $\tau$ flavor restricted to neutrinos and antineutrinos.
For processes belonging to the $\Delta L=\pm1$ classes, the generalized flavor $\Delta F_L=1$ if the flavor index $z$ coincides with either $x$ or $y$, whereas $\Delta F_L=3$ if $z$ differs from both. All processes in the $\Delta L=\pm3$ classes likewise have $\Delta F_L=3$.
Processes involving same charged leptons and kaons but different neutrino or antineutrino states are grouped in the same column due to their similar experimental signatures. Neglecting neutrino and antineutrino flavors, we identify 26 distinct four-body nucleon decay modes when considering all possible combinations of charged leptons $e^\pm$ and $\mu^\pm$:
$p\to e^+ e^+ (e^-,\mu^-)K^0$, $p\to \mu^+ \mu^+ (e^-,\mu^-)K^0$, $p\to e^+ \mu^+ (e^-,\mu^-)K^0$,
$p\to \hat\nu\hat\nu (e^+,\mu^+) K^0$, 
$p\to (e^- e^+,\mu^-\mu^+, e^-\mu^+, \mu^- e^+)\hat\nu K^+$, $p\to \hat\nu\hat\nu\hat\nu K^+$, together with their isospin counterparts obtained by simultaneously interchanging $p\leftrightarrow n$, $K^+\leftrightarrow K^0$, and $\ell^+\leftrightarrow\ell^-$,
where $\hat\nu$ collectively denotes neutrinos and antineutrinos.

Within the LEFT framework, the $\Delta F_L=3$ processes can be mediated at leading order only by dim-9 operators.
In contrast, the $\Delta F_L=1$ processes can also receive contributions from lower-dimensional BNV operators, with field configurations such as $l sqq$, $l sqq\partial$, and $l sqq\partial^2$, through the exchange of additional SM mediators. 
In this work, we focus on the contributions from the dim-9 LEFT operators. 
We begin by constructing the relevant dim-9 BNV operators containing a strange quark that mediate all the processes listed in \cref{tab:n2lllK}, following the general classification of triple-quark structures given in~\cite{Liao:2025vlj}. 
Although the complete dim-9 LEFT operator basis was presented in~\cite{Li:2020tsi}, adapting it to our analysis would be rather cumbersome. We therefore construct an alternative, explicit operator basis with well-defined chiral irrep properties, facilitating chiral matching and subsequent calculations of the four-body nucleon decays considered in this work. 

Within the LEFT framework, all dim-9 operators consisting of three leptons and three quarks can be recast into a lepton-quark factorized form, $(l_1l_2)(l_3 q_1)(q_2 q_3)$, with the help of Fierz transformations.
Following Ref.~\cite{Liao:2025vlj}, 
these dim-9 operators can be expressed in terms of the following four structures 
\begin{subequations}
\begin{align}
{\cal O}_{a}^{uvw} =\,& \overline{\Psi_{a}}\calN_{uvw}^{\tL\tL},
&
{\cal O}_{b}^{uvw} =\,&\overline{\Psi_{b}}\calN_{uvw}^{\tR\tL}, 
\\
{\cal O}_{c}^{uvw} =\,& \overline{\Psi_{c,\mu}}
\calN_{uvw}^{\tL\tR,\mu},
&
{\cal O}_{d}^{uvw} =\,& \overline{\Psi_{d,\mu\nu}}
\calN_{uvw}^{\tL\tL,\mu\nu}, 
\label{eq:LEFT3qO}
\end{align} 
\end{subequations}
together with their chirality-flipped partners (chiral partners in short), obtained by interchanging $\tL \leftrightarrow \tR$. 
Here, $u,v,w=u,d,s$ denote light quark-flavor indices, 
and the $\Psi$ fields represent appropriate combinations of three lepton fields.  
The $\calN$ objects are constructed from three chiral quark fields and transform according to definite irreducible representations (irreps) of the QCD chiral group $\rm SU(3)_\tL\otimes SU(3)_\tR$: 
\begin{subequations}
\label{eq:Nyzw}
\begin{align}
\calN_{uvw}^{\tL\tL} \equiv\,&  q_{\tL, u}^\alpha \big(\overline{ q_{\tL, v}^{\beta \C} }\, q_{\tL, w}^\gamma \big)\epsilon_{\alpha \beta \gamma} \;\in\; \pmb{8}_\tL \otimes  \pmb{1}_\tR , 
\\
\calN_{uvw}^{\tR\tL} \equiv\,& q_{\tR, u}^\alpha \big(\overline{ q_{\tL, v}^{\beta \C} }\, q_{\tL,w}^\gamma\big)\epsilon_{\alpha \beta \gamma} \;\in\; 
\bar{\pmb{3}}_\tL \otimes \pmb{3}_\tR , 
\\
\calN_{uvw}^{\tL\tR,\mu} \equiv\,& q_{\tL,\{u}^\alpha \big(\overline{ q_{\tL, v\}}^{\beta \C} }\, \gamma^\mu q_{\tR,w}^\gamma\big)\epsilon_{\alpha \beta \gamma}
\;\in\; \pmb{6}_\tL \otimes \pmb{3}_\tR ,
\\
\calN_{uvw}^{\tL\tL,\mu\nu} \equiv\,&  q_{\tL, \{u}^\alpha \big(\overline{ q_{\tL, v}^{\beta \C} }\, \sigma^{\mu\nu} q_{\tL, w\} }^\gamma \big)\epsilon_{\alpha \beta \gamma} \;\in\; \pmb{10}_\tL \otimes \pmb{1}_\tR, 
\end{align}
\end{subequations}  
together with their chiral partners with  $\tL \leftrightarrow \tR$. Here, $q_{\tL(\tR),v}^\C$ denotes the charge-conjugated field of the left- (right-)
handed chiral quark field $q_{\tL(\tR),v}$ with flavor $v$.
The curly brackets indicate symmetrization over all flavor indices of fields with the same chirality, viz.,
$A_{\{u}B_{v\}} \equiv (1/2)(A_{u}B_{v}+A_{v}B_{u})$ and 
$A_{\{u}B_v C_{w\}} \equiv (1/6) [A_{u}B_{v}C_{w} + \text{5 perm. of } (u,v,w)]$.

In our analysis, the relevant triple-quark building blocks involve three flavor configurations: $uus$, $dds$ and $uds$.
The $uus$ and $dds$ configurations contribute to proton and neutron decays, respectively, while the $uds$ configuration contributes to both. The resulting three-lepton-plus-kaon decay modes are summarized in \cref{tab:n2lllK}.
Based on the triple-quark conventions defined in \cref{eq:Nyzw} and the relevant lepton configurations, the independent triple-quark components can be chosen as follows: 
\begin{subequations}
\label{eq:Nqqs}
\begin{align}
\ell^+\ell^+\ell^-~\&~ \ell^+\hat\nu\hat\nu:&~ 
\calN_{usu}^{\tL\tL(\tR\tR)};\,
\calN_{usu}^{\tR\tL(\tL\tR)};
\nn\\
&~\calN_{uus}^{\tL\tR(\tR\tL),\mu},\,
\calN_{usu}^{\tL\tR(\tR\tL),\mu};
\calN_{uus}^{\tL\tL(\tR\tR),\mu\nu}.
\\
\ell^+\ell^-\ell^-~\&~
\ell^-\hat\nu\hat\nu:&~
\calN_{dds}^{\tL\tL(\tR\tR)};\,
\calN_{dds}^{\tR\tL(\tL\tR)};
\nn\\
&~\calN_{dds}^{\tL\tR(\tR\tL),\mu},\,
\calN_{dsd}^{\tL\tR(\tR\tL),\mu};\,
\calN_{dds}^{\tL\tL(\tR\tR),\mu\nu}.
\\
\ell^+\ell^-\hat\nu~\&~
\hat\nu\hat\nu\hat\nu:&~ 
\calN_{dsu}^{\tL\tL(\tR\tR)},\,
\calN_{sud}^{\tL\tL(\tR\tR)};
\nn\\
&~
\calN_{uds}^{\tR\tL(\tL\tR)},\,
\calN_{dsu}^{\tR\tL(\tL\tR)},\,
\calN_{sud}^{\tR\tL(\tL\tR)};
\nn\\
&~\calN_{dsu}^{\tL\tR(\tR\tL),\mu},\,
\calN_{usd}^{\tL\tR(\tR\tL),\mu},\,\calN_{uds}^{\tL\tR(\tR\tL),\mu};
\nn\\
&~\calN_{uds}^{\tL\tL(\tR\tR),\mu\nu}.
\end{align}
\end{subequations}
As can be seen, for each of $uus$ and $dds$ quark configurations, there are 1, 1, 2, and 1 independent triple-quark structures corresponding, respectively, to the four irreps listed in \cref{eq:Nyzw}. 
For the $uds$ configuration, the corresponding numbers are 2, 3, 3, and 1. 
Note that, in \cref{eq:Nqqs}, the structure $\calN_{uds}^{\tL\tL(\tR\tR)}$ is absent,
as it can be expressed as a linear combination of
$\calN_{dsu}^{\tL\tL(\tR\tR)}$ and $\calN_{sud}^{\tL\tL(\tR\tR)}$~\cite{Liao:2025vlj}.

In terms of $\calN$s in \cref{eq:Nqqs}, each dim-9 BNV operator involving three leptons can be expressed in the form:
\begin{align}
\calO \sim (\overline{l_x}\Gamma_1 l_y)(\overline{l_z}\Gamma_2 \calN_{uvw}^{\Gamma_3}),
\end{align}
where $l_{i} \in\{\ell_{\tL(\tR)},\,\ell_{\tL(\tR)}^\C,\,\nu_{\tL},\,\nu_{\tL}^\C\}$, with $i=e,\mu,\tau$, denotes a generic chiral lepton field.
Here, $\ell_{\tL(\tR)}^\C$ and $\nu_{\tL}^\C$ represent the charge-conjugated fields of the chiral charged leptons $\ell_{\tL(\tR)}$ and the left-handed neutrino $\nu_{\tL}$, respectively. 
Following our previous construction of dim-9 operators involving only up and down quarks in~\cite{Liao:2025wxk}, we use Fierz transformations to group leptons of the same type into a single current and adopt this convention throughout. 
The Dirac gamma structures $\Gamma_{1,2,3}$ are determined once all six chiral fields are specified.  

For each chiral field configuration associated with the process classes in \cref{tab:n2lllK}, we construct the independent operators following the prescriptions described above. The resulting operator bases are summarized in \cref{tab:dim9ope_Lp1A,tab:dim9ope_Lp1B} for $\Delta L=+1$, \cref{tab:dim9ope_Lm1A,tab:dim9ope_Lm1B} for $\Delta L=-1$, and \cref{tab:dim9ope_L3} for $\Delta L=\pm3$ processes.  
We use a tilde over $\calO$ to distinguish these operators from those without a strange quark in~\cite{Liao:2025wxk}.
In each table, the ``chiral irrep'' column lists the irreducible representations of the relevant operators under the chiral group $\rm SU(3)_\tL\otimes SU(3)_\tR$, while the ``\#'' column gives the number of independent operators for general $n_{\ell}$ charged leptons and $n_\nu$ neutrinos; the numbers in angle brackets correspond to $n_\nu=3$ ($\nu_e,\nu_\mu,\nu_\tau$) and $n_\ell=2$ ($e,\mu$).
We have cross-checked our constructions by using the package {\tt Basisgen}~\cite{Criado:2019ugp} to enumerate the independent operators for each field configuration.

For each pair of processes related by $\nu\leftrightarrow\bar\nu$, indicated by (x) and (x$^\prime$) in \cref{tab:n2lllK} (x = a, b, c, d, e), we choose their operator bases to be related by the simultaneous interchanges of $\tL\leftrightarrow\tR$ and $\nu_\tL\leftrightarrow\nu_\tL^\C$.
For example, the $\Delta L=+1$ processes $n\to\bar\nu_x\bar\nu_y\ell_z^- K^+$ and the $\Delta L=-3$ processes  $n\to \nu_x\nu_y\ell_z^- K^+$ each admit nine generic operators (without counting lepton flavors), whose parametrizations follow the above convention, as shown in \cref{tab:dim9ope_Lp1A} and \cref{tab:dim9ope_L3}, respectively.
Similar constructions apply to the remaining four pairs of processes with (b, b$^\prime$), (c, c$^\prime$), (d, d$^\prime$), and (e, e$^\prime$). 
As will be seen later, such a ``symmetric" construction simplifies the chiral matching and decay-width calculations significantly.
From \cref{tab:dim9ope_Lp1A,tab:dim9ope_Lp1B,tab:dim9ope_Lm1A,tab:dim9ope_Lm1B,tab:dim9ope_L3}, 
the relevant operators for each flavor-specific process can be easily identified.
For instance, 
each of the four modes $p\to e^+ e^+(e^-,\mu^-)K^0$ and $p\to \mu^+ \mu^+(e^-,\mu^-)K^0$ is described by 16 independent operators, whereas each of the two modes $p\to e^+ \mu^+(e^-,\mu^-)K^0$ involves 34 independent operators. 

In the SMEFT framework, these dim-9 LEFT operators can arise from SMEFT operators of various dimensions~\cite{Weinberg:1980bf,Hambye:2017qix,Liao:2020jmn,Li:2020xlh,Harlander:2023psl,Heeck:2026dmh}. In the $(\Delta L,\Delta F_L)=(1,1)$ and $(-1,1)$ sectors, the corresponding operators can be generated at lowest order by dim-6 and dim-7 SMEFT interactions, respectively, upon integrating out heavy SM fields. 
By contrast, due to their nontrivial flavor structures, the dim-9 LEFT operators in the $(\Delta L,\Delta F_L)=(-1,3)$ and $(3,3)$ sectors require SMEFT operators of dim-9 and dim-11 or higher odd dimensions, respectively, while those in the $(1,3)$ and $(-3,3)$ sectors require dim-10  and dim-12 or higher even dimensions, respectively~\cite{Liao:2025wxk}. In Appendix \ref{app:smeft_case}, we illustrate the matching onto our LEFT bases using the dim-9 SMEFT operators involving three leptons and three quarks. From a UV-complete perspective, these dim-9 operators can arise in various models; see Ref.\,~\cite{Heeck:2026dmh} and references therein for examples.

\begin{table}[t]
\center
\resizebox{\linewidth}{!}{
\renewcommand{\arraystretch}{1.25}
\begin{tabular}{l l l l}
\hline   
Name & Operator & Irrep & \#
\\\clineB{1-4}{1.5}
\multicolumn{4}{l}{\color{magenta}$\Delta L = + 1$: $p\to\ell_x^+ \ell_y^+ \ell_z^-K^0$}
\\\cline{1-4}
$\tO_{\ell\ell\bar\ell ,xyz}^{\tV\tR\tV\tL\tS\tL}$
& $(\overline{\ell_{\tL x}^\C}\gamma_{\mu}\ell_{\tR y})
(\overline{\ell_{\tL z}}\gamma^\mu\calN_{usu}^{\tL\tL}) $
&$\pmb{8}_\tL\otimes{\pmb{1}}_\tR$
& $n_\ell^3\,\langle 8\rangle$ 
\\\cline{1-4}%
$\tO_{\ell\ell\bar\ell ,xyz}^{\tV\tR\tV\tL\tS\tR}$
& $(\overline{\ell_{\tL x}^\C}\gamma_{\mu}\ell_{\tR y})
(\overline{\ell_{\tL z}}\gamma^\mu\calN_{usu}^{\tL\tR}) $    
&$\pmb{3}_\tL\otimes{\bar{\pmb{3}}}_\tR$
& $n_\ell^3\,\langle 8 \rangle$
\\
$\tO_{\ell\ell\bar\ell ,xyz}^{\tV\tR\tS\tR\tV\tL1}$
& $(\overline{\ell_{\tL x}^\C}\gamma_\mu\ell_{\tR y})
(\overline{\ell_{\tL z}}\calN_{usu}^{\tR\tL,\mu}) $    
&$\pmb{3}_\tL\otimes{\pmb{6}}_\tR$
& $n_\ell^3\,\langle 8 \rangle$
\\\cline{1-4}%
$\tO_{\ell\ell\bar\ell ,xyz}^{\tV\tR\tS\tR\tV\tL2}$
& $(\overline{\ell_{\tL x}^\C}\gamma_{\mu}\ell_{\tR y})
(\overline{\ell_{\tL z}}\calN_{uus}^{\tR\tL,\mu}) $   
&$\pmb{3}_\tL\otimes\pmb{6}_\tR$
& $n_\ell^3\,\langle 8 \rangle$
\\\cline{1-4}%
$\tO_{\ell\ell\bar\ell ,xyz}^{\tS\tL\tS\tR\tS\tR}$
& $(\overline{\ell_{\tL x}^\C}\ell_{\tL y})
(\overline{\ell_{\tL z}} \calN_{usu}^{\tR\tR}) $    
&$\pmb{1}_\tL\otimes\pmb{8}_\tR$
&$\frac{1}{2}n_\ell^2 (n_\ell+1)\,\langle 6 \rangle$
\\\cline{1-4}%
$\tO_{\ell\ell\bar\ell ,xyz}^{\tS\tL\tS\tL\tS\tL}$
&$(\overline{\ell_{\tL x}^\C}\ell_{\tL y})
(\overline{\ell_{\tR z}} \calN_{usu}^{\tL\tL}) $    
&$\pmb{8}_\tL\otimes\pmb{1}_\tR$
&$\frac{1}{2}n_\ell^2 (n_\ell+1)\,\langle 6 \rangle$
\\
$\tO_{\ell\ell\bar\ell ,xyz}^{\tT\tL\tT\tL\tS\tL}$
& $(\overline{\ell_{\tL x}^\C}\sigma_{\mu\nu}\ell_{\tL y})
(\overline{\ell_{\tR z}} \sigma^{\mu\nu}\calN_{usu}^{\tL\tL}) $    
&$\pmb{8}_\tL\otimes\pmb{1}_\tR$
&$\frac{1}{2}n_\ell^2 (n_\ell-1)\,\langle 2 \rangle$
\\
$\tO_{\ell\ell\bar\ell ,xyz}^{\tT\tL\tS\tL\tT\tL}$
&$(\overline{\ell_{\tL x}^\C}\sigma_{\mu\nu}\ell_{\tL y})
(\overline{\ell_{\tR z}}\calN_{uus}^{\tL\tL,\mu\nu}) $ 
&$\pmb{10}_\tL\otimes\pmb{1}_\tR$
&$\frac{1}{2}n_\ell^2 (n_\ell-1)\,\langle 2 \rangle$
\\\cline{1-4}%
$\tO_{\ell\ell\bar\ell ,xyz}^{\tS\tL\tS\tR\tS\tL}$
&$(\overline{\ell_{\tL x}^\C} \ell_{\tL y})
(\overline{\ell_{\tL z}}\calN_{usu}^{\tR\tL}) $    
&$\bar{\pmb{3}}_\tL\otimes\pmb{3}_\tR$
&$\frac{1}{2}n_\ell^2 (n_\ell+1)\,\langle 6 \rangle$
\\
$\tO_{\ell\ell\bar\ell ,xyz}^{\tT\tL\tV\tL\tV\tR1}$
&$(\overline{\ell_{\tL x}^\C}\sigma_{\mu\nu} \ell_{\tL y})
(\overline{\ell_{\tL z}}\gamma^\mu\calN_{usu}^{\tL\tR,\nu}) $    
&$\pmb{6}_\tL \otimes\pmb{3}_\tR$
&$\frac{1}{2}n_\ell^2 (n_\ell-1)\,\langle 2\rangle$
\\\cline{1-4}%
$\tO_{\ell\ell\bar\ell ,xyz}^{\tT\tL\tV\tL\tV\tR2}$
&$(\overline{\ell_{\tL x}^\C}\sigma_{\mu\nu} \ell_{\tL y})
(\overline{\ell_{\tL z}}\gamma^\mu \calN_{uus}^{\tL\tR,\nu}) $    
&$\pmb{6}_\tL \otimes\pmb{3}_\tR$
&$\frac{1}{2}n_\ell^2 (n_\ell-1)\,\langle 2\rangle$
\\\cline{1-4}%
$\tO_{\ell\ell\bar\ell ,xyz}^{\tS\tL\tS\tL\tS\tR}$
&$(\overline{\ell_{\tL x}^\C}\ell_{\tL y})
(\overline{\ell_{\tR z}}\calN_{usu}^{\tL\tR}) $    
&$\pmb{3}_\tL\otimes\bar{\pmb{3}}_\tR$
&$\frac{1}{2}n_\ell^2 (n_\ell+1)\,\langle 6 \rangle$
\\
$\tO_{\ell\ell\bar\ell ,xyz}^{\tT\tL\tT\tL\tS\tR}$
&$(\overline{\ell_{\tL x}^\C}\sigma_{\mu\nu}\ell_{\tL y})
(\overline{\ell_{\tR z}}\sigma^{\mu\nu}\calN_{usu}^{\tL\tR}) $    
&$\pmb{3}_\tL\otimes\bar{\pmb{3}}_\tR$
&$\frac{1}{2}n_\ell^2 (n_\ell-1)\,\langle 2 \rangle$
\\\clineB{1-4}{1.5}
$\tO_{\ell\ell\bar\ell ,xyz}^{\tV\tL\tV\tR\tS\tR}$
&$(\overline{\ell_{\tR x}^\C}\gamma_{\mu}\ell_{\tL y})
(\overline{\ell_{\tR z}}\gamma^\mu\calN_{usu}^{\tR\tR}) $    
&$\pmb{1}_\tL\otimes\pmb{8}_\tR$
&$n_\ell^3\,\langle 8 \rangle$
\\\cline{1-4}%
$\tO_{\ell\ell\bar\ell ,xyz}^{\tV\tL\tV\tR\tS\tL}$
& $(\overline{\ell_{\tR x}^\C}\gamma_{\mu}\ell_{\tL y})
(\overline{\ell_{\tR z}}\gamma^\mu\calN_{usu}^{\tR\tL}) $    
&$\bar{\pmb{3}}_\tL\otimes\pmb{3}_\tR$
&$n_\ell^3\,\langle 8 \rangle$
\\
$\tO_{\ell\ell\bar\ell,xyz}^{\tV\tL\tS\tL\tV\tR1}$
&$(\overline{\ell_{\tR x}^\C}\gamma_\mu\ell_{\tL y})
(\overline{\ell_{\tR z}}\calN_{usu}^{\tL\tR,\mu}) $    
&$\pmb{6}_\tL\otimes\pmb{3}_\tR$
&$n_\ell^3\,\langle 8 \rangle$
\\\cline{1-4}%
$\tO_{\ell\ell\bar\ell ,xyz}^{\tV\tL\tS\tL\tV\tR2}$
&$(\overline{\ell_{\tR x}^\C}\gamma_{\mu}\ell_{\tL y})
(\overline{\ell_{\tR z}}\calN_{uus}^{\tL\tR,\mu}) $   
&$\pmb{6}_\tL\otimes{\pmb{3}}_\tR$
&$n_\ell^3\,\langle 8 \rangle$
\\\cline{1-4}
$\tO_{\ell\ell\bar\ell ,xyz}^{\tS\tR\tS\tL\tS\tL}$
&$(\overline{\ell_{\tR x}^\C}\ell_{\tR y})
(\overline{\ell_{\tR z}} \calN_{usu}^{\tL\tL}) $    
&$\pmb{8}_\tL\otimes\pmb{1}_\tR$
&$\frac{1}{2}n_\ell^2 (n_\ell+1)\,\langle 6 \rangle$ 
\\\cline{1-4}%
$\tO_{\ell\ell\bar\ell ,xyz}^{\tS\tR\tS\tR\tS\tR}$
& $(\overline{\ell_{\tR x}^\C}\ell_{\tR y})
(\overline{\ell_{\tL z}} \calN_{usu}^{\tR\tR}) $    
&$\pmb{1}_\tL\otimes\pmb{8}_\tR$
&$\frac{1}{2}n_\ell^2 (n_\ell+1)\,\langle 6 \rangle$
\\
$\tO_{\ell\ell\bar\ell ,xyz}^{\tT\tR\tT\tR\tS\tR}$
& $(\overline{\ell_{\tR x}^\C}\sigma_{\mu\nu}\ell_{\tR y})
(\overline{\ell_{\tL z}} \sigma^{\mu\nu}\calN_{usu}^{\tR\tR}) $    
&$\pmb{1}_\tL\otimes\pmb{8}_\tR$
&$\frac{1}{2}n_\ell^2 (n_\ell-1)\,\langle 2 \rangle$
\\
$\tO_{\ell\ell\bar\ell ,xyz}^{\tT\tR\tS\tR\tT\tR}$
&$(\overline{\ell_{\tR x}^\C}\sigma_{\mu\nu}\ell_{\tR y})
(\overline{\ell_{\tL z}}\calN_{uus}^{\tR\tR,\mu\nu}) $
&$\pmb{1}_\tL\otimes\pmb{10}_\tR$
&$\frac{1}{2}n_\ell^2 (n_\ell-1)\,\langle 2 \rangle$
\\\cline{1-4}%
$\tO_{\ell\ell\bar\ell ,xyz}^{\tS\tR\tS\tL\tS\tR}$
& $(\overline{\ell_{\tR x}^\C} \ell_{\tR y})
(\overline{\ell_{\tR z}}\calN_{usu}^{\tL\tR}) $    
&$\pmb{3}_\tL\otimes\bar{\pmb{3}}_\tR$
&$\frac{1}{2}n_\ell^2 (n_\ell+1)\,\langle 6 \rangle$
\\
$\tO_{\ell\ell\bar\ell ,xyz}^{\tT\tR\tV\tR\tV\tL1}$
& $(\overline{\ell_{\tR x}^\C}\sigma_{\mu\nu} \ell_{\tR y})
(\overline{\ell_{\tR z}}\gamma^\mu\calN_{usu}^{\tR\tL,\nu}) $    
&$\pmb{3}_\tL\otimes\pmb{6}_\tR$
&$\frac{1}{2}n_\ell^2 (n_\ell-1)\,\langle 2 \rangle$
\\\cline{1-4}%
$\tO_{\ell\ell\bar\ell ,xyz}^{\tT\tR\tV\tR\tV\tL2}$
&$(\overline{\ell_{\tR x}^\C}\sigma_{\mu\nu} \ell_{\tR y})
(\overline{\ell_{\tR z}}\gamma^\mu \calN_{uus}^{\tR\tL,\nu}) $    
&$\pmb{3}_\tL\otimes\pmb{6}_\tR$
&$\frac{1}{2}n_\ell^2 (n_\ell-1)\,\langle 2 \rangle$
\\\cline{1-4}%
$\tO_{\ell\ell\bar\ell ,xyz}^{\tS\tR\tS\tR\tS\tL}$
&$(\overline{\ell_{\tR x}^\C}\ell_{\tR y})
(\overline{\ell_{\tL z}}\calN_{usu}^{\tR\tL}) $    
&$\bar{\pmb{3}}_\tL\otimes\pmb{3}_\tR$
&$\frac{1}{2}n_\ell^2 (n_\ell+1)\,\langle 6 \rangle$
\\
$\tO_{\ell\ell\bar\ell ,xyz}^{\tT\tR\tT\tR\tS\tL}$
&$(\overline{\ell_{\tR x}^\C}\sigma_{\mu\nu}\ell_{\tR y})
(\overline{\ell_{\tL z}}\sigma^{\mu\nu}\calN_{usu}^{\tR\tL}) $    
&$\bar{\pmb{3}}_\tL\otimes\pmb{3}_\tR$
&$\frac{1}{2}n_\ell^2(n_\ell-1)\,\langle 2 \rangle$
\\\clineB{1-4}{1.5}
\multicolumn{4}{l}{\color{magenta}$\Delta L = + 1$: $p\to\nu_x \bar\nu_y \ell_z^+K^0$}
\\\hline
$\tO_{\bar\nu \nu \ell,xyz}^{\tV\tL\tV\tR\tS\tR}$
&$(\overline{\nu_{\tL x}}\gamma_\mu \nu_{\tL y})
(\overline{\ell_{\tL z}^\C}\gamma^\mu \calN_{usu}^{\tR\tR}) $
&$\pmb{1}_\tL\otimes\pmb{8}_\tR$
&$n_\ell n_\nu^2 \,\langle 18 \rangle$ 
\\\cline{1-4}%
$\tO_{\bar\nu \nu \ell,xyz}^{\tV\tL\tV\tR\tS\tL}$
&$(\overline{\nu_{\tL x}}\gamma_\mu \nu_{\tL y})
(\overline{\ell_{\tL z}^\C}\gamma^\mu \calN_{usu}^{\tR\tL}) $
&$\bar{\pmb{3}}_\tL\otimes\pmb{3}_\tR$
&$n_\ell n_\nu^2 \,\langle 18 \rangle$ 
\\
$\tO_{\bar\nu \nu \ell,xyz}^{\tV\tL\tS\tL\tV\tR1}$
&$(\overline{\nu_{\tL x}}\gamma_\mu \nu_{\tL y})
(\overline{\ell_{\tL z}^\C}\calN_{usu}^{\tL\tR,\mu}) $
&$\pmb{6}_\tL\otimes\pmb{3}_\tR$
&$n_\ell n_\nu^2 \,\langle 18 \rangle$ 
\\\cline{1-4}%
$\tO_{\bar\nu \nu \ell,xyz}^{\tV\tL\tS\tL\tV\tR2}$
&$(\overline{\nu_{\tL x}}\gamma_\mu \nu_{\tL y})
(\overline{\ell_{\tL z}^\C}\calN_{uus}^{\tL\tR,\mu}) $
&$\pmb{6}_\tL\otimes\pmb{3}_\tR$
&$n_\ell n_\nu^2  \,\langle 18 \rangle$ 
\\\cline{1-4}%
$\tO_{\bar\nu \nu \ell,xyz}^{\tV\tL\tV\tL\tS\tL}$
&$(\overline{\nu_{\tL x}}\gamma_\mu\nu_{\tL y})
(\overline{\ell_{\tR z}^\C}\gamma^\mu\calN_{usu}^{\tL\tL}) $
&$\pmb{8}_\tL\otimes\pmb{1}_\tR$
&$n_\ell n_\nu^2 \,\langle 18 \rangle$
\\\cline{1-4}%
$\tO_{\bar\nu \nu \ell,xyz}^{\tV\tL\tV\tL\tS\tR}$
&$(\overline{\nu_{\tL x}}\gamma_\mu\nu_{\tL y})
(\overline{\ell_{\tR z}^\C}\gamma^\mu\calN_{usu}^{\tL\tR}) $
&$\pmb{3}_\tL \otimes\bar{\pmb{3}}_\tR$
&$n_\ell n_\nu^2 \,\langle 18 \rangle$ 
\\
$\tO_{\bar\nu \nu \ell,xyz}^{\tV\tL\tS\tR\tV\tL1}$ 
&$(\overline{\nu_{\tL x}}\gamma_{\mu}\nu_{\tL y})
(\overline{\ell_{\tR z}^\C}\calN_{usu}^{\tR\tL,\mu}) $
&$\pmb{3}_\tL\otimes\pmb{6}_\tR$
&$n_\ell n_\nu^2 \,\langle 18 \rangle$ 
\\\cline{1-4}%
$\tO_{\bar\nu \nu \ell,xyz}^{\tV\tL\tS\tR\tV\tL2}$ 
&$(\overline{\nu_{\tL x}}\gamma_\mu \nu_{\tL y})
(\overline{\ell_{\tR z}^\C}\calN_{uus}^{\tR\tL,\mu}) $
&$\pmb{3}_\tL\otimes\pmb{6}_\tR$
&$n_\ell n_\nu^2 \,\langle 18 \rangle$
\\\clineB{1-4}{1.5}
\multicolumn{4}{l}{\color{magenta}$\Delta L = +1 $: $n\to\bar\nu_x \bar\nu_y \ell_z^- K^+$}
\\\hline%
$\tO_{\nu\nu\bar\ell,xyz}^{\tS\tL\tS\tR\tS\tR}$
&$(\overline{\nu_{\tL x}^\C} \nu_{\tL y})
(\overline{\ell_{\tL z}}\calN_{dds}^{\tR\tR}) $
&$\pmb{1}_\tL\otimes\pmb{8}_\tR$
&$\frac{1}{2}n_\ell(n_\nu^2+n_\nu) \,\langle 12 \rangle$ 
\\\hline%
$\tO_{\nu\nu\bar\ell,xyz}^{\tS\tL\tS\tL\tS\tL}$
&$(\overline{\nu_{\tL x}^\C} \nu_{\tL y})
(\overline{\ell_{\tR z}}\calN_{dds}^{\tL\tL}) $
&$\pmb{8}_\tL\otimes\pmb{1}_\tR$
&$\frac{1}{2}n_\ell(n_\nu^2+n_\nu)\,\langle 12 \rangle$ 
\\
$\tO_{\nu\nu\bar\ell,xyz}^{\tT\tL\tT\tL\tS\tL}$
&$(\overline{\nu_{\tL x}^\C}\sigma_{\mu\nu} \nu_{\tL y})
(\overline{\ell_{\tR z}}\sigma^{\mu\nu}\calN_{dds}^{\tL\tL}) $
&$\pmb{8}_\tL\otimes\pmb{1}_\tR$
&$\frac{1}{2}n_\ell n_\nu (n_\nu-1)\,\langle 6 \rangle$ 
\\
$\tO_{\nu\nu\bar\ell,xyz}^{\tT\tL\tS\tL\tT\tL}$
&$(\overline{\nu_{\tL x}^\C}\sigma_{\mu\nu} \nu_{\tL y})
(\overline{\ell_{\tR z}}\calN_{dds}^{\tL\tL,\mu\nu}) $ 
&$\pmb{10}_\tL\otimes\pmb{1}_\tR$
&$\frac{1}{2}n_\ell n_\nu (n_\nu-1)\,\langle 6 \rangle$ 
\\\hline%
$\tO_{\nu\nu\bar\ell,xyz}^{\tS\tL\tS\tR\tS\tL}$
&$(\overline{\nu_{\tL x}^\C} \nu_{\tL y})
(\overline{\ell_{\tL z}} \calN_{dds}^{\tR\tL}) $
&$\bar{\pmb{3}}_\tL\otimes\pmb{3}_\tR$
&$\frac{1}{2}n_\ell(n_\nu^2+n_\nu) \,\langle 12 \rangle$
\\
$\tO_{\nu\nu\bar\ell,xyz}^{\tT\tL\tV\tL\tV\tR1}$
&$(\overline{\nu_{\tL x}^\C} \sigma_{\mu\nu}\nu_{\tL y})
(\overline{\ell_{\tL z}}\gamma^\mu\calN_{dsd}^{\tL\tR,\nu}) $  
&$\pmb{6}_\tL\otimes\pmb{3}_\tR$
&$\frac{1}{2}n_\ell n_\nu(n_\nu-1) \,\langle 6 \rangle$
\\\hline%
$\tO_{\nu\nu\bar\ell,xyz}^{\tT\tL\tV\tL\tV\tR2}$
&$(\overline{\nu_{\tL x}^\C} \sigma_{\mu\nu}\nu_{\tL y})
(\overline{\ell_{\tL z}}\gamma^\mu\calN_{dds}^{\tL\tR,\nu}) $ 
&$\pmb{6}_\tL\otimes\pmb{3}_\tR$
&$\frac{1}{2}n_\ell n_\nu(n_\nu-1) \,\langle 6 \rangle$ 
\\\hline%
$\tO_{\nu\nu\bar\ell,xyz}^{\tS\tL\tS\tL\tS\tR}$
&$(\overline{\nu_{\tL x}^\C} \nu_{\tL y})
(\overline{\ell_{\tR z}}\calN_{dds}^{\tL\tR})$
&$\pmb{3}_\tL\otimes\bar{\pmb{3}}_\tR$
&$\frac{1}{2}n_\ell(n_\nu^2+n_\nu) \,\langle 12 \rangle$ 
\\
$\tO_{\nu\nu\bar\ell,xyz}^{\tT\tL\tT\tL\tS\tR}$
&$(\overline{\nu_{\tL x}^\C} \sigma_{\mu\nu}\nu_{\tL y})
(\overline{\ell_{\tR z}} \sigma^{\mu\nu}\calN_{dds}^{\tL\tR}) $
&$\pmb{3}_\tL\otimes\bar{\pmb{3}}_\tR$
&$\frac{1}{2}n_\ell n_\nu(n_\nu-1) \,\langle 6 \rangle$
\\\hline
\end{tabular} }
\caption{
The dim-9 $\Delta L=+1$ BNV LEFT operators involving a single strange quark. 
The chiral irrep. column lists the irreducible representations under the chiral group $\rm SU(3)_\tL\otimes SU(3)_\tR$. 
The \# column counts the number of independent operators for general $n_{\ell}$ charged leptons and $n_\nu$ neutrinos, and the numbers in the angle brackets are for $n_\nu=3$ and $n_\ell=2$. }
\label{tab:dim9ope_Lp1A}
\end{table}

\begin{table}[t]
\center
\resizebox{\linewidth}{!}{
\renewcommand{\arraystretch}{1.3}
\begin{tabular}{l l l l}
\hline   
Name & Operator & Irrep & \#
\\\clineB{1-4}{1.5}
\multicolumn{4}{l}{\color{magenta}$\Delta L = + 1$: $p\to\ell_x^- \ell_y^+ \bar\nu_z K^+$, 
$n\to\ell_x^- \ell_y^+ \bar\nu_z K^0$}
\\\hline
$\tO_{\bar\ell \ell\nu,xyz}^{\tV\tR\tV\tR\tS\tR1}$
&$(\overline{\ell_{\tR x}}\gamma_\mu \ell_{\tR y})
(\overline{\nu_{\tL z}^\C}\gamma^\mu \calN_{dsu}^{\tR\tR}) $
&$\pmb{1}_\tL\otimes\pmb{8}_\tR$
&$n_\ell^2 n_\nu\,\langle 12 \rangle$ 
\\
$\tO_{\bar\ell \ell\nu,xyz}^{\tV\tR\tV\tR\tS\tR2}$
&$(\overline{\ell_{\tR x}}\gamma_\mu \ell_{\tR y})
(\overline{\nu_{\tL z}^\C}\gamma^\mu \calN_{sud}^{\tR\tR}) $
&$\pmb{1}_\tL\otimes\pmb{8}_\tR$
&$n_\ell^2 n_\nu\,\langle 12 \rangle$
\\\hline
$\tO_{\bar\ell \ell\nu,xyz}^{\tV\tL\tV\tR\tS\tR1}$
&$(\overline{\ell_{\tL x}}\gamma_\mu \ell_{\tL y})
(\overline{\nu_{\tL z}^\C}\gamma^\mu \calN_{dsu}^{\tR\tR}) $  
&$\pmb{1}_\tL\otimes\pmb{8}_\tR$
&$n_\ell^2 n_\nu\,\langle 12 \rangle$ 
\\
$\tO_{\bar\ell \ell\nu,xyz}^{\tV\tL\tV\tR\tS\tR2}$
&$(\overline{\ell_{\tL x}}\gamma_\mu \ell_{\tL y})
(\overline{\nu_{\tL z}^\C}\gamma^\mu \calN_{sud}^{\tR\tR}) $  
&$\pmb{1}_\tL\otimes\pmb{8}_\tR$
&$n_\ell^2 n_\nu\,\langle 12 \rangle$ 
\\\hline
$\tO_{\bar\ell \ell\nu,xyz}^{\tV\tR\tV\tR\tS\tL1}$
&$(\overline{\ell_{\tR x}}\gamma_\mu \ell_{\tR y})(\overline{\nu_{\tL z}^\C}\gamma^\mu\calN_{uds}^{\tR\tL})$  
&$\bar{\pmb{3}}_\tL\otimes\pmb{3}_\tR$
&$n_\ell^2 n_\nu\,\langle 12 \rangle$
\\
$\tO_{\bar\ell \ell\nu,xyz}^{\tV\tR\tS\tL\tV\tR1}$
&$(\overline{\ell_{\tR x}}\gamma_\mu \ell_{\tR y})
(\overline{\nu_{\tL z}^\C} \calN_{dsu}^{\tL\tR,\mu}) $  
&$\pmb{6}_\tL\otimes\pmb{3}_\tR$
&$n_\ell^2 n_\nu\,\langle 12 \rangle$ 
\\\hline%
$\tO_{\bar\ell \ell\nu,xyz}^{\tV\tR\tV\tR\tS\tL2}$
&$(\overline{\ell_{\tR x}}\gamma_\mu \ell_{\tR y})
(\overline{\nu_{\tL z}^\C}\gamma^\mu \calN_{dsu}^{\tR\tL}) $ 
&$\bar{\pmb{3}}_\tL\otimes\pmb{3}_\tR$
&$n_\ell^2 n_\nu\,\langle 12 \rangle$ 
\\
$\tO_{\bar\ell \ell\nu,xyz}^{\tV\tR\tS\tL\tV\tR2}$
&$(\overline{\ell_{\tR x}}\gamma_\mu \ell_{\tR y})
(\overline{\nu_{\tL z}^\C} \calN_{usd}^{\tL\tR,\mu}) $ 
&$\pmb{6}_\tL\otimes\pmb{3}_\tR$
&$n_\ell^2 n_\nu\,\langle 12 \rangle$ 
\\\hline%
$\tO_{\bar\ell \ell\nu,xyz}^{\tV\tR\tV\tR\tS\tL3}$
&$(\overline{\ell_{\tR x}}\gamma_\mu \ell_{\tR y})
(\overline{\nu_{\tL z}^\C}\gamma^\mu \calN_{sud}^{\tR\tL}) $
&$\bar{\pmb{3}}_\tL\otimes\pmb{3}_\tR$
&$n_\ell^2 n_\nu\,\langle 12 \rangle$ 
\\
$\tO_{\bar\ell \ell\nu,xyz}^{\tV\tR\tS\tL\tV\tR3}$
&$(\overline{\ell_{\tR x}}\gamma_\mu \ell_{\tR y})
(\overline{\nu_{\tL z}^\C} \calN_{uds}^{\tL\tR,\mu}) $ 
&$\pmb{6}_\tL\otimes\pmb{3}_\tR$
&$n_\ell^2 n_\nu\,\langle 12 \rangle$ 
\\\hline
$\tO_{\bar\ell \ell\nu,xyz}^{\tV\tL\tV\tR\tS\tL1}$
&$(\overline{\ell_{\tL x}}\gamma_\mu\ell_{\tL y})
(\overline{\nu_{\tL z}^\C}\gamma^\mu\calN_{uds}^{\tR\tL})$
&$\bar{\pmb{3}}_\tL\otimes\pmb{3}_\tR$
&$n_\ell^2 n_\nu\,\langle 12 \rangle$ 
\\
$\tO_{\bar\ell \ell\nu,xyz}^{\tV\tL\tS\tL\tV\tR1}$
&$(\overline{\ell_{\tL x}}\gamma_\mu \ell_{\tL y})
(\overline{\nu_{\tL z}^\C} \calN_{dsu}^{\tL\tR,\mu}) $
&$\pmb{6}_\tL\otimes\pmb{3}_\tR$
&$n_\ell^2 n_\nu\,\langle 12 \rangle$
\\\hline%
$\tO_{\bar\ell \ell\nu,xyz}^{\tV\tL\tV\tR\tS\tL2}$
&$(\overline{\ell_{\tL x}}\gamma_\mu \ell_{\tL y})
(\overline{\nu_{\tL z}^\C} \gamma^\mu\calN_{dsu}^{\tR\tL}) $
&$\bar{\pmb{3}}_\tL\otimes\pmb{3}_\tR$
&$n_\ell^2 n_\nu\,\langle 12 \rangle$ 
\\
$\tO_{\bar\ell \ell\nu,xyz}^{\tV\tL\tS\tL\tV\tR2}$
&$(\overline{\ell_{\tL x}}\gamma_\mu \ell_{\tL y})
(\overline{\nu_{\tL z}^\C} \calN_{usd}^{\tL\tR,\mu}) $
&$\pmb{6}_\tL\otimes\pmb{3}_\tR$
&$n_\ell^2 n_\nu\,\langle 12 \rangle$ 
\\\hline%
$\tO_{\bar\ell \ell\nu,xyz}^{\tV\tL\tV\tR\tS\tL3}$
&$(\overline{\ell_{\tL x}}\gamma_\mu \ell_{\tL y})
(\overline{\nu_{\tL z}^\C} \gamma^\mu\calN_{sud}^{\tR\tL}) $
&$\bar{\pmb{3}}_\tL\otimes\pmb{3}_\tR$
&$n_\ell^2 n_\nu\,\langle 12 \rangle$ 
\\
$\tO_{\bar\ell \ell\nu,xyz}^{\tV\tL\tS\tL\tV\tR3}$
&$(\overline{\ell_{\tL x}}\gamma_\mu \ell_{\tL y})
(\overline{\nu_{\tL z}^\C} \calN_{uds}^{\tL\tR,\mu}) $
&$\pmb{6}_\tL\otimes\pmb{3}_\tR$
&$n_\ell^2 n_\nu\,\langle 12 \rangle$  
\\\hline
$\tO_{\bar\ell \ell\nu,xyz}^{\tS\tL\tS\tL\tS\tL1}$
&$(\overline{\ell_{\tR x}}\ell_{\tL y})
(\overline{\nu_{\tL z}^\C}\calN_{dsu}^{\tL\tL})$ 
&$\pmb{8}_\tL\otimes\pmb{1}_\tR$
&$n_\ell^2 n_\nu\,\langle 12 \rangle$ 
\\
$\tO_{\bar\ell \ell\nu,xyz}^{\tT\tL\tT\tL\tS\tL1}$
&$(\overline{\ell_{\tR x}} \sigma_{\mu\nu}\ell_{\tL y})
(\overline{\nu_{\tL z}^\C} \sigma^{\mu\nu}\calN_{dsu}^{\tL\tL}) $ 
&$\pmb{8}_\tL\otimes\pmb{1}_\tR$
&$n_\ell^2 n_\nu\,\langle 12 \rangle$ 
\\
$\tO_{\bar\ell \ell\nu,xyz}^{\tT\tL\tS\tL\tT\tL}$
&$(\overline{\ell_{\tR x}} \sigma_{\mu\nu}\ell_{\tL y})
(\overline{\nu_{\tL z}^\C} \calN_{uds}^{\tL\tL,\mu\nu}) $   
&$\pmb{10}_\tL\otimes\pmb{1}_\tR$
&$n_\ell^2 n_\nu\,\langle 12 \rangle$
\\%
$\tO_{\bar\ell \ell\nu,xyz}^{\tS\tL\tS\tL\tS\tL2}$
&$(\overline{\ell_{\tR x}} \ell_{\tL y})
(\overline{\nu_{\tL z}^\C}\calN_{sud}^{\tL\tL})$ 
&$\pmb{8}_\tL\otimes\pmb{1}_\tR$
&$n_\ell^2 n_\nu\,\langle 12 \rangle$ 
\\
$\tO_{\bar\ell \ell\nu,xyz}^{\tT\tL\tT\tL\tS\tL2}$
&$(\overline{\ell_{\tR x}} \sigma_{\mu\nu}\ell_{\tL y})
(\overline{\nu_{\tL z}^\C} \sigma^{\mu\nu}\calN_{sud}^{\tL\tL}) $ 
&$\pmb{8}_\tL\otimes\pmb{1}_\tR$
&$n_\ell^2 n_\nu\,\langle 12 \rangle$  
\\\hline
$\tO_{\bar\ell\ell\nu,xyz}^{\tS\tR\tS\tL\tS\tL1}$
&$(\overline{\ell_{\tL x}} \ell_{\tR y})
(\overline{\nu_{\tL z}^\C} \calN_{dsu}^{\tL\tL}) $ 
&$\pmb{8}_\tL\otimes\pmb{1}_\tR$
&$n_\ell^2 n_\nu\,\langle 12 \rangle$ 
\\
$\tO_{\bar\ell \ell\nu,xyz}^{\tS\tR\tS\tL\tS\tL2}$
&$(\overline{\ell_{\tL x}} \ell_{\tR y})
(\overline{\nu_{\tL z}^\C} \calN_{sud}^{\tL\tL}) $ 
&$\pmb{8}_\tL\otimes\pmb{1}_\tR$
&$n_\ell^2 n_\nu\,\langle 12 \rangle$
\\\hline
$\tO_{\bar\ell\ell\nu,xyz}^{\tS\tL\tS\tL\tS\tR1}$ 
&$(\overline{\ell_{\tR x}} \ell_{\tL y})
(\overline{\nu_{\tL z}^\C}\calN_{uds}^{\tL\tR}) $  
&$\pmb{3}_\tL \otimes\bar{\pmb{3}}_\tR$
&$n_\ell^2 n_\nu\,\langle 12 \rangle$ 
\\
$\tO_{\bar\ell\ell\nu,xyz}^{\tT\tL\tT\tL\tS\tR1}$ 
&$(\overline{\ell_{\tR x}} \sigma_{\mu\nu}\ell_{\tL y})
(\overline{\nu_{\tL z}^\C}\sigma^{\mu\nu}\calN_{uds}^{\tL\tR}) $  
&$\pmb{3}_\tL \otimes\bar{\pmb{3}}_\tR$
&$n_\ell^2 n_\nu\,\langle 12 \rangle$
\\\hline%
$\tO_{\bar\ell \ell\nu,xyz}^{\tS\tL\tS\tL\tS\tR2}$
&$(\overline{\ell_{\tR x}} \ell_{\tL y})
(\overline{\nu_{\tL z}^\C} \calN_{dsu}^{\tL\tR}) $ 
&$\pmb{3}_\tL\otimes\bar{\pmb{3}}_\tR$
&$n_\ell^2 n_\nu\,\langle 12 \rangle$ 
\\
$\tO_{\bar\ell \ell\nu,xyz}^{\tT\tL\tT\tL\tS\tR2}$
&$(\overline{\ell_{\tR x}} \sigma_{\mu\nu}\ell_{\tL y})
(\overline{\nu_{\tL z}^\C} \sigma^{\mu\nu}\calN_{dsu}^{\tL\tR}) $ 
&$\pmb{3}_\tL\otimes\bar{\pmb{3}}_\tR$
&$n_\ell^2 n_\nu\,\langle 12 \rangle$ 
\\\hline%
$\tO_{\bar\ell\ell\nu,xyz}^{\tS\tL\tS\tL\tS\tR3}$
&$(\overline{\ell_{\tR x}} \ell_{\tL y})
(\overline{\nu_{\tL z}^\C} \calN_{sud}^{\tL\tR}) $ 
&$\pmb{3}_\tL\otimes\bar{\pmb{3}}_\tR$
&$n_\ell^2 n_\nu\,\langle 12 \rangle$ 
\\
$\tO_{\bar\ell \ell\nu,xyz}^{\tT\tL\tT\tL\tS\tR3}$
&$(\overline{\ell_{\tR x}} \sigma_{\mu\nu}\ell_{\tL y})
(\overline{\nu_{\tL z}^\C} \sigma^{\mu\nu}\calN_{sud}^{\tL\tR}) $ 
&$\pmb{3}_\tL\otimes\bar{\pmb{3}}_\tR$
&$n_\ell^2 n_\nu\,\langle 12 \rangle$ 
\\\hline
$\tO_{\bar\ell \ell\nu,xyz}^{\tS\tR\tS\tL\tS\tR1}$ 
&$(\overline{\ell_{\tL x}} \ell_{\tR y})
(\overline{\nu_{\tL z}^\C} \calN_{uds}^{\tL\tR}) $  
&$\pmb{3}_\tL \otimes\bar{\pmb{3}}_\tR$
&$n_\ell^2 n_\nu\,\langle 12 \rangle$
\\
$\tO_{\bar\ell \ell\nu,xyz}^{\tT\tR\tV\tR\tV\tL1}$ 
&$(\overline{\ell_{\tL x}}\sigma_{\mu\nu} \ell_{\tR y})
(\overline{\nu_{\tL z}^\C} \gamma^\mu\calN_{dsu}^{\tR\tL,\nu}) $ 
&$\pmb{3}_\tL\otimes\pmb{6}_\tR$
&$n_\ell^2 n_\nu\,\langle 12 \rangle$  
\\\hline%
$\tO_{\bar\ell \ell\nu,xyz}^{\tS\tR\tS\tL\tS\tR2}$
&$(\overline{\ell_{\tL x}} \ell_{\tR y})
(\overline{\nu_{\tL z}^\C} \calN_{dsu}^{\tL\tR}) $ 
&$\pmb{3}_\tL\otimes\bar{\pmb{3}}_\tR$
&$n_\ell^2 n_\nu\,\langle 12 \rangle$ 
\\
$\tO_{\bar\ell \ell\nu,xyz}^{\tT\tR\tV\tR\tV\tL2}$ 
&$(\overline{\ell_{\tL x}}\sigma_{\mu\nu} \ell_{\tR y})
(\overline{\nu_{\tL z}^\C} \gamma^\mu\calN_{usd}^{\tR\tL,\nu}) $ 
&$\pmb{3}_\tL\otimes\pmb{6}_\tR$
&$n_\ell^2 n_\nu\,\langle 12 \rangle$
\\\hline%
$\tO_{\bar\ell \ell\nu,xyz}^{\tS\tR\tS\tL\tS\tR3}$
&$(\overline{\ell_{\tL x}} \ell_{\tR y})
(\overline{\nu_{\tL z}^\C} \calN_{sud}^{\tL\tR}) $ 
&$\pmb{3}_\tL\otimes\bar{\pmb{3}}_\tR$
&$n_\ell^2 n_\nu\,\langle 12 \rangle$ 
\\
$\tO_{\bar\ell \ell\nu,xyz}^{\tT\tR\tV\tR\tV\tL3}$ 
&$(\overline{\ell_{\tL x}}\sigma_{\mu\nu} \ell_{\tR y})
(\overline{\nu_{\tL z}^\C} \gamma^\mu\calN_{uds}^{\tR\tL,\nu}) $ 
&$\pmb{3}_\tL\otimes\pmb{6}_\tR$
&$n_\ell^2 n_\nu\,\langle 12 \rangle$   
\\\clineB{1-4}{1.5}
\multicolumn{4}{l}{\color{magenta}$\Delta L = + 1 $: $p\to \bar\nu_x\bar\nu_y \nu_z K^+$, 
$n\to \bar\nu_x\bar\nu_y \nu_z K^0$}
\\\hline
$\tO_{\nu\nu\bar\nu,xyz}^{\tS\tL\tS\tR\tS\tR1}$
&$(\overline{\nu_{\tL x}^\C}\nu_{\tL y})(\overline{\nu_{\tL z}}\calN_{dsu}^{\tR\tR})$
&$\pmb{1}_\tL\otimes\pmb{8}_\tR$
&$\frac{1}{2}n_\nu^2(n_\nu+1)\,\langle 18 \rangle$
\\
$\tO_{\nu\nu\bar\nu,xyz}^{\tS\tL\tS\tR\tS\tR2}$
&$(\overline{\nu_{\tL x}^\C}\nu_{\tL y})(\overline{\nu_{\tL z}}\calN_{sud}^{\tR\tR})$
&$\pmb{1}_\tL\otimes\pmb{8}_\tR$
&$\frac{1}{2}n_\nu^2(n_\nu+1)\,\langle 18 \rangle$
\\\hline
$\tO_{\nu\nu\bar\nu,xyz}^{\tS\tL\tS\tR\tS\tL1}$
& $(\overline{\nu_{\tL x}^\C}\nu_{\tL y})(\overline{\nu_{\tL z}}\calN_{uds}^{\tR\tL})$
&$\bar{\pmb{3}}_\tL\otimes\pmb{3}_\tR$
&$\frac{1}{2}n_\nu^2(n_\nu+1)\,\langle 18 \rangle$ 
\\
$\tO_{\nu\nu\bar\nu,xyz}^{\tT\tL\tV\tL\tV\tR1}$
&$(\overline{\nu_{\tL x}^\C}\sigma_{\mu\nu}\nu_{\tL y})(\overline{\nu_{\tL z}}\gamma^\mu \calN_{dsu}^{\tL\tR,\nu})$
&$\pmb{6}_\tL\otimes\pmb{3}_\tR$
&$\frac{1}{2}n_\nu^2(n_\nu-1)\,\langle 9 \rangle$ 
\\\hline%
$\tO_{\nu\nu\bar\nu,xyz}^{\tS\tL\tS\tR\tS\tL2}$
&$(\overline{\nu_{\tL x}^\C}\nu_{\tL y})(\overline{\nu_{\tL z}}\calN_{dsu}^{\tR\tL})$
&$\bar{\pmb{3}}_\tL\otimes\pmb{3}_\tR$
&$\frac{1}{2}n_\nu^2(n_\nu+1)\,\langle 18 \rangle$  
\\
$\tO_{\nu\nu\bar\nu,xyz}^{\tT\tL\tV\tL\tV\tR2}$
&$(\overline{\nu_{\tL x}^\C}\sigma_{\mu\nu}\nu_{\tL y})(\overline{\nu_{\tL z}}\gamma^\mu \calN_{usd}^{\tL\tR,\nu})$
&$\pmb{6}_\tL\otimes\pmb{3}_\tR$
&$\frac{1}{2}n_\nu^2(n_\nu-1)\,\langle 9 \rangle$ 
\\\hline%
$\tO_{\nu\nu\bar\nu,xyz}^{\tS\tL\tS\tR\tS\tL3}$
&$(\overline{\nu_{\tL x}^\C}\nu_{\tL y})(\overline{\nu_{\tL z}}\calN_{sud}^{\tR\tL})$
&$\bar{\pmb{3}}_\tL\otimes\pmb{3}_\tR$
&$\frac{1}{2}n_\nu^2(n_\nu+1)\,\langle 18 \rangle$ 
\\
$\tO_{\nu\nu\bar\nu,xyz}^{\tT\tL\tV\tL\tV\tR3}$
&$(\overline{\nu_{\tL x}^\C}\sigma_{\mu\nu}\nu_{\tL y})(\overline{\nu_{\tL z}}\gamma^\mu \calN_{uds}^{\tL\tR,\nu})$
&$\pmb{6}_\tL\otimes\pmb{3}_\tR$
&$\frac{1}{2}n_\nu^2(n_\nu-1)\,\langle 9 \rangle$
\\\hline
\end{tabular} }
\caption{Continuation of \cref{tab:dim9ope_Lp1A}.}
\label{tab:dim9ope_Lp1B}
\end{table}

\begin{table}[t]
\center
\resizebox{\linewidth}{!}{
\renewcommand{\arraystretch}{1.3}
\begin{tabular}{l l l l}
\hline   
Name & Operator & Irrep & \#
\\\clineB{1-4}{1.5}
\multicolumn{4}{l}{\color{magenta}$\Delta L = - 1$: $p\to \nu_x\nu_y\ell_z^+K^0$}
\\\hline
$\tO_{\bar\nu\bar\nu\ell,xyz}^{\tS\tR\tS\tL\tS\tL}$
&$(\overline{\nu_{\tL x}}\nu_{\tL y}^\C)(\overline{\ell_{\tL z}^\C}\calN_{usu}^{\tL\tL})$
&$\pmb{8}_\tL\otimes\pmb{1}_\tR$
&$\frac{1}{2}n_\ell (n_\nu^2+n_\nu)\,\langle 12 \rangle$ 
\\\cline{1-4}%
$\tO_{\bar\nu\bar\nu\ell,xyz}^{\tS\tR\tS\tR\tS\tR}$
&$(\overline{\nu_{\tL x}}\nu_{\tL y}^\C)(\overline{\ell_{\tR z}^\C}\calN_{usu}^{\tR\tR})$
&$\pmb{1}_\tL\otimes\pmb{8}_\tR$
&$\frac{1}{2}n_\ell (n_\nu^2+n_\nu)\,\langle 12 \rangle$
\\
$\tO_{\bar\nu\bar\nu\ell,xyz}^{\tT\tR\tT\tR\tS\tR}$
&$(\overline{\nu_{\tL x}}\sigma_{\mu\nu}\nu_{\tL y}^\C)(\overline{\ell_{\tR z}^\C}\sigma^{\mu\nu} \calN_{usu}^{\tR\tR})$
&$\pmb{1}_\tL\otimes\pmb{8}_\tR$
&$\frac{1}{2}n_\ell n_\nu(n_\nu-1)\,\langle 6 \rangle$  
\\
$\tO_{\bar\nu\bar\nu\ell,xyz}^{\tT\tR\tS\tR\tT\tR}$ 
&$(\overline{\nu_{\tL x}}\sigma_{\mu\nu}\nu_{\tL y}^\C)(\overline{\ell_{\tR z}^\C}\calN_{uus}^{\tR\tR,\mu\nu})$
&$\pmb{1}_\tL\otimes\pmb{10}_\tR$
&$ \frac{1}{2}n_\ell n_\nu(n_\nu-1)\,\langle 6 \rangle$ 
\\\cline{1-4}%
$\tO_{\bar\nu\bar\nu\ell,xyz}^{\tS\tR\tS\tL\tS\tR}$
&$(\overline{\nu_{\tL x}}\nu_{\tL y}^\C)(\overline{\ell_{\tL z}^\C}\calN_{usu}^{\tL\tR})$
&$\pmb{3}_\tL\otimes\bar{\pmb{3}}_\tR$
&$\frac{1}{2}n_\ell (n_\nu^2+n_\nu)\,\langle 12 \rangle$ 
\\
$\tO_{\bar\nu\bar\nu\ell,xyz}^{\tT\tR\tV\tR\tV\tL1}$
&$(\overline{\nu_{\tL x}}\sigma_{\mu\nu}\nu_{\tL y}^\C)(\overline{\ell_{\tL z}^\C}\gamma^\mu \calN_{usu}^{\tR\tL,\nu})$
&$\pmb{3}_\tL\otimes\pmb{6}_\tR$
&$\frac{1}{2}n_\ell n_\nu(n_\nu-1)\,\langle 6 \rangle$ 
\\\cline{1-4}%
$\tO_{\bar\nu\bar\nu\ell,xyz}^{\tT\tR\tV\tR\tV\tL2}$
&$(\overline{\nu_{\tL x}}\sigma_{\mu\nu}\nu_{\tL y}^\C)(\overline{\ell_{\tL z}^\C}\gamma^\mu \calN_{uus}^{\tR\tL,\nu})$
&$\pmb{3}_\tL\otimes\pmb{6}_\tR$
&$\frac{1}{2}n_\ell n_\nu(n_\nu-1)\,\langle 6 \rangle$ 
\\\cline{1-4}%
$\tO_{\bar\nu\bar\nu\ell,xyz}^{\tS\tR\tS\tR\tS\tL}$
&$(\overline{\nu_{\tL x}}\nu_{\tL y}^\C)(\overline{\ell_{\tR z}^\C} \calN_{usu}^{\tR\tL})$
&$\bar{\pmb{3}}_\tL\otimes\pmb{3}_\tR$
&$\frac{1}{2} n_\ell (n_\nu^2+n_\nu)\,\langle 12 \rangle$ 
\\
$\tO_{\bar\nu\bar\nu\ell,xyz}^{\tT\tR\tT\tR\tS\tL}$ 
&$(\overline{\nu_{\tL x}}\sigma_{\mu\nu}\nu_{\tL y}^\C)(\overline{\ell_{\tR z}^\C} \sigma^{\mu\nu} \calN_{usu}^{\tR\tL})$
&$\bar{\pmb{3}}_\tL\otimes\pmb{3}_\tR$
&$\frac{1}{2}n_\ell n_\nu(n_\nu-1)\,\langle 6 \rangle$
\\\clineB{1-4}{1.5}
\multicolumn{4}{l}{\color{magenta}$\Delta L = -1 $: $n\to \ell_x^-\ell_y^- \ell_z^+ K^+$}
\\\hline%
$\tO_{\bar\ell \bar\ell\ell,xyz}^{\tV\tR\tV\tL\tS\tL}$
&$(\overline{\ell_{\tR x}}\gamma_\mu \ell_{\tL y}^\C)
(\overline{\ell_{\tR z}^\C}\gamma^\mu \calN_{dds}^{\tL\tL})
$  
&$\pmb{8}_\tL\otimes\pmb{1}_\tR$
&$n_\ell^3\,\langle 8 \rangle$ 
\\\hline%
$\tO_{\bar\ell \bar\ell\ell,xyz}^{\tV\tR\tV\tL\tS\tR}$
&$(\overline{\ell_{\tR x}}\gamma_\mu \ell_{\tL y}^\C)
(\overline{\ell_{\tR z}^\C}\gamma^\mu \calN_{dds}^{\tL\tR})
$  
&$\pmb{3}_\tL\otimes\bar{\pmb{3}}_\tR$
&$n_\ell^3\,\langle 8 \rangle$ 
\\
$\tO_{\bar\ell \bar\ell\ell,xyz}^{\tV\tR\tS\tR\tV\tL1}$
&$(\overline{\ell_{\tR x}}\gamma_\mu \ell_{\tL y}^\C)
(\overline{\ell_{\tR z}^\C}\calN_{dsd}^{\tR\tL,\mu})
$ 
&$\pmb{3}_\tL\otimes\pmb{6}_\tR$
&$n_\ell^3\,\langle 8 \rangle$ 
\\\hline%
$\tO_{\bar\ell \bar\ell\ell,xyz}^{\tV\tR\tS\tR\tV\tL2}$
&$(\overline{\ell_{\tR x}}\gamma_\mu \ell_{\tL y}^\C)
(\overline{\ell_{\tR z}^\C}\calN_{dds}^{\tR\tL,\mu})
$
&$\pmb{3}_\tL\otimes\pmb{6}_\tR$
&$n_\ell^3\,\langle 8 \rangle$ 
\\\hline%
$\tO_{\bar\ell \bar\ell\ell,xyz}^{\tS\tL\tS\tR\tS\tR}$
&$(\overline{\ell_{\tR x}} \ell_{\tR y}^\C)
(\overline{\ell_{\tR z}^\C} \calN_{dds}^{\tR\tR}) $
&$\pmb{1}_\tL\otimes\pmb{8}_\tR$
&$\frac{1}{2}n_\ell^2(n_\ell+1)\,\langle 6 \rangle$ 
\\\hline%
$\tO_{\bar\ell \bar\ell\ell,xyz}^{\tS\tL\tS\tL\tS\tL}$
&$(\overline{\ell_{\tR x}} \ell_{\tR y}^\C)
(\overline{\ell_{\tL z}^\C}\calN_{dds}^{\tL\tL}) $
&$\pmb{8}_\tL\otimes\pmb{1}_\tR$
&$\frac{1}{2}n_\ell^2 (n_\ell+1)\,\langle 6 \rangle$ 
\\
$\tO_{\bar\ell \bar\ell\ell,xyz}^{\tT\tL\tT\tL\tS\tL}$
&$(\overline{\ell_{\tR x}} \sigma_{\mu\nu}\ell_{\tR y}^\C)
(\overline{\ell_{\tL z}^\C}\sigma^{\mu\nu}\calN_{dds}^{\tL\tL}) $
&$\pmb{8}_\tL\otimes\pmb{1}_\tR$
&$\frac{1}{2}n_\ell^2 (n_\ell-1)\,\langle 2 \rangle$ 
\\
$\tO_{\bar\ell \bar\ell\ell,xyz}^{\tT\tL\tS\tL\tT\tL}$
&$(\overline{\ell_{\tR x}} \sigma_{\mu\nu}\ell_{\tR y}^\C)
(\overline{\ell_{\tL z}^\C}\calN_{dds}^{\tL\tL,\mu\nu}) $ 
&$\pmb{10}_\tL\otimes\pmb{1}_\tR$
&$\frac{1}{2}n_\ell^2 (n_\ell-1)\,\langle 2 \rangle$   
\\\hline%
$\tO_{\bar\ell \bar\ell\ell,xyz}^{\tS\tL\tS\tR\tS\tL}$
&$(\overline{\ell_{\tR x}} \ell_{\tR y}^\C)
(\overline{\ell_{\tR z}^\C}\calN_{dds}^{\tR\tL}) $
&$\bar{\pmb{3}}_\tL\otimes\pmb{3}_\tR$
&$\frac{1}{2}n_\ell^2(n_\ell+1)\,\langle 6 \rangle$ 
\\
$\tO_{\bar\ell \bar\ell\ell,xyz}^{\tT\tL\tV\tL\tV\tR1}$
&$(\overline{\ell_{\tR x}} \sigma_{\mu\nu}\ell_{\tR y}^\C)
(\overline{\ell_{\tR z}^\C}\gamma^\mu\calN_{dsd}^{\tL\tR,\nu}) $ 
&$\pmb{6}_\tL\otimes\pmb{3}_\tR$
&$\frac{1}{2}n_\ell^2(n_\ell-1)\,\langle 2 \rangle$ 
\\\hline%
$\tO_{\bar\ell \bar\ell\ell,xyz}^{\tT\tL\tV\tL\tV\tR2}$
&$(\overline{\ell_{\tR x}} \sigma_{\mu\nu}\ell_{\tR y}^\C)
(\overline{\ell_{\tR z}^\C}\gamma^\mu\calN_{dds}^{\tL\tR,\nu}) $  
&$\pmb{6}_\tL\otimes\pmb{3}_\tR$
&$\frac{1}{2}n_\ell^2(n_\ell-1)\,\langle 2 \rangle$ 
\\\hline%
$\tO_{\bar\ell \bar\ell\ell,xyz}^{\tS\tL\tS\tL\tS\tR}$
&$(\overline{\ell_{\tR x}} \ell_{\tR y}^\C)
(\overline{\ell_{\tL z}^\C} \calN_{dds}^{\tL\tR}) $
&$\pmb{3}_\tL\otimes\bar{\pmb{3}}_\tR$
&$\frac{1}{2}n_\ell^2(n_\ell+1)\,\langle 6 \rangle$ 
\\
$\tO_{\bar\ell \bar\ell\ell,xyz}^{\tT\tL\tT\tL\tS\tR}$
&$(\overline{\ell_{\tR x}}\sigma_{\mu\nu} \ell_{\tR y}^\C)
(\overline{\ell_{\tL z}^\C} \sigma^{\mu\nu}\calN_{dds}^{\tL\tR}) $
&$\pmb{3}_\tL\otimes\bar{\pmb{3}}_\tR$
&$\frac{1}{2}n_\ell^2(n_\ell-1)\,\langle 2 \rangle$
\\\clineB{1-4}{1.5}
$\tO_{\bar\ell \bar\ell\ell,xyz}^{\tV\tL\tV\tR\tS\tR}$
&$(\overline{\ell_{\tL x}}\gamma_\mu \ell_{\tR y}^\C)
(\overline{\ell_{\tL z}^\C}\gamma^\mu \calN_{dds}^{\tR\tR})$  
&$\pmb{1}_\tL\otimes\pmb{8}_\tR$
&$n_\ell^3\,\langle 8 \rangle$ 
\\\hline%
$\tO_{\bar\ell \bar\ell\ell,xyz}^{\tV\tL\tV\tR\tS\tL}$
& $(\overline{\ell_{\tL x}}\gamma_\mu \ell_{\tR y}^\C)
(\overline{\ell_{\tL z}^\C}\gamma^\mu \calN_{dds}^{\tR\tL})$
&$\bar{\pmb{3}}_\tL\otimes\pmb{3}_\tR$
&$n_\ell^3\,\langle 8 \rangle$ 
\\
$\tO_{\bar\ell \bar\ell\ell,xyz}^{\tV\tL\tS\tL\tV\tR1}$
&$(\overline{\ell_{\tL x}}\gamma_\mu \ell_{\tR y}^\C)
(\overline{\ell_{\tL z}^\C}\calN_{dsd}^{\tL\tR,\mu}) $  
&$\pmb{6}_\tL\otimes\pmb{3}_\tR$
&$n_\ell^3\,\langle 8 \rangle$ 
\\\hline%
$\tO_{\bar\ell \bar\ell\ell,xyz}^{\tV\tL\tS\tL\tV\tR2}$
&$(\overline{\ell_{\tL x}}\gamma_\mu \ell_{\tR y}^\C)
(\overline{\ell_{\tL z}^\C}\calN_{dds}^{\tL\tR,\mu}) $  
&$\pmb{6}_\tL\otimes\pmb{3}_\tR$
&$n_\ell^3\,\langle 8 \rangle$ 
\\\hline%
$\tO_{\bar\ell \bar\ell\ell,xyz}^{\tS\tR\tS\tL\tS\tL}$
&$(\overline{\ell_{\tL x}} \ell_{\tL y}^\C)
(\overline{\ell_{\tL z}^\C} \calN_{dds}^{\tL\tL}) $
&$\pmb{8}_\tL\otimes\pmb{1}_\tR$
&$\frac{1}{2}n_\ell^2(n_\ell+1)\,\langle 6 \rangle$ 
\\\hline%
$\tO_{\bar\ell \bar\ell\ell,xyz}^{\tS\tR\tS\tR\tS\tR}$
&$(\overline{\ell_{\tL x}} \ell_{\tL y}^\C)
(\overline{\ell_{\tR z}^\C}\calN_{dds}^{\tR\tR}) $
&$\pmb{1}_\tL\otimes\pmb{8}_\tR$
&$\frac{1}{2}n_\ell^2 (n_\ell+1)\,\langle 6 \rangle$ 
\\
$\tO_{\bar\ell \bar\ell\ell,xyz}^{\tT\tR\tT\tR\tS\tR}$
&$(\overline{\ell_{\tL x}} \sigma_{\mu\nu}\ell_{\tL y}^\C)
(\overline{\ell_{\tR z}^\C}\sigma^{\mu\nu}\calN_{dds}^{\tR\tR}) $
&$\pmb{1}_\tL\otimes\pmb{8}_\tR$
&$\frac{1}{2}n_\ell^2 (n_\ell-1)\,\langle 2 \rangle$ 
\\
$\tO_{\bar\ell \bar\ell\ell,xyz}^{\tT\tR\tS\tR\tT\tR}$
&$(\overline{\ell_{\tL x}} \sigma_{\mu\nu}\ell_{\tL y}^\C)
(\overline{\ell_{\tR z}^\C}\calN_{dds}^{\tR\tR,\mu\nu}) $
&$\pmb{1}_\tL\otimes\pmb{10}_\tR$
&$\frac{1}{2}n_\ell^2 (n_\ell-1)\,\langle 2 \rangle$
\\\hline%
$\tO_{\bar\ell \bar\ell\ell,xyz}^{\tS\tR\tS\tL\tS\tR}$
&$(\overline{\ell_{\tL x}} \ell_{\tL y}^\C)
(\overline{\ell_{\tL z}^\C}\calN_{dds}^{\tL\tR}) $
&$\pmb{3}_\tL\otimes\bar{\pmb{3}}_\tR$
&$\frac{1}{2}n_\ell^2(n_\ell+1)\,\langle 6 \rangle$ 
\\
$\tO_{\bar\ell \bar\ell\ell,xyz}^{\tT\tR\tV\tR\tV\tL1}$
&$(\overline{\ell_{\tL x}} \sigma_{\mu\nu}\ell_{\tL y}^\C)
(\overline{\ell_{\tL z}^\C}\gamma^\mu\calN_{dsd}^{\tR\tL,\nu}) $ 
&$\pmb{3}_\tL\otimes\pmb{6}_\tR$
&$\frac{1}{2}n_\ell^2(n_\ell-1)\,\langle 2 \rangle$
\\\hline%
$\tO_{\bar\ell \bar\ell\ell,xyz}^{\tT\tR\tV\tR\tV\tL2}$
&$(\overline{\ell_{\tL x}} \sigma_{\mu\nu}\ell_{\tL y}^\C)
(\overline{\ell_{\tL z}^\C}\gamma^\mu\calN_{dds}^{\tR\tL,\nu}) $ 
&$\pmb{3}_\tL\otimes\pmb{6}_\tR$
&$\frac{1}{2}n_\ell^2(n_\ell-1)\,\langle 2 \rangle$ 
\\\hline%
$\tO_{\bar\ell \bar\ell\ell,xyz}^{\tS\tR\tS\tR\tS\tL}$
&$(\overline{\ell_{\tL x}} \ell_{\tL y}^\C)
(\overline{\ell_{\tR z}^\C} \calN_{dds}^{\tR\tL}) $
&$\bar{\pmb{3}}_\tL\otimes\pmb{3}_\tR$
&$\frac{1}{2}n_\ell^2(n_\ell+1)\,\langle 6 \rangle$ 
\\
$\tO_{\bar\ell \bar\ell\ell,xyz}^{\tT\tR\tT\tR\tS\tL}$
&$(\overline{\ell_{\tL x}}\sigma_{\mu\nu} \ell_{\tL y}^\C)
(\overline{\ell_{\tR z}^\C} \sigma^{\mu\nu}\calN_{dds}^{\tR\tL}) $
&$\bar{\pmb{3}}_\tL\otimes\pmb{3}_\tR$
&$\frac{1}{2}n_\ell^2(n_\ell-1)\,\langle 2 \rangle$ 
\\\clineB{1-4}{1.5}
\multicolumn{4}{l}{\color{magenta}$\Delta L = -1$: $n\to\nu_x \bar\nu_y \ell_z^- K^+$}
\\\cline{1-4}
$\tO_{\bar\nu \nu \bar\ell,xyz}^{\tV\tL\tV\tR\tS\tR}$
&$(\overline{\nu_{\tL x}}\gamma_\mu \nu_{\tL y})
(\overline{\ell_{\tR z}}\gamma^\mu \calN_{dds}^{\tR\tR}) $
&$\pmb{1}_\tL\otimes\pmb{8}_\tR$
&$n_\ell n_\nu^2 \,\langle 18 \rangle$ 
\\\hline%
$\tO_{\bar\nu \nu \bar\ell,xyz}^{\tV\tL\tV\tR\tS\tL}$
&$(\overline{\nu_{\tL x}}\gamma_\mu \nu_{\tL y})
(\overline{\ell_{\tR z}}\gamma^\mu \calN_{dds}^{\tR\tL}) $
&$\bar{\pmb{3}}_\tL\otimes\pmb{3}_\tR$
&$n_\ell n_\nu^2 \,\langle 18 \rangle$ 
\\
$\tO_{\bar\nu \nu \bar\ell,xyz}^{\tV\tL\tS\tL\tV\tR1}$
&$(\overline{\nu_{\tL x}}\gamma_\mu \nu_{\tL y})
(\overline{\ell_{\tR z}}\calN_{dsd}^{\tL\tR,\mu})$
&$\pmb{6}_\tL\otimes\pmb{3}_\tR$
&$n_\ell n_\nu^2 \,\langle 18 \rangle$ 
\\\hline%
$\tO_{\bar\nu \nu \bar\ell,xyz}^{\tV\tL\tS\tL\tV\tR2}$
&$(\overline{\nu_{\tL x}}\gamma_\mu \nu_{\tL y})
(\overline{\ell_{\tR z}}\calN_{dds}^{\tL\tR,\mu})$
&$\pmb{6}_\tL\otimes\pmb{3}_\tR$
&$n_\ell n_\nu^2 \,\langle 18 \rangle$ 
\\\hline
$\tO_{\bar\nu \nu \bar\ell,xyz}^{\tV\tL\tV\tL\tS\tL}$
&$(\overline{\nu_{\tL x}}\gamma_\mu \nu_{\tL y})
(\overline{\ell_{\tL z}}\gamma^\mu \calN_{dds}^{\tL\tL}) $
&$\pmb{8}_\tL\otimes\pmb{1}_\tR$
&$n_\ell n_\nu^2 \,\langle 18 \rangle$ 
\\\hline%
$\tO_{\bar\nu \nu \bar\ell,xyz}^{\tV\tL\tV\tL\tS\tR}$
&$(\overline{\nu_{\tL x}}\gamma_\mu \nu_{\tL y})
(\overline{\ell_{\tL z}}\gamma^\mu \calN_{dds}^{\tL\tR}) $
&$\pmb{3}_\tL\otimes\bar{\pmb{3}}_\tR$
&$n_\ell n_\nu^2 \,\langle 18 \rangle$ 
\\
$\tO_{\bar\nu \nu \bar\ell,xyz}^{\tV\tL\tS\tR\tV\tL1}$
&$(\overline{\nu_{\tL x}}\gamma_\mu \nu_{\tL y})
(\overline{\ell_{\tL z}}\calN_{dsd}^{\tR\tL,\mu}) $ 
&$\pmb{3}_\tL\otimes\pmb{6}_\tR$
&$n_\ell n_\nu^2 \,\langle 18 \rangle$
\\\hline%
$\tO_{\bar\nu \nu \bar\ell,xyz}^{\tV\tL\tS\tR\tV\tL2}$
&$(\overline{\nu_{\tL x}}\gamma_\mu \nu_{\tL y})
(\overline{\ell_{\tL z}}\calN_{dds}^{\tR\tL,\mu}) $  
&$\pmb{3}_\tL\otimes\pmb{6}_\tR$
&$n_\ell n_\nu^2 \,\langle 18 \rangle$ 
\\\hline
\end{tabular} }
\caption{The dim-9 $\Delta L=-1$ BNV LEFT operators involving a single strange quark.}
\label{tab:dim9ope_Lm1A}
\end{table}

\begin{table}[t]
\center
\resizebox{\linewidth}{!}{
\renewcommand{\arraystretch}{1.3}
\begin{tabular}{l l l l}
\hline   
Name & Operator & Irrep &\#
\\
\clineB{1-4}{1.5}
\multicolumn{4}{l}{\color{magenta}$\Delta L = - 1 $: $p\to\ell_x^- \ell_y^+ \nu_z K^+$,
$n\to\ell_x^- \ell_y^+ \nu_z K^0$}
\\\hline
$\tO_{\bar\ell \ell \bar\nu,xyz}^{\tV\tL\tV\tL\tS\tL1}$
&$(\overline{\ell_{\tL x}}\gamma_\mu \ell_{\tL y})
(\overline{\nu_{\tL z}}\gamma^\mu \calN_{dsu}^{\tL\tL}) $
&$\pmb{8}_\tL\otimes\pmb{1}_\tR$
&$n_\ell^2 n_\nu\,\langle 12 \rangle$ 
\\
$\tO_{\bar\ell \ell \bar\nu,xyz}^{\tV\tL\tV\tL\tS\tL2}$
&$(\overline{\ell_{\tL x}}\gamma_\mu \ell_{\tL y})
(\overline{\nu_{\tL z}}\gamma^\mu \calN_{sud}^{\tL\tL}) $
&$\pmb{8}_\tL\otimes\pmb{1}_\tR$
&$n_\ell^2 n_\nu\,\langle 12 \rangle$ 
\\\hline%
$\tO_{\bar\ell \ell \bar\nu,xyz}^{\tV\tR\tV\tL\tS\tL1}$
& $(\overline{\ell_{\tR x}}\gamma_\mu \ell_{\tR y})
(\overline{\nu_{\tL z}}\gamma^\mu \calN_{dsu}^{\tL\tL}) $
&$\pmb{8}_\tL\otimes\pmb{1}_\tR$
&$n_\ell^2 n_\nu\,\langle 12 \rangle$ 
\\
$\tO_{\bar\ell \ell \bar\nu,xyz}^{\tV\tR\tV\tL\tS\tL2}$
&$(\overline{\ell_{\tR x}}\gamma_\mu \ell_{\tR y})
(\overline{\nu_{\tL z}}\gamma^\mu \calN_{sud}^{\tL\tL}) $
&$\pmb{8}_\tL\otimes\pmb{1}_\tR$
&$n_\ell^2 n_\nu\,\langle 12 \rangle$ 
\\\hline
$\tO_{\bar\ell \ell \bar\nu,xyz}^{\tV\tL\tV\tL\tS\tR1}$
&$(\overline{\ell_{\tL x}}\gamma_\mu \ell_{\tL y})
(\overline{\nu_{\tL z}}\gamma^\mu\calN_{uds}^{\tL\tR}) $  
&$\pmb{3}_\tL\otimes\bar{\pmb{3}}_\tR$
&$n_\ell^2 n_\nu\,\langle 12 \rangle$ 
\\
$\tO_{\bar\ell \ell \bar\nu,xyz}^{\tV\tL\tS\tR\tV\tL1}$
& $(\overline{\ell_{\tL x}}\gamma_\mu \ell_{\tL y})
(\overline{\nu_{\tL z}}\calN_{dsu}^{\tR\tL,\mu}) $  
&$\pmb{3}_\tL\otimes\pmb{6}_\tR$
&$n_\ell^2 n_\nu\,\langle 12 \rangle$
\\\hline%
$\tO_{\bar\ell \ell \bar\nu,xyz}^{\tV\tL\tV\tL\tS\tR2}$
& $(\overline{\ell_{\tL x}}\gamma_\mu \ell_{\tL y})
(\overline{\nu_{\tL z}}\gamma^\mu \calN_{dsu}^{\tL\tR}) $
&$\pmb{3}_\tL\otimes\bar{\pmb{3}}_\tR$
&$n_\ell^2 n_\nu\,\langle 12 \rangle$
\\
$\tO_{\bar\ell \ell \bar\nu,xyz}^{\tV\tL\tS\tR\tV\tL2}$
&$(\overline{\ell_{\tL x}}\gamma_\mu \ell_{\tL y})
(\overline{\nu_{\tL z}}\calN_{usd}^{\tR\tL,\mu}) $  
&$\pmb{3}_\tL\otimes\pmb{6}_\tR$
&$n_\ell^2 n_\nu\,\langle 12 \rangle$
\\\hline%
$\tO_{\bar\ell \ell \bar\nu,xyz}^{\tV\tL\tV\tL\tS\tR3}$
&$(\overline{\ell_{\tL x}}\gamma_\mu \ell_{\tL y})
(\overline{\nu_{\tL z}}\gamma^\mu \calN_{sud}^{\tL\tR}) $
&$\pmb{3}_\tL\otimes\bar{\pmb{3}}_\tR$
&$n_\ell^2 n_\nu\,\langle 12 \rangle$ 
\\
$\tO_{\bar\ell \ell \bar\nu,xyz}^{\tV\tL\tS\tR\tV\tL3}$
&$(\overline{\ell_{\tL x}}\gamma_\mu \ell_{\tL y})
(\overline{\nu_{\tL z}}\calN_{uds}^{\tR\tL,\mu}) $  
&$\pmb{3}_\tL\otimes\pmb{6}_\tR$
&$n_\ell^2 n_\nu\,\langle 12 \rangle$
\\\hline
$\tO_{\bar\ell \ell \bar\nu,xyz}^{\tV\tR\tV\tL\tS\tR1}$
&$(\overline{\ell_{\tR x}}\gamma_\mu \ell_{\tR y})
(\overline{\nu_{\tL z}}\gamma^\mu\calN_{uds}^{\tL\tR}) $  
&$\pmb{3}_\tL\otimes\bar{\pmb{3}}_\tR$
&$n_\ell^2 n_\nu\,\langle 12 \rangle$
\\
$\tO_{\bar\ell \ell \bar\nu,xyz}^{\tV\tR\tS\tR\tV\tL1}$
&$(\overline{\ell_{\tR x}}\gamma_\mu \ell_{\tR y})
(\overline{\nu_{\tL z}}\calN_{dsu}^{\tR\tL,\mu}) $  
&$\pmb{3}_\tL\otimes\pmb{6}_\tR$
&$n_\ell^2 n_\nu\,\langle 12 \rangle$ 
\\\hline%
$\tO_{\bar\ell \ell \bar\nu,xyz}^{\tV\tR\tV\tL\tS\tR2}$
&$(\overline{\ell_{\tR x}}\gamma_\mu \ell_{\tR y})
(\overline{\nu_{\tL z}}\gamma^\mu \calN_{dsu}^{\tL\tR}) $  
&$\pmb{3}_\tL\otimes\bar{\pmb{3}}_\tR$
&$n_\ell^2 n_\nu\,\langle 12 \rangle$ 
\\
$\tO_{\bar\ell \ell \bar\nu,xyz}^{\tV\tR\tS\tR\tV\tL2}$
&$(\overline{\ell_{\tR x}}\gamma_\mu \ell_{\tR y})
(\overline{\nu_{\tL z}}\calN_{usd}^{\tR\tL,\mu}) $  
&$\pmb{3}_\tL\otimes\pmb{6}_\tR$
&$n_\ell^2 n_\nu\,\langle 12 \rangle$ 
\\\hline%
$\tO_{\bar\ell \ell \bar\nu,xyz}^{\tV\tR\tV\tL\tS\tR3}$
&$(\overline{\ell_{\tR x}}\gamma_\mu \ell_{\tR y})
(\overline{\nu_{\tL z}}\gamma^\mu \calN_{sud}^{\tL\tR}) $
&$\pmb{3}_\tL\otimes\bar{\pmb{3}}_\tR$
&$n_\ell^2 n_\nu\,\langle 12 \rangle$ 
\\
$\tO_{\bar\ell \ell \bar\nu,xyz}^{\tV\tR\tS\tR\tV\tL3}$
&$(\overline{\ell_{\tR x}}\gamma_\mu \ell_{\tR y})
(\overline{\nu_{\tL z}}\calN_{uds}^{\tR\tL,\mu}) $  
&$\pmb{3}_\tL\otimes\pmb{6}_\tR$
&$n_\ell^2 n_\nu\,\langle 12 \rangle$ 
\\\hline
$\tO_{\bar\ell \ell \bar\nu,xyz}^{\tS\tR\tS\tR\tS\tR1}$
&$(\overline{\ell_{\tL x}}\ell_{\tR y})
(\overline{\nu_{\tL z}}\calN_{dsu}^{\tR\tR})$ 
&$\pmb{1}_\tL\otimes\pmb{8}_\tR$
&$n_\ell^2 n_\nu\,\langle 12 \rangle$ 
\\
$\tO_{\bar\ell \ell \bar\nu,xyz}^{\tT\tR\tT\tR\tS\tR1}$
&$(\overline{\ell_{\tL x}}\sigma_{\mu\nu}\ell_{\tR y})
(\overline{\nu_{\tL z}}\sigma^{\mu\nu} \calN_{dsu}^{\tR\tR}) $  
&$\pmb{1}_\tL\otimes\pmb{8}_\tR$
&$n_\ell^2 n_\nu\,\langle 12 \rangle$ 
\\
$\tO_{\bar\ell \ell \bar\nu,xyz}^{\tT\tR\tS\tR\tT\tR}$
& $(\overline{\ell_{\tL x}}\sigma_{\mu\nu}\ell_{\tR y})
(\overline{\nu_{\tL z}}\calN_{uds}^{\tR\tR,\mu\nu}) $ 
&$\pmb{1}_\tL\otimes\pmb{10}_\tR$
&$n_\ell^2 n_\nu\,\langle 12 \rangle$ 
\\%
$\tO_{\bar\ell \ell \bar\nu,xyz}^{\tS\tR\tS\tR\tS\tR2}$
&$(\overline{\ell_{\tL x}}\ell_{\tR y})
(\overline{\nu_{\tL z}}\calN_{sud}^{\tR\tR})$ 
&$\pmb{1}_\tL\otimes\pmb{8}_\tR$
&$n_\ell^2 n_\nu\,\langle 12 \rangle$ 
\\
$\tO_{\bar\ell \ell \bar\nu,xyz}^{\tT\tR\tT\tR\tS\tR2}$
&$(\overline{\ell_{\tL x}}\sigma_{\mu\nu}\ell_{\tR y})
(\overline{\nu_{\tL z}}\sigma^{\mu\nu} \calN_{sud}^{\tR\tR}) $ 
&$\pmb{1}_\tL\otimes\pmb{8}_\tR$
&$n_\ell^2 n_\nu\,\langle 12 \rangle$ 
\\\hline
$\tO_{\bar\ell \ell \bar\nu,xyz}^{\tS\tL\tS\tR\tS\tR1}$
& $(\overline{\ell_{\tR x}}\ell_{\tL y})
(\overline{\nu_{\tL z}}\calN_{dsu}^{\tR\tR}) $ 
&$\pmb{1}_\tL\otimes\pmb{8}_\tR$
&$n_\ell^2 n_\nu\,\langle 12 \rangle$ 
\\
$\tO_{\bar\ell \ell \bar\nu,xyz}^{\tS\tL\tS\tR\tS\tR2}$
&$(\overline{\ell_{\tR x}}\ell_{\tL y})
(\overline{\nu_{\tL z}}\calN_{sud}^{\tR\tR})$
&$\pmb{1}_\tL\otimes\pmb{8}_\tR$
&$n_\ell^2 n_\nu\,\langle 12 \rangle$ 
\\\hline
$\tO_{\bar\ell \ell \bar\nu,xyz}^{\tS\tR\tS\tR\tS\tL1}$ 
&$(\overline{\ell_{\tL x}} \ell_{\tR y})
(\overline{\nu_{\tL z}}\calN_{uds}^{\tR\tL})$
&$\bar{\pmb{3}}_\tL\otimes\pmb{3}_\tR$
&$n_\ell^2 n_\nu\,\langle 12 \rangle$ 
\\
$\tO_{\bar\ell \ell \bar\nu,xyz}^{\tT\tR\tT\tR\tS\tL1}$ 
&$(\overline{\ell_{\tL x}} \sigma_{\mu\nu}\ell_{\tR y})
(\overline{\nu_{\tL z}}\sigma^{\mu\nu}\calN_{uds}^{\tR\tL}) $  
&$\bar{\pmb{3}}_\tL\otimes\pmb{3}_\tR$
&$n_\ell^2 n_\nu\,\langle 12 \rangle$ 
\\\hline%
$\tO_{\bar\ell \ell \bar\nu,xyz}^{\tS\tR\tS\tR\tS\tL2}$
&$(\overline{\ell_{\tL x}}\ell_{\tR y})
(\overline{\nu_{\tL z}}\calN_{dsu}^{\tR\tL})$
&$\bar{\pmb{3}}_\tL\otimes\pmb{3}_\tR$
&$n_\ell^2 n_\nu\,\langle 12 \rangle$ 
\\
$\tO_{\bar\ell \ell \bar\nu,xyz}^{\tT\tR\tT\tR\tS\tL2}$
&$(\overline{\ell_{\tL x}}\sigma_{\mu\nu}\ell_{\tR y})
(\overline{\nu_{\tL z}}\sigma^{\mu\nu}\calN_{dsu}^{\tR\tL}) $  
&$\bar{\pmb{3}}_\tL\otimes\pmb{3}_\tR$
&$n_\ell^2 n_\nu\,\langle 12 \rangle$ 
\\\hline%
$\tO_{\bar\ell \ell \bar\nu,xyz}^{\tS\tR\tS\tR\tS\tL3}$
&$(\overline{\ell_{\tL x}}\ell_{\tR y})
(\overline{\nu_{\tL z}}\calN_{sud}^{\tR\tL})$
&$\bar{\pmb{3}}_\tL\otimes\pmb{3}_\tR$
&$n_\ell^2 n_\nu\,\langle 12 \rangle$ 
\\
$\tO_{\bar\ell \ell \bar\nu,xyz}^{\tT\tR\tT\tR\tS\tL3}$
&$(\overline{\ell_{\tL x}}\sigma_{\mu\nu}\ell_{\tR y})
(\overline{\nu_{\tL z}}\sigma^{\mu\nu}\calN_{sud}^{\tR\tL}) $  
&$\bar{\pmb{3}}_\tL\otimes\pmb{3}_\tR$
&$n_\ell^2 n_\nu\,\langle 12 \rangle$ 
\\\hline
$\tO_{\bar\ell \ell \bar\nu,xyz}^{\tS\tL\tS\tR\tS\tL1}$ 
&$(\overline{\ell_{\tR x}} \ell_{\tL y})
(\overline{\nu_{\tL z}}\calN_{uds}^{\tR\tL})$
&$\bar{\pmb{3}}_\tL\otimes\pmb{3}_\tR$
&$n_\ell^2 n_\nu\,\langle 12 \rangle$
\\
$\tO_{\bar\ell \ell \bar\nu,xyz}^{\tT\tL\tV\tL\tV\tR1}$ 
&$(\overline{\ell_{\tR x}}\sigma_{\mu\nu} \ell_{\tL y})
(\overline{\nu_{\tL z}}\gamma^\mu\calN_{dsu}^{\tL\tR,\nu}) $  
&$\pmb{6}_\tL\otimes\pmb{3}_\tR$
&$n_\ell^2 n_\nu\,\langle 12 \rangle$ 
\\\hline%
$\tO_{\bar\ell \ell \bar\nu,xyz}^{\tS\tL\tS\tR\tS\tL2}$
&$(\overline{\ell_{\tR x}}\ell_{\tL y})
(\overline{\nu_{\tL z}}\calN_{dsu}^{\tR\tL})$
& $\bar{\pmb{3}}_\tL\otimes\pmb{3}_\tR$
&$n_\ell^2 n_\nu\,\langle 12 \rangle$ 
\\
$\tO_{\bar\ell \ell \bar\nu,xyz}^{\tT\tL\tV\tL\tV\tR2}$ 
&$(\overline{\ell_{\tR x}}\sigma_{\mu\nu}\ell_{\tL y})
(\overline{\nu_{\tL z}}\gamma^\mu\calN_{usd}^{\tL\tR,\nu}) $
&$\pmb{6}_\tL\otimes\pmb{3}_\tR$
&$n_\ell^2 n_\nu\,\langle 12 \rangle$ 
\\\hline%
$\tO_{\bar\ell \ell \bar\nu,xyz}^{\tS\tL\tS\tR\tS\tL3}$
&$(\overline{\ell_{\tR x}}\ell_{\tL y})
(\overline{\nu_{\tL z}}\calN_{sud}^{\tR\tL})$
&$\bar{\pmb{3}}_\tL\otimes\pmb{3}_\tR$
&$n_\ell^2 n_\nu\,\langle 12 \rangle$ 
\\
$\tO_{\bar\ell \ell \bar\nu,xyz}^{\tT\tL\tV\tL\tV\tR3}$ 
&$(\overline{\ell_{\tR x}}\sigma_{\mu\nu}\ell_{\tL y})
(\overline{\nu_{\tL z}}\gamma^\mu\calN_{uds}^{\tL\tR,\nu}) $  
&$\pmb{6}_\tL\otimes\pmb{3}_\tR$
&$n_\ell^2 n_\nu\,\langle 12 \rangle$ 
\\\clineB{1-4}{1.5}
\multicolumn{4}{l}{\color{magenta}$\Delta L = - 1 $: $p\to \nu_x\nu_y \bar\nu_z K^+$,
$n\to \nu_x\nu_y \bar\nu_z K^0$
}
\\\hline%
$\tO_{\bar\nu\bar\nu\nu,xyz}^{\tS\tR\tS\tL\tS\tL1}$
&$(\overline{\nu_{\tL x}}\nu_{\tL y}^\C)(\overline{\nu_{\tL z}^\C}\calN_{dsu}^{\tL\tL})$
&$\pmb{8}_\tL\otimes\pmb{1}_\tR$
&$\frac{1}{2}n_\nu^2(n_\nu+1)\,\langle 18 \rangle$
\\
$\tO_{\bar\nu\bar\nu\nu,xyz}^{\tS\tR\tS\tL\tS\tL2}$
&$(\overline{\nu_{\tL x}}\nu_{\tL y}^\C)(\overline{\nu_{\tL z}^\C}\calN_{sud}^{\tL\tL})$
&$\pmb{8}_\tL\otimes\pmb{1}_\tR$
&$\frac{1}{2}n_\nu^2(n_\nu+1)\,\langle 18 \rangle$ 
\\\hline%
$\tO_{\bar\nu\bar\nu\nu,xyz}^{\tS\tR\tS\tL\tS\tR1}$
&$(\overline{\nu_{\tL x}}\nu_{\tL y}^\C)(\overline{\nu_{\tL z}^\C}\calN_{uds}^{\tL\tR})$
&$\pmb{3}_\tL\otimes\bar{\pmb{3}}_\tR$
&$\frac{1}{2}n_\nu^2(n_\nu+1)\,\langle 18 \rangle$ 
\\
$\tO_{\bar\nu\bar\nu\nu,xyz}^{\tT\tR\tV\tR\tV\tL1}$
&$(\overline{\nu_{\tL x}}\sigma_{\mu\nu}\nu_{\tL y}^\C)(\overline{\nu_{\tL z}^\C}\gamma^\mu \calN_{dsu}^{\tR\tL,\nu})$
&$\pmb{3}_\tL\otimes\pmb{6}_\tR$
&$\frac{1}{2}n_\nu^2(n_\nu-1)\,\langle 9 \rangle$ 
\\\hline%
$\tO_{\bar\nu\bar\nu\nu,xyz}^{\tS\tR\tS\tL\tS\tR2}$
&$(\overline{\nu_{\tL x}}\nu_{\tL y}^\C)(\overline{\nu_{\tL z}^\C}\calN_{dsu}^{\tL\tR})$
&$\pmb{3}_\tL\otimes\bar{\pmb{3}}_\tR$
&$\frac{1}{2}n_\nu^2(n_\nu+1)\,\langle 18 \rangle$ 
\\
$\tO_{\bar\nu\bar\nu\nu,xyz}^{\tT\tR\tV\tR\tV\tL2}$
&$(\overline{\nu_{\tL x}}\sigma_{\mu\nu}\nu_{\tL y}^\C)(\overline{\nu_{\tL z}^\C}\gamma^\mu \calN_{usd}^{\tR\tL,\nu})$
&$\pmb{3}_\tL\otimes\pmb{6}_\tR$
&$\frac{1}{2}n_\nu^2(n_\nu-1)\,\langle 9 \rangle$
\\\hline%
$\tO_{\bar\nu\bar\nu\nu,xyz}^{\tS\tR\tS\tL\tS\tR3}$
&$(\overline{\nu_{\tL x}}\nu_{\tL y}^\C)(\overline{\nu_{\tL z}^\C}\calN_{sud}^{\tL\tR})$
&$\pmb{3}_\tL\otimes\bar{\pmb{3}}_\tR$
&$\frac{1}{2}n_\nu^2(n_\nu+1)\,\langle 18 \rangle$ 
\\
$\tO_{\bar\nu\bar\nu\nu,xyz}^{\tT\tR\tV\tR\tV\tL3}$
&$(\overline{\nu_{\tL x}}\sigma_{\mu\nu}\nu_{\tL y}^\C)(\overline{\nu_{\tL z}^\C}\gamma^\mu \calN_{uds}^{\tR\tL,\nu})$
&$\pmb{3}_\tL\otimes\pmb{6}_\tR$
&$\frac{1}{2}n_\nu^2(n_\nu-1)\,\langle 9 \rangle$
\\\hline
\end{tabular} }
\caption{Continuation of \cref{tab:dim9ope_Lm1A}.}
\label{tab:dim9ope_Lm1B}
\end{table}
\clearpage

\begin{table}[t]
\center
\resizebox{\linewidth}{!}{
\renewcommand{\arraystretch}{1.3}
\begin{tabular}{l l l l}
\hline   
Name & Operator & Irrep & \#
\\\clineB{1-4}{1.5}
\multicolumn{4}{l}{\color{magenta}$\Delta L = +3 $: 
$p\to \bar\nu_x\bar \nu_y \ell_z^+K^0$}
\\\hline%
$\tO_{\nu\nu\ell,xyz}^{\tS\tL\tS\tR\tS\tR}$
& $(\overline{\nu_{\tL x}^\C}\nu_{\tL y})(\overline{\ell_{\tR z}^\C}\calN_{usu}^{\tR\tR})$
&$\pmb{1}_\tL\otimes\pmb{8}_\tR$
&$\frac{1}{2}n_\ell (n_\nu^2+n_\nu)\,\langle 12\rangle$ 
\\\hline%
$\tO_{\nu\nu\ell,xyz}^{\tS\tL\tS\tL\tS\tL}$
&$(\overline{\nu_{\tL x}^\C}\nu_{\tL y})(\overline{\ell_{\tL z}^\C}\calN_{usu}^{\tL\tL})$
&$\pmb{8}_\tL\otimes\pmb{1}_\tR$
&$\frac{1}{2}n_\ell (n_\nu^2+n_\nu)\,\langle 12 \rangle$  
\\
$\tO_{\nu\nu\ell,xyz}^{\tT\tL\tT\tL\tS\tL}$
&$(\overline{\nu_{\tL x}^\C}\sigma_{\mu\nu}\nu_{\tL y})(\overline{\ell_{\tL z}^\C}\sigma^{\mu\nu} \calN_{usu}^{\tL\tL})$
&$\pmb{8}_\tL\otimes\pmb{1}_\tR$
&$\frac{1}{2}n_\ell n_\nu(n_\nu-1)\,\langle 6 \rangle$ 
\\
$\tO_{\nu\nu\ell,xyz}^{\tT\tL\tS\tL\tT\tL}$
&$(\overline{\nu_{\tL x}^\C}\sigma_{\mu\nu}\nu_{\tL y})(\overline{\ell_{\tL z}^\C}\calN_{uus}^{\tL\tL,\mu\nu})$
&$\pmb{10}_\tL\otimes\pmb{1}_\tR$
&$\frac{1}{2}n_\ell n_\nu(n_\nu-1)\,\langle 6 \rangle$ 
\\\hline%
$\tO_{\nu\nu\ell,xyz}^{\tS\tL\tS\tR\tS\tL}$
&$(\overline{\nu_{\tL x}^\C}\nu_{\tL y})(\overline{\ell_{\tR z}^\C}\calN_{usu}^{\tR\tL})$
&$\bar{\pmb{3}}_\tL\otimes\pmb{3}_\tR$
&$\frac{1}{2}n_\ell (n_\nu^2+n_\nu)\,\langle 12 \rangle$ 
\\
$\tO_{\nu\nu\ell,xyz}^{\tT\tL\tV\tL\tV\tR1}$ 
&$(\overline{\nu_{\tL x}^\C}\sigma_{\mu\nu}\nu_{\tL y})(\overline{\ell_{\tR z}^\C}\gamma^\mu \calN_{usu}^{\tL\tR,\nu})$
&$\pmb{6}_\tL\otimes\pmb{3}_\tR$
&$\frac{1}{2}n_\ell n_\nu(n_\nu-1)\,\langle 6 \rangle$ 
\\\hline%
$\tO_{\nu\nu\ell,xyz}^{\tT\tL\tV\tL\tV\tR2}$ 
&$(\overline{\nu_{\tL x}^\C}\sigma_{\mu\nu}\nu_{\tL y})(\overline{\ell_{\tR z}^\C}\gamma^\mu \calN_{uus}^{\tL\tR,\nu})$
&$\pmb{6}_\tL\otimes\pmb{3}_\tR$
&$\frac{1}{2}n_\ell n_\nu(n_\nu-1)\,\langle 6 \rangle$ 
\\\hline%
$\tO_{\nu\nu\ell,xyz}^{\tS\tL\tS\tL\tS\tR}$ 
&$(\overline{\nu_{\tL x}^\C}\nu_{\tL y})(\overline{\ell_{\tL z}^\C} \calN_{usu}^{\tL\tR})$
&$\pmb{3}_\tL\otimes\bar{\pmb{3}}_\tR$
&$\frac{1}{2}n_\ell (n_\nu^2+n_\nu)\,\langle 12 \rangle$ 
\\
$\tO_{\nu\nu\ell,xyz}^{\tT\tL\tT\tL\tS\tR}$ 
&$(\overline{\nu_{\tL x}^\C}\sigma_{\mu\nu}\nu_{\tL y})(\overline{\ell_{\tL z}^\C} \sigma^{\mu\nu} \calN_{usu}^{\tL\tR})$
&$\pmb{3}_\tL\otimes\bar{\pmb{3}}_\tR$
&$\frac{1}{2}n_\ell n_\nu(n_\nu-1)\,\langle 6 \rangle$ 
\\\clineB{1-4}{1.5}
\multicolumn{4}{l}{\color{magenta}$\Delta L = +3$:
$p\to\bar\nu_x\bar\nu_y\bar\nu_z K^+$,
$n\to\bar\nu_x\bar\nu_y\bar\nu_z K^0$}
\\\hline%
$\tO_{\nu\nu\nu,xyz}^{\tS\tL\tS\tL\tS\tL1}$
&$(\overline{\nu_{\tL x}^\C}\nu_{\tL [y})(\overline{\nu_{\tL z]}^\C}\calN_{dsu}^{\tL\tL})$
&$\pmb{8}_\tL\otimes\pmb{1}_\tR$
&$\frac{1}{3} n_\nu (n_\nu^2-1)\,\langle 8 \rangle$ 
\\
$\tO_{\nu\nu\nu,xyz}^{\tS\tL\tS\tL\tS\tL2}$
&$(\overline{\nu_{\tL x}^\C}\nu_{\tL [y})(\overline{\nu_{\tL z]}^\C}\calN_{sud}^{\tL\tL})$
&$\pmb{8}_\tL\otimes\pmb{1}_\tR$
&$\frac{1}{3} n_\nu (n_\nu^2-1)\,\langle 8 \rangle$ 
\\
$\tO_{\nu\nu\nu,xyz}^{\tT\tL\tS\tL\tT\tL}$
&$(\overline{\nu_{\tL [x}^\C}\sigma_{\mu\nu}\nu_{\tL y})(\overline{\nu_{\tL z]}^\C}\calN_{uds}^{\tL\tL,\mu\nu})$
&$\pmb{10}_\tL\otimes\pmb{1}_\tR$
&$\frac{1}{6}(n_\nu^3-3n_\nu^2+2n_\nu)\,\langle 1 \rangle$
\\\hline%
$\tO_{\nu\nu\nu,xyz}^{\tS\tL\tS\tL\tS\tR1}$
&$(\overline{\nu^\C_{\tL x}}\nu_{\tL [y})(\overline{\nu^\C_{\tL z]}}\calN_{uds}^{\tL\tR})$
&$\pmb{3}_\tL\otimes\bar{\pmb{3}}_\tR$ 
&$\frac{1}{3} n_\nu (n_\nu^2-1)\,\langle 8 \rangle$ 
\\\hline%
$\tO_{\nu\nu\nu,xyz}^{\tS\tL\tS\tL\tS\tR2}$
&$(\overline{\nu_{\tL x}^\C}\nu_{\tL [y})(\overline{\nu_{\tL z]}^\C}\calN_{dsu}^{\tL\tR})$
&$\pmb{3}_\tL\otimes\bar{\pmb{3}}_\tR$
&$\frac{1}{3}n_\nu (n_\nu^2-1)\,\langle 8 \rangle$
\\\hline%
$\tO_{\nu\nu\nu,xyz}^{\tS\tL\tS\tL\tS\tR3}$
&$(\overline{\nu_{\tL x}^\C}\nu_{\tL [y})(\overline{\nu_{\tL z]}^\C}\calN_{sud}^{\tL\tR})$
&$\pmb{3}_\tL\otimes\bar{\pmb{3}}_\tR$
& $\frac{1}{3} n_\nu (n_\nu^2-1)\,\langle 8 \rangle$ 
\\\clineB{1-4}{1.5}
\multicolumn{4}{l}{\color{magenta}$\Delta L = -3 $: $n\to\nu_x \nu_y \ell_z^- K^+$}
\\\hline%
$\tO_{\bar\nu\bar\nu\bar\ell,xyz}^{\tS\tR\tS\tL\tS\tL}$
&$(\overline{\nu_{\tL x}} \nu_{\tL y}^\C)
(\overline{\ell_{\tR z}}\calN_{dds}^{\tL\tL}) $
&$\pmb{8}_\tL\otimes\pmb{1}_\tR$
&$\frac{1}{2}n_\ell(n_\nu^2+n_\nu) \,\langle 12 \rangle$ 
\\\hline%
$\tO_{\bar\nu\bar\nu\bar\ell,xyz}^{\tS\tR\tS\tR\tS\tR}$
&$(\overline{\nu_{\tL x}} \nu_{\tL y}^\C)
(\overline{\ell_{\tL z}}\calN_{dds}^{\tR\tR}) $
&$\pmb{1}_\tL\otimes\pmb{8}_\tR$
&$\frac{1}{2}n_\ell(n_\nu^2+n_\nu)\,\langle 12 \rangle$ 
\\ $\tO_{\bar\nu\bar\nu\bar\ell,xyz}^{\tT\tR\tT\tR\tS\tR}$
&$(\overline{\nu_{\tL x}}\sigma_{\mu\nu} \nu_{\tL y}^\C)
(\overline{\ell_{\tL z}}\sigma^{\mu\nu}\calN_{dds}^{\tR\tR}) $
&$\pmb{1}_\tL\otimes\pmb{8}_\tR$
&$\frac{1}{2}n_\ell n_\nu (n_\nu-1)\,\langle 6 \rangle$ 
\\ $\tO_{\bar\nu\bar\nu\bar\ell,xyz}^{\tT\tR\tS\tR\tT\tR}$
&$(\overline{\nu_{\tL x}}\sigma_{\mu\nu} \nu_{\tL y}^\C)
(\overline{\ell_{\tL z}}\calN_{dds}^{\tR\tR,\mu\nu}) $
&$\pmb{1}_\tL\otimes\pmb{10}_\tR$
&$\frac{1}{2} n_\ell n_\nu (n_\nu-1)\,\langle 6 \rangle$ 
\\\hline%
$\tO_{\bar\nu\bar\nu\bar\ell,xyz}^{\tS\tR\tS\tL\tS\tR}$
&$(\overline{\nu_{\tL x}} \nu_{\tL y}^\C)
(\overline{\ell_{\tR z}} \calN_{dds}^{\tL\tR}) $
&$\pmb{3}_\tL\otimes\bar{\pmb{3}}_\tR$
&$\frac{1}{2}n_\ell(n_\nu^2+n_\nu) \,\langle 12 \rangle$ 
\\
$\tO_{\bar\nu\bar\nu\bar\ell,xyz}^{\tT\tR\tV\tR\tV\tL1}$
&$(\overline{\nu_{\tL x}} \sigma_{\mu\nu}\nu_{\tL y}^\C)
(\overline{\ell_{\tR z}}\gamma^\mu\calN_{dsd}^{\tR\tL,\nu}) $  
&$\pmb{3}_\tL\otimes\pmb{6}_\tR$
&$\frac{1}{2}n_\ell n_\nu(n_\nu-1) \,\langle 6 \rangle$ 
\\\hline%
$\tO_{\bar\nu\bar\nu\bar\ell,xyz}^{\tT\tR\tV\tR\tV\tL2}$
&$(\overline{\nu_{\tL x}} \sigma_{\mu\nu}\nu_{\tL y}^\C)
(\overline{\ell_{\tR z}}\gamma^\mu\calN_{dds}^{\tR\tL,\nu}) $  
&$\pmb{3}_\tL\otimes\pmb{6}_\tR$
&$\frac{1}{2}n_\ell n_\nu(n_\nu-1) \,\langle 6 \rangle$ 
\\\hline%
$\tO_{\bar\nu\bar\nu\bar\ell,xyz}^{\tS\tR\tS\tR\tS\tL}$
&$(\overline{\nu_{\tL x}} \nu_{\tL y}^\C)
(\overline{\ell_{\tL z}} \calN_{dds}^{\tR\tL}) $
&$\bar{\pmb{3}}_\tL\otimes\pmb{3}_\tR$
&$\frac{1}{2}n_\ell(n_\nu^2+n_\nu) \,\langle 12 \rangle$ 
\\
$\tO_{\bar\nu\bar\nu\bar\ell,xyz}^{\tT\tR\tT\tR\tS\tL}$
&$(\overline{\nu_{\tL x}} \sigma_{\mu\nu}\nu_{\tL y}^\C)
(\overline{\ell_{\tL z}} \sigma^{\mu\nu}\calN_{dds}^{\tR\tL}) $
&$\bar{\pmb{3}}_\tL\otimes\pmb{3}_\tR$
&$\frac{1}{2}n_\ell n_\nu(n_\nu-1) \,\langle 6 \rangle$ 
\\\clineB{1-4}{1.5}
\multicolumn{4}{l}{\color{magenta}$\Delta L = -3 $: $p\to\nu_x\nu_y\nu_z K^+$,
$n\to\nu_x\nu_y\nu_z K^0$
}
\\\hline%
$\tO_{\bar\nu\bar\nu\bar\nu,xyz}^{\tS\tR\tS\tR\tS\tR1}$
&$(\overline{\nu_{\tL x}}\nu_{\tL [y}^\C)(\overline{\nu_{\tL z]}}\calN_{dsu}^{\tR\tR})$
&$\pmb{1}_\tL\otimes\pmb{8}_\tR$
&$\frac{1}{3} n_\nu (n_\nu^2-1)\,\langle 8 \rangle$ 
\\ $\tO_{\bar\nu\bar\nu\bar\nu,xyz}^{\tS\tR\tS\tR\tS\tR2}$
&$(\overline{\nu_{\tL x}}\nu_{\tL [y}^\C)(\overline{\nu_{\tL z]}}\calN_{sud}^{\tR\tR})$
&$\pmb{1}_\tL\otimes\pmb{8}_\tR$
&$\frac{1}{3} n_\nu (n_\nu^2-1)\,\langle 8 \rangle$
\\
$\tO_{\bar\nu\bar\nu\bar\nu,xyz}^{\tT\tR\tS\tR\tT\tR}$
&$(\overline{\nu_{\tL [x}}\sigma_{\mu\nu}\nu_{\tL y}^\C)(\overline{\nu_{\tL z]}}\calN_{uds}^{\tR\tR,\mu\nu})$
&$\pmb{1}_\tL\otimes\pmb{10}_\tR$
&$\frac{1}{6}(n_\nu^3-3n_\nu^2+2n_\nu)\,\langle 1\rangle$
\\\hline%
$\tO_{\bar\nu\bar\nu\bar\nu,xyz}^{\tS\tR\tS\tR\tS\tL1}$
&$(\overline{\nu_{\tL x}}\nu_{\tL [y}^\C)(\overline{\nu_{\tL z]}}\calN_{uds}^{\tR\tL})$
&$\bar{\pmb{3}}_\tL\otimes\pmb{3}_\tR$ 
&$\frac{1}{3} n_\nu (n_\nu^2-1)\,\langle 8 \rangle$ 
\\\hline%
$\tO_{\bar\nu\bar\nu\bar\nu,xyz}^{\tS\tR\tS\tR\tS\tL2}$
&$(\overline{\nu_{\tL x}}\nu_{\tL [y}^\C)(\overline{\nu_{\tL z]}}\calN_{dsu}^{\tR\tL})$ 
&$\bar{\pmb{3}}_\tL\otimes\pmb{3}_\tR$ 
&$\frac{1}{3} n_\nu (n_\nu^2-1)\,\langle 8 \rangle$ 
\\\hline%
$\tO_{\bar\nu\bar\nu\bar\nu,xyz}^{\tS\tR\tS\tR\tS\tL3}$
&$(\overline{\nu_{\tL x}}\nu_{\tL [y}^\C)(\overline{\nu_{\tL z]}}\calN_{sud}^{\tR\tL})$ 
&$\bar{\pmb{3}}_\tL\otimes\pmb{3}_\tR$ 
&$\frac{1}{3} n_\nu (n_\nu^2-1)\,\langle 8 \rangle$ 
\\\hline
\end{tabular} }
\caption{The dim-9 $\Delta L=\pm3$ BNV LEFT operators involving a single strange quark.
Note that antisymmetrization over lepton flavors is imposed for operators involving three left-handed neutrino fields or their charge conjugates, $A_{[y}B_{z]} \equiv (1/2)(A_y B_z-A_z B_y)$ and 
$A_{[x}B_y C_{z]} \equiv (1/6) [A_x B_y C_z - A_x B_z C_y + A_y B_z C_x -A_y B_x C_z+ A_z B_x C_y-A_z B_y C_x]$, thereby ensuring that only  independent operators are counted in each field configuration.  As observed, there is no contribution from dim-9 LEFT operators to the proton or neutron decays into three neutrinos or antineutrinos with identical flavors. }
\label{tab:dim9ope_L3}
\end{table}

\section{Chiral perturbation theory for nucleon decays}
\label{sec:chpt}

\begin{table}[!t]
\centering
\renewcommand{\arraystretch}{1.15}
\resizebox{\linewidth}{!}{
\begin{tabular}{l}
\\\hline
Spurion fields
\\\hline
\color{magenta}$\Delta L = +1$:
$p\to\ell_x^+\ell_y^+\ell_z^-K^0$ 
\\
$\makecell[l]{ \calP_{usu}^{\tL\tL}= C^{ \tS\tL\tS\tL\tS\tL}_{\ell\ell\bar\ell,xyz}(\overline{\ell_{\tL x}^\C}\ell_{\tL y})\overline{\ell_{\tR z}}
+ C^{\tS\tR\tS\tL\tS\tL}_{\ell\ell\bar\ell,xyz}(\overline{\ell_{\tR x}^\C}\ell_{\tR y})\overline{\ell_{\tR z}}
\\
\qquad\quad~
+ C^{\tV\tR\tV\tL\tS\tL}_{\ell\ell\bar\ell,xyz}(\overline{\ell_{\tL x}^\C}\gamma_\mu\ell_{\tR y})\overline{\ell_{\tL z}}\gamma^\mu
+ C^{ \tT\tL\tT\tL\tS\tL}_{\ell\ell\bar\ell,xyz}(\overline{\ell_{\tL x}^\C}\sigma_{\mu\nu}\ell_{\tL y})\overline{\ell_{\tR z}}\sigma^{\mu\nu},}$
\\
$\makecell[l]{ \calP_{usu}^{\tL\tR}=
C^{ \tS\tL\tS\tL\tS\tR}_{\ell\ell\bar\ell,xyz}(\overline{\ell_{\tL x}^\C}\ell_{\tL y})\overline{\ell_{\tR z}}
+ C^{\tS\tR\tS\tL\tS\tR}_{\ell\ell\bar\ell,xyz}(\overline{\ell_{\tR x}^\C}\ell_{\tR y})\overline{\ell_{\tR z}}
\\
\qquad\quad~+ C^{\tV\tR\tV\tL\tS\tR}_{\ell\ell\bar\ell,xyz}(\overline{\ell_{\tL x}^\C}\gamma_\mu\ell_{\tR y})\overline{\ell_{\tL z}}\gamma^\mu
+ C^{ \tT\tL\tT\tL\tS\tR}_{\ell\ell\bar\ell,xyz}(\overline{\ell_{\tL x}^\C}\sigma_{\mu\nu}\ell_{\tL y})\overline{\ell_{\tR z}}\sigma^{\mu\nu},}$
\\
$ \calP_{usu}^{\tL\tR,\mu}=
\frac{1}{2}
C^{\tV\tL\tS\tL\tV\tR1}_{\ell\ell\bar\ell  ,xyz}(\overline{\ell_{\tR x}^\C}\gamma^{\mu} \ell_{\tL y})\overline{\ell_{\tR z}}
+ \frac{1}{2}
C^{\tT\tL\tV\tL\tV\tR1}_{\ell\ell\bar\ell  ,xyz}(\overline{\ell_{\tL x}^\C}\sigma^{\nu\mu} \ell_{\tL y})\overline{\ell_{\tL z}}\gamma_\nu$, 
\\
$\calP_{uus}^{\tL\tR,\mu}=
C^{\tV\tL\tS\tL\tV\tR2}_{\ell\ell\bar\ell,xyz}(\overline{\ell_{\tR x}^\C}\gamma^{\mu} \ell_{\tL y})\overline{\ell_{\tR z}}
+ C^{\tT\tL\tV\tL\tV\tR2}_{\ell\ell\bar\ell,xyz}(\overline{\ell_{\tL x}^\C}\sigma^{\nu\mu} \ell_{\tL y})\overline{\ell_{\tL z}}\gamma_\nu$,
\\
$\calP_{uus}^{\tL\tL,\mu\nu}= \frac{1}{3}
C^{ \tT\tL\tS\tL\tT\tL}_{\ell\ell\bar\ell,xyz}(\overline{\ell_{\tL x}^\C}
\sigma_{\mu\nu}\ell_{\tL y})
\overline{\ell_{\tR z}}$,
\\
$+\tL\leftrightarrow \tR$.
\\\hline
\color{magenta}$\Delta L = +1$:
$p \to {\nu}_x\bar{\nu}_y{\ell}_z^+K^0$
\\
$\calP_{usu}^{\tL\tL}= C^{\tV\tL\tV\tL\tS\tL}_{\bar\nu\nu\ell,xyz}(\overline{\nu_{\tL x}}\gamma_\mu\nu_{\tL y})\overline{\ell_{\tR z}^\C}\gamma^\mu$,\,
$\calP_{usu}^{\tR\tR}= C^{\tV\tL\tV\tR\tS\tR}_{\bar\nu\nu\ell,xyz}(\overline{\nu_{\tL x}}\gamma_\mu\nu_{\tL y})\overline{\ell_{\tL z}^\C}\gamma^\mu$,
\\
$\calP_{usu}^{\tL\tR}= C^{\tV\tL\tV\tL\tS\tR}_{\bar\nu\nu\ell,xyz}(\overline{\nu_{\tL x}}\gamma_\mu\nu_{\tL y})\overline{\ell_{\tR z}^\C}\gamma^\mu$,\,
$\calP_{usu}^{\tR\tL}= C^{\tV\tL\tV\tR\tS\tL}_{\bar\nu\nu\ell,xyz}(\overline{\nu_{\tL x}}\gamma_\mu\nu_{\tL y})\overline{\ell_{\tL z}^\C}\gamma^\mu $,
\\
$\calP_{usu}^{\tL\tR,\mu}=\frac{1}{2}
C^{\tV\tL\tS\tL\tV\tR1}_{\bar\nu\nu\ell,xyz}(\overline{\nu_{\tL x}}\gamma^\mu\nu_{\tL y})\overline{\ell_{\tL z}^\C}$
,\,
$\calP_{usu}^{\tR\tL,\mu}=\frac{1}{2} C^{\tV\tL\tS\tR\tV\tL1}_{\bar\nu\nu\ell,xyz}(\overline{\nu_{\tL x}}\gamma^\mu\nu_{\tL y})\overline{\ell_{\tR z}^\C} $, 
\\
$\calP_{uus}^{\tL\tR,\mu}=
C^{\tV\tL\tS\tL\tV\tR2}_{\bar\nu\nu\ell,xyz}(\overline{\nu_{\tL x}}\gamma^\mu\nu_{\tL y})\overline{\ell_{\tL z}^\C}$
,\,~~
$\calP_{uus}^{\tR\tL,\mu}= C^{\tV\tL\tS\tR\tV\tL2}_{\bar\nu\nu\ell,xyz}(\overline{\nu_{\tL x}}\gamma^\mu\nu_{\tL y})\overline{\ell_{\tR z}^\C} $.
\\\hline
\color{magenta}$\Delta L = +1$:
$n \to \bar{\nu}_x\bar{\nu}_y\ell_z^-K^+$ (a)
\\
$\calP_{dds}^{\tL\tL}= C^{\tS\tL\tS\tL\tS\tL}_{\nu\nu\bar\ell,xyz}(\overline{\nu_{\tL x}^\C}\nu_{\tL y})\overline{\ell_{\tR z}}
+ C^{\tT\tL\tT\tL\tS\tL}_{\nu\nu\bar\ell,xyz}(\overline{\nu_{\tL x}^\C}\sigma_{\mu\nu}\nu_{\tL y})\overline{\ell_{\tR z}}\sigma^{\mu\nu}$,
\\
$\calP_{dds}^{\tL\tR}=
C^{\tS\tL\tS\tL\tS\tR}_{\nu\nu\bar\ell,xyz}(\overline{\nu_{\tL x}^\C}\nu_{\tL y})\overline{\ell_{\tR z}}
+ C^{\tT\tL\tT\tL\tS\tR}_{\nu\nu\bar\ell,xyz}(\overline{\nu_{\tL x}^\C}\sigma_{\mu\nu}\nu_{\tL y})\overline{\ell_{\tR z}}\sigma^{\mu\nu}$,
\\
$\calP_{dds}^{\tR\tR}=C^{\tS\tL\tS\tR\tS\tR}_{\nu\nu\bar\ell,xyz}(\overline{\nu_{\tL x}^\C}\nu_{\tL y})\overline{\ell_{\tL z}}$,\,
$\calP_{dds}^{\tR\tL}= C^{\tS\tL\tS\tR\tS\tL}_{\nu\nu\bar\ell,xyz}(\overline{\nu_{\tL x}^\C}\nu_{\tL y})\overline{\ell_{\tL z}}$,
\\
$\calP_{dsd}^{\tL\tR,\mu}= \frac{1}{2}
C^{\tT\tL\tV\tL\tV\tR1}_{\nu\nu\bar\ell,xyz}(\overline{\nu_{\tL x}^\C}
\sigma^{\nu\mu}\nu_{\tL y})
\overline{\ell_{\tL z}}\gamma_\nu$,
\\
$\calP_{dds}^{\tL\tR,\mu}= C^{\tT\tL\tV\tL\tV\tR2}_{\nu\nu\bar\ell,xyz}(\overline{\nu_{\tL x}^\C}
\sigma^{\nu\mu}\nu_{\tL y})
\overline{\ell_{\tL z}}\gamma_\nu$,
\\
$\calP_{dds}^{\tL\tL,\mu\nu}=\frac{1}{3}
C^{\tT\tL\tS\tL\tT\tL}_{\nu\nu\bar\ell,xyz}(\overline{\nu_{\tL x}^\C}
\sigma_{\mu\nu}\nu_{\tL y})
\overline{\ell_{\tR z}}$.
\\\hline
\color{magenta}$\Delta L = +1$:
$p \to {\ell}_x^-{\ell}_y^+\bar{\nu}_zK^+,\,
n \to {\ell}_x^-{\ell}_y^+\bar{\nu}_zK^0$  (b)
\\
$\makecell[l]{\calP_{dsu}^{\tL\tL}=C^{\tS\tL\tS\tL\tS\tL1}_{\bar\ell\ell\nu,xyz}(\overline{\ell_{\tR x}}\ell_{\tL y})\overline{\nu_{\tL z}^\C}
+C^{\tS\tR\tS\tL\tS\tL1}_{\bar\ell\ell\nu,xyz}(\overline{\ell_{\tL x}}\ell_{\tR y})\overline{\nu_{\tL z}^\C}
\\
\qquad\quad~+C^{\tT\tL\tT\tL\tS\tL1}_{\bar\ell\ell\nu,xyz}(\overline{\ell_{\tR x}}\sigma_{\mu\nu}\ell_{\tL y})\overline{\nu_{\tL z}^\C}\sigma^{\mu\nu},}$
\\
$\makecell[l]{\calP_{sud}^{\tL\tL}=C^{\tS\tL\tS\tL\tS\tL2}_{\bar\ell\ell\nu,xyz}(\overline{\ell_{\tR x}}\ell_{\tL y})\overline{\nu_{\tL z}^\C}
+C^{\tS\tR\tS\tL\tS\tL2}_{\bar\ell\ell\nu,xyz}(\overline{\ell_{\tL x}}\ell_{\tR y})\overline{\nu_{\tL z}^\C}
\\
\qquad\quad~+C^{\tT\tL\tT\tL\tS\tL2}_{\bar\ell\ell\nu,xyz}(\overline{\ell_{\tR x}}\sigma_{\mu\nu}\ell_{\tL y})\overline{\nu_{\tL z}^\C}\sigma^{\mu\nu},}$
\\
$\calP_{dsu}^{\tR\tR}=C^{\tV\tR\tV\tR\tS\tR1}_{\bar\ell\ell\nu,xyz}(\overline{\ell_{\tR x}}\gamma_\mu\ell_{\tR y})\overline{\nu_{\tL z}^\C}\gamma^\mu
+C^{\tV\tL\tV\tR\tS\tR1}_{\bar\ell\ell\nu,xyz}(\overline{\ell_{\tL x}}\gamma_\mu\ell_{\tL y})\overline{\nu_{\tL z}^\C}\gamma^\mu$,
\\
$\calP_{sud}^{\tR\tR}=C^{\tV\tR\tV\tR\tS\tR2}_{\bar\ell\ell\nu,xyz}(\overline{\ell_{\tR x}}\gamma_\mu\ell_{\tR y})\overline{\nu_{\tL z}^\C}\gamma^\mu
+C^{\tV\tL\tV\tR\tS\tR2}_{\bar\ell\ell\nu,xyz}(\overline{\ell_{\tL x}}\gamma_\mu\ell_{\tL y})\overline{\nu_{\tL z}^\C}\gamma^\mu$,
\\
$ \makecell[l]{
\calP_{uds}^{\tL\tR}=C^{\tS\tL\tS\tL\tS\tR1}_{\bar\ell\ell\nu,xyz}(\overline{\ell_{\tR x}}\ell_{\tL y})\overline{\nu_{\tL z}^\C}
+C^{\tS\tR\tS\tL\tS\tR1}_{\bar\ell\ell\nu,xyz}(\overline{\ell_{\tL x}}\ell_{\tR y})\overline{\nu_{\tL z}^\C}
\\
\qquad\quad~
+C^{\tT\tL\tT\tL\tS\tR1}_{\bar\ell\ell\nu,xyz}(\overline{\ell_{\tR x}}\sigma_{\mu\nu}\ell_{\tL y})\overline{\nu_{\tL z}^\C}\sigma^{\mu\nu},}$
\\
$ \makecell[l]{\calP_{dsu}^{\tL\tR}=C^{\tS\tL\tS\tL\tS\tR2}_{\bar\ell\ell\nu,xyz}(\overline{\ell_{\tR x}}\ell_{\tL y})\overline{\nu_{\tL z}^\C}
+C^{\tS\tR\tS\tL\tS\tR2}_{\bar\ell\ell\nu,xyz}(\overline{\ell_{\tL x}}\ell_{\tR y})\overline{\nu_{\tL z}^\C}
\\
\qquad\quad~
+C^{\tT\tL\tT\tL\tS\tR2}_{\bar\ell\ell\nu,xyz}(\overline{\ell_{\tR x}}\sigma_{\mu\nu}\ell_{\tL y})\overline{\nu_{\tL z}^\C}\sigma^{\mu\nu},}$
\\
$ \makecell[l]{\calP_{sud}^{\tL\tR}=C^{\tS\tL\tS\tL\tS\tR3}_{\bar\ell\ell\nu,xyz}(\overline{\ell_{\tR x}}\ell_{\tL y})\overline{\nu_{\tL z}^\C}
+C^{\tS\tR\tS\tL\tS\tR3}_{\bar\ell\ell\nu,xyz}(\overline{\ell_{\tL x}}\ell_{\tR y})\overline{\nu_{\tL z}^\C}
\\
\qquad\quad~
+C^{\tT\tL\tT\tL\tS\tR3}_{\bar\ell\ell\nu,xyz}(\overline{\ell_{\tR x}}\sigma_{\mu\nu}\ell_{\tL y})\overline{\nu_{\tL z}^\C}\sigma^{\mu\nu},}$
\\
$\calP_{uds}^{\tR\tL}=C^{\tV\tR\tV\tR\tS\tL1}_{\bar\ell\ell\nu,xyz}(\overline{\ell_{\tR x}}\gamma_\mu\ell_{\tR y})\overline{\nu_{\tL z}^\C}\gamma^\mu
+C^{\tV\tL\tV\tR\tS\tL1}_{\bar\ell\ell\nu,xyz}(\overline{\ell_{\tL x}}\gamma_\mu\ell_{\tL y})\overline{\nu_{\tL z}^\C}\gamma^\mu$,
\\
$\calP_{dsu}^{\tR\tL}=
C^{\tV\tR\tV\tR\tS\tL2}_{\bar\ell\ell\nu,xyz}
(\overline{\ell_{\tR x}}\gamma_\mu\ell_{\tR y})
\overline{\nu_{\tL z}^\C}\gamma^\mu
+C^{\tV\tL\tV\tR\tS\tL2}_{\bar\ell\ell\nu,xyz}
(\overline{\ell_{\tL x}}\gamma_\mu\ell_{\tL y})
\overline{\nu_{\tL z}^\C}\gamma^\mu$,
\\
$\calP_{sud}^{\tR\tL}=
C^{\tV\tR\tV\tR\tS\tL3}_{\bar\ell\ell\nu,xyz}
(\overline{\ell_{\tR x}}\gamma_\mu\ell_{\tR y})
\overline{\nu_{\tL z}^\C}\gamma^\mu
+C^{\tV\tL\tV\tR\tS\tL3}_{\bar\ell\ell\nu,xyz}
(\overline{\ell_{\tL x}}\gamma_\mu\ell_{\tL y})
\overline{\nu_{\tL z}^\C}\gamma^\mu$,
\\
$\calP_{dsu}^{\tL\tR,\mu}=\frac{1}{2}
C^{\tV\tR\tS\tL\tV\tR1}_{\bar\ell\ell\nu,xyz}
(\overline{\ell_{\tR x}}\gamma^\mu\ell_{\tR y})
\overline{\nu_{\tL z}^\C}
+\frac{1}{2}
C^{\tV\tL\tS\tL\tV\tR1}_{\bar\ell\ell\nu,xyz}
(\overline{\ell_{\tL x}}\gamma^\mu\ell_{\tL y})
\overline{\nu_{\tL z}^\C}$,
\\
$\calP_{usd}^{\tL\tR,\mu}=\frac{1}{2}
C^{\tV\tR\tS\tL\tV\tR2}_{\bar\ell\ell\nu,xyz}
(\overline{\ell_{\tR x}}\gamma^\mu\ell_{\tR y})
\overline{\nu_{\tL z}^\C}
+\frac{1}{2}
C^{\tV\tL\tS\tL\tV\tR2}_{\bar\ell\ell\nu,xyz}
(\overline{\ell_{\tL x}}\gamma^\mu\ell_{\tL y})
\overline{\nu_{\tL z}^\C}$,
\\
$\calP_{uds}^{\tL\tR,\mu}=\frac{1}{2}
C^{\tV\tR\tS\tL\tV\tR3}_{\bar\ell\ell\nu,xyz}
(\overline{\ell_{\tR x}}\gamma^\mu\ell_{\tR y})
\overline{\nu_{\tL z}^\C}
+\frac{1}{2}
C^{\tV\tL\tS\tL\tV\tR3}_{\bar\ell\ell\nu,xyz}
(\overline{\ell_{\tL x}}\gamma^\mu\ell_{\tL y})
\overline{\nu_{\tL z}^\C}$,
\\
$\calP_{dsu}^{\tR\tL,\mu}=\frac{1}{2} 
C^{\tT\tR\tV\tR\tV\tL1}_{\bar\ell\ell\nu,xyz}
(\overline{\ell_{\tL x}}\sigma^{\nu\mu}\ell_{\tR y})
\overline{\nu_{\tL z}^\C}\gamma_\nu$,
\\
$\calP_{usd}^{\tR\tL,\mu}=\frac{1}{2} 
C^{\tT\tR\tV\tR\tV\tL2}_{\bar\ell\ell\nu,xyz}
(\overline{\ell_{\tL x}}\sigma^{\nu\mu}\ell_{\tR y})
\overline{\nu_{\tL z}^\C}\gamma_\nu$,
\\
$\calP_{uds}^{\tR\tL,\mu}=\frac{1}{2}
C^{\tT\tR\tV\tR\tV\tL3}_{\bar\ell\ell\nu,xyz}
(\overline{\ell_{\tL x}}\sigma^{\nu\mu}\ell_{\tR y})
\overline{\nu_{\tL z}^\C}\gamma_\nu$,
\\
$\calP_{uds}^{\tL\tL,\mu\nu}=\frac{1}{6}
C^{\tT\tL\tS\tL\tT\tL}_{\bar\ell\ell\nu,xyz}
(\overline{\ell_{\tR x}}\sigma_{\mu\nu}\ell_{\tL y})
\overline{\nu_{\tL z}^\C}$.
\\\hline
\color{magenta}$\Delta L = +1$:
$p \to \bar{\nu}_x\bar{\nu}_y{\nu}_zK^+,\, 
n \to \bar{\nu}_x\bar{\nu}_y{\nu}_z K^0$ (c)
\\
$\calP_{dsu}^{\tR\tR}= C^{\tS\tL\tS\tR\tS\tR1}_{\nu\nu\bar\nu,xyz}
(\overline{\nu_{\tL x}^\C}\nu_{\tL y})
\overline{\nu_{\tL z}}$,\,
$\calP_{sud}^{\tR\tR}= 
C^{\tS\tL\tS\tR\tS\tR2}_{\nu\nu\bar\nu,xyz}
(\overline{\nu_{\tL x}^\C}\nu_{\tL y})
\overline{\nu_{\tL z}}$,
\\
$\calP_{uds}^{\tR\tL}= 
C^{\tS\tL\tS\tR\tS\tL1}_{\nu\nu\bar\nu,xyz}
(\overline{\nu_{\tL x}^\C}\nu_{\tL y})
\overline{\nu_{\tL z}}$,\,
$\calP_{dsu}^{\tR\tL}= 
C^{\tS\tL\tS\tR\tS\tL2}_{\nu\nu\bar\nu,xyz}
(\overline{\nu_{\tL x}^\C}\nu_{\tL y})
\overline{\nu_{\tL z}}$,
\\
$\calP_{sud}^{\tR\tL}= 
C^{\tS\tL\tS\tR\tS\tL3}_{\nu\nu\bar\nu,xyz}
(\overline{\nu_{\tL x}^\C}\nu_{\tL y})
\overline{\nu_{\tL z}}$,
\\
$\calP_{dsu}^{\tL\tR,\mu}=\frac{1}{2}
C^{\tT\tL\tV\tL\tV\tR1}_{\nu\nu\bar\nu,xyz}
(\overline{\nu_{\tL x}^\C} \sigma^{\nu\mu}\nu_{\tL y})
\overline{\nu_{\tL z}}\gamma_\nu$,
\\
$\calP_{usd}^{\tL\tR,\mu}=\frac{1}{2}
C^{\tT\tL\tV\tL\tV\tR2}_{\nu\nu\bar\nu,xyz}(\overline{\nu_{\tL x}^\C} \sigma^{\nu\mu}\nu_{\tL y})
\overline{\nu_{\tL z}}\gamma_\nu$,
\\
$\calP_{uds}^{\tL\tR,\mu}=\frac{1}{2}
C^{\tT\tL\tV\tL\tV\tR3}_{\nu\nu\bar\nu,xyz}(\overline{\nu_{\tL x}^\C} \sigma^{\nu\mu}\nu_{\tL y})
\overline{\nu_{\tL z}}\gamma_\nu$.
\\\hline
\end{tabular} }
\caption{Spurion fields in the $\Delta L=+1$ sector.}
\label{tab:spurions1}
\end{table}

\begin{table}[!t]
\centering
\renewcommand{\arraystretch}{1.4}
\resizebox{\linewidth}{!}{
\begin{tabular}{l}
\\\hline
Spurion fields
\\\hline
\color{magenta}$\Delta L = -1$:
$p \to {\nu}_x{\nu}_y\ell_z^+K^0$ (d)
\\
$\calP_{usu}^{\tL\tL}=
C^{\tS\tR\tS\tL\tS\tL}_{\bar\nu\bar\nu\ell,xyz}
(\overline{\nu_{\tL x}}\nu_{\tL y}^\C)
\overline{\ell_{\tL z}^\C}$,\,
$\calP_{usu}^{\tL\tR}= 
C^{\tS\tR\tS\tL\tS\tR}_{\bar\nu\bar\nu\ell,xyz}
(\overline{\nu_{\tL x}}\nu_{\tL y}^\C)
\overline{\ell_{\tL z}^\C}$,
\\
$\calP_{usu}^{\tR\tR}= 
C^{\tS\tR\tS\tR\tS\tR}_{\bar\nu\bar\nu\ell,xyz}
(\overline{\nu_{\tL x}}\nu_{\tL y}^\C) \overline{\ell_{\tR z}^\C}
+ C^{\tT\tR\tT\tR\tS\tR}_{\bar\nu\bar\nu\ell,xyz}
(\overline{\nu_{\tL x}}\sigma_{\mu\nu}\nu_{\tL y}^\C)
\overline{\ell_{\tR z}^\C}\sigma^{\mu\nu}$,
\\
$\calP_{usu}^{\tR\tL}=
C^{\tS\tR\tS\tR\tS\tL}_{\bar\nu\bar\nu\ell,xyz}
(\overline{\nu_{\tL x}}\nu_{\tL y}^\C)
\overline{\ell_{\tR z}^\C}
+C^{\tT\tR\tT\tR\tS\tL}_{\bar\nu\bar\nu\ell,xyz}
(\overline{\nu_{\tL x}}\sigma_{\mu\nu}\nu_{\tL y}^\C)
\overline{\ell_{\tR z}^\C}\sigma^{\mu\nu}$,
\\
$\calP_{usu}^{\tR\tL,\mu}=\frac{1}{2} 
C^{\tT\tR\tV\tR\tV\tL1}_{\bar\nu\bar\nu\ell,xyz}(\overline{\nu_{\tL x}}\sigma^{\nu\mu}\nu_{\tL y}^\C)
\overline{\ell_{\tL z}^\C}\gamma_\nu$,
\\
$\calP_{uus}^{\tR\tL,\mu}= 
C^{\tT\tR\tV\tR\tV\tL2}_{\bar\nu\bar\nu\ell,xyz}
(\overline{\nu_{\tL x}}\sigma^{\nu\mu}\nu_{\tL y}^\C)
\overline{\ell_{\tL z}^\C}\gamma_\nu$,
\\
$\calP_{uus}^{\tR\tR,\mu\nu}= \frac{1}{3}
C^{\tT\tR\tS\tR\tT\tR}_{\bar\nu\bar\nu\ell,xyz}
(\overline{\nu_{\tL x}}\sigma_{\mu\nu}\nu_{\tL y}^\C)
\overline{\ell_{\tR z}^\C}$.
\\\hline
\color{magenta}$\Delta L = -1$:  
$n \to \ell_x^-\ell_y^-\ell_z^+K^+$ 
\\
$\makecell[l]{ \calP_{dds}^{\tL\tL}= C^{\tS\tL\tS\tL\tS\tL}_{\bar\ell\bar\ell\ell,xyz}
(\overline{\ell_{\tR x}}\ell_{\tR y}^\C)
\overline{\ell_{\tL z}^\C}
+C^{\tS\tR\tS\tL\tS\tL}_{\bar\ell\bar\ell\ell,xyz}(\overline{\ell_{\tL x}}\ell_{\tL y}^\C)
\overline{\ell_{\tL z}^\C}
\\
\qquad\quad~
+C^{\tV\tR\tV\tL\tS\tL}_{\bar\ell\bar\ell\ell,xyz}
(\overline{\ell_{\tR x}}\gamma_\mu\ell_{\tL y}^\C)
\overline{\ell_{\tR z}^\C}\gamma^\mu
+ C^{\tT\tL\tT\tL\tS\tL}_{\bar\ell\bar\ell\ell,xyz}(\overline{\ell_{\tR x}}\sigma_{\mu\nu}\ell_{\tR y}^\C)
\overline{\ell_{\tL z}^\C}\sigma^{\mu\nu},}$
\\
$\makecell[l]{ \calP_{dds}^{\tL\tR}=
C^{ \tS\tL\tS\tL\tS\tR}_{\bar\ell\bar\ell\ell,xyz}
(\overline{\ell_{\tR x}}\ell_{\tR y}^\C)
\overline{\ell_{\tL z}^\C}
+C^{\tS\tR\tS\tL\tS\tR}_{\bar\ell\bar\ell\ell,xyz}
(\overline{\ell_{\tL x}}\ell_{\tL y}^\C)
\overline{\ell_{\tL z}^\C}
\\
\qquad\quad
+C^{\tV\tR\tV\tL\tS\tR}_{\bar\ell\bar\ell\ell,xyz}(\overline{\ell_{\tR x}}\gamma_\mu\ell_{\tL y}^\C)
\overline{\ell_{\tR z}^\C}\gamma^\mu
+C^{ \tT\tL\tT\tL\tS\tR}_{\bar\ell\bar\ell\ell,xyz}
(\overline{\ell_{\tR x}}\sigma_{\mu\nu}\ell_{\tR y}^\C)
\overline{\ell_{\tL z}^\C}\sigma^{\mu\nu},}$
\\
$ \calP_{dsd}^{\tL\tR,\mu}=\frac{1}{2}
C^{\tV\tL\tS\tL\tV\tR1}_{\bar\ell\bar\ell\ell,xyz}
(\overline{\ell_{\tL x}}\gamma^{\mu}\ell_{\tR y}^\C)
\overline{\ell_{\tL z}^\C}
+\frac{1}{2}
C^{\tT\tL\tV\tL\tV\tR1}_{\bar\ell\bar\ell\ell,xyz}(\overline{\ell_{\tR x}}\sigma^{\nu\mu}\ell_{\tR y}^\C)
\overline{\ell_{\tR z}^\C}\gamma_\nu$,
\\
$ \calP_{dds}^{\tL\tR,\mu}=
C^{\tV\tL\tS\tL\tV\tR2}_{\bar\ell\bar\ell\ell,xyz}
(\overline{\ell_{\tL x}}\gamma^{\mu}\ell_{\tR y}^\C)
\overline{\ell_{\tL z}^\C}
+C^{\tT\tL\tV\tL\tV\tR2}_{\bar\ell\bar\ell\ell,xyz}
(\overline{\ell_{\tR x}}\sigma^{\nu\mu} \ell_{\tR y}^\C)
\overline{\ell_{\tR z}^\C}\gamma_\nu$,
\\
$ \calP_{dds}^{\tL\tL,\mu\nu}= \frac{1}{3}
C^{ \tT\tL\tS\tL\tT\tL}_{\bar\ell\bar\ell\ell,xyz}
(\overline{\ell_{\tR x}}\sigma_{\mu\nu}\ell_{\tR y}^\C)
\overline{\ell_{\tL z}^\C}$,
\\
$+\tL\leftrightarrow\tR$.
\\\hline
\color{magenta}$\Delta L = -1$:
$n \to {\nu}_x\bar{\nu}_y {\ell}_z^-K^+$
\\
$\calP_{dds}^{\tL\tL}= C^{\tV\tL\tV\tL\tS\tL}_{\bar\nu\nu\bar\ell,xyz}
(\overline{\nu_{\tL x}}\gamma_\mu\nu_{\tL y})
\overline{\ell_{\tL z}}\gamma^\mu$,\,
$\calP_{dds}^{\tR\tR}= C^{\tV\tL\tV\tR\tS\tR}_{\bar\nu\nu\bar\ell,xyz}
(\overline{\nu_{\tL x}}\gamma_\mu\nu_{\tL y})
\overline{\ell_{\tR z}}\gamma^\mu$,
\\
$\calP_{dds}^{\tL\tR}= C^{\tV\tL\tV\tL\tS\tR}_{\bar\nu\nu\bar\ell,xyz}
(\overline{\nu_{\tL x}}\gamma_\mu\nu_{\tL y})
\overline{\ell_{\tL z}}\gamma^\mu$,\,
$\calP_{dds}^{\tR\tL}= C^{\tV\tL\tV\tR\tS\tL}_{\bar\nu\nu\bar\ell,xyz}
(\overline{\nu_{\tL x}}\gamma_\mu\nu_{\tL y})
\overline{\ell_{\tR z}}\gamma^\mu$,
\\
$\calP_{dsd}^{\tL\tR,\mu}=\frac{1}{2}
C^{\tV\tL\tS\tL\tV\tR1}_{\bar\nu\nu\bar\ell,xyz}
(\overline{\nu_{\tL x}}\gamma^\mu\nu_{\tL y})
\overline{\ell_{\tR z}}$,\,
$\calP_{dsd}^{\tR\tL,\mu}=\frac{1}{2}
C^{\tV\tL\tS\tR\tV\tL1}_{\bar\nu\nu\bar\ell,xyz}(\overline{\nu_{\tL x}}\gamma^\mu\nu_{\tL y})
\overline{\ell_{\tL z}}$,
\\
$\calP_{dds}^{\tL\tR,\mu}=
C^{\tV\tL\tS\tL\tV\tR2}_{\bar\nu\nu\bar\ell,xyz}(\overline{\nu_{\tL x}}\gamma^\mu\nu_{\tL y})\overline{\ell_{\tR z}}$,\,
~~$\calP_{dds}^{\tR\tL,\mu}=
C^{\tV\tL\tS\tR\tV\tL2}_{\bar\nu\nu\bar\ell,xyz}(\overline{\nu_{\tL x}}\gamma^\mu\nu_{\tL y})
\overline{\ell_{\tL z}}$.
\\\hline
\color{magenta}$\Delta L = + 3$:  
$p\to\bar{\nu}_x\bar{\nu}_y\bar{\nu}_z K^+,\,
n\to\bar{\nu}_x\bar{\nu}_y\bar{\nu}_z K^0$ (e)
\\
$\calP_{dsu}^{\tL\tL}= C^{\tS\tL\tS\tL\tS\tL1}_{\nu\nu\nu,xyz}
(\overline{\nu_{\tL x}^\C}\nu_{\tL[y})
\overline{\nu_{\tL z]}^\C}$,\,
$\calP_{sud}^{\tL\tL}= C^{\tS\tL\tS\tL\tS\tL2}_{\nu\nu\nu,xyz}
(\overline{\nu_{\tL x}^\C}\nu_{\tL[y})
\overline{\nu_{\tL z]}^\C}$,
\\%
$\calP_{uds}^{\tL\tR}=
C^{\tS\tL\tS\tL\tS\tR1}_{\nu\nu\nu,xyz}(\overline{\nu_{\tL x}^\C}\nu_{\tL[y})
\overline{\nu_{\tL z]}^\C} $,\,
$\calP_{dsu}^{\tL\tR}=
C^{\tS\tL\tS\tL\tS\tR2}_{\nu\nu\nu,xyz}(\overline{\nu_{\tL x}^\C}\nu_{\tL[y})
\overline{\nu_{\tL z]}^\C}$,
\\
$\calP_{sud}^{\tL\tR}=
C^{\tS\tL\tS\tL\tS\tR3}_{\nu\nu\nu,xyz}(\overline{\nu_{\tL x}^\C}\nu_{\tL[y})
\overline{\nu_{\tL z]}^\C}$,
\\
$\calP_{uds}^{\tL\tL,\mu\nu}=\frac{1}{6} 
C^{\tT\tL\tS\tL\tT\tL}_{\nu\nu\nu,xyz}(\overline{\nu_{\tL[x}^\C}\sigma^{\mu\nu}\nu_{\tL y})
\overline{\nu_{\tL z]}^\C}$.
\\\hline
\end{tabular} }
\caption{Spurion fields in the $\Delta L=-1$ and $\Delta L=+3$ sectors.}
\label{tab:spurions2}
\end{table}

To calculate the nucleon decay amplitudes induced by the LEFT dim-9 operators constructed in the previous section, a systematic and convenient way is to work within the baryon ChPT framework and perform a chiral matching to translate the quark degrees of freedom into hadronic building blocks such as nucleons and pseudoscalar mesons~\cite{Jenkins:1990jv,Bijnens:1985kj,Claudson:1981gh,Liao:2025vlj,Liao:2025sqt,Ma:2026hiy}.
A leading-order chiral matching formalism for general triple-quark operators was recently developed in~\cite{Liao:2025vlj}. 
In this work, we apply this formalism to calculate four-body nucleon decays. In doing so, we first identify the relevant spurion fields required for the chiral construction. 
In terms of $\calN_i$ in \cref{eq:Nqqs}, all dim-9 operators in \cref{tab:dim9ope_Lp1A,tab:dim9ope_Lp1B,tab:dim9ope_Lm1A,tab:dim9ope_Lm1B,tab:dim9ope_L3} related to four-body nucleon decays with a kaon can be written as
\begin{align}
{\cal L}_{sq^2l^3}^{\slashed{B}} =\,   
& \calP_{usu}^{\tL\tL} \calN_{usu}^{\tL\tL}
+\calP_{dds}^{\tL\tL} \calN_{dds}^{\tL\tL}
+ \calP_{dsu}^{\tL\tL} \calN_{dsu}^{\tL\tL}
+ \calP_{sud}^{\tL\tL} \calN_{sud}^{\tL\tL}
\nn\\%
&+ \calP_{usu}^{\tL\tR} \calN_{usu}^{\tL\tR}
+\calP_{dds}^{\tL\tR} \calN_{dds}^{\tL\tR}
+ \calP_{uds}^{\tL\tR} \calN_{uds}^{\tL\tR}
\nn\\
& + \calP_{dsu}^{\tL\tR} \calN_{dsu}^{\tL\tR}
+ \calP_{sud}^{\tL\tR} \calN_{sud}^{\tL\tR}
\nn\\%
& + \calP_{uus,\mu}^{\tL\tR} \calN_{uus}^{\tL\tR,\mu}
+ \calP_{usu,\mu}^{\tL\tR} \calN_{usu}^{\tL\tR,\mu}
\nn\\
& + \calP_{dds,\mu}^{\tL\tR} \calN_{dds}^{\tL\tR,\mu}
+ \calP_{dsd,\mu}^{\tL\tR} \calN_{dsd}^{\tL\tR,\mu}
\nn\\
&+ \calP_{dsu,\mu}^{\tL\tR} \calN_{dsu}^{\tL\tR,\mu}
+ \calP_{usd,\mu}^{\tL\tR} \calN_{usd}^{\tL\tR,\mu}
+ \calP_{uds,\mu}^{\tL\tR} \calN_{uds}^{\tL\tR,\mu}
\nn\\%
&+\calP_{uus,\mu\nu}^{\tL\tL} \calN_{uus}^{\tL\tL,\mu\nu}
+\calP_{dds,\mu\nu}^{\tL\tL} \calN_{dds}^{\tL\tL,\mu\nu}
\nn
\\
&+\calP_{uds,\mu\nu}^{\tL\tL} \calN_{uds}^{\tL\tL,\mu\nu}
+\tL\leftrightarrow \tR.
\label{eq:Lagq3l3}
\end{align}
Here, $\calP_i$ are the so-called spurion fields, defined as linear combinations of the products of the non-quark component of each operator and their corresponding WCs.
The relevant spurion fields derived from these dim-9 operators are summarized in \cref{tab:spurions1,tab:spurions2}, where $C_i^j$ denotes the WC associated with the operator $\tO_i^j$.
For brevity, we omit the results for processes marked with a prime in \cref{tab:n2lllK} [denoted as (a$^\prime$), (b$^\prime$), (c$^\prime$), (d$^\prime$), (e$^\prime$)]. 
These processes are directly related to their unprimed counterparts [(a), (b), (c) ,(d), (e)] via the interchange $\nu\leftrightarrow\bar\nu$.
The spurion fields for (i$^\prime$)-class processes can be directly obtained from those of (i)-class processes in \cref{tab:spurions1,tab:spurions2}  by applying the simultaneous interchange of $\tL\leftrightarrow\tR$ and $\nu_\tL\leftrightarrow\nu_\tL^\C$.

The leading-order chiral realizations of all triple-quark interactions relevant to our analysis are given in Eqs.~(12) and (13) of Ref.~\cite{Liao:2025vlj}.
Note that in the chiral Lagrangian of Eqs.~(12) and (13) in Ref.~\cite{Liao:2025vlj}, the indices $x,y,z$ in the spurion fields
$\calP_{xyz}^{\tL\tR(\tR\tL),\mu}$ and $\calP_{xyz}^{\tL\tL(\tR\tR),\mu\nu}$
are implicitly summed over.
Therefore, before inserting $\calP_{xyz}^{\tL\tR(\tR\tL),\mu}$ and
$\calP_{xyz}^{\tL\tL(\tR\tR),\mu\nu}$ into the chiral Lagrangian,
the spurion fields must first be symmetrized with respect to the indices
$x,y$ for $\calP_{xyz}^{\tL\tR(\tR\tL),\mu}$ and $x,y,z$ for
$\calP_{xyz}^{\tL\tL(\tR\tR),\mu\nu}$~\cite{Fan:2025xhi}. For example, for the mode $n\to\bar\nu_x\bar\nu_y\ell^-_z K^+$,
we have
\begin{align}
\calP_{dsd}^{\tL\tR,\mu}=\,&\calP_{sdd}^{\tL\tR,\mu}
=\frac{1}{2}C^{\tT\tL\tV\tL\tV\tR1}_{\nu\nu\bar\ell,xyz}
(\overline{\nu_{\tL x}^\C}\sigma^{\nu\mu}\nu_{\tL y})\overline{\ell_{\tL z}}\gamma_\nu,
\label{eq:symm}
\\
\calP_{dds}^{\tL\tR,\mu\nu}=\,&\calP_{dsd}^{\tL\tR,\mu\nu}
=\calP_{sdd}^{\tL\tR,\mu\nu}
=\frac{1}{3}C^{\tT\tL\tS\tL\tT\tL}_{\nu\nu\bar\ell,xyz}
(\overline{\nu_{\tL x}^\C}\sigma^{\mu\nu}\nu_{\tL y})\overline{\ell_{\tR z}}.
\nn
\end{align}
The similar symmetrization should also be performed over all the other $\calP_{xyz}^{\tL\tR(\tR\tL),\mu}$ and
$\calP_{xyz}^{\tL\tL(\tR\tR),\mu\nu}$.
For brevity, only the independent spurion fields are shown in \cref{tab:spurions1,tab:spurions2}.  

By expanding the pseudoscalar matrix to the zeroth order in the meson fields, the effective Lagrangian in \cref{eq:Lagq3l3} yields the following four-point vertices involving three leptons and a hyperon~\cite{Fan:2025xhi}: 
\begin{align}
{\cal L}_{\calP B}
=\,&
\big[ c_1 \calP_{usu}^{\tL\tR} + c_2 \calP_{usu}^{\tL\tL} 
+c_{3\chi}(\calP_{usu}^{\tL\tR,\mu}-\calP_{uus}^{\tL\tR,\mu}) i\tilde{\partial}_\mu \big] \Sigma^+_\tL 
\nn\\
& +\big[c_1 \calP_{dds}^{\tL\tR} + c_2 \calP_{dds}^{\tL\tL} 
- c_{3\chi}(\calP_{dsd}^{\tL\tR,\mu}-\calP_{dds}^{\tL\tR,\mu})
i\tilde{\partial}_\mu \big] \Sigma^-_\tL
\nn\\
&+\frac{1}{\sqrt{6}}\big[
c_1 ( \calP_{uds}^{\tL\tR} + \calP_{dsu}^{\tL\tR}-2\calP_{sud}^{\tL\tR})
\nn\\
&+ c_2 (\calP_{dsu}^{\tL\tL} - 2 \calP_{sud}^{\tL\tL}) 
+3c_{3\chi}(\calP_{dsu}^{\tL\tR,\mu}-\calP_{usd}^{\tL\tR,\mu})
i \tilde{\partial}_\mu \big] \Lambda^0_\tL
\nn\\
& 
+\frac{1}{\sqrt{2}}\big[
c_1(\calP_{uds}^{\tL\tR}-\calP_{dsu}^{\tL\tR})
- c_2 \calP_{dsu}^{\tL\tL} 
\nn\\
&-c_{3\chi}(\calP_{dsu}^{\tL\tR,\mu}
+ \calP_{usd}^{\tL\tR,\mu} 
- 2\calP_{uds}^{\tL\tR,\mu}) 
i\tilde{\partial}_\mu \big] \Sigma^0_\tL 
\nn\\
&- \tL \leftrightarrow \tR. 
\label{eq:LPN}
\end{align} 
where $c_{3\chi} \equiv c_3\Lambda_\chi^{-1}$, with $c_{1,2,3}$ and $c_4$ below the hadronic low energy constants (LECs)~\cite{Liao:2025vlj}, and $\Lambda_\chi\approx1.2~\text{GeV}$ denotes the chiral symmetry breaking scale.
Similarly, expanding the pseudoscalar matrix to the first order in the meson fields yields the following five-point vertices involving three leptons, a nucleon, and a kaon:
\begin{align}
{\cal L}_{\calP{\tN}K} 
=\,&\Big\{ 
\frac{\bar{K}^0}{\sqrt{2}}\big[ 
c_1 \calP_{usu}^{\tL\tR} 
- c_2 \calP_{usu}^{\tL\tL} 
\nn\\
&- c_{3\chi}
( \calP_{usu}^{\tL\tR,\mu}
+\calP_{uus}^{\tL\tR,\mu})i\tilde{\partial}_\mu
\big]p_\tL
\nn\\
&-\sqrt{2}c_{4\chi} \calP_{uus}^{\tL\tL,\mu\nu }\hat{\Gamma}^{\tL}_{\mu\nu\alpha\beta}i\partial^\alpha p_\tL i\partial^\beta \bar{K}^0
\nn\\
& + \frac{K^-}{\sqrt{2}}\big[
c_1 \calP_{dds}^{\tL\tR} 
- c_2 \calP_{dds}^{\tL\tL} 
\nn\\
&+ c_{3\chi}(\calP_{dsd}^{\tL\tR,\mu}
+\calP_{dds}^{\tL\tR,\mu})i\tilde{\partial}_\mu
\big]n_\tL 
\nn\\
&+\sqrt{2}c_{4\chi} \calP_{dds}^{\tL\tL,\mu\nu }\hat{\Gamma}^{\tL}_{\mu\nu\alpha\beta}i\partial^\alpha n_\tL i\partial^\beta {K}^-
\nn\\
&+ \frac{K^-}{\sqrt{2}} \big[ 
c_1 ( \calP_{uds}^{\tL\tR}+ \calP_{sud}^{\tL\tR}) 
+ c_2 \calP_{sud}^{\tL\tL} 
\nn\\
&- c_{3\chi} (\calP_{dsu}^{\tL\tR,\mu}
-2\calP_{usd}^{\tL\tR,\mu}
-\calP_{uds}^{\tL\tR,\mu}) i \tilde{\partial}_\mu 
\big] p_\tL 
\nn\\
&+\sqrt{2}c_{4\chi} \calP_{uds}^{\tL\tL,\mu\nu }\hat{\Gamma}^{\tL}_{\mu\nu\alpha\beta}i\partial^\alpha p_\tL i\partial^\beta {K}^-
\nn\\
&+\frac{\bar{K}^0}{\sqrt{2}}  
\big[ 
c_1 (\calP_{dsu}^{\tL\tR}+ \calP_{sud}^{\tL\tR} ) 
- c_2 (\calP_{dsu}^{\tL\tL}-\calP_{sud}^{\tL\tL})
\nn\\
&- c_{3\chi} (2\calP_{dsu}^{\tL\tR,\mu}
- \calP_{usd}^{\tL\tR,\mu}
+ \calP_{uds}^{\tL\tR,\mu}) i \tilde{\partial}_\mu  
\big]n_\tL 
\nn\\
&
-\sqrt{2}c_{4\chi} \calP_{uds}^{\tL\tL,\mu\nu }\hat{\Gamma}^{\tL}_{\mu\nu\alpha\beta}i\partial^\alpha n_\tL i\partial^\beta \bar{K}^0
\Big\} {i \over F_0}
\nn\\
&+ \tL \leftrightarrow \tR.
\label{eq:LPNM}    
\end{align}
where $c_{4\chi}\equiv c_4 \Lambda_\chi^{-2}$, and $F_0=f_{\pi}/\sqrt{2}$ with the pion decay constant $f_{\pi}=130.41(20)~\text{MeV}$~\cite{ParticleDataGroup:2024cfk}.

In \cref{eq:LPN,eq:LPNM}, the derivative operator is defined as $\tilde{\partial}_\mu = \partial_\mu - \tfrac{1}{4}\gamma_\mu 
\slashed{\partial}$. The projector related to the $\bm{10}_{\tL(\tR)}\otimes \bm{1}_{\tR(\tL)}$ irreps is defined as 
$\hat\Gamma_{\mu\nu\alpha\beta}^{\tL,\tR}  \equiv
{1\over 24} \left( 
2 \{\sigma_{\mu\nu},\sigma_{\alpha\beta}\}
- [\sigma_{\mu\nu},\sigma_{\alpha\beta}]
\right) P_{\tL,\tR}$.
The LECs $c_1=-0.01257(111)\,{\text{GeV}}^3$ and $c_2=0.01269(107)\,{\text{GeV}}^3$ are taken from lattice QCD calculations~\cite{Yoo:2021gql}. For the LEC $c_{3}$ and $c_{4}$, only the naive dimensional analysis estimations are available, with a value of $c_3=0.011\,{\text{GeV}}^3, c_4=0.007\,{\text{GeV}}^3$~\cite{Liao:2025vlj}.  For later convenience, we define the ratios $\kappa_2\equiv c_2/c_1$, $\kappa_3 \equiv c_3/c_1$ and $\kappa_4 \equiv c_4/c_1$ to organize our results. 

Equation (\ref{eq:LPN}) will contribute to the four-body nucleon decays into three leptons plus a kaon via noncontact diagrams, while \cref{eq:LPNM} contribute through contact diagrams.    
The noncontact contributions also involve the standard baryon ChPT interactions. The relevant terms are those three-point vertices containing a nucleon, a hyperon, and a kaon~\cite{Fan:2025xhi}:
\begin{align}
{\cal L}_{\bar{B}\texttt{N}K}\supset\, & 
-\frac{D+3F}{2\sqrt{3}F_0} 
\big( 
\overline{\Lambda^0} \gamma^\mu \gamma_5 p \partial_\mu K^- 
+\overline{\Lambda^0} \gamma^\mu \gamma_5 n \partial_\mu \bar K^0
\big)
\nn\\
&+ \frac{D-F}{2F_0} 
\big[
\overline{\Sigma^0} \gamma^\mu \gamma_5 p \partial_\mu K^-
-\overline{\Sigma^0} \gamma^\mu \gamma_5 n \partial_\mu \bar K^0 
\nn\\
&+\sqrt{2}(\overline{\Sigma^+}\gamma^\mu \gamma_5 p \partial_\mu \bar K^0 
+\overline{\Sigma^-}\gamma^\mu \gamma_5 n \partial_\mu K^-) \big] 
\nn\\
\equiv\,&\sum_{\tN,B,M} { C_{\tN BM}\over F_0 }
\overline{B}\gamma_\mu \gamma_5 \tN\, \partial^\mu \bar M,   
\label{eq:N2BM}
\end{align}
where the dimensionless coefficient $C_{\texttt{N}BM}$s are 
\begin{align}
C_{p\Lambda^0K^+}=\,&-\frac{D+3F}{2\sqrt{3}}, 
&
C_{n\Lambda^0K^0}=\,&-\frac{D+3F}{2\sqrt{3}},
\nn\\
C_{p\Sigma^0K^+}=\,&\frac{D-F}{2},
&
C_{n\Sigma^0K^0}=\,&-\frac{D-F}{2},
\nn\\
C_{p\Sigma^+K^0}=\,&\frac{D-F}{\sqrt{2}}
&
C_{n\Sigma^-K^+}=\,&\frac{D-F}{\sqrt{2}}.
\end{align}
For later numerical evaluation, we use the 
lattice results for the axial couplings $D=0.730(11)$ and $F=0.447^{+0.006}_{-0.007}$~\cite{Bali:2022qja}. 

As the higher-order chiral matching results are not yet available, we use the leading-order Lagrangian terms to provide the first estimates of these four-body decays. Based on the expected convergence of ChPT, we anticipate that higher-order corrections are subdominant.

\section{Calculation of decay width}
\label{sec:decay_width}

\begin{figure}[t]
\centering
\includegraphics[width=0.5\textwidth]{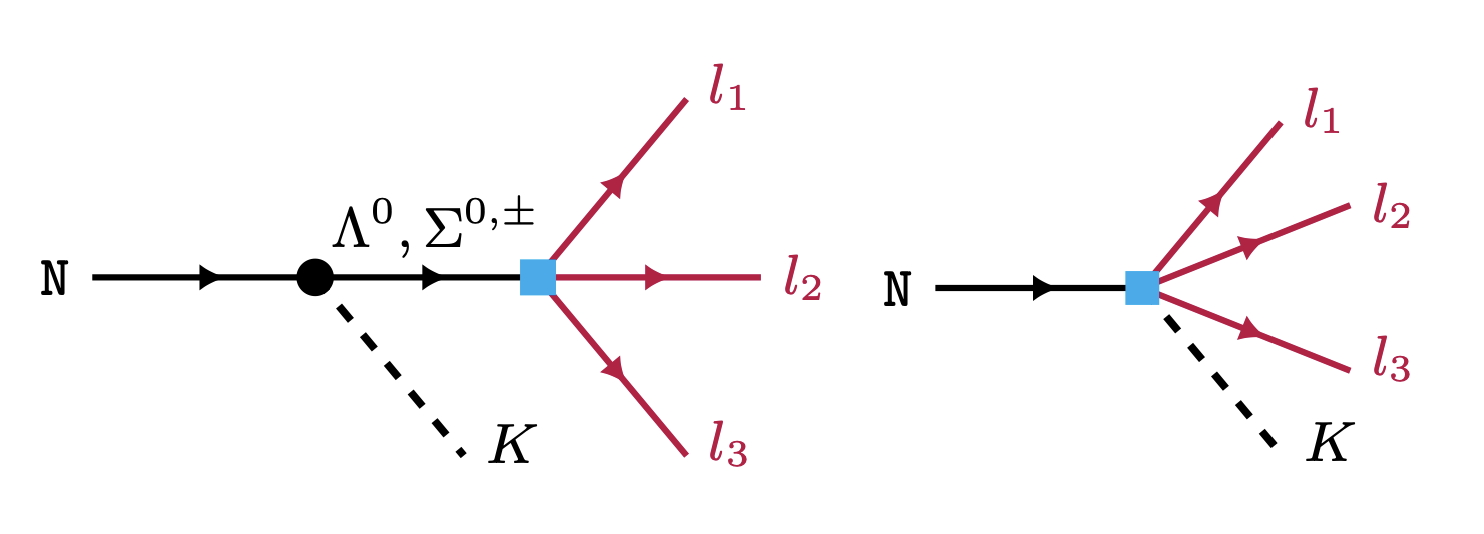}
\caption{Diagrams for four-body nucleon decays into three leptons plus a kaon. The cyan square (black blob) represents the insertion of a BNV (usual) chiral interaction vertex.}
\label{fig:diagram_N2lllK}
\end{figure}

Having obtained all relevant interaction vertices, we proceed to calculate the decay widths. 
At leading order, each four-body mode $\tN(p)\to l_1(p_1) l_2(p_2) l_3 (p_3) M(p_4)$ is mediated by two Feynman diagrams, as shown in \cref{fig:diagram_N2lllK}. 
Considering the spurion fields collected in \cref{tab:spurions1,tab:spurions2}, the noncontact diagram (left) involves conventional QCD strong interaction vertices from \cref{eq:N2BM} connected to four-point BNV vertices from \cref{eq:LPN}, while the contact diagram (right) is governed by five-point vertices from \cref{eq:LPNM}.

The decay amplitude for each specific mode is quite tedious due to numerous LEFT operators that contribute, and we summarize these expressions in \cref{app:decay_width_amp}. 
For the amplitude $\mathcal{M}$ relevant to the decay process
$\tN(p)\to l_1(p_1)\,l_2(p_2)\,l_3(p_3)\,M(p_4)$,
the decay width is given by
\begin{align}
\Gamma
=\frac{1}{S}
\frac{1}{2m_\tN}
\int \overline{|\mathcal{M}|^2}\,d\Phi^{(4)},
\label{eq:dGamma}
\end{align}
where $\overline{|\mathcal{M}|^2}$ denotes 
the spin summed- and averaged-matrix element squared.
For $m$ identical final-state particles, the factor $1/S=1/m!$ is used to remove overcounting in the phase space integration.
The Lorentz-invariant four-body phase-space measure is defined as 
\begin{align}
d\Phi^{(4)}
=
(2\pi)^4\,
\delta^{(4)}
\Big(p-\sum_{i=1}^{4}p_i\Big)
\prod_{i=1}^{4}
\frac{d^3p_i}{(2\pi)^3\,2E_i}.
\end{align}

For the numerical evaluation of the decay widths, 
we adopt the central values of the particle masses from Particle Data Group~\cite{ParticleDataGroup:2024cfk} and the parameters $F_0,D,F,c_{1,2}$ given above.
The uncertainties in the hadronic LECs have a negligible impact on the resulting decay widths.
In our final results, we retain the ratios $\kappa_{3,4}=c_{3,4}/c_1$ as free parameters because the LECs $c_{3,4}$ remain undetermined. 
We compute the four-body phase space integration using 
an efficient Monte Carlo procedure, with the technical details described in \cref{app:4body_PS}.
The complete expressions for the decay widths in terms of dim-9 LEFT WCs are presented in \cref{app:decay_width_WCs}.
The Monte Carlo procedure also allows us to study the momentum distributions of the final-state charged particles, which may provide additional information for distinguishing among different operators once these decay modes are observed experimentally. For each decay mode, the momentum distributions associated with a given operator can be straightforwardly obtained by binning the simulated events in momentum, with the event weights [see \cref{eq:GammaMC}] determined by the corresponding squared matrix element given in \cref{app:decay_width_amp}.

\section{Numerical constraints}
\label{sec:constraints}

Having established the general expressions for all relevant four-body nucleon decays into a $K^+$ or $K^0$ in terms of the dim-9 LEFT WCs, we now derive the corresponding experimental constraints. 
To date, there have been no dedicated searches for nucleon decays into three leptons and a kaon.
Nevertheless, all the modes considered
here can be constrained by reinterpreting existing Super-K data, following the approach adopted for nucleon decays into an invisible particle in Refs.\,\cite{Fan:2025xhi,Ma:2025mjy,Liang:2026vqk}.

Since final-state neutrinos and antineutrinos are invisible to the detector, the signatures of these four-body decays are determined entirely by the final-state charged leptons, the visible decay products of the kaon, and the de-excitation products of the daughter nucleus.
For the decay modes containing at most one charged lepton, $p\to\hat\nu\hat\nu(e^+,\mu^+)K^0$,
$n\to\hat\nu\hat\nu(e^-,\mu^-)K^+$,
$p\to\hat\nu\hat\nu\hat\nu K^+$, and
$n\to\hat\nu\hat\nu\hat\nu K^0$,  
the visible-particle content is
identical to that targeted by dedicated searches for the corresponding two-body nucleon decays, including $p\to(e^+,\mu^+)K^0$~\cite{Super-Kamiokande:2005lev,Super-Kamiokande:2022egr},
$n\to(e^-,\mu^-) K^+$~\cite{Frejus:1991ben}, 
$p\to\hat\nu K^+$~\cite{Super-Kamiokande:2014otb}, and
$n\to\hat\nu K^0$~\cite{Super-Kamiokande:2025ibz}.
The experimental data sets can therefore be reanalyzed to derive stringent bounds on these four-body modes. The details are presented in~\cref{subsec:twobody}.  

For decay modes containing at least two charged leptons, 
there are no two-body nucleon decays with similar experimental signatures. 
However, we note that the recent Super-K search for the two three-body modes $p\to (e^+,\mu^+)\pi^0\pi^0$~\cite{Kamiokande:2026zwu} provides an alternative opportunity to constrain these modes.   
At the Cherenkov-ring level, the decays $p\to (e^+,\mu^+)\pi^0\pi^0$ produce three to five reconstructed rings, yielding topologies similar to those of the modes considered here, since both the charged
leptons and the visible kaon decay products can produce Cherenkov rings.
We therefore reinterpret this data set to constrain 
$p\to\ell_x^+\ell_y^+\ell_z^-K^0$,
$n\to\ell_x^-\ell_y^-\ell_z^+K^+$,
$p\to\ell_x^+\ell_y^-\hat\nu K^+$, and
$n\to\ell_x^-\ell_y^+\hat\nu K^0$ ($x,y,z=e,\mu$), as detailed in \cref{subsec:recastC2}.  

\subsection{Recast constraints from Super-K two-body nucleon decay searches}
\label{subsec:twobody}

In Ref.~\cite{Liang:2026vqk}, 
we reanalyzed the Super-K data sets originally collected for the two-body searches $p\to\ell^+K^0$~\cite{Super-Kamiokande:2005lev,Super-Kamiokande:2022egr},
$n\to\ell^+\pi^-$~\cite{Super-Kamiokande:2017gev}, and
$p\to \hat\nu K^+$~\cite{Super-Kamiokande:2014otb},    
to constrain the exotic three-body nucleon
decays involving an invisible dark photon $X$: $p\to X \ell^+K^0$, $n\to X\ell^-K^+$, and
$p\to X\hat\nu K^+$, respectively.  
Since there is no dedicated Super-K search for $n\to\ell^- K^+$, we use the $n\to\ell^+\pi^-$ data set to constrain $n\to X\ell^-K^+$ at the Cherenkov-ring level. This is justified because, when focusing on the kaon leptonic decay channel $K^+\to \mu^+\bar\nu_\mu$, the two processes produce similar Cherenkov-ring signatures.  
For the three decay modes with at most one charged lepton in the final state,  
$p\to\hat\nu \hat\nu \ell^+K^0$, $n\to\hat\nu\hat\nu\ell^-K^+$, and
$p\to\hat\nu\hat\nu\hat\nu K^+$, there is a one-to-one correspondence with the corresponding three-body dark-photon modes upon replacing the invisible particle $X$ with a neutrino pair $\hat\nu\hat\nu$. 
We therefore adopt the same analysis strategy to derive constraints on these four-body decay modes, including the treatment of the initial-state nuclear effects, event samples, selection criteria,
ring-identification efficiencies, and statistical procedures. 
The non-kinematic efficiencies are retained from the three-body analysis, while the kinematic efficiencies are reevaluated for each four-body decay mode for each specific operator as detailed in~\cite{Liang:2026vqk}.
Note that since $K^+$ decays at rest, the experimental bound on $p\to \hat\nu K^+$~\cite{Super-Kamiokande:2014otb} can be directly applied to $p\to\hat\nu\hat\nu\hat\nu K^+$.  

For the decay mode $n\to\hat\nu\hat\nu\hat\nu K^0$, the corresponding dark-photon mode $n\to X\hat\nu K^0$ was not analyzed in Ref.~\cite{Liang:2026vqk}.
We therefore reinterpret the recent Super-K search for $n\to\hat\nu K^0$~\cite{Super-Kamiokande:2025ibz}, which is based on $0.401~{\rm Mton\cdot yr}$ of SK-I--V data, to constrain this mode.
In this search, the kaon is the only visible particle, identified through its decay products, including $\pi^0\pi^0$ and $\pi^+\pi^-$. 
The final 90\,\%\,CL lower partial lifetime bound is 
$\Gamma^{-1}(n\to\hat\nu K^0)
>7.8\times10^{32}~\text{yr}$. 
Here, we focus on the $K_S^0\to2\pi^0$ channel, which provides the dominant contribution to the published sensitivity.
The four photons produced in the subsequent decay ($K_S^0\to\pi^0\pi^0\to4\gamma$) are used to reconstruct its invariant mass $W_K$ and momentum $P_K$. After applying the standard non-kinematic selections, candidate events are required to satisfy both $200<P_K<500~\text{MeV}/c$ and $300<W_K<700~\text{MeV}/c^2$. 
A spectrum fit was then performed for the surviving events on the $W_K$ distributions.

The four-body decay $n\to\hat\nu\hat\nu\hat\nu K^0$ yields the same visible final state. Although the additional invisible neutrinos modify the kaon momentum spectrum, the reconstructed $W_K$ distribution is expected to have a similar shape to that in the two-body decay, since the same $K^0$ is reconstructed. 
Consequently, the selection efficiencies associated with the non-kinematic cuts and the $W_K$ fit are assumed to be the same for the two- and four-body decays,
with the only difference in acceptance arising from the kaon-momentum requirement. We therefore reevaluate the efficiency of the momentum requirement for the four-body decay, $\varepsilon_\text{4body}$, and use the resulting efficiency ratio, $\varepsilon_\text{4body}/\varepsilon_\text{2body}$, to establish the limit, $\Gamma^{-1}_\text{4body}
= (\varepsilon_\text{4body}/ \varepsilon_\text{2body})
\Gamma^{-1}_\text{2body}$, 
where the efficiency in the two-body decay $\varepsilon_\text{2body}=0.775$ is taken from Fig.~2 of Ref.~\cite{Super-Kamiokande:2025ibz}. With a constant squared matrix element, we find that $\varepsilon_\text{4body}=0.47$, leading to 
$\Gamma^{-1}(n\to\hat\nu\hat\nu\hat\nu K^0)
>4.8\times10^{32}~\text{yr}$ at 90\,\% CL. 

For each decay mode, the momentum dependence of the squared  matrix element, $|\mathcal{M}|^2$, varies among the contributing operators, leading to different reweightings of the four-body phase space.
This, in turn, modifies the kinematic
distributions of visible particles and selection efficiencies, resulting in operator-dependent lifetime limits.  
For each of the six decay modes with at most one charged lepton,  $p\to\hat\nu \hat\nu (e^+,\mu^+)K^0$, $n\to\hat\nu\hat\nu(e^-,\mu^-)K^+$, $p\to\hat\nu\hat\nu\hat\nu K^+$, and $n\to\hat\nu\hat\nu\hat\nu K^0$, 
we evaluate the lifetime limits for each contributing operator individually, following the methodology in~\cite{Liang:2026vqk} and the discussion above.
For each decay mode, the resulting lifetime limits vary by at most a factor of 1.2 across all contributing operators. We therefore adopt the most conservative (i.e., weakest) limit as the final bound for each mode. 
The final lower limits on the lifetimes of the six decay modes are summarized at the bottom of \cref{fig:recast_limits}. They reach $3.2(10)\times10^{32}~\text{yr}$ for $p\to \hat\nu\hat\nu e^+(\mu^+)K^0$, $1.5(5)\times10^{32}~\text{yr}$ for $n\to \hat\nu\hat\nu e^-(\mu^-)K^+$, and $3.5\times10^{32}~\text{yr}$ for $n\to \hat\nu\hat\nu\hat\nu K^0$. 
We find that these constraints are comparable to those obtained for three-body decays involving a dark photon with mass $m_X \simeq 0.1~\text{GeV}$ in~\cite{Liang:2026vqk}. This is because the neutrino pair can carry a sizable invariant mass, leading to softer visible-particle spectra than in the case of a massless dark photon.

\subsection{Recast constraints from Super-K searches for \tf{$p\to (e^+,\mu^+)\pi^0\pi^0$}{p2l+pi0pi0}}
\label{subsec:recastC2}

\begin{figure}[t]
\centering
\includegraphics[width=\linewidth]{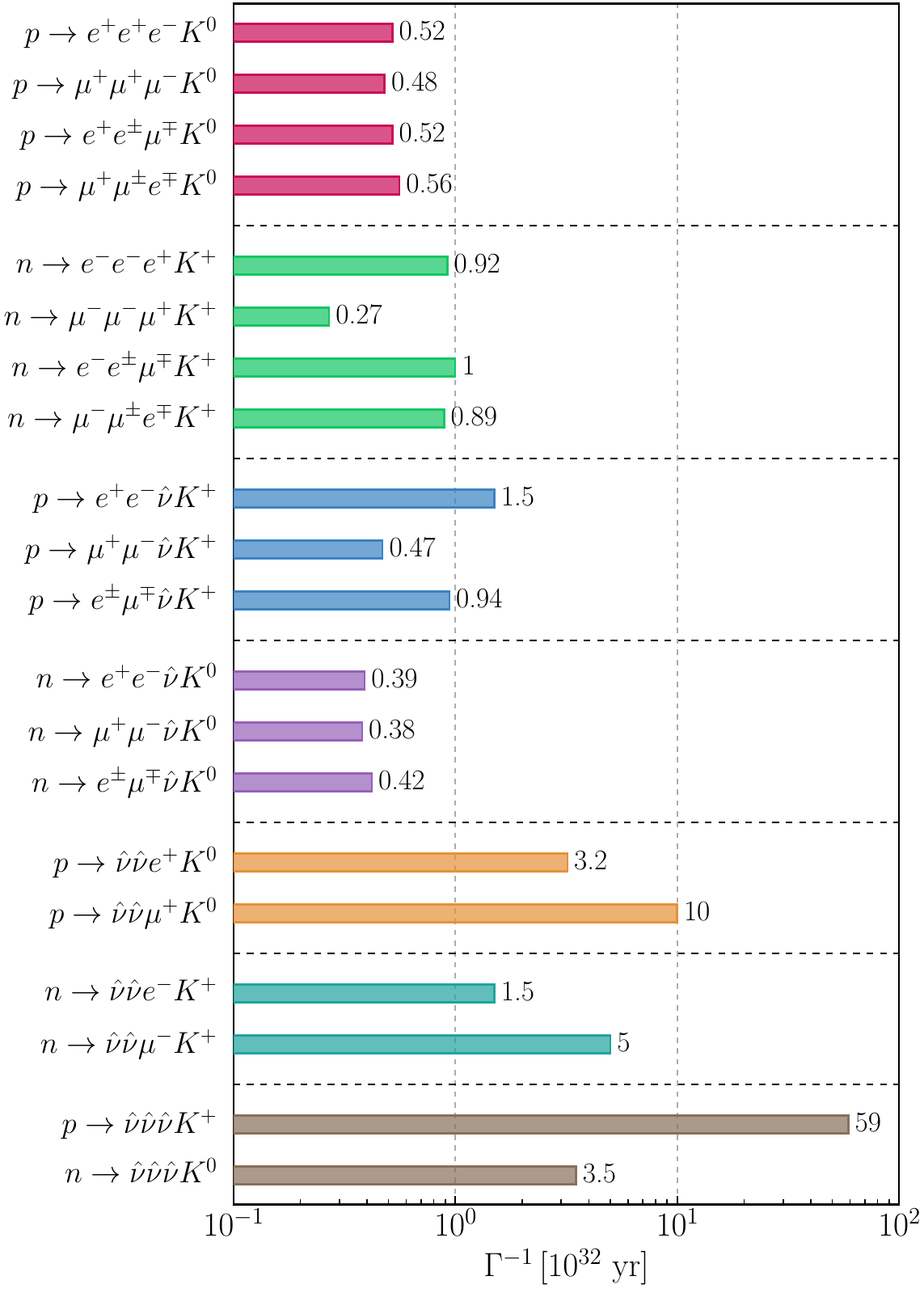}
\caption{Recasting lower limits on the partial lifetimes of 26 four-body nucleon decay modes into three leptons and a kaon.}
\label{fig:recast_limits}
\end{figure}

We now turn to the recent
Super-K searches for $p\to(e^+,\mu^+)\pi^0\pi^0$, which is based on the full
SK I--V exposure of $0.401~{\rm Mton\cdot yr}$~\cite{Kamiokande:2026zwu}, to constrain the four-body decay modes involving at least two charged leptons.
The Super-K analysis applies sequential event selection criteria: events with their reconstructed vertex fully contained inside the fiducial volume (C1); a requirement of three to five reconstructed Cherenkov rings (C2); and subsequent optimized particle-identification, Michel-electron, neutron-tagging, and kinematic selections tailored to the search for $p\to(e^+,\mu^+)\pi^0\pi^0$ (C3--C7). Since the selections in C3--C7 are not directly applicable to our four-body decay modes, we use the data after C2 but before C3.
The corresponding observed event count, $n=6974$, and the expected atmospheric-neutrino background, $b_0=6383$, are extracted from the bottom panels of Fig.~4 in Ref.~\cite{Kamiokande:2026zwu}.

In signal simulation, initial-state nuclear effects and final-state intranuclear interactions are treated as in Ref.~\cite{Liang:2026vqk}. 
The number of Cherenkov rings, $N_{\text{rings}}$, is estimated by assigning one ring to each final-state particle whose momentum exceeds its Cherenkov threshold in water: $p_{\text{th}}\simeq0.58$, $120$, and $159~\text{MeV}/c$ for $e^\pm/\gamma$, $\mu^\pm$, and $\pi^\pm$, respectively.
Potential ring merging effects are neglected in this counting. 
For the final-state $K^0$ meson in the decays $p\to \ell^+_x\ell^+_y\ell^-_z K^0$ and $n\to \ell^+_x\ell^+_y\hat\nu K^0$, we consider only the $K_S^0$ component, since the long-lived $K_L^0$ typically does not produce the required prompt Cherenkov-ring topology. Accordingly, we include a factor of $1/2$ in the signal efficiency to account for the $K_S^0$ fraction.
We consider the two dominant decay modes, $K_S^0\to\pi^+\pi^-$ and $\pi^0\pi^0$, with branching ratios of 0.692 and 0.307, respectively, followed by 
the subsequent decays $\pi^0\to\gamma\gamma$ for the latter mode.
For the final-state $K^+$ meson in the decays $n\to \ell^-_x\ell^-_y\ell^+_z K^+$ and $p\to \ell^+_x\ell^+_y\hat\nu K^+$, 
we focus on the two dominant decay modes, $K^+\to\mu^+\nu_\mu$ and $K^+\to\pi^+\pi^0$, with branching ratios of 0.636 and 0.207, respectively, 
thereby obtaining conservative estimates of the signal efficiencies and resulting limits.
The resulting final states from each four-body decay are then subjected to the C2 requirement, $3\leq N_{\text{rings}}\leq5$, with a flat C1 efficiency of $0.97$ adopted from~\cite{Kamiokande:2026zwu}. 

To derive the partial lifetime limits, 
we marginalize the Poisson likelihood over a Gaussian prior for the background, with mean $b_0=6383$ and standard deviation $\sigma_b=\sqrt{b_0}$. 
Following the discussion around Eq.~(5.4) in~\cite{Liang:2026vqk}, 
this yields a $90\,\%$ CL upper limit of $s_{90}=7.4\times10^2$ signal events.  
Here, only the statistical uncertainty in the background is taken into account. The partial lifetime limit for each decay mode is then given by 
\begin{equation}
\Gamma^{-1}
=
\frac{\varepsilon(\text{C1}\cdot\text{C2})\,\lambda_{\tN}}{s_{90}},
\end{equation}
where $\lambda_{\tN}$ is the nucleon exposure.
Based on the total exposure of $0.401~{\rm Mton\cdot yr}$ reported in Ref.~\cite{Kamiokande:2026zwu}, 
the proton and neutron exposures are  $\lambda_p=1.34\times10^{35}~{\rm proton\cdot yr}$ and $\lambda_n=1.07\times10^{35}~{\rm neutron\cdot yr}$, respectively. $\varepsilon(\text{C1}\cdot\text{C2})$ denotes the combined signal efficiency for the C1 and C2 selection criteria, evaluated separately for each contributing operator in each decay mode.

Following the same procedure as in \cref{subsec:twobody},  
we evaluate the signal efficiency and the lifetime limit for each contributing operator separately for each of the 12 triple-charged-lepton and 8 double-charged-lepton decay modes. 
We then adopt the most conservative limit for each mode as its final bound. The complete results are summarized in the top rows of \cref{fig:recast_limits}. As before, the resulting lifetime limits can vary by  up to a factor of 1.2 among different operators.
These limits range from $3\times 10^{31}~\text{yr}$ for $n\to\mu^-\mu^-\mu^+ K^+$ to $1.5\times 10^{32}~\text{yr}$ for $p\to e^+ e^-\hat\nu K^+$,
with $\varepsilon(\text{C1}\cdot\text{C2})=0.81$ in the latter mode.
This high efficiency in $p\to e^+ e^-\hat\nu K^+$ reflects the low Cherenkov threshold for $e^\pm$s and the favorable ring multiplicity. The two positrons produce two rings, while $K^+\to\mu^+\nu_\mu$ contributes one additional ring if the muon is above its Cherenkov threshold, and $K^+\to\pi^+\pi^0$ yields two to three additional rings. Consequently, this decay mode predominantly satisfies the C2 requirement of three to five rings.

In summary, for all 20 four-body nucleon decay modes involving at least two charged leptons, as well as the two proton decay modes $p\to \hat\nu\hat\nu (e^+, \mu^+)K^0$, our recast limits are more stringent than the inclusive bounds, 
$\Gamma^{-1}(\tN\to e^++\text{anything})>0.6\times10^{30}~\text{yr}$~\cite{Learned:1979gp} and
$\Gamma^{-1}(\tN\to\mu^++\text{anything})>1.2\times10^{31}~\text{yr}$~\cite{Cherry:1981uq},
by up to two orders of magnitude.
These results highlight the importance of reanalyzing existing experimental data to constrain nucleon decay modes that have not yet been directly searched for.
Nevertheless, the sensitivity of these recasts is limited by the use of selection criteria not optimized for the $\tN\to l_x l_y l_z+K$ topology, leaving substantial residual backgrounds. Dedicated multitrack searches, such as a search for $\tN\to l_x l_y l_z+\pi^+\pi^-$ with $m_{\pi^+\pi^-}\simeq m_K$, could substantially improve the limits. These considerations motivate dedicated searches for these four-body decays $\tN\to l_x l_y l_z+K$ at Super-K and JUNO.

\begin{figure}[!t]
\centering
\includegraphics[width=0.94\linewidth]{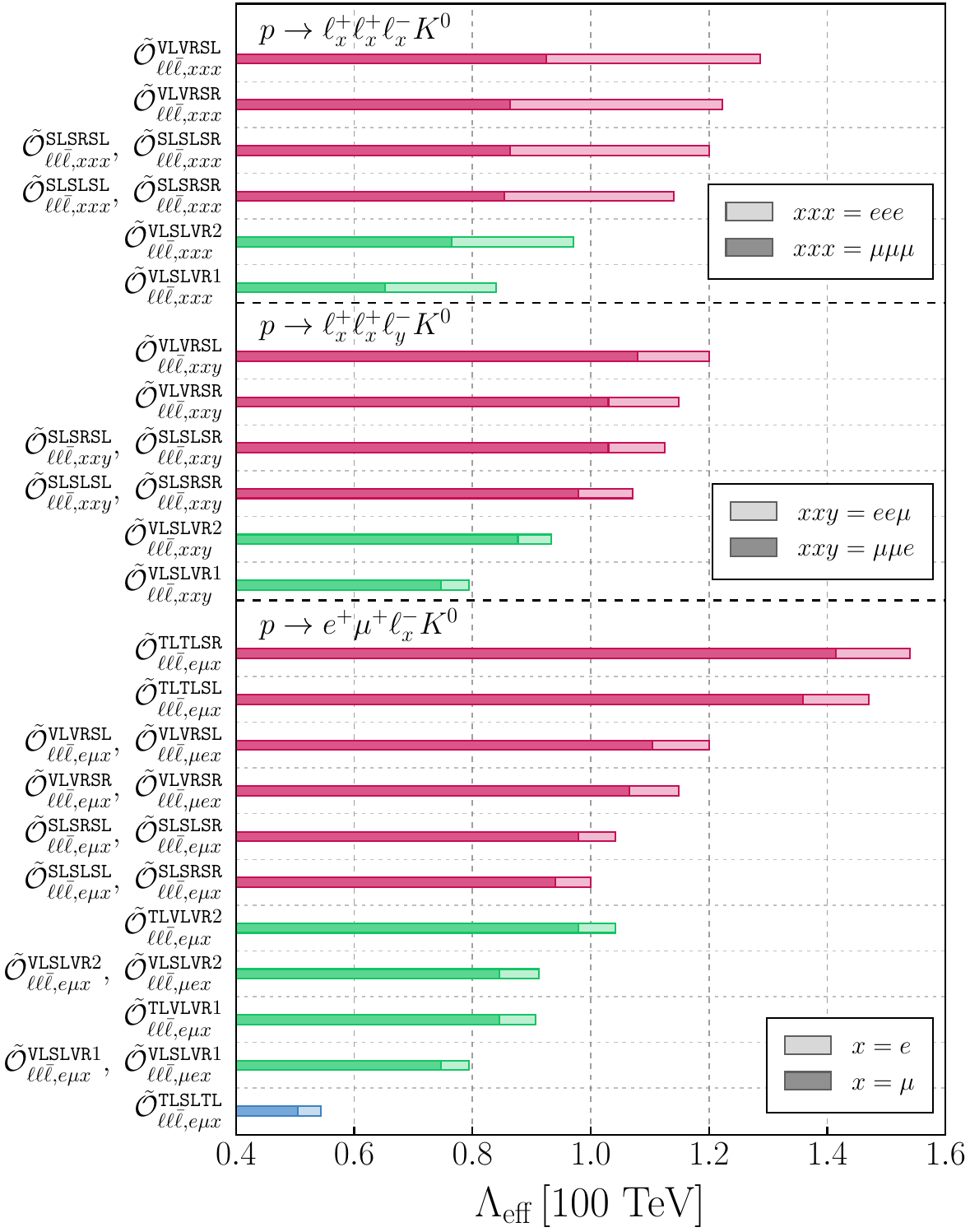}
\\
\vspace{0.3em}
\includegraphics[width=0.94\linewidth]{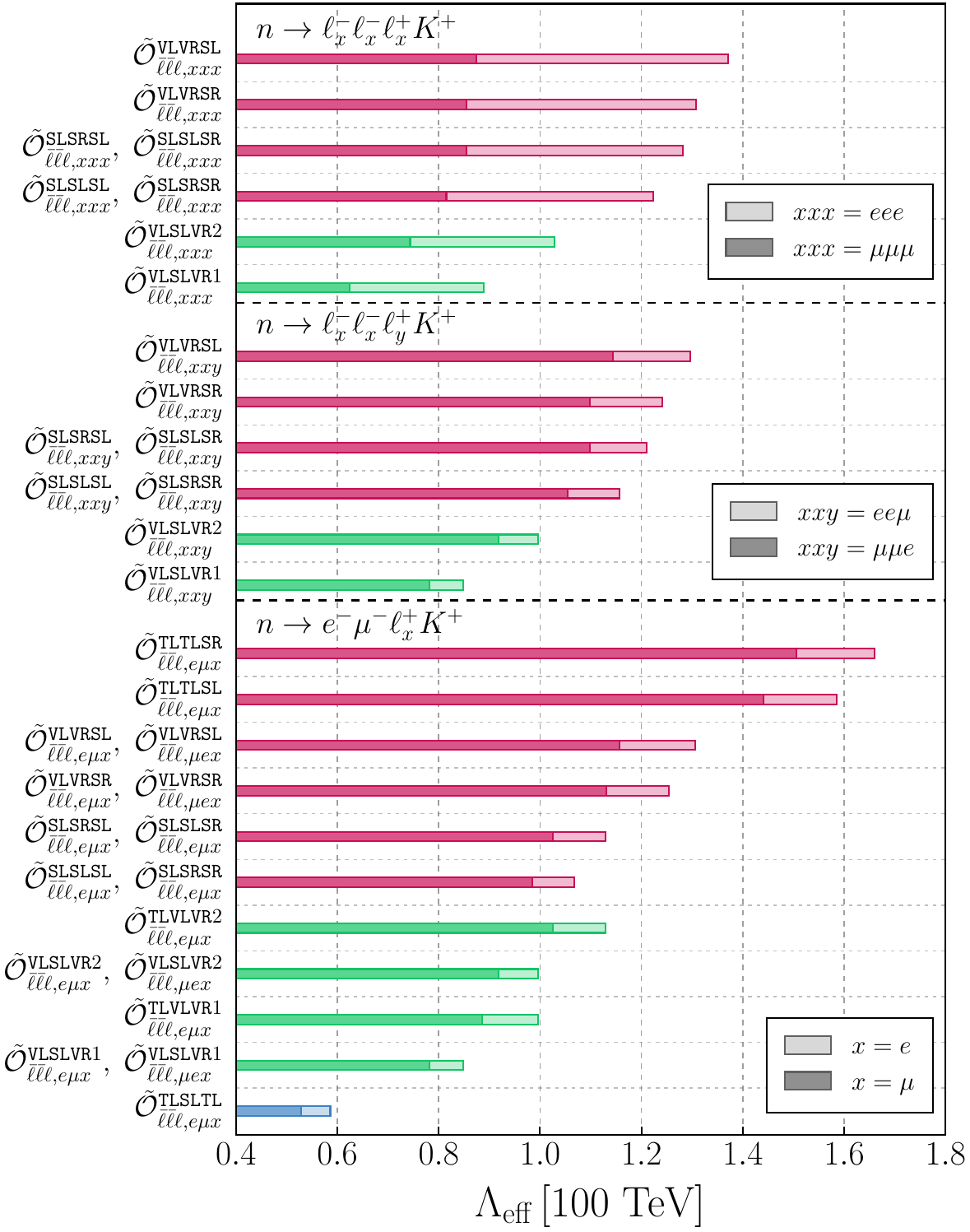}
\caption{Constraints on the effective scales associated with operators contributing to $p\to\ell_x^+ \ell_y^+ \ell_z^-K^0$ and $n\to\ell_x^- \ell_y^- \ell_z^+K^+$, using the recasting limits for the corresponding modes in~\cref{fig:recast_limits}.}
\label{fig:Lambda_eff_N2lllK}
\end{figure}

\subsection{Constraints on the Wilson coefficients}

\begin{figure}[t]
\centering
\includegraphics[width=0.94\linewidth]{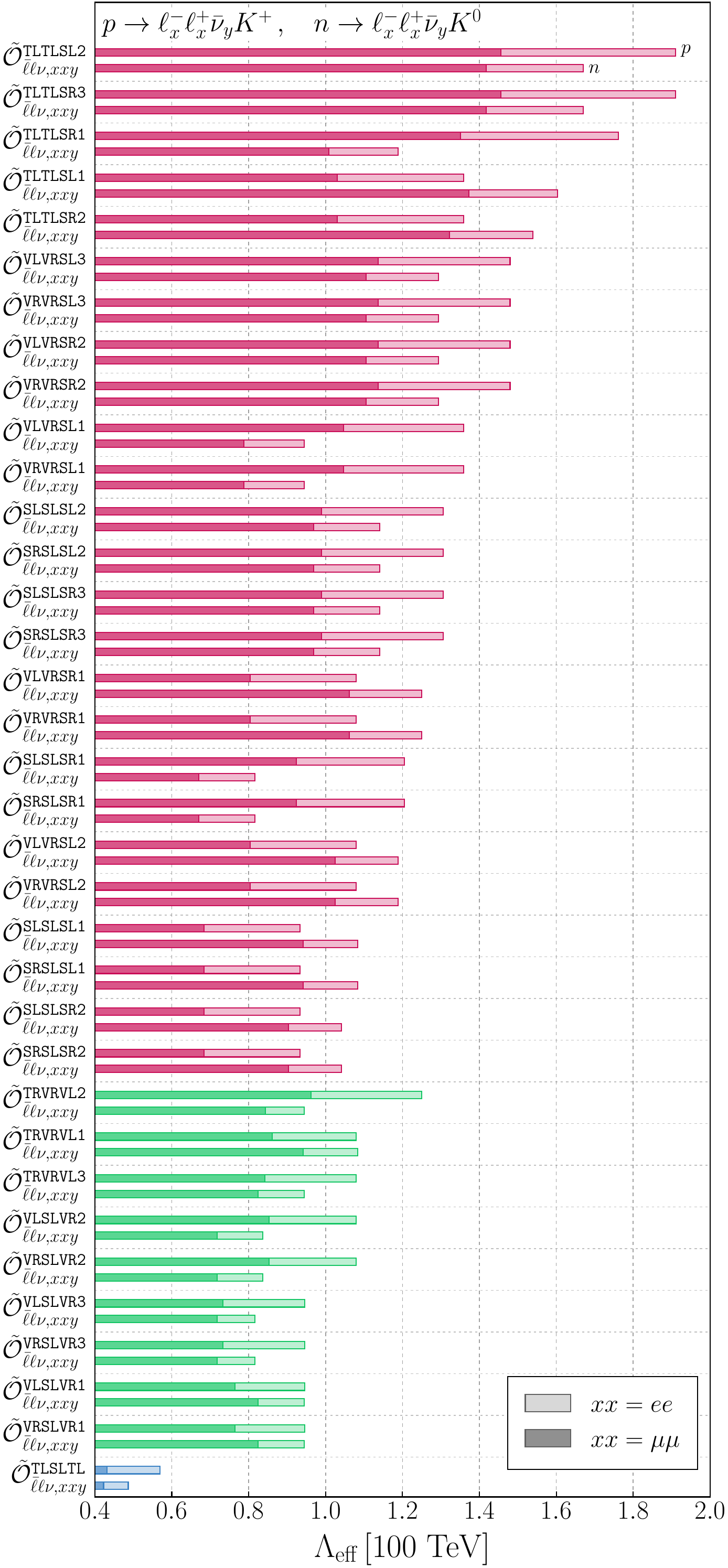}
\caption{
Constraints on the effective scales associated with operators contributing to $\tN\to (e^+ e^-, \mu^+\mu^-) \bar{\nu}_y K$, using the recasting limits 
$\Gamma^{-1}(p\to e^+ e^- \hat{\nu} K^+)>1.5\times 10^{32}~\text{yr}$,
$\Gamma^{-1}(p\to \mu^+ \mu^- \hat{\nu} K^+)>4.7\times 10^{31}~\text{yr}$, 
$\Gamma^{-1}(n\to e^+ e^- \hat{\nu} K^0)>3.9\times 10^{31}~\text{yr}$, 
and $\Gamma^{-1}(n\to \mu^+ \mu^- \hat{\nu} K^0)>3.8\times 10^{31}~\text{yr}$ from \cref{fig:recast_limits}.}
\label{fig:Lambda_eff_N2llvK1}
\end{figure}

\begin{figure}[t]
\centering
\includegraphics[width=0.94\linewidth]{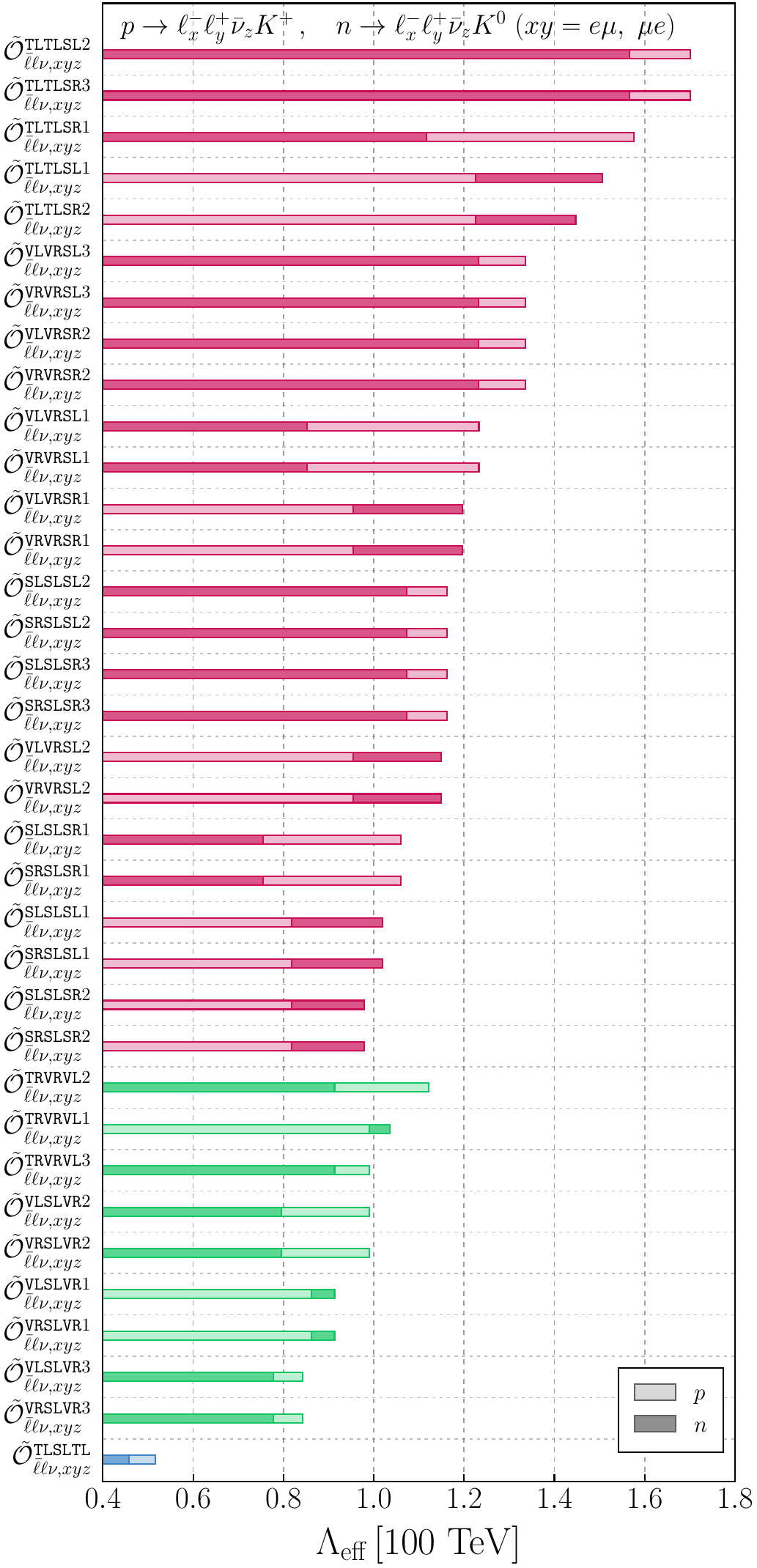}
\caption{Constraints on the effective scales  associated with operators contributing to $\tN\to e^{\pm}\mu^{\mp} \bar{\nu}_z K$, using the recasting limits $\Gamma^{-1}(p\to e^{\pm}\mu^{\mp} \hat{\nu} K^+)>9.4\times 10^{31}~\text{yr}$ and $\Gamma^{-1}(n\to e^{\pm}\mu^{\mp} \hat{\nu} K^0)>4.2\times 10^{31}~\text{yr}$ from \cref{fig:recast_limits}.}
\label{fig:Lambda_eff_N2llvK2}
\end{figure}

In this subsection, we derive constraints on the WCs of the dim-9 operators. 
For each operator, we express the constraint in terms of an effective scale, defined through its WC via $\Lambda_{\text{eff}}\equiv |C_i^j|^{-1/5}$. 
The operators mediating the decay-mode pairs labeled with and without a prime in \cref{tab:n2lllK} are related by the simultaneous interchanges of $\tL\leftrightarrow\tR$ and $\nu_\tL\leftrightarrow\nu_\tL^\C$.
Consequently, identical lower limits on the partial lifetimes of these paired modes yield identical constraints on the corresponding effective scales. We therefore present results only for the unprimed modes. 

\begin{figure}[t]
\centering
\includegraphics[width=0.94\linewidth]{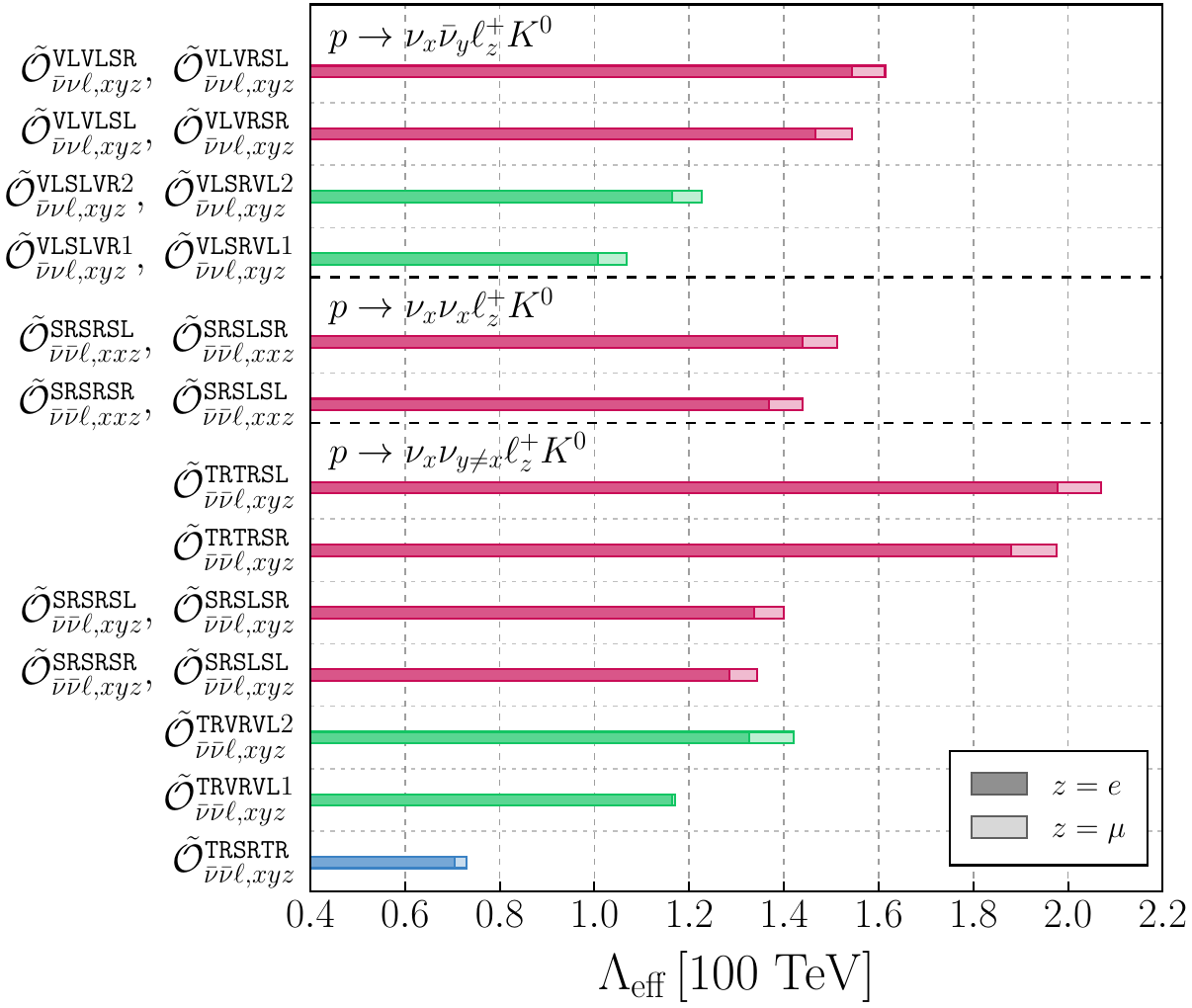}
\\
\vspace{0.3em}
\includegraphics[width=0.94\linewidth]{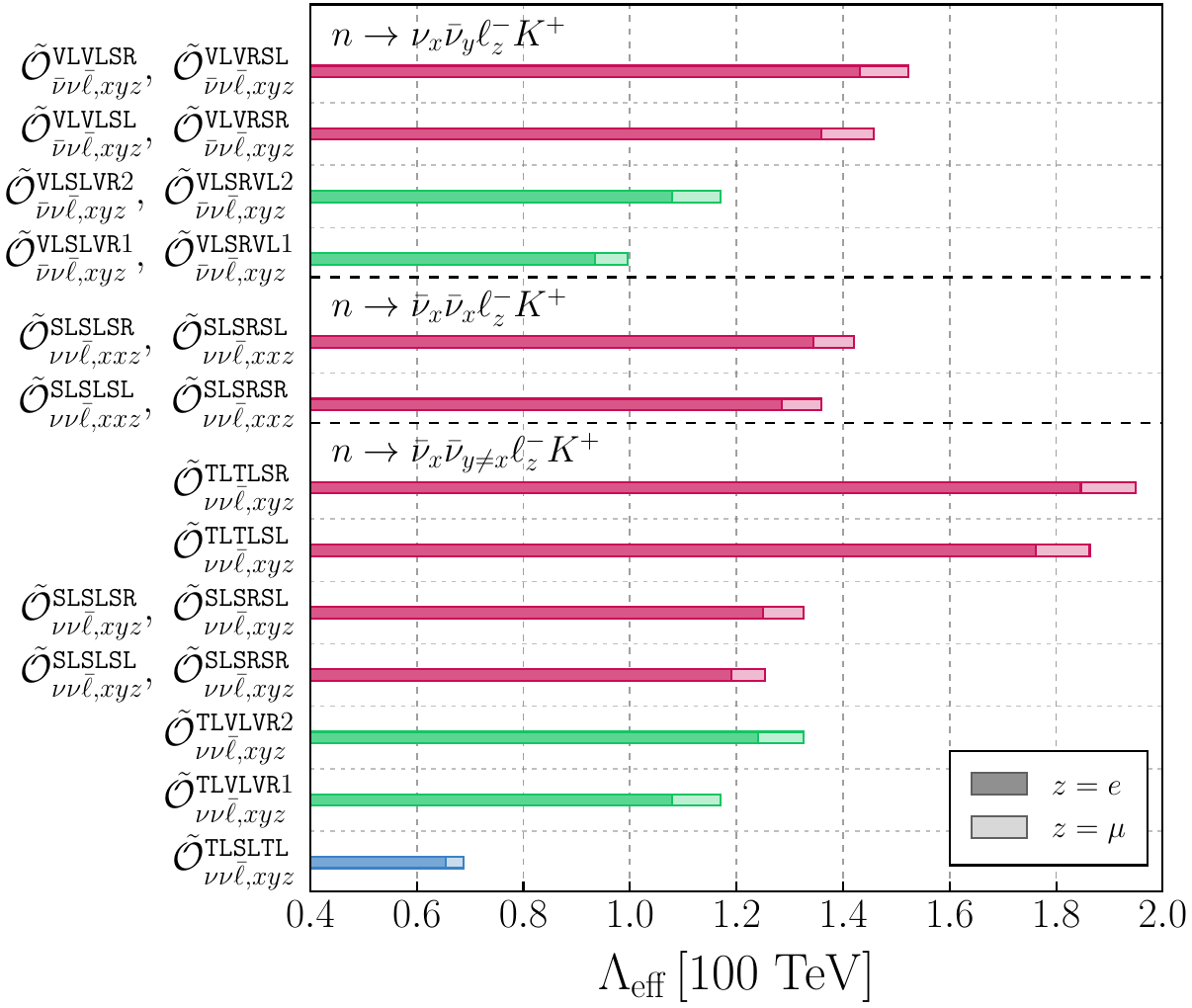}
\caption{Constraints on the effective scales associated with operators contributing to  
$p\to\hat{\nu}_x\hat{\nu}_y(e^+,\mu^+)K^0$ and 
$n\to\hat{\nu}_x\hat{\nu}_y(e^-,\mu^-)K^+$, 
using the recasting limits 
$\Gamma^{-1}(p\to \hat\nu\hat\nu e^+ K^0)>3.2\times 10^{32}~\text{yr}$,
$\Gamma^{-1}(p\to \hat\nu\hat\nu \mu^+ K^0)>1.0\times 10^{33}~\text{yr}$, 
$\Gamma^{-1}(n\to \hat\nu\hat\nu e^- K^+)>1.5\times 10^{32}~\text{yr}$, 
and $\Gamma^{-1}(n\to \hat\nu\hat\nu \mu^- K^+)>5.0\times 10^{32}~\text{yr}$ from~\cref{fig:recast_limits}.}
\label{fig:Lambda_eff_N2vvlK}
\end{figure}

\begin{figure}[t]
\centering
\includegraphics[width=0.94\linewidth]{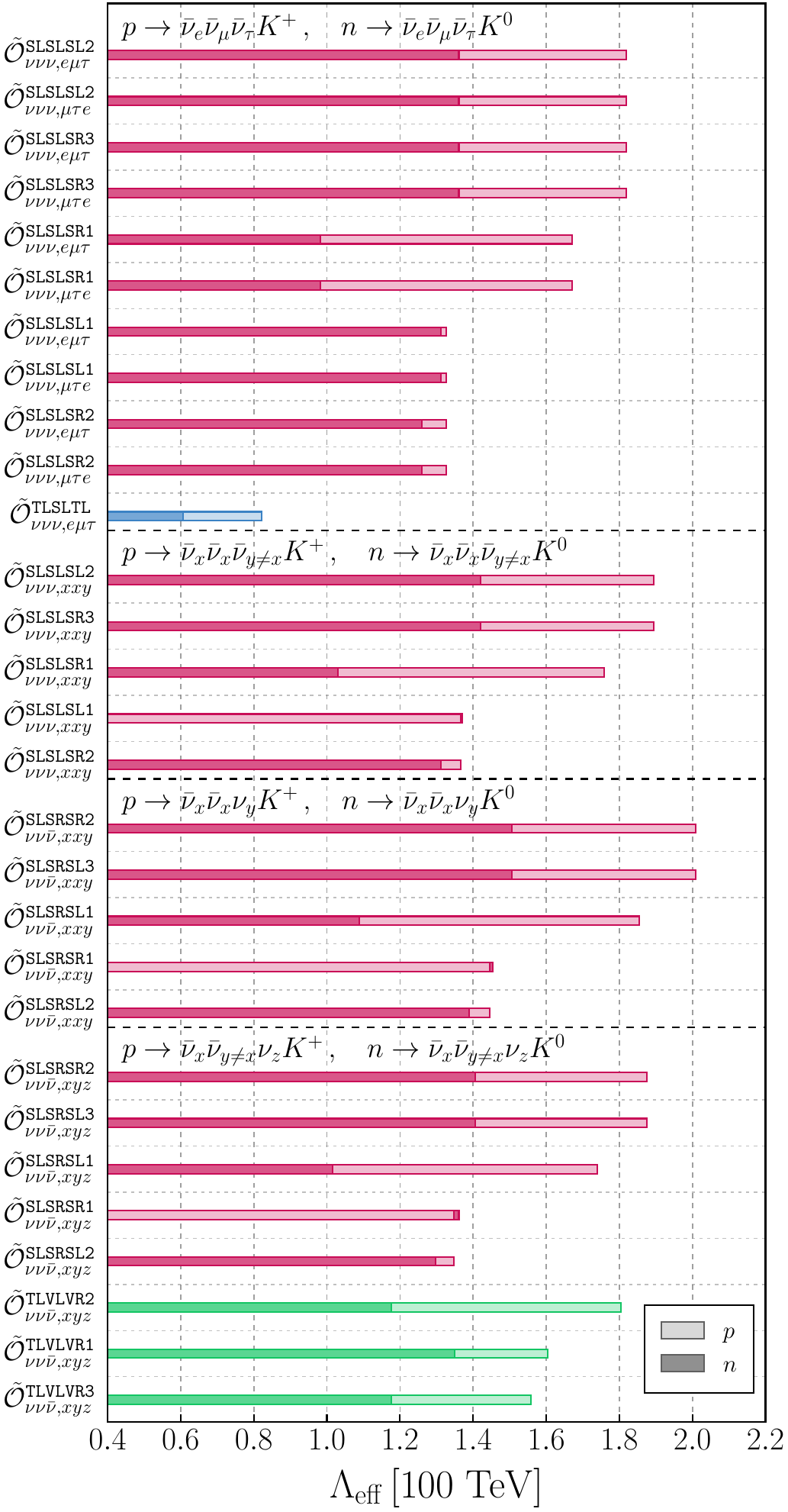}
\caption{Constraints on the effective scales  associated with operators contributing to $\tN\to \hat{\nu}_x \hat{\nu}_y \hat{\nu}_z K$, using the bounds $\Gamma^{-1}(p\to \hat{\nu} \hat{\nu} \hat{\nu} K^+)>5.9\times 10^{33}~\text{yr}$ and $\Gamma^{-1}(n\to \hat{\nu} \hat{\nu} \hat{\nu} K^0)>3.5\times 10^{32}~\text{yr}$ from~\cref{fig:recast_limits}.}
\label{fig:Lambda_eff_N2vvvK}
\end{figure}

Using the derived lifetime limits in~\cref{fig:recast_limits} and the decay width expressions summarized in \cref{app:decay_width_WCs}, we show the resulting constraints on the effective scales in \cref{fig:Lambda_eff_N2lllK} for the triple-charged-lepton modes,  
\cref{fig:Lambda_eff_N2llvK1,fig:Lambda_eff_N2llvK2} for the double-charged-lepton modes, \cref{fig:Lambda_eff_N2vvlK} for the one-charged-lepton modes, and 
\cref{fig:Lambda_eff_N2vvvK} for the triple-antineutrino modes.
The red bars denote operators transforming under the chiral irreps $\pmb{8}_{\tL(\tR)}\otimes \pmb{1}_{\tR(\tL)}$ and $\pmb{3}_{\tL(\tR)}\otimes \bar{\pmb{3}}_{\tR(\tL)}$, the green bars denote those transforming under $\pmb{3}_{\tL(\tR)}\otimes \pmb{6}_{\tR(\tL)}$, and the blue bars denote those transforming under $\pmb{10}_{\tL(\tR)}\otimes \pmb{1}_{\tR(\tL)}$. For operators belonging to the latter two irreps, we set $\kappa_3=\kappa_4=1$ when deriving the effective-scale constraints. A more precise determination of the LECs $c_{3,4}$ would rescale the obtained limits by factors of $\kappa_{3,4}^{1/5}$, respectively.

As shown in these plots, the effective scales are constrained to the range of approximately 40--200~TeV, demonstrating the high sensitivity of nucleon decay searches to BNV new physics. Compared with the $\Lambda_{\text{eff}}$ constraints from purely triple-lepton modes in~\cite{Liao:2025wxk}, these bounds are typically weaker by a factor of 2--5, owing to both the additional phase-space suppression in four-body decays and the weaker lifetime bounds obtained in our analysis.

\section{Summary}
\label{sec:summary}

In this work, we have conducted a thorough investigation of four-body BNV nucleon decays into three leptons and a kaon, mediated by local dim-9 BNV LEFT operators involving a strange quark. We construct a complete operator basis classified according to the irreducible representations of the QCD chiral group. By matching these operators onto chiral perturbation theory, we calculate the decay widths for 26 distinct four-body decay modes (without accounting for neutrino flavors) and express them entirely in terms of the LEFT WCs.

Furthermore, employing existing Super-K data from conventional two-body nucleon decay searches and searches for the three-body modes $p\to (e^+,\mu^+)\pi^0\pi^0$, we perform a systematic reinterpretation of these data to establish the first stringent limits on the four-body decay modes considered in this work. 
We obtain partial lifetime limits of approximately $\calO(10^{31\text{--}34})~\text{yr}$ as summarized in~\cref{fig:recast_limits}, which are more stringent than the corresponding inclusive limits by up to two orders of magnitude. 
Finally, we translate the lifetime limits into constraints on the effective scales associated with the dim-9 operators, with the sensitivity reaching approximately 40--200 TeV.    

In light of ongoing and upcoming large-fiducial-mass neutrino experiments, such as Hyper-Kamiokande (Hyper-K)~\cite{Hyper-Kamiokande:2018ofw} and JUNO~\cite{JUNO:2022qgr}, 
our results provide a useful benchmark for future searches for these four-body nucleon decays involving a kaon.
Dedicated searches optimized for four-body decay topologies could substantially improve upon our conservative recast limits through more effective background suppression strategies.
Super-K and Hyper-K are particularly promising for modes involving $K^0$, given their large exposures and multi-ring reconstruction capabilities, which can help identify the characteristic Cherenkov-ring topologies arising from the dominant pion decay channels of $K^0$.
JUNO offers complementary sensitivity. Unlike water-Cherenkov detectors, JUNO can detect energy deposited by particles below the Cherenkov threshold in water, including $K^+$ mesons and low-momentum muons. The sequential energy deposits from the $K^+$, its decay muon, and the subsequent Michel positron provide a distinctive triple-coincidence signature~\cite{JUNO:2022qgr}. This capability is particularly advantageous for modes such as $n\to\mu^+\mu^-\mu^-K^+$, where the limited phase space leaves several final-state muons below the Cherenkov threshold in water.  

\acknowledgments
This work was supported 
by Grants 
No.\,NSFC-12305110 
and No.\,NSFC-12035008.

\appendix
\section{Matching dim-9 SMEFT six-fermion operators onto the LEFT}
\label{app:smeft_case}

From the top-down EFT approach, for a given UV scenario, it will first generate the SMEFT interactions at the new physics scale after integrating out the heavy states, which further run down to the electroweak scale and match onto the LEFT. 
To illustrate the connection of our LEFT formalism to  those operators in the SMEFT, 
we focus on the seven  dim-9 $\Delta B=-\Delta L=1$ SMEFT operators involving three leptons and three quarks~\cite{Liao:2020jmn} without involving any derivative,
\begin{align}
{\cal O}_{eeeddd}=\,&
\epsilon_{\alpha\beta\gamma}(\overline{e}e^\C)(\overline{e^\C}d^\alpha)(\overline{d^{\C,\beta}} d^\gamma),
\nn\\
{\cal O}_{eLLddd}=\,&
\epsilon_{\alpha\beta\gamma}(\overline{e}L)(\overline{L}d^\alpha)(\overline{d^{\C,\beta}} d^\gamma),
\nn\\
{\cal O}_{LeLdud}=\,&
\epsilon_{\alpha\beta\gamma}\epsilon_{ij}(\overline{L^i}e)(\overline{L^j} d^\alpha)(\overline{u^{\C,\beta}} d^\gamma),
\nn\\
{\cal O}_{LLeudd}=\,&
\epsilon_{\alpha\beta\gamma}\epsilon_{ij}(\overline{L^i}\sigma_{\mu\nu}L^{\C,j})(\overline{e^\C}u^\alpha)(\overline{d^{\C,\beta}}\sigma^{\mu\nu}d^\gamma),
\nn\\
{\cal O}_{eQLded}=\,&
\epsilon_{\alpha\beta\gamma}(\overline{e}Q^\alpha)(\overline{L}d^\beta)(\overline{e^\C}d^\gamma),
\nn\\
{\cal O}_{LQLdLd}=\,&
\epsilon_{\alpha\beta\gamma}\delta_{ik}\delta_{jl}(\overline{L^{\C,i}} Q^{j\alpha})(\overline{L^k} d^\beta)(\overline{L^l} d^\gamma),
\nn\\
{\cal O}_{LeLdQQ}=\,&
\epsilon_{\alpha\beta\gamma}\delta_{ik}\delta_{jl}(\overline{L^i}e)(\overline{L^j}d^\alpha)(\overline{Q^{\C,k\beta}} Q^{l\gamma}).
\end{align}
For the operators $\calO_{eeeddd}$, $\calO_{eLLddd}$, and $\calO_{LLeudd}$ exhibiting flavor symmetries among right-handed down-type quarks, we retain only one  representative for each independent flavor assignment in the matching. For example, we use $\calO_{eeeddd}^{xyz121}=-\calO_{eeeddd}^{xyz112}$ and include only $\calO_{eeeddd}^{xyz112}$ in the matching.
Here, the generation indices 1 and 2 correspond to the first- and second-generation quarks, respectively.
The same convention for choosing independent flavor assignments is also applied to identical lepton fields.

To simply our discussion, we neglect quark flavor-to-mass eigenstate rotations and match these operators onto our LEFT bases at the electroweak scale. We denote the WCs of the SMEFT operators by replacing $\calO_i^j$ with $C_i^j$. After expressing the lepton and quark doublets in terms of their physical components and extensively applying various Fierz transformations, we match the seven SMEFT operators onto the corresponding subsets of LEFT operators in~\cref{tab:dim9ope_Lm1A,tab:dim9ope_Lm1B}.
The resulting nonvanishing LEFT WCs are summarized as follows.
\begin{itemize}
\item ${\cal O}_{eeeddd}$:
\begin{align}
C_{\bar{\ell}\bar{\ell}\ell,xyz}^{\tS\tL\tS\tR\tS\tR}=C_{eeeddd}^{xyz112}.
\end{align}
\item ${\cal O}_{eLLddd}$:
\begin{align}     
C_{\bar{\ell}\bar{\ell}\ell,xyz}^{\tV\tL\tV\tR\tS\tR}&= -\frac{1}{2}C_{eLLddd}^{yzx112};
\nn\\
C_{\bar{\nu}\nu\bar{\ell},xyz}^{\tV\tL\tV\tR\tS\tR}&=-\frac{1}{2}C_{eLLddd}^{zyx112}.
\end{align}
\item ${\cal O}_{LeLdud}$:
\begin{align}
C_{\bar{\ell}\ell\bar{\nu},xyz}^{\tS\tR\tS\tR\tS\tR1}&=\frac{1}{2}C_{LeLdud}^{zyx112}+C_{LeLdud}^{xyz112},
\nn\\
C_{\bar{\ell}\ell\bar{\nu},xyz}^{\tT\tR\tT\tR\tS\tR1}&=\frac{1}{8}C_{LeLdud}^{zyx112},
\nn\\
C_{\bar{\ell}\ell\bar{\nu},xyz}^{\tS\tR\tS\tR\tS\tR2}&=-\frac{1}{2}C_{LeLdud}^{zyx211}-C_{LeLdud}^{xyz211},
\nn\\
C_{\bar{\ell}\ell\bar{\nu},xyz}^{\tT\tR\tT\tR\tS\tR2}&=-\frac{1}{8}C_{LeLdud}^{zyx211}.
\end{align}
\item ${\cal O}_{LLeudd}$:
\begin{align}   
C_{\bar{\ell}\ell\bar{\nu},xyz}^{\tS\tR\tS\tR\tS\tR1}&=-2(C_{LLeudd}^{xzy 112}+C_{LLeudd}^{zxy 112}),
\nn\\
C_{\bar{\ell}\ell\bar{\nu},xyz}^{\tT\tR\tT\tR\tS\tR1}&=-\frac{1}{6}(C_{LLeudd}^{xzy 112}+C_{LLeudd}^{zxy 112}),
\nn\\
C_{\bar{\ell}\ell\bar{\nu},xyz}^{\tT\tR\tS\tR\tT\tR}&=C_{LLeudd}^{xzy 112}+C_{LLeudd}^{zxy 112},
\nn\\
C_{\bar{\ell}\ell\bar{\nu},xyz}^{\tS\tR\tS\tR\tS\tR2}&=2(C_{LLeudd}^{xzy 112}+C_{LLeudd}^{zxy 112}),
\nn\\
C_{\bar{\ell}\ell\bar{\nu},xyz}^{\tT\tR\tT\tR\tS\tR2}&=\frac{1}{6}(C_{LLeudd}^{xzy 112}+C_{LLeudd}^{zxy 112}).
\end{align}
\item ${\cal O}_{eQLded}$:
\begin{align}   
C^{\tV\tR\tV\tL\tS\tR}_{\bar{\ell}\bar{\ell}\ell,xyz}&=\frac{1}{4}(C_{eQLded}^{x1y1z2}-C_{eQLded}^{x1y2z1})
\nn\\
C^{\tV\tR\tS\tR\tV\tL1}_{\bar{\ell}\bar{\ell}\ell,xyz}&=\frac{1}{2}(C_{eQLded}^{x1y1z2}+C_{eQLded}^{x1y2z1}),
\nn\\
C^{\tV\tR\tS\tR\tV\tL2}_{\bar{\ell}\bar{\ell}\ell,xyz}&=\frac{1}{2}C_{eQLded}^{x2y1z1};
\nn\\%
C^{\tV\tR\tV\tL\tS\tR1}_{\bar{\ell}\ell\bar{\nu},xyz}&=\frac{1}{4}(C_{eQLded}^{x1z1y2}-C_{eQLded}^{x1z2y1}),
\nn\\
C^{\tV\tR\tS\tR\tV\tL1}_{\bar{\ell}\ell\bar{\nu},xyz}&=-\frac{1}{2}(C_{eQLded}^{x1z1y2}+C_{eQLded}^{x1z2y1}).
\end{align}
\item ${\cal O}_{LQLdLd}$: 
\begin{align}
C_{\bar{\ell}\bar{\ell}\ell,xyz}^{\tS\tR\tS\tL\tS\tR}&=-\frac{1}{2(1+\delta_{xy})}(C_{LQLdLd}^{z1x1y2}-C_{LQLdLd}^{z1x2y1}
+ x\leftrightarrow y),
\nn\\
C_{\bar{\ell}\bar{\ell}\ell,xyz}^{\tT\tR\tV\tR\tV\tL1}&=-\frac{i}{4}(C_{LQLdLd}^{z1x1y2}+C_{LQLdLd}^{z1x2y1}
- x\leftrightarrow y)
\nn\\
C_{\bar{\ell}\bar{\ell}\ell,xyz}^{\tT\tR\tV\tR\tV\tL2}&=-\frac{i}{4}(C_{LQLdLd}^{z2x1y1}-C_{LQLdLd}^{z2y1x1});
\nn\\%
C_{\bar{\nu}\nu\bar{\ell},xyz}^{\tV\tL\tV\tL\tS\tR}&=-\frac{1}{4}(C_{LQLdLd}^{y1x1z2}-C_{LQLdLd}^{y1x2z1}),
\nn\\
C_{\bar{\nu}\nu\bar{\ell},xyz}^{\tV\tL\tS\tR\tV\tL1}&=-\frac{1}{2}(C_{LQLdLd}^{y1x1z2}+C_{LQLdLd}^{y1x2z1}),
\nn\\
C_{\bar{\nu}\nu\bar{\ell},xyz}^{\tV\tL\tS\tR\tV\tL2}&=-\frac{1}{2}C_{LQLdLd}^{y2x1z1};
\nn\\%
C_{\bar{\ell}\ell\bar{\nu},xyz}^{\tV\tL\tV\tL\tS\tR1}&=-\frac{1}{4}(C_{LQLdLd}^{y1x1z2}-C_{LQLdLd}^{y1x2z1}),
\nn\\
C_{\bar{\ell}\ell\bar{\nu},xyz}^{\tV\tL\tS\tR\tV\tL1}&=-\frac{1}{2}(C_{LQLdLd}^{y1x1z2}+C_{LQLdLd}^{y1x2z1});
\nn\\%
C_{\bar{\nu}\bar{\nu}\nu,xyz}^{\tS\tR\tS\tL\tS\tR1}&=-\frac{1}{2(1+\delta_{xy})}(C_{LQLdLd}^{z1x1y2}-C_{LQLdLd}^{z1x2y1}
+ x\leftrightarrow y),
\nn\\
C_{\bar{\nu}\bar{\nu}\nu,xyz}^{\tT\tR\tV\tR\tV\tL1}&=-\frac{i}{4}(C_{LQLdLd}^{z1x1y2}+C_{LQLdLd}^{z1x2y1}
- x\leftrightarrow y),
\end{align}
\item ${\cal O}_{LeLdQQ}$:
\begin{align}  
C_{\bar{\ell}\bar{\ell}\ell,xyz}^{\tS\tR\tS\tR\tS\tL}&=-\frac{1}{2(1+\delta_{xy})}(C_{LeLdQQ}^{xzy112}-C_{LeLdQQ}^{xzy121}
+ x\leftrightarrow y),
\nn\\
C_{\bar{\ell}\bar{\ell}\ell,xyz}^{\tT\tR\tT\tR\tS\tL}&=\frac{1}{8}(C_{LeLdQQ}^{xzy112}-C_{LeLdQQ}^{xzy121}
-x\leftrightarrow y);
\nn\\
C_{\bar{\ell}\ell\bar{\nu},xyz}^{\tS\tR\tS\tR\tS\tL2}&=C_{LeLdQQ}^{xyz121}+\frac{1}{2}C_{LeLdQQ}^{zyx112},
\nn\\
C_{\bar{\ell}\ell\bar{\nu},xyz}^{\tT\tR\tT\tR\tS\tL2}&=\frac{1}{8}C_{LeLdQQ}^{zyx112},
\nn\\
C_{\bar{\ell}\ell\bar{\nu},xyz}^{\tS\tR\tS\tR\tS\tL3}&=-C_{LeLdQQ}^{xyz211}
-\frac{1}{2}C_{LeLdQQ}^{zyx211},
\nn\\
C_{\bar{\ell}\ell\bar{\nu},xyz}^{\tT\tR\tT\tR\tS\tL3}&=-\frac{1}{8}C_{LeLdQQ}^{zyx211}.
\end{align}
\end{itemize}

\section{Monte Carlo four-body phase-space integration}
\label{app:4body_PS}

In this section, we describe the Monte Carlo procedure used in our work to generate
the final-state kinematics and evaluate the decay widths numerically.
For the decay process
$\tN(p)\to l_1(p_1)\,l_2(p_2)\,l_3(p_3)\,M(p_4)$,
we follow the standard recursive treatment in~\cite{Barger:1987nn} and cluster the final-state particles as $\{1+[2+(34)]\}$. 
The four-body phase space then can be decomposed into a product of three two-body phase-space factors as follows:
\begin{align}
d\Phi^{(4)}
=\,&
\frac{dM_{234}^2}{2\pi}\,
\frac{dM_{34}^2}{2\pi}\,
d\Phi^{(2)}(p\to p_1,P_{234})
\nn\\
&\times
d\Phi^{(2)}(P_{234}\to p_2,P_{34})
\nn\\
&\times
d\Phi^{(2)}(P_{34}\to p_3,p_4),
\label{eq:PS4decomp}
\end{align}
where
$P_{34}=p_3+p_4$ and $P_{234}=p_2+p_3+p_4$.
The invariant masses vary within
\begin{align}
m_3+m_4
&\leq M_{34}\leq M_{234}-m_2,
\nn\\
m_2+m_3+m_4
&\leq M_{234}\leq m_\tN-m_1.
\end{align}

In the rest frame of the parent system, each two-body phase space can be expressed as
\begin{align}
d\Phi^{(2)}(Q\to q_1,q_2)
&=
\frac{1}{16\pi^2}
\frac{q^*}{M_Q}\,d\Omega^*,
\label{eq:PS2}\\
q^*
&=
\frac{\lambda^{1/2}
\big(M_Q^2,q_1^2,q_2^2\big)}
{2M_Q},
\end{align}
where $M_Q=\sqrt{Q^2}$, and $q^*=|\pmb{q}^*|$ is the magnitude of the three-momentum of either final-state particle in the parent rest frame.
Throughout, quantities evaluated in the parent rest frame are denoted by a star. 
The triangle function is defined as
\begin{align}
\lambda(x,y,z)
=
x^2+y^2+z^2-2xy-2yz-2zx.
\end{align}

We now apply \cref{eq:PS2} to the three two-body phase-space factors in
\cref{eq:PS4decomp}.
For notational convenience, we introduce the nested invariant masses
\begin{align}
\tilde m_1=m_\tN,~
\tilde m_2=M_{234},~
\tilde m_3=M_{34},~
\tilde m_4=m_4.
\end{align}
With this notation, the corresponding three-momentum magnitudes for the three successive two-body subsystems can be written compactly as
\begin{align}
p_j^* =
\frac{\lambda^{1/2}
\big(\tilde m_j^2,m_j^2,\tilde m_{j+1}^2\big)}
{2\tilde m_j},
\qquad j=1,2,3.
\end{align}
For each three-momentum $\pmb{p}_j^*$,
its direction in the corresponding parent rest frame is specified by the polar and azimuthal angles
$(\theta_j^*,\phi_j^*)$.
Thus, the four-body phase-space integration is fully described by eight kinematic
variables: the two nested invariant masses $M_{234}$ and $M_{34}$, together with the three solid angles
$\Omega_1^*$, $\Omega_2^*$, and $\Omega_3^*$.

For the Monte Carlo integration, eight random numbers $\{x_{1,2,3}, y_{1,2,3}, z_{2,3}\}\in[0,1]$ are
mapped onto these kinematic variables according to
\begin{align}
\phi_i^*
&=
2\pi\,x_i,~i=1,2,3,
\nn\\
\cos\theta_i^*
&= 1-2y_i,~i=1,2,3,
\nn\\
\tilde m_i
&=
\sigma_i +(m_\tN-\sigma_1)\,z_i,~i=2,3.
\label{eq:RNmapping}
\end{align}
where $\sigma_i = m_i+m_{i+1}+\cdots+m_4$.
The random numbers $z_2$ and $z_3$ are ordered such that
$z_2\geq z_3$, which ensures the kinematic condition
$\tilde m_3\leq\tilde m_2-m_2$.

With the above parametrization, the decay width is calculated as
\begin{align}
\Gamma
&\simeq
\frac{1}{N_\text{sim}}
\sum_{j=1}^{N_\text{sim}}w_j,
\nn\\
w_j
&=
\frac{1}{2}
\left(
\frac{m_\tN-\sigma_1}{4\pi^2}
\right)^2
\frac{p_1^*p_2^*p_3^*}{4\pi m_\tN}\,
\frac{1}{S}\frac{1}{2m_\tN}\,
\overline{|\mathcal{M}|^2}^{(j)},
\label{eq:GammaMC}
\end{align}
where $N_\text{sim}$ denotes the total number of simulated decay events, which we take to be $\calO(10^5)$ in our calculation. 
For $z_2< z_3$ in the sampling procedure, we interchange $z_2\leftrightarrow z_3$  to improve sampling efficiency, with the prefactor $1/2$ included to account for the resulting double counting.

For each generated phase-space point, the four-momenta are constructed
recursively in the $P_{34}$ and $P_{234}$ rest frames and then boosted
successively to the nucleon rest frame, where
$\overline{|\mathcal{M}|^2}$ is evaluated.
Specifically, in each parent rest frame the three-momentum of particle
$j$ can be written as
\begin{align}
    \pmb{p}_j^*=p_j^*(\sin\theta_j^*\cos\phi_j^*,
\sin\theta_j^*\sin\phi_j^*,\cos\theta_j^*),
\end{align}
whose energy is $E_j^*=\sqrt{(p_j^*)^2+m_j^2}$,
while the recoiling
subsystem carries the opposite three-momentum.
For a given four-momentum $p_3^*$, constructed from the random numbers $x_3,\,y_3,\,z_3$ according to \cref{eq:RNmapping}, 
we first boost $p_3^*$ and its corresponding recoiling momentum $p_4^*$ from the $P_{34}$ rest frame to the $P_{234}$ rest frame.
Together with $p_2^*$, constructed from the random numbers $x_2,\,y_2,\,z_2$, these momenta in $P_{234}$ rest frame are then further boosted to the nucleon rest frame,  
where $p_1=p_1^*$ is constructed directly from $x_1,\,y_1$.
Each boost from the rest frame of a subsystem $Q^*=(M_Q,\pmb{0})$ to the frame in which the
subsystem carries a four-momentum $Q=(E_Q,\pmb{Q})$ is the standard
Lorentz transformation
\begin{align}
E_k=\,&\gamma\left(E_k^*+\beta\,\hat{\pmb{n}}\cdot\pmb{k}^*\right),
\nn\\
\pmb{k}=\,&\pmb{k}^*
+\left[(\gamma-1)\,(\hat{\pmb{n}}\cdot\pmb{k}^*)
+\gamma\beta\,E_k^*\right]\hat{\pmb{n}},
\end{align}
where $\hat{\pmb{n}}=\pmb{Q}/|\pmb{Q}|$ denotes the boost direction,
and $\gamma=E_Q/M_Q$, $\beta=|\pmb{Q}|/E_Q$.
For our case, the relevant subsystems are $P_{34}$ and $P_{234}$. The first boost takes $Q_{1}^*=(\tilde m_3,\pmb{0})$ to $Q_{1}=(\sqrt{\tilde m_3^2 + (\pmb{p}_2^*)^2 },-\pmb{p}_2^*)$, 
and the second boost takes $Q_{2}^*=(\tilde m_2,\pmb{0})$ to $Q_{2}=(\sqrt{\tilde m_2^2 + (\pmb{p}_1^*)^2 },-\pmb{p}_1^*)$. 

\section{Nucleon decay amplitudes} 
\label{app:decay_width_amp}

There are two diagrams contributing to $\tN(p)\to l_1(p_1) l_2(p_2) l_3 (p_3) M(p_4)$ shown in \cref{fig:diagram_N2lllK}. 
To write the amplitude in a compact way, we denote the partial amplitude related to the hadronic part of the noncontact diagram by 
\begin{align}
U_B \equiv  \frac{\slashed{p}_B+m_B}{p_B^2-m_B^2}(m_\tN+\slashed{p}_B)\gamma_5 u_\tN,  
\end{align}
where $B=\Sigma^\pm,\Sigma^0,\Lambda^0$ represents the baryon mediator, whose mass and momentum are denoted by $m_B$ and  $p_B=p_1+p_2+p_3=p-p_4$, respectively. 
The spinor for the initial nucleon $\tN$ is denoted by $u_\tN$. For the amplitudes associated with operators in the irreps $\pmb{3}_{\tL(\tR)}\otimes \pmb{6}_{\tR(\tL)}$, we also employ the abbreviations $\tilde p^\mu = p^\mu - \frac{1}{4}\gamma^\mu\slashed{p}$ and $\tilde p_B^\mu = p_B^\mu - \frac{1}{4}\gamma^\mu\slashed{p}_B$ with $B=\Sigma_{\pm,0},\Lambda_0$.

In addition, since multiple LEFT operators can contribute to the same four-fermion operators in chiral matching, it is convenient to abbreviate the linear combinations of WCs related to each hadronic operator in each decay mode by a a single symbol. In the following, we use the symbols $\calK_i^j$ and $\calC_i^j$ to represent the WC combinations related to the vertices in noncontact and contact diagrams of \cref{fig:diagram_N2lllK}, respectively.  
In the non-contact contributions, the abbreviated WC combinations $\calK_i^j$ depend on the baryon mediators $\Sigma^{\pm}$, $\Sigma^0$, and $\Lambda^0$, and we explicitly indicate them as superscripts.

In the following, for each nucleon decay mode listed in \cref{tab:spurions1,tab:spurions2}, we provide the relevant WC combinations and the resulting matrix elements. 

\subsection{\tf{$\Delta L = +1$: 
$p\to\ell_x^+\ell_y^+\ell_z^-K^0$}{DeltaL=+1:p2lplplmK0}}
The relevant WC combinations are:
\begin{align}
\calK(\calC)^{\tS\tL\tS\tL\tS}_{\ell\ell\bar\ell,xyz}
&\equiv C^{\tS\tL\tS\tL\tS\tR}_{\ell\ell\bar\ell,xyz}
\pm \kappa_2 C^{\tS\tL\tS\tL\tS\tL}_{\ell\ell\bar\ell,xyz},
\nn\\
\calK(\calC)^{\tS\tR\tS\tL\tS}_{\ell\ell\bar\ell,xyz}
&\equiv 
C^{\tS\tR\tS\tL\tS\tR}_{\ell\ell\bar\ell,xyz}
\pm \kappa_2 C^{\tS\tR\tS\tL\tS\tL}_{\ell\ell\bar\ell,xyz},
\nn\\
\calK(\calC)^{\tV\tR\tV\tL\tS}_{\ell\ell\bar\ell,xyz}
&\equiv
C^{\tV\tR\tV\tL\tS\tR}_{\ell\ell\bar\ell,xyz}
\pm \kappa_2 C^{\tV\tR\tV\tL\tS\tL}_{\ell\ell\bar\ell,xyz},
\nn\\
\calK(\calC)^{\tT\tL\tT\tL\tS}_{\ell\ell\bar\ell,xyz}
&\equiv 
C^{\tT\tL\tT\tL\tS\tR}_{\ell\ell\bar\ell,xyz}
\pm \kappa_2 C^{\tT\tL\tT\tL\tS\tL}_{\ell\ell\bar\ell,xyz},
\nn\\%
\calK(\calC)^{\tV\tL\tS\tL\tV\tR}_{\ell\ell\bar\ell,xyz}&\equiv
\kappa_{3\chi} (C^{\tV\tL\tS\tL\tV\tR1}_{\ell\ell\bar\ell,xyz}/2
\mp C^{\tV\tL\tS\tL\tV\tR2}_{\ell\ell\bar\ell,xyz}),
\nn\\
\calK(\calC)^{\tT\tL\tV\tL\tV\tR}_{\ell\ell\bar\ell,xyz}&\equiv
\kappa_{3\chi}
(C^{\tT\tL\tV\tL\tV\tR1}_{\ell\ell\bar\ell,xyz}/2
\mp C^{\tT\tL\tV\tL\tV\tR2}_{\ell\ell\bar\ell,xyz}),
\end{align}  
and their chiral partners with $\tL\leftrightarrow\tR$.
The upper and lower signs in each line are related to the symbol $\calK$ and  $\calC$, respectively.
The amplitudes for the related decay modes are given as follows.

\noindent$\bullet~p(p)\to e^+(p_1)\mu^+(p_2)\ell_x^-(p_3)K^0(p_4)$ with $x=e,\mu$:
\begin{align}
\mathcal{M} =\,&
\Big\{
[\overline{u_1} 
(\calK_{\ell\ell\bar\ell,e\mu x}^{\tS\tL\tS\tL\tS} P_\tL
+\calK_{\ell\ell\bar\ell,e\mu x}^{\tS\tR\tS\tL\tS} P_\tR) v_2]
(\overline{u_3}P_\tL U_{\Sigma^+})
\nn\\
&+\calK_{\ell\ell\bar\ell,e\mu x}^{\tT\tL\tT\tL\tS}
(\overline{u_1} \sigma_{\mu\nu}P_\tL v_2)
(\overline{u_3} \sigma^{\mu\nu}P_\tL U_{\Sigma^+})
\nn\\
&+
\calK_{\ell\ell\bar\ell,e\mu x}^{\tT\tL\tV\tL\tV\tR}
(\overline{u_1} \sigma_{\nu\mu}P_\tL v_2)
(\overline{u_3} \gamma^\nu \tilde{p}_{\Sigma^+}^\mu P_\tL U_{\Sigma^+})
\nn\\
&+
[\overline{u_1}\gamma_\mu (\calK_{\ell\ell\bar\ell,e\mu x}^{\tV\tR\tV\tL\tS} P_\tR - \calK_{\ell\ell\bar\ell,\mu e x}^{\tV\tR\tV\tL\tS}P_\tL )v_2]
(\overline{u_3}\gamma^\mu P_\tL U_{\Sigma^+})
\nn\\
&+ [\overline{u_1} \gamma_{\mu}( \calK_{\ell\ell\bar\ell,e\mu x}^{\tV\tL\tS\tL\tV\tR}P_\tL - \calK_{\ell\ell\bar\ell,\mu e x}^{\tV\tL\tS\tL\tV\tR} P_\tR) v_2]
\nn\\
&\times (\overline{u_3} \tilde{p}_{\Sigma^+}^\mu P_\tL U_{\Sigma^+})-\tL\leftrightarrow\tR
\Big\} \frac{ic_1 C_{p\Sigma^+K^0} }{F_0}
\nn\\%
&+\Big\{
[\overline{u_1} 
(\calC_{\ell\ell\bar\ell,e\mu x}^{\tS\tL\tS\tL\tS} P_\tL
+\calC_{\ell\ell\bar\ell,e\mu x}^{\tS\tR\tS\tL\tS} P_\tR) v_2]
(\overline{u_3} P_\tL u_p)
\nn\\
&+\calC_{\ell\ell\bar\ell,e\mu x}^{\tT\tL\tT\tL\tS}
(\overline{u_1} \sigma_{\mu\nu} P_\tL v_2)
(\overline{u_3} \sigma^{\mu\nu} P_\tL u_p)
\nn\\
&-
\calC_{\ell\ell\bar\ell,e\mu x}^{\tT\tL\tV\tL\tV\tR}
(\overline{u_1} \sigma_{\nu\mu}P_\tL v_2)
(\overline{u_3} \gamma^\nu \tilde p^\mu P_\tL u_p)
\nn\\
&+\Big[
\calC_{\ell\ell\bar\ell,e\mu x}^{\tV\tR\tV\tL\tS}
(\overline{u_1}\gamma_\mu P_\tR v_2)
(\overline{u_3}\gamma^\mu P_\tL u_p)
-\calC_{\ell\ell\bar\ell,e\mu x}^{\tV\tL\tS\tL\tV\tR}
\nn\\
&\times (\overline{u_1} \gamma_{\mu}P_\tL v_2)
(\overline{u_3} \tilde p^\mu P_\tL u_p)
-(e,p_1)\leftrightarrow (\mu,p_2)\Big]
\nn\\
&+\frac{2}{3}\kappa_{4\chi}
C^{\tT\tL\tS\tL\tT\tL}_{\ell\ell\bar\ell,e\mu x}
[\overline{u_1}\sigma_{\mu\nu} P_\tL v_2] 
[\overline{u_3}\hat{\Gamma}^{\tL}_{\mu\nu\alpha\beta} u_p] p^\alpha p_4^\beta
\nn\\
&+\tL\leftrightarrow\tR \Big\}
\frac{i c_1}{\sqrt{2}F_0}.
\end{align}
where $u_{1}=v_{1}^\C$.\\
\noindent$\bullet~
p(p)\to\ell_x^+(p_1)\ell_x^+(p_2)\ell_y^-(p_3)K^0(p_4)$ with $x,y=e,\mu$:
\begin{align}
\mathcal{M} =\,& 
\Big\{\big[
[\overline{u_1} 
(\calK_{\ell\ell\bar\ell,xxy}^{\tS\tL\tS\tL\tS} P_\tL
+\calK_{\ell\ell\bar\ell,xxy}^{\tS\tR\tS\tL\tS} P_\tR) v_2]
(\overline{u_3} P_\tL U_{\Sigma^+})
\nn\\
&+\calK_{\ell\ell\bar\ell,xxy}^{\tV\tR\tV\tL\tS}
(\overline{u_1}\gamma_\mu P_\tR v_2)
(\overline{u_3}\gamma^\mu P_\tL U_{\Sigma^+}) 
\nn\\
&+\calK_{\ell\ell\bar\ell,xxy}^{\tV\tL\tS\tL\tV\tR}
(\overline{u_1} \gamma_{\mu}P_\tL v_2)
(\overline{u_3} \tilde p_{\Sigma^+}^\mu P_\tL U_{\Sigma^+})
\nn\\
&
-(p_1\leftrightarrow p_2)\big]-\tL\leftrightarrow\tR
\Big\}\frac{ic_1}{F_0}C_{p\Sigma^+K^0}
\nn\\%
&+\Big\{\big[
[\overline{u_1} 
(\calC_{\ell\ell\bar\ell,xxy}^{\tS\tL\tS\tL\tS} P_\tL
+\calC_{\ell\ell\bar\ell,xxy}^{\tS\tR\tS\tL\tS} P_\tR) v_2]
(\overline{u_3} P_\tL u_p)
\nn\\
&
+\calC_{\ell\ell\bar\ell,xxy}^{\tV\tR\tV\tL\tS}
(\overline{u_1}\gamma_\mu P_\tR v_2)
(\overline{u_3}\gamma^\mu P_\tL u_p)
\nn\\
&-\calC_{\ell\ell\bar\ell,xxy}^{\tV\tL\tS\tL\tV\tR}
(\overline{u_1} \gamma_{\mu}P_\tL v_2)
(\overline{u_3} \tilde p^\mu P_\tL u_p)
\nn\\
&
-(p_1\leftrightarrow p_2)\big]+\tL\leftrightarrow\tR
\Big\}\frac{ic_1}{\sqrt{2}F_0},
\end{align}
where $u_{1,2}=v_{1,2}^\C$.

\subsection{\tf{$\Delta L = +1$: $p\to\nu_x\bar\nu_y\ell_z^+K^0$}{p2vvblpK0}}
The relevant WC combinations are:
\begin{align}
\calK(\calC)^{\tV\tL\tV\tL\tS}_{\bar\nu \nu \ell,xyz}
&\equiv C_{\bar\nu \nu \ell,xyz}^{\tV\tL\tV\tL\tS\tR}
\pm\kappa_2C_{\bar\nu \nu \ell,xyz}^{\tV\tL\tV\tL\tS\tL},
\nn\\
\calK(\calC)^{\tV\tL\tV\tR\tS}_{\bar\nu \nu \ell,xyz}
&\equiv C_{\bar\nu \nu \ell,xyz}^{\tV\tL\tV\tR\tS\tL}
\pm\kappa_2C_{\bar\nu \nu \ell,xyz}^{\tV\tL\tV\tR\tS\tR},
\nn\\
\calK(\calC)_{\bar\nu \nu \ell,xyz}^{\tV\tL\tS\tL\tV\tR}
&\equiv\kappa_{3\chi}(C_{\bar\nu \nu \ell,xyz}^{\tV\tL\tS\tL\tV\tR1}/2
\mp C_{\bar\nu \nu \ell,xyz}^{\tV\tL\tS\tL\tV\tR2}),
\nn\\
\calK(\calC)_{\bar\nu \nu \ell,xyz}^{\tV\tL\tS\tR\tV\tL}
&\equiv\kappa_{3\chi}
(C_{\bar\nu \nu \ell,xyz}^{\tV\tL\tS\tR\tV\tL1}/2
\mp C_{\bar\nu \nu \ell,xyz}^{\tV\tL\tS\tR\tV\tL2}).
\end{align}

\noindent$\bullet~
p(p)\to\nu_x(p_1)\bar\nu_y(p_2)\ell_z^+(p_3)K^0(p_4)$ with $x,y=e,\mu,\tau$ and $z=e,\mu$:
\begin{align}
\mathcal{M} =\,&
\frac{ic_1}{F_0}C_{p\Sigma^+ K^0} 
(\overline{u_1} \gamma_\mu P_\tL v_2)
\nn\\
&\times \Big\{\overline{u_3} \big[
\gamma^\mu ( \calK_{\bar\nu\nu\ell,xyz}^{\tV\tL\tV\tL\tS} P_\tL 
-\calK_{\bar\nu\nu\ell,xyz}^{\tV\tL\tV\tR\tS} P_\tR )
\\
&+\tilde p^\mu_{\Sigma^+}
( \calK_{\bar\nu\nu\ell,xyz}^{\tV\tL\tS\tL\tV\tR} P_\tL
- \calK_{\bar\nu\nu\ell,xyz}^{\tV\tL\tS\tR\tV\tL} P_\tR )\big]U_{\Sigma^+} \Big\}
\nn\\
&+(\overline{u_1} \gamma_\mu P_\tL v_2)
\Big\{\overline{u_3} \big[
\gamma^\mu ( \calC_{\bar\nu\nu\ell,xyz}^{\tV\tL\tV\tL\tS} P_\tL 
+\calC_{\bar\nu\nu\ell,xyz}^{\tV\tL\tV\tR\tS} P_\tR )
\nn\\
&-\tilde p^\mu
( \calC_{\bar\nu\nu\ell,xyz}^{\tV\tL\tS\tL\tV\tR} P_\tL
+ \calC_{\bar\nu\nu\ell,xyz}^{\tV\tL\tS\tR\tV\tL} P_\tR )\big]u_p\Big\}\frac{ic_1}{\sqrt{2}F_0},
\nn
\end{align}
where $u_3=v_3^\C$.

\subsection{\tf{$\Delta L = +1$: $n\to\bar\nu_x\bar\nu_y\ell_z^-K^+$ (a)}{n2vbvblmKp(a)}}

The relevant WC combinations are:
\begin{align}
\calK(\calC)^{\tS\tL\tS\tL\tS}_{\nu\nu\bar\ell,xyz}
&\equiv C^{\tS\tL\tS\tL\tS\tR}_{\nu\nu\bar\ell,xyz}
\pm\kappa_2 C^{\tS\tL\tS\tL\tS\tL}_{\nu\nu\bar\ell,xyz},
\nn\\
\calK(\calC)^{\tT\tL\tT\tL\tS}_{\nu\nu\bar\ell,xyz}
&\equiv C^{\tT\tL\tT\tL\tS\tR}_{\nu\nu\bar\ell,xyz}
\pm\kappa_2 C^{\tT\tL\tT\tL\tS\tL}_{\nu\nu\bar\ell,xyz},
\nn\\
\calK(\calC)^{\tS\tL\tS\tR\tS}_{\nu\nu\bar\ell,xyz}
&\equiv C^{\tS\tL\tS\tR\tS\tL}_{\nu\nu\bar\ell,xyz}
\pm\kappa_2 C^{\tS\tL\tS\tR\tS\tR}_{\nu\nu\bar\ell,xyz},
\nn\\
\calK(\calC)^{\tT\tL\tV\tL\tV\tR}_{\nu\nu\bar\ell,xyz}
&\equiv \kappa_{3\chi}
(C^{\tT\tL\tV\tL\tV\tR1}_{\nu\nu\bar\ell,xyz}/2
\mp C^{\tT\tL\tV\tL\tV\tR2}_{\nu\nu\bar\ell,xyz}).
\end{align}

\noindent$\bullet~n(p)\to \bar\nu_x(p_1) \bar\nu_y(p_2) \ell_z^-(p_3)K^+(p_4)$ with $x,y=e,\mu,\tau$ and $z=e,\mu$:
\begin{align}
\mathcal{M} =\, &\, \Big\{
(\overline{u_1} P_\tL v_2)
[\overline{u_3}
(\calK_{\nu\nu\bar\ell,xyz}^{\tS\tL\tS\tL\tS} P_\tL-\calK_{\nu\nu\bar\ell,xyz}^{\tS\tL\tS\tR\tS}P_\tR
) U_{\Sigma^-}]
\nn\\
&- \calK_{\nu\nu\bar\ell,xyz}^{\tT\tL\tV\tL\tV\tR}
(\overline{u_1} \sigma_{\nu\mu}P_\tL v_2)
(\overline{u_3} \gamma^\nu \tilde p^\mu_{\Sigma^-} P_\tL U_{\Sigma^-})
\nn\\
& + \calK_{\nu\nu\bar\ell,xyz}^{\tT\tL\tT\tL\tS}
(\overline{u_1} \sigma_{\mu\nu}P_\tL v_2)
(\overline{u_3}\sigma^{\mu\nu} P_\tL U_{\Sigma^-})
\nn\\
&-\delta_{x,y}(p_1\leftrightarrow p_2)
\Big\} \frac{ic_1}{F_0} C_{n\Sigma^- K^+}
\nn\\
&+\Big\{ (\overline{u_1} P_\tL v_2)
[\overline{u_3}
(\calC_{\nu\nu\bar\ell,xyz}^{\tS\tL\tS\tR\tS}P_\tR
+\calC_{\nu\nu\bar\ell,xyz}^{\tS\tL\tS\tL\tS} P_\tL) u_n]
\nn\\
&+\calC_{\nu\nu\bar\ell,xyz}^{\tT\tL\tV\tL\tV\tR}
(\overline{u_1} \sigma_{\nu\mu}P_\tL v_2)
(\overline{u_3} \gamma^\nu \tilde p^\mu P_\tL u_n)
\nn\\
& + \calC_{\nu\nu\bar\ell,xyz}^{\tT\tL\tT\tL\tS}
(\overline{u_1} \sigma_{\mu\nu}P_\tL v_2)
(\overline{u_3}\sigma^{\mu\nu} P_\tL u_n)
\nn\\
&-\frac{2}{3}\kappa_{4\chi} 
C^{\tT\tL\tS\tL\tT\tL}_{\nu\nu\bar\ell,xyz}
(\overline{u_1}\sigma_{\mu\nu} P_\tL v_2) 
(\overline{u_3}\hat{\Gamma}^{\tL}_{\mu\nu\alpha\beta} u_n) p^\alpha p_4^\beta 
\nn\\
&-\delta_{x,y}(p_1\leftrightarrow p_2)\Big\}\frac{ic_1}{\sqrt{2}F_0},
\end{align}
where $u_{1,2}=v_{1,2}^\C$.

\subsection{\tf{$\Delta L = +1$: 
$p\to\ell_x^-\ell_y^+\bar\nu_zK^+,\,
n\to\ell_x^-\ell_y^+\bar\nu_z K^0$ (b)}{N2lmlpvbK(b)}}

The relevant WC combinations are:
\begin{align}
\calC^{\tS\tL\tS\tL\tS ,p}_{\bar\ell\ell\nu,xyz}
\equiv\,& \frac{1}{\sqrt{2}}
(C^{\tS\tL\tS\tL\tS\tR3}_{\bar\ell\ell\nu,xyz}
+C^{\tS\tL\tS\tL\tS\tR1}_{\bar\ell\ell\nu,xyz}
+\kappa_2 C^{\tS\tL\tS\tL\tS\tL2}_{\bar\ell\ell\nu,xyz}),
\nn\\
\calC^{\tT\tL\tT\tL\tS ,p}_{\bar\ell\ell\nu,xyz}
\equiv\,& \frac{1}{\sqrt{2}}
(C^{\tT\tL\tT\tL\tS\tR3}_{\bar\ell\ell\nu,xyz}
+C^{\tT\tL\tT\tL\tS\tR1}_{\bar\ell\ell\nu,xyz}
+\kappa_2 C^{\tT\tL\tT\tL\tS\tL2}_{\bar\ell\ell\nu,xyz}),
\nn\\
\calC^{\tS\tR\tS\tL\tS ,p}_{\bar\ell\ell\nu,xyz}
\equiv\,& \frac{1}{\sqrt{2}}
(C^{\tS\tR\tS\tL\tS\tR3}_{\bar\ell\ell\nu,xyz}
+C^{\tS\tR\tS\tL\tS\tR1}_{\bar\ell\ell\nu,xyz}
 +\kappa_2 C^{\tS\tR\tS\tL\tS\tL2}_{\bar\ell\ell\nu,xyz}),
\nn\\
\calC^{\tV\tR\tV\tR\tS ,p}_{\bar\ell\ell\nu,xyz}
\equiv\,& \frac{1}{\sqrt{2}}
(C^{\tV\tR\tV\tR\tS\tL3}_{\bar\ell\ell\nu,xyz}
+C^{\tV\tR\tV\tR\tS\tL1}_{\bar\ell\ell\nu,xyz}
+\kappa_2 C^{\tV\tR\tV\tR\tS\tR2}_{\bar\ell\ell\nu,xyz}),
\nn\\
\calC^{\tV\tL\tV\tR\tS,p}_{\bar\ell\ell\nu,xyz}
\equiv\,& \frac{1}{\sqrt{2}}
(C^{\tV\tL\tV\tR\tS\tL3}_{\bar\ell\ell\nu,xyz}
+C^{\tV\tL\tV\tR\tS\tL1}_{\bar\ell\ell\nu,xyz}
+\kappa_2 C^{\tV\tL\tV\tR\tS\tR2}_{\bar\ell\ell\nu,xyz}),
\nn\\
\calC^{\tT\tR\tV\tR\tV\tL,p}_{\bar\ell\ell\nu,xyz}
\equiv\,&  -\frac{\kappa_{3\chi}}{2\sqrt{2}}
(C^{\tT\tR\tV\tR\tV\tL1}_{\bar\ell\ell\nu,xyz}
-C^{\tT\tR\tV\tR\tV\tL3}_{\bar\ell\ell\nu,xyz}
-2C^{\tT\tR\tV\tR\tV\tL2}_{\bar\ell\ell\nu,xyz}),
\nn\\
\calC^{\tV\tR\tS\tL\tV\tR,p}_{\bar\ell\ell\nu,xyz}
\equiv\,& -\frac{\kappa_{3\chi}}{2\sqrt{2}}
(C^{\tV\tR\tS\tL\tV\tR1}_{\bar\ell\ell\nu,xyz}
-C^{\tV\tR\tS\tL\tV\tR3}_{\bar\ell\ell\nu,xyz}
-2C^{\tV\tR\tS\tL\tV\tR2}_{\bar\ell\ell\nu,xyz}),
\nn\\
\calC^{\tV\tL\tS\tL\tV\tR,p}_{\bar\ell\ell\nu,xyz}
\equiv\,&  -\frac{\kappa_{3\chi}}{2\sqrt{2}}
(C^{\tV\tL\tS\tL\tV\tR1}_{\bar\ell\ell\nu,xyz}
-C^{\tV\tL\tS\tL\tV\tR3}_{\bar\ell\ell\nu,xyz}
-2C^{\tV\tL\tS\tL\tV\tR2}_{\bar\ell\ell\nu,xyz}),
\nn\\
\calC^{\tS\tL\tS\tL\tS ,n}_{\bar\ell\ell\nu,xyz}
\equiv\,& \frac{1}{\sqrt{2}}
\big[C^{\tS\tL\tS\tL\tS\tR2}_{\bar\ell\ell\nu,xyz}
+C^{\tS\tL\tS\tL\tS\tR3}_{\bar\ell\ell\nu,xyz}
\nn\\
&+\kappa_2 (C^{\tS\tL\tS\tL\tS\tL2}_{\bar\ell\ell\nu,xyz}
-C^{\tS\tL\tS\tL\tS\tL1}_{\bar\ell\ell\nu,xyz})\big],
\nn\\
\calC^{\tT\tL\tT\tL\tS ,n}_{\bar\ell\ell\nu,xyz}
\equiv\,& \frac{1}{\sqrt{2}}
\big[C^{\tT\tL\tT\tL\tS\tR2}_{\bar\ell\ell\nu,xyz}
+C^{\tT\tL\tT\tL\tS\tR3}_{\bar\ell\ell\nu,xyz}
\nn\\
&+\kappa_2 (C^{\tT\tL\tT\tL\tS\tL2}_{\bar\ell\ell\nu,xyz}
-C^{\tT\tL\tT\tL\tS\tL1}_{\bar\ell\ell\nu,xyz})\big],
\nn\\
\calC^{\tS\tR\tS\tL\tS ,n}_{\bar\ell\ell\nu,xyz}
\equiv\,& \frac{1}{\sqrt{2}}
\big[C^{\tS\tR\tS\tL\tS\tR2}_{\bar\ell\ell\nu,xyz}
+C^{\tS\tR\tS\tL\tS\tR3}_{\bar\ell\ell\nu,xyz}
\nn\\
&+\kappa_2 (C^{\tS\tR\tS\tL\tS\tL2}_{\bar\ell\ell\nu,xyz}
-C^{\tS\tR\tS\tL\tS\tL1}_{\bar\ell\ell\nu,xyz})\big],
\nn\\
\calC^{\tV\tR\tV\tR\tS ,n}_{\bar\ell\ell\nu,xyz}
\equiv\,& \frac{1}{\sqrt{2}}
\big[C^{\tV\tR\tV\tR\tS\tL2}_{\bar\ell\ell\nu,xyz}
+C^{\tV\tR\tV\tR\tS\tL3}_{\bar\ell\ell\nu,xyz}
\nn\\
&+\kappa_2 (C^{\tV\tR\tV\tR\tS\tR2}_{\bar\ell\ell\nu,xyz}
-C^{\tV\tR\tV\tR\tS\tR1}_{\bar\ell\ell\nu,xyz})\big],
\nn\\
\calC^{\tV\tL\tV\tR\tS ,n}_{\bar\ell\ell\nu,xyz}
\equiv\,& \frac{1}{\sqrt{2}}
\big[C^{\tV\tL\tV\tR\tS\tL2}_{\bar\ell\ell\nu,xyz}
+C^{\tV\tL\tV\tR\tS\tL3}_{\bar\ell\ell\nu,xyz}
\nn\\
&+\kappa_2 (C^{\tV\tL\tV\tR\tS\tR2}_{\bar\ell\ell\nu,xyz}
-C^{\tV\tL\tV\tR\tS\tR1}_{\bar\ell\ell\nu,xyz})\big],
\nn\\
\calC^{\tT\tR\tV\tR\tV\tL ,n}_{\bar\ell\ell\nu,xyz}
\equiv\,& -\frac{\kappa_{3\chi}}{2\sqrt{2}}
(2C^{\tT\tR\tV\tR\tV\tL1}_{\bar\ell\ell\nu,xyz}+C^{\tT\tR\tV\tR\tV\tL3}_{\bar\ell\ell\nu,xyz}-C^{\tT\tR\tV\tR\tV\tL2}_{\bar\ell\ell\nu,xyz}),
\nn\\
\calC^{\tV\tR\tS\tL\tV\tR ,n}_{\bar\ell\ell\nu,xyz}
\equiv\,& -\frac{\kappa_{3\chi}}{2\sqrt{2}}
(2C^{\tV\tR\tS\tL\tV\tR1}_{\bar\ell\ell\nu,xyz}
+C^{\tV\tR\tS\tL\tV\tR3}_{\bar\ell\ell\nu,xyz}
-C^{\tV\tR\tS\tL\tV\tR2}_{\bar\ell\ell\nu,xyz}),
\nn\\
\calC^{\tV\tL\tS\tL\tV\tR ,n}_{\bar\ell\ell\nu,xyz}
\equiv\,& -\frac{\kappa_{3\chi}}{2\sqrt{2}}
(2C^{\tV\tL\tS\tL\tV\tR1}_{\bar\ell\ell\nu,xyz}
+C^{\tV\tL\tS\tL\tV\tR3}_{\bar\ell\ell\nu,xyz}
-C^{\tV\tL\tS\tL\tV\tR2}_{\bar\ell\ell\nu,xyz}),
\nn\\
\calK^{\tS\tL\tS\tL\tS,\Lambda^0}_{\bar\ell\ell\nu,xyz}
\equiv\,& \frac{1}{\sqrt{6}}
\big[C^{\tS\tL\tS\tL\tS\tR1}_{\bar\ell\ell\nu,xyz}
+C^{\tS\tL\tS\tL\tS\tR2}_{\bar\ell\ell\nu,xyz}
-2C^{\tS\tL\tS\tL\tS\tR3}_{\bar\ell\ell\nu,xyz}
\nn\\
&+\kappa_2 (C^{\tS\tL\tS\tL\tS\tL1}_{\bar\ell\ell\nu,xyz}
-2C^{\tS\tL\tS\tL\tS\tL2}_{\bar\ell\ell\nu,xyz})\big],
\nn\\
\calK^{\tT\tL\tT\tL\tS,\Lambda^0}_{\bar\ell\ell\nu,xyz}
\equiv\,& \frac{1}{\sqrt{6}}
\big[C^{\tT\tL\tT\tL\tS\tR1}_{\bar\ell\ell\nu,xyz}
+C^{\tT\tL\tT\tL\tS\tR2}_{\bar\ell\ell\nu,xyz}
-2C^{\tT\tL\tT\tL\tS\tR3}_{\bar\ell\ell\nu,xyz}
\nn\\
&+\kappa_2 (C^{\tT\tL\tT\tL\tS\tL1}_{\bar\ell\ell\nu,xyz}
-2C^{\tT\tL\tT\tL\tS\tL2}_{\bar\ell\ell\nu,xyz})\big],
\nn\\
\calK^{\tS\tR\tS\tL\tS,\Lambda^0}_{\bar\ell\ell\nu,xyz}
\equiv\,&\frac{1}{\sqrt{6}}
\big[C^{\tS\tR\tS\tL\tS\tR1}_{\bar\ell\ell\nu,xyz}
+C^{\tS\tR\tS\tL\tS\tR2}_{\bar\ell\ell\nu,xyz}
-2C^{\tS\tR\tS\tL\tS\tR3}_{\bar\ell\ell\nu,xyz}
\nn\\
&+\kappa_2 (C^{\tS\tR\tS\tL\tS\tL1}_{\bar\ell\ell\nu,xyz}
-2C^{\tS\tR\tS\tL\tS\tL2}_{\bar\ell\ell\nu,xyz})\big],
\nn\\
\calK^{\tV\tR\tV\tR\tS,\Lambda^0}_{\bar\ell\ell\nu,xyz}
\equiv\,& \frac{1}{\sqrt{6}}
\big[C^{\tV\tR\tV\tR\tS\tL1}_{\bar\ell\ell\nu,xyz}
+C^{\tV\tR\tV\tR\tS\tL2}_{\bar\ell\ell\nu,xyz}
-2C^{\tV\tR\tV\tR\tS\tL3}_{\bar\ell\ell\nu,xyz}
\nn\\
&+\kappa_2 (C^{\tV\tR\tV\tR\tS\tR1}_{\bar\ell\ell\nu,xyz}
-2C^{\tV\tR\tV\tR\tS\tR2}_{\bar\ell\ell\nu,xyz})\big],
\nn\\
\calK^{\tV\tL\tV\tR\tS,\Lambda^0}_{\bar\ell\ell\nu,xyz}
\equiv\,& \frac{1}{\sqrt{6}}
\big[C^{\tV\tL\tV\tR\tS\tL1}_{\bar\ell\ell\nu,xyz}
+C^{\tV\tL\tV\tR\tS\tL2}_{\bar\ell\ell\nu,xyz}
-2C^{\tV\tL\tV\tR\tS\tL3}_{\bar\ell\ell\nu,xyz}
\nn\\
&+\kappa_2 (C^{\tV\tL\tV\tR\tS\tR1}_{\bar\ell\ell\nu,xyz}
-2C^{\tV\tL\tV\tR\tS\tR2}_{\bar\ell\ell\nu,xyz}
)\big],
\nn\\
\calK^{\tT\tR\tV\tR\tV\tL,\Lambda^0}_{\bar\ell\ell\nu,xyz}
\equiv\,& -({\sqrt{6}\kappa_{3\chi}}/4)
(C^{\tT\tR\tV\tR\tV\tL2}_{\bar\ell\ell\nu,xyz}
-C^{\tT\tR\tV\tR\tV\tL1}_{\bar\ell\ell\nu,xyz}),
\nn\\
\calK^{\tV\tR\tS\tL\tV\tR,\Lambda^0}_{\bar\ell\ell\nu,xyz}
\equiv\,&-({\sqrt{6}\kappa_{3\chi}}/4)
(C^{\tV\tR\tS\tL\tV\tR2}_{\bar\ell\ell\nu,xyz}
-C^{\tV\tR\tS\tL\tV\tR1}_{\bar\ell\ell\nu,xyz}),
\nn\\
\calK^{\tV\tL\tS\tL\tV\tR,\Lambda^0}_{\bar\ell\ell\nu,xyz}
\equiv\,& -({\sqrt{6}\kappa_{3\chi}}/4)
(C^{\tV\tL\tS\tL\tV\tR2}_{\bar\ell\ell\nu,xyz}
-C^{\tV\tL\tS\tL\tV\tR1}_{\bar\ell\ell\nu,xyz}), 
\\
\calK^{\tS\tL\tS\tL\tS,\Sigma^0}_{\bar\ell\ell\nu,xyz}
\equiv\,& \frac{1}{\sqrt{2}}
(C^{\tS\tL\tS\tL\tS\tR1}_{\bar\ell\ell\nu,xyz}
-C^{\tS\tL\tS\tL\tS\tR2}_{\bar\ell\ell\nu,xyz}
 -\kappa_2 C^{\tS\tL\tS\tL\tS\tL1}_{\bar\ell\ell\nu,xyz}),
\nn\\
\calK^{\tT\tL\tT\tL\tS,\Sigma^0}_{\bar\ell\ell\nu,xyz}
\equiv\,& \frac{1}{\sqrt{2}}
(C^{\tT\tL\tT\tL\tS\tR1}_{\bar\ell\ell\nu,xyz}
-C^{\tT\tL\tT\tL\tS\tR2}_{\bar\ell\ell\nu,xyz}
-\kappa_2 C^{\tT\tL\tT\tL\tS\tL1}_{\bar\ell\ell\nu,xyz}),
\nn\\
\calK^{\tS\tR\tS\tL\tS,\Sigma^0}_{\bar\ell\ell\nu,xyz}
\equiv\,& \frac{1}{\sqrt{2}}
(C^{\tS\tR\tS\tL\tS\tR1}_{\bar\ell\ell\nu,xyz}
-C^{\tS\tR\tS\tL\tS\tR2}_{\bar\ell\ell\nu,xyz}
 -\kappa_2 C^{\tS\tR\tS\tL\tS\tL1}_{\bar\ell\ell\nu,xyz}),
\nn\\
\calK^{\tV\tR\tV\tR\tS,\Sigma^0}_{\bar\ell\ell\nu,xyz}
\equiv\,& \frac{1}{\sqrt{2}}
(C^{\tV\tR\tV\tR\tS\tL1}_{\bar\ell\ell\nu,xyz}
-C^{\tV\tR\tV\tR\tS\tL2}_{\bar\ell\ell\nu,xyz}
-\kappa_2 C^{\tV\tR\tV\tR\tS\tR1}_{\bar\ell\ell\nu,xyz}),
\nn\\
\calK^{\tV\tL\tV\tR\tS,\Sigma^0}_{\bar\ell\ell\nu,xyz}
\equiv\,& \frac{1}{\sqrt{2}}
(C^{\tV\tL\tV\tR\tS\tL1}_{\bar\ell\ell\nu,xyz}
-C^{\tV\tL\tV\tR\tS\tL2}_{\bar\ell\ell\nu,xyz}
-\kappa_2 C^{\tV\tL\tV\tR\tS\tR1}_{\bar\ell\ell\nu,xyz}),
\nn\\
\calK^{\tT\tR\tV\tR\tV\tL,\Sigma^0}_{\bar\ell\ell\nu,xyz}
\equiv\,& \frac{\kappa_{3\chi}}{2\sqrt{2}}
(2C^{\tT\tR\tV\tR\tV\tL3}_{\bar\ell\ell\nu,xyz}
-C^{\tT\tR\tV\tR\tV\tL2}_{\bar\ell\ell\nu,xyz}
-C^{\tT\tR\tV\tR\tV\tL1}_{\bar\ell\ell\nu,xyz}),
\nn\\
\calK^{\tV\tR\tS\tL\tV\tR,\Sigma^0}_{\bar\ell\ell\nu,xyz}
\equiv\,& \frac{\kappa_{3\chi}}{2\sqrt{2}}
(2C^{\tV\tR\tS\tL\tV\tR3}_{\bar\ell\ell\nu,xyz}
-C^{\tV\tR\tS\tL\tV\tR2}_{\bar\ell\ell\nu,xyz}
-C^{\tV\tR\tS\tL\tV\tR1}_{\bar\ell\ell\nu,xyz}),
\nn\\
\calK^{\tV\tL\tS\tL\tV\tR,\Sigma^0}_{\bar\ell\ell\nu,xyz}
\equiv\,& \frac{\kappa_{3\chi}}{2\sqrt{2}}
(2C^{\tV\tL\tS\tL\tV\tR3}_{\bar\ell\ell\nu,xyz}
-C^{\tV\tL\tS\tL\tV\tR2}_{\bar\ell\ell\nu,xyz}
-C^{\tV\tL\tS\tL\tV\tR1}_{\bar\ell\ell\nu,xyz}).
\nn
\end{align}

\noindent$\bullet~p(p)\to\ell_x^-(p_1)\ell_y^+(p_2)\bar\nu_z(p_3)K^+(p_4)$ with $x,y=e,\mu$ and $z=e,\mu,\tau$:
\begin{align}
\mathcal{M} =\, &
\sum_{B=\Lambda^0,\Sigma^0}\Big\{
\calK_{\bar\ell \ell \nu,xyz}^{\tT\tL\tT\tL\tS,B}
(\overline{u_1} \sigma_{\mu\nu}P_\tL v_2)
(\overline{u_3}\sigma^{\mu\nu}P_\tL U_B)
\nn\\
&+[\overline{u_1}
(\calK_{\bar\ell \ell \nu,xyz}^{\tS\tL\tS\tL\tS,B} P_\tL 
+ \calK_{\bar\ell\ell\nu,xyz}^{\tS\tR\tS\tL\tS,B} P_\tR)
v_2]
[\overline{u_3}P_\tL U_B ]
\nn\\
&- [\overline{u_1}\gamma_\mu 
(\calK_{\bar\ell \ell \nu,xyz}^{\tV\tL\tV\tR\tS,B} P_\tL
+\calK_{\bar\ell \ell \nu,xyz}^{\tV\tR\tV\tR\tS,B} P_\tR) 
v_2]
(\overline{u_3}\gamma^\mu P_\tR U_B)
\nn\\
&- \calK_{\bar\ell \ell \nu,xyz}^{\tT\tR\tV\tR\tV\tL,B}
(\overline{u_1} \sigma_{\nu\mu}P_\tR v_2)
(\overline{u_3}\gamma^\nu \tilde p^\mu_B P_\tR U_B)
\nn\\
&+[\overline{u_1} \gamma_{\mu}
(\calK_{\bar\ell \ell \nu,xyz}^{\tV\tL\tS\tL\tV\tR,B}P_\tL
+ \calK_{\bar\ell \ell \nu,xyz}^{\tV\tR\tS\tL\tV\tR,B} P_\tR) v_2]
\nn\\
&\times [\overline{u_3} \tilde p^\mu_B P_\tL U_B ] \Big\}\frac{ic_1}{F_0}C_{pBK^+}
\\
&+\frac{ic_1}{F_0} \Big\{
[\overline{u_1}( \calC_{\bar\ell \ell \nu,xyz}^{\tS\tL\tS\tL\tS,p} P_\tL 
+\calC_{\bar\ell\ell\nu,xyz}^{\tS\tR\tS\tL\tS,p} P_\tR)v_2]
(\overline{u_3}P_\tL u_p)
\nn\\
&+ [\overline{u_1}\gamma_\mu 
(\calC_{\bar\ell \ell \nu,xyz}^{\tV\tL\tV\tR\tS,p} P_\tL
+\calC_{\bar\ell \ell \nu,xyz}^{\tV\tR\tV\tR\tS,p} P_\tR) 
v_2]
(\overline{u_3}\gamma^\mu P_\tR u_p)
\nn\\
&+\calC_{\bar\ell \ell \nu,xyz}^{\tT\tL\tT\tL\tS,p}
(\overline{u_1} \sigma_{\mu\nu}P_\tL v_2)
(\overline{u_3}\sigma^{\mu\nu}P_\tL u_p)
\nn\\
&+[\overline{u_1} \gamma_{\mu}
(\calC_{\bar\ell\ell\nu,xyz}^{\tV\tL\tS\tL\tV\tR,p}P_\tL
+\calC_{\bar\ell\ell\nu,xyz}^{\tV\tR\tS\tL\tV\tR,p}P_\tR)
v_2]
(\overline{u_3} \tilde p^\mu P_\tL u_p)
\nn\\
&+\calC_{\bar\ell \ell \nu,xyz}^{\tT\tR\tV\tR\tV\tL,p}
(\overline{u_1} \sigma_{\nu\mu}P_\tR v_2)
(\overline{u_3}\gamma^\nu \tilde p^\mu P_\tR u_p)
\nn\\
&-\frac{\sqrt{2}}{6}\kappa_{4\chi}
C^{\tT\tL\tS\tL\tT\tL}_{\bar\ell\ell\nu,xyz}
(\overline{u_1}\sigma_{\mu\nu} P_\tL v_2)
(\overline{u_3}\hat{\Gamma}^{\tL}_{\mu\nu\alpha\beta} u_p) p^\alpha p_4^\beta
\Big\},
\nn
\end{align}
where $u_3=v_3^\C$.

\noindent$\bullet~n(p)\to\ell_x^-(p_1)\ell_y^+(p_2)\bar\nu_z(p_3)K^0(p_4)$ with $x,y=e,\mu$ and $z=e,\mu,\tau$:
\begin{align}
\mathcal{M} =\, &\,
\sum_{B=\Lambda^0,\Sigma^0}\Big\{
\calK_{\bar\ell \ell \nu,xyz}^{\tT\tL\tT\tL\tS,B}
(\overline{u_1} \sigma_{\mu\nu}P_\tL v_2)
(\overline{u_3}\sigma^{\mu\nu}P_\tL U_B)
\nn\\
&+[\overline{u_1}( \calK_{\bar\ell \ell \nu,xyz}^{\tS\tL\tS\tL\tS,B} P_\tL 
+\calK_{\bar\ell \ell \nu,xyz}^{\tS\tR\tS\tL\tS,B} P_\tR) v_2]
(\overline{u_3}P_\tL U_B)
\nn\\
&- [\overline{u_1}\gamma_\mu 
(\calK_{\bar\ell \ell \nu,xyz}^{\tV\tL\tV\tR\tS,B} P_\tL
+\calK_{\bar\ell \ell \nu,xyz}^{\tV\tR\tV\tR\tS,B} P_\tR) 
v_2]
(\overline{u_3}\gamma^\mu P_\tR U_B)
\nn\\
&-\calK_{\bar\ell \ell \nu,xyz}^{\tT\tR\tV\tR\tV\tL,B}
(\overline{u_1} \sigma_{\nu\mu}P_\tR v_2)
(\overline{u_3}\gamma^\nu \tilde p^\mu_B P_\tR U_B)
\nn\\
& +[\overline{u_1} \gamma_{\mu}
(\calK_{\bar\ell\ell\nu,xyz}^{\tV\tL\tS\tL\tV\tR,B}P_\tL
+\calK_{\bar\ell\ell\nu,xyz}^{\tV\tR\tS\tL\tV\tR,B} P_\tR)
v_2]
\nn\\
&\times
(\overline{u_3} \tilde p^\mu_B P_\tL U_B)
\Big\}\frac{ic_1}{F_0}C_{nBK^0}
\\
&+\frac{ic_1}{F_0}\Big\{
[\overline{u_1}( 
\calC_{\bar\ell\ell\nu,xyz}^{\tS\tL\tS\tL\tS,n} P_\tL 
+ \calC_{\bar\ell\ell\nu,xyz}^{\tS\tR\tS\tL\tS,n} P_\tR)v_2]
(\overline{u_3}P_\tL u_n)
\nn\\
&+ [\overline{u_1}\gamma_\mu 
(\calC_{\bar\ell \ell \nu,xyz}^{\tV\tL\tV\tR\tS,n} P_\tL
+\calC_{\bar\ell \ell \nu,xyz}^{\tV\tR\tV\tR\tS,n} P_\tR) 
v_2]
(\overline{u_3}\gamma^\mu P_\tR u_n)
\nn\\
&+\calC_{\bar\ell \ell \nu,xyz}^{\tT\tL\tT\tL\tS,n}
(\overline{u_1} \sigma_{\mu\nu}P_\tL v_2)
(\overline{u_3}\sigma^{\mu\nu}P_\tL u_n)
\nn\\
&+[\overline{u_1} \gamma_{\mu}
(\calC_{\bar\ell \ell \nu,xyz}^{\tV\tL\tS\tL\tV\tR,n}P_\tL
+\calC_{\bar\ell \ell \nu,xyz}^{\tV\tR\tS\tL\tV\tR,n} P_\tR)
v_2]
(\overline{u_3} \tilde p^\mu P_\tL u_n)
\nn\\
&+\calC_{\bar\ell \ell \nu,xyz}^{\tT\tR\tV\tR\tV\tL,n}
(\overline{u_1} \sigma_{\nu\mu}P_\tR v_2)
(\overline{u_3}\gamma^\nu \tilde p^\mu P_\tR u_n)
\nn\\
&+\frac{\sqrt{2}}{6}\kappa_{4\chi}
C^{\tT\tL\tS\tL\tT\tL}_{\bar\ell\ell\nu,xyz}
(\overline{u_1}\sigma_{\mu\nu} P_\tL v_2)
(\overline{u_3}\hat{\Gamma}^{\tL}_{\mu\nu\alpha\beta} u_n) p^\alpha p_4^\beta
\Big\},
\nn
\end{align}
where $u_3=v_3^\C$.

\subsection{\tf{$\Delta L = +1$: $p\to\bar\nu_x\bar\nu_y\nu_zK^+,\,n\to\bar\nu_x\bar\nu_y\nu_z K^0$ (c)}{p2vbvbvKp(c)}}

The relevant WC combinations are:
\begin{align}
\calK^{\tS\tL\tS\tR\tS,\Lambda^0}_{\nu\nu\bar\nu,xyz}
\equiv\,&\frac{1}{\sqrt{6}}\big[C^{\tS\tL\tS\tR\tS\tL1}_{\nu\nu\bar\nu,xyz}+C^{\tS\tL\tS\tR\tS\tL2}_{\nu\nu\bar\nu,xyz}
-2C^{\tS\tL\tS\tR\tS\tL3}_{\nu\nu\bar\nu,xyz}
\nn\\
&+\kappa_2 (C^{\tS\tL\tS\tR\tS\tR1}_{\nu\nu\bar\nu,xyz}
-2C^{\tS\tL\tS\tR\tS\tR2}_{\nu\nu\bar\nu,xyz})\big],
\nn\\
\calK^{\tT\tL\tV\tL\tV\tR,\Lambda^0}_{\nu\nu\bar\nu,xyz}
\equiv\,&-(\sqrt{6}\kappa_{3\chi}/4)(C^{\tT\tL\tV\tL\tV\tR2}_{\nu\nu\bar\nu,xyz}
-C^{\tT\tL\tV\tL\tV\tR1}_{\nu\nu\bar\nu,xyz}
),
\nn\\
\calK^{\tS\tL\tS\tR\tS,\Sigma^0}_{\nu\nu\bar\nu,xyz}
\equiv\,&\frac{1}{\sqrt{2}}(C^{\tS\tL\tS\tR\tS\tL1}_{\nu\nu\bar\nu,xyz}-C^{\tS\tL\tS\tR\tS\tL2}_{\nu\nu\bar\nu,xyz}
-\kappa_2 C^{\tS\tL\tS\tR\tS\tR1}_{\nu\nu\bar\nu,xyz}
),
\nn\\
\calK^{\tT\tL\tV\tL\tV\tR,\Sigma^0}_{\nu\nu\bar\nu,xyz}
\equiv\,&\frac{\kappa_{3\chi}}{2\sqrt{2}}(2C^{\tT\tL\tV\tL\tV\tR3}_{\nu\nu\bar\nu,xyz}-C^{\tT\tL\tV\tL\tV\tR2}_{\nu\nu\bar\nu,xyz}-C^{\tT\tL\tV\tL\tV\tR1}_{\nu\nu\bar\nu,xyz}).
\nn\\%
\calC^{\tS\tL\tS\tR\tS ,p}_{\nu\nu\bar\nu,xyz}
\equiv\,&\frac{1}{\sqrt{2}}(
C^{\tS\tL\tS\tR\tS\tL1}_{\nu\nu\bar\nu,xyz}
+C^{\tS\tL\tS\tR\tS\tL3}_{\nu\nu\bar\nu,xyz}
+\kappa_2 C^{\tS\tL\tS\tR\tS\tR2}_{\nu\nu\bar\nu,xyz}),
\nn\\
\calC^{\tT\tL\tV\tL\tV\tR ,p}_{\nu\nu\bar\nu,xyz}
\equiv\,&-\frac{\kappa_{3\chi}}{2\sqrt{2}}
(C^{\tT\tL\tV\tL\tV\tR1}_{\nu\nu\bar\nu,xyz}
-2C^{\tT\tL\tV\tL\tV\tR2}_{\nu\nu\bar\nu,xyz}
-C^{\tT\tL\tV\tL\tV\tR3}_{\nu\nu\bar\nu,xyz}),
\nn\\
\calC^{\tS\tL\tS\tR\tS ,n}_{\nu\nu\bar\nu,xyz}
\equiv\,&\frac{1}{\sqrt{2}}\big[C^{\tS\tL\tS\tR\tS\tL2}_{\nu\nu\bar\nu,xyz}
+C^{\tS\tL\tS\tR\tS\tL3}_{\nu\nu\bar\nu,xyz}
\nn\\
&-\kappa_2 
(C^{\tS\tL\tS\tR\tS\tR1}_{\nu\nu\bar\nu,xyz}
-C^{\tS\tL\tS\tR\tS\tR2}_{\nu\nu\bar\nu,xyz})\big],
\\
\calC^{\tT\tL\tV\tL\tV\tR ,n}_{\nu\nu\bar\nu,xyz}
\equiv\,&-\frac{\kappa_{3\chi}}{2\sqrt{2}}(2C^{\tT\tL\tV\tL\tV\tR1}_{\nu\nu\bar\nu,xyz}
-C^{\tT\tL\tV\tL\tV\tR2}_{\nu\nu\bar\nu,xyz}
+C^{\tT\tL\tV\tL\tV\tR3}_{\nu\nu\bar\nu,xyz}).
\nn
\end{align}
\noindent$\bullet~p(p)\to\bar\nu_x(p_1)\bar\nu_{y}(p_2)\nu_z(p_3)K^+(p_4)$ with $x,y,z=e,\mu,\tau$:
\begin{align}
\mathcal{M} =\, &\,\Big\{
\sum_{B=\Lambda^0,\Sigma^0}C_{pBK^+}
[-\calK_{\nu\nu\bar\nu,xyz}^{\tS\tL\tS\tR\tS,B}
(\overline{u_1} P_\tL v_2)
(\overline{u_3} P_\tR U_B)
\nn\\
& +\calK_{\nu\nu\bar\nu,xyz}^{\tT\tL\tV\tL\tV\tR,B}
(\overline{u_1} \sigma_{\nu\mu}P_\tL v_2)
(\overline{u_3}\gamma^\nu \tilde p_B^\mu P_\tL U_B)]
\nn\\
&+\calC_{\nu\nu\bar\nu,xyz}^{\tT\tL\tV\tL\tV\tR,p}
(\overline{u_1} \sigma_{\nu\mu}P_\tL v_2)
(\overline{u_3}\gamma^\nu \tilde p^\mu P_\tL u_p) 
\nn\\
& +\calC_{\nu\nu\bar\nu,xyz}^{\tS\tL\tS\tR\tS,p}
(\overline{u_1} P_\tL v_2)
(\overline{u_3} P_\tR u_p)
\nn\\
&-\delta_{x,y}(p_1\leftrightarrow p_2)\Big\}\frac{ic_1}{F_0},
\end{align}
where $u_{1,2}=v_{1,2}^\C$.

\noindent$\bullet~n(p)\to\bar\nu_x(p_1)\bar\nu_{y}(p_2)\nu_z(p_3)K^0(p_4)$ with $x,y,z=e,\mu,\tau$:
\begin{align}
\mathcal{M} =\, &\,\Big\{
\sum_{B=\Lambda^0,\Sigma^0}C_{nBK^0}
[-\calK_{\nu\nu\bar\nu,xyz}^{\tS\tL\tS\tR\tS,B}
(\overline{u_1} P_\tL v_2) 
(\overline{u_3} P_\tR U_B)
\nn\\
&+\calK_{\nu\nu\bar\nu,xyz}^{\tT\tL\tV\tL\tV\tR,B}
(\overline{u_1} \sigma_{\nu\mu}P_\tL v_2)
(\overline{u_3}\gamma^\nu \tilde p_B^\mu P_\tL U_B)]
\nn\\
&+\calC_{\nu\nu\bar\nu,xyz}^{\tT\tL\tV\tL\tV\tR,n}
(\overline{u_1} \sigma_{\nu\mu}P_\tL v_2)
(\overline{u_3}\gamma^\nu \tilde p^\mu P_\tL u_n)
\nn\\
&+\calC_{\nu\nu\bar\nu,xyz}^{\tS\tL\tS\tR\tS,n}
(\overline{u_1} P_\tL v_2)\,(\overline{u_3} P_\tR u_n)  
\nn\\
&-\delta_{x,y}(p_1\leftrightarrow p_2)\Big\}\frac{ic_1}{F_0},
\end{align}
where $u_{1,2}=v_{1,2}^\C$.

\subsection{\tf{$\Delta L = -1$: 
$p\to\nu_x\nu_y\ell_z^+K^0$ (d)}{p2vvlpK0(d)}}

The relevant WC combinations are:
\begin{align}
\calK(\calC)^{\tS\tR\tS\tL\tS}_{\bar\nu\bar\nu\ell,xyz}
&\equiv C^{\tS\tR\tS\tL\tS\tR}_{\bar\nu\bar\nu\ell,xyz}
\pm\kappa_2 C^{\tS\tR\tS\tL\tS\tL}_{\bar\nu\bar\nu\ell,xyz},
\nn\\
\calK(\calC)^{\tS\tR\tS\tR\tS}_{\bar\nu\bar\nu\ell,xyz}
&\equiv C^{\tS\tR\tS\tR\tS\tL}_{\bar\nu\bar\nu\ell,xyz}
\pm\kappa_2 C^{\tS\tR\tS\tR\tS\tR}_{\bar\nu\bar\nu\ell,xyz},
\nn\\
\calK(\calC)^{\tT\tR\tT\tR\tS}_{\bar\nu\bar\nu\ell,xyz}
&\equiv C^{\tT\tR\tT\tR\tS\tL}_{\bar\nu\bar\nu\ell,xyz}
\pm\kappa_2 C^{\tT\tR\tT\tR\tS\tR}_{\bar\nu\bar\nu\ell,xyz},
\nn\\
\calK(\calC)^{\tT\tR\tV\tR\tV\tL}_{\bar\nu\bar\nu\ell,xyz}
&\equiv{\kappa_{3\chi}}(C^{\tT\tR\tV\tR\tV\tL1}_{\bar\nu\bar\nu\ell,xyz}/2
\mp C^{\tT\tR\tV\tR\tV\tL2}_{\bar\nu\bar\nu\ell,xyz}).
\end{align}

\noindent$\bullet~p(p)\to \nu_x(p_1) \nu_y(p_2) \ell_z^+(p_3)K^0(p_4)$ with $x,y=e,\mu,\tau$ and $z=e,\mu$:
\begin{align}
\mathcal{M} =\, &\Big\{
(\overline{u_1} P_\tR v_2)
[\overline{u_3}
(\calK_{\bar\nu\bar\nu\ell,xyz}^{\tS\tR\tS\tL\tS}P_\tL-\calK_{\bar\nu\bar\nu\ell,xyz}^{\tS\tR\tS\tR\tS} P_\tR
) U_{\Sigma^+}]
\nn\\
&- \calK_{\bar\nu\bar\nu\ell,xyz}^{\tT\tR\tV\tR\tV\tL}
(\overline{u_1}\sigma_{\nu\mu}P_\tR v_2)
(\overline{u_3}\gamma^\nu\tilde p^\mu_{\Sigma^+} P_\tR U_{\Sigma^+})
\nn\\
&- \calK_{\bar\nu\bar\nu\ell,xyz}^{\tT\tR\tT\tR\tS}
(\overline{u_1} \sigma_{\mu\nu}P_\tR v_2)
(\overline{u_3}\sigma^{\mu\nu} P_\tR U_{\Sigma^+})
\Big\}\frac{ic_1}{F_0}C_{p\Sigma^+K^0}
\nn\\
&+\frac{ic_1}{\sqrt{2}F_0}\Big\{
(\overline{u_1} P_\tR v_2)
[\overline{u_3}
(\calC_{\bar\nu\bar\nu\ell,xyz}^{\tS\tR\tS\tL\tS}P_\tL
+\calC_{\bar\nu\bar\nu\ell,xyz}^{\tS\tR\tS\tR\tS} P_\tR)u_p]
\nn\\
&- \calC_{\bar\nu\bar\nu\ell,xyz}^{\tT\tR\tV\tR\tV\tL}
(\overline{u_1} \sigma_{\nu\mu}P_\tR v_2)
(\overline{u_3} \gamma^\nu \tilde p^\mu P_\tR u_p)
\nn\\
&+\calC_{\bar\nu\bar\nu\ell,xyz}^{\tT\tR\tT\tR\tS}
(\overline{u_1} \sigma_{\mu\nu}P_\tR v_2)
(\overline{u_3}\sigma^{\mu\nu} P_\tR u_p)
\nn\\
&+\frac{2}{3}\kappa_{4\chi}C^{\tT\tR\tS\tR\tT\tR}_{\bar\nu\bar\nu\ell,xyz}
(\overline{u_1}\sigma_{\mu\nu} P_\tR v_2) 
(\overline{u_3}\hat{\Gamma}^{\tR}_{\mu\nu\alpha\beta} u_p)
p^\alpha p_4^\beta\Big\}
\nn\\
&-\delta_{x,y}(p_1\leftrightarrow p_2).
\end{align}
where $v_{1,2}=u_{1,2}^\C$ and $u_3=v_3^\C$.

\subsection{\tf{$\Delta L = -1$: 
$n\to\ell_x^-\ell_y^-\ell_z^+K^+$}{n2lmlmmpKp}}

The relevant WC combinations are:
\begin{align}
\calK(\calC)^{\tS\tL\tS\tL\tS}_{\bar\ell\bar\ell\ell,xyz}
&\equiv C^{\tS\tL\tS\tL\tS\tR}_{\bar\ell\bar\ell\ell,xyz}
\pm\kappa_2 C^{\tS\tL\tS\tL\tS\tL}_{\bar\ell\bar\ell\ell,xyz},
\nn\\
\calK(\calC)^{\tS\tR\tS\tL\tS}_{\bar\ell\bar\ell\ell,xyz}&\equiv C^{\tS\tR\tS\tL\tS\tR}_{\bar\ell\bar\ell\ell,xyz}
\pm\kappa_2 C^{\tS\tR\tS\tL\tS\tL}_{\bar\ell\bar\ell\ell,xyz},
\nn\\
\calK(\calC)^{\tV\tR\tV\tL\tS}_{\bar\ell\bar\ell\ell,xyz}
&\equiv C^{\tV\tR\tV\tL\tS\tR}_{\bar\ell\bar\ell\ell,xyz}
\pm\kappa_2 C^{\tV\tR\tV\tL\tS\tL}_{\bar\ell\bar\ell\ell,xyz},
\nn\\
\calK(\calC)^{\tT\tL\tT\tL\tS}_{\bar\ell\bar\ell\ell,xyz}
&\equiv C^{\tT\tL\tT\tL\tS\tR}_{\bar\ell\bar\ell\ell,xyz}
\pm\kappa_2 C^{\tT\tL\tT\tL\tS\tL}_{\bar\ell\bar\ell\ell,xyz},
\nn\\%
\calK(\calC)^{\tV\tL\tS\tL\tV\tR}_{\bar\ell\bar\ell\ell,xyz}
&\equiv \kappa_{3\chi}
(C^{\tV\tL\tS\tL\tV\tR1}_{\bar\ell\bar\ell\ell,xyz}/2
\mp
C^{\tV\tL\tS\tL\tV\tR2}_{\bar\ell\bar\ell\ell,xyz}),
\nn\\
\calK(\calC)^{\tT\tL\tV\tL\tV\tR}_{\bar\ell\bar\ell\ell,xyz}&
\equiv \kappa_{3\chi}
(C^{\tT\tL\tV\tL\tV\tR1}_{\bar\ell\bar\ell\ell,xyz}/2
\mp
C^{\tT\tL\tV\tL\tV\tR2}_{\bar\ell\bar\ell\ell,xyz}),
\end{align}
and their chiral partners with $\tL\leftrightarrow\tR$.

\noindent$\bullet~n(p)\to e^-(p_1)\mu^-(p_2)\ell_x^+(p_3)K^+(p_4)$ with $x=e,\mu$:
\begin{align}
\mathcal{M} =\, &
\Big\{
[\overline{u_1} 
(\calK_{\bar\ell\bar\ell\ell,e\mu x}^{\tS\tL\tS\tL\tS} P_\tL
+\calK_{\bar\ell\bar\ell\ell,e\mu x}^{\tS\tR\tS\tL\tS} P_\tR) v_2]
(\overline{u_3}P_\tL U_{\Sigma^-})
\nn\\
&+\calK_{\bar\ell\bar\ell\ell,e\mu x}^{\tT\tL\tT\tL\tS}
(\overline{u_1} \sigma_{\mu\nu}P_\tL v_2)
(\overline{u_3} \sigma^{\mu\nu}P_\tL U_{\Sigma^-})
\nn\\
&-\calK_{\bar\ell\bar\ell\ell,e\mu x}^{\tT\tL\tV\tL\tV\tR}
(\overline{u_1}\sigma_{\nu\mu}P_\tL v_2)
(\overline{u_3}\gamma^\nu\tilde{p}_{\Sigma^-}^\mu P_\tL U_{\Sigma^-})
\nn\\
&+[\calK_{\bar\ell\bar\ell\ell,e\mu x}^{\tV\tR\tV\tL\tS}(\overline{u_1}\gamma_\mu P_\tR v_2)
(\overline{u_3}\gamma^\mu P_\tL U_{\Sigma^-})
\nn\\
&-\calK_{\bar\ell\bar\ell\ell,e\mu x}^{\tV\tL\tS\tL\tV\tR}
(\overline{u_1} \gamma_{\mu}P_\tL v_2)
(\overline{u_3} \tilde{p}_{\Sigma^-}^\mu P_\tL U_{\Sigma^-})
\nn\\
&-(e,p_1)\leftrightarrow (\mu,p_2) ] 
-\tL\leftrightarrow\tR
\Big\} \frac{ic_1}{F_0}C_{n\Sigma^-K^+}
\nn\\
&+\Big\{
[\overline{u_1} 
(\calC_{\bar\ell\bar\ell\ell,e\mu x}^{\tS\tL\tS\tL\tS} P_\tL
+\calC_{\bar\ell\bar\ell\ell,e\mu x}^{\tS\tR\tS\tL\tS} P_\tR) v_2]
(\overline{u_3} P_\tL u_n)
\nn\\
&+\calC_{\bar\ell\bar\ell\ell,e\mu x}^{\tT\tL\tT\tL\tS}
(\overline{u_1} \sigma_{\mu\nu} P_\tL v_2)
(\overline{u_3} \sigma^{\mu\nu} P_\tL u_n)
\nn\\
&+\calC_{\bar\ell\bar\ell\ell,e\mu x}^{\tT\tL\tV\tL\tV\tR}
(\overline{u_1} \sigma_{\nu\mu}P_\tL v_2)
(\overline{u_3} \gamma^\nu \tilde p^\mu P_\tL u_n)
\nn\\
&+[\calC_{\bar\ell\bar\ell\ell,e\mu x}^{\tV\tL\tS\tL\tV\tR}
(\overline{u_1} \gamma_{\mu}P_\tL v_2)
(\overline{u_3} \tilde p^\mu P_\tL u_n)
+\calC_{\bar\ell\bar\ell\ell,e\mu x}^{\tV\tR\tV\tL\tS}
\nn\\
&\times(\overline{u_1}\gamma_\mu P_\tR v_2)
(\overline{u_3}\gamma^\mu P_\tL u_n)
-(e,p_1)\leftrightarrow (\mu,p_2)]
\nn\\
&-\frac{2}{3}\kappa_{4\chi}
C^{\tT\tL\tS\tL\tT\tL}_{\bar\ell\bar\ell\ell,e\mu x}
(\overline{u_1}\sigma_{\mu\nu} P_\tL v_2) 
(\overline{u_3}\hat{\Gamma}^{\tL}_{\mu\nu\alpha\beta} u_n) p^\alpha p_4^\beta
\nn\\
&+\tL\leftrightarrow\tR \Big\}\frac{ic_1}{\sqrt{2}F_0},
\end{align}
where $v_{1,2}=u_{1,2}^\C$ and $u_3=v_3^\C$.

\noindent$\bullet~n(p)\to\ell_x^-(p_1)\ell_x^-(p_2)\ell_y^+(p_3)K^+(p_4)$ with $x,y=e,\mu$:
\begin{align}
\mathcal{M} =\, &\Big\{
\Big[
[\overline{u_1} 
(\calK_{\bar\ell\bar\ell\ell,xxy}^{\tS\tL\tS\tL\tS} P_\tL
+\calK_{\bar\ell\bar\ell\ell,xxy}^{\tS\tR\tS\tL\tS} P_\tR) v_2]
(\overline{u_3} P_\tL U_{\Sigma^-})
\nn\\
&- \calK_{\bar\ell\bar\ell\ell,xxy}^{\tV\tL\tS\tL\tV\tR}
(\overline{u_1} \gamma_{\mu}P_\tL v_2)
(\overline{u_3} \tilde p_{\Sigma^-}^\mu P_\tL U_{\Sigma^-})
\nn\\
&+\calK_{\bar\ell\bar\ell\ell,xxy}^{\tV\tR\tV\tL\tS}
(\overline{u_1}\gamma_\mu P_\tR v_2)
(\overline{u_3}\gamma^\mu P_\tL U_{\Sigma^-}) 
-(p_1\leftrightarrow p_2)\Big]
\nn\\
&-\tL\leftrightarrow\tR
\Big\}\frac{ic_1}{F_0}C_{n\Sigma^-K^+}
\nn\\
&+\Big\{\Big[
[\overline{u_1} 
(\calC_{\bar\ell\bar\ell\ell,xxy}^{\tS\tL\tS\tL\tS} P_\tL
+\calC_{\bar\ell\bar\ell\ell,xxy}^{\tS\tR\tS\tL\tS} P_\tR) v_2]
(\overline{u_3} P_\tL u_n)
\nn\\
&+ \calC_{\bar\ell\bar\ell\ell,xxy}^{\tV\tL\tS\tL\tV\tR}
(\overline{u_1} \gamma_{\mu}P_\tL v_2)
(\overline{u_3} \tilde p^\mu P_\tL u_n)
\nn\\
&+\calC_{\bar\ell\bar\ell\ell,xxy}^{\tV\tR\tV\tL\tS}
(\overline{u_1}\gamma_\mu P_\tR v_2)
(\overline{u_3}\gamma^\mu P_\tL u_n) -(p_1\leftrightarrow p_2)\Big]
\nn\\
&
+\tL\leftrightarrow\tR
\Big\}\frac{ic_1}{\sqrt{2}F_0},
\end{align}
where $v_{1,2}=u_{1,2}^\C$ and $u_3=v_3^\C$.

\subsection{\tf{$\Delta L = -1$: 
$n\to\nu_x\bar\nu_y\ell_z^-K^+$}{n2vvblmKp}}

The relevant WC combinations are:
\begin{align}
\calK(\calC)^{\tV\tL\tV\tL\tS}_{\bar\nu\nu\bar\ell,xyz}
&\equiv C^{\tV\tL\tV\tL\tS\tR}_{\bar\nu\nu\bar\ell,xyz}
\pm \kappa_2 C^{\tV\tL\tV\tL\tS\tL}_{\bar\nu\nu\bar\ell,xyz},
\nn\\
\calK(\calC)^{\tV\tL\tV\tR\tS}_{\bar\nu\nu\bar\ell,xyz}
&\equiv C^{\tV\tL\tV\tR\tS\tL}_{\bar\nu\nu\bar\ell,xyz}
\pm\kappa_2 C^{\tV\tL\tV\tR\tS\tR}_{\bar\nu\nu\bar\ell,xyz},
\nn\\%
\calK(\calC)^{\tV\tL\tS\tL\tV\tR}_{\bar\nu\nu\bar\ell,xyz}
&\equiv \kappa_{3\chi}
(C^{\tV\tL\tS\tL\tV\tR1}_{\bar\nu\nu\bar\ell,xyz}/2
\mp
C^{\tV\tL\tS\tL\tV\tR2}_{\bar\nu\nu\bar\ell,xyz}),
\nn\\
\calK(\calC)^{\tV\tL\tS\tR\tV\tL}_{\bar\nu\nu\bar\ell,xyz}
&\equiv \kappa_{3\chi}
(C^{\tV\tL\tS\tR\tV\tL1}_{\bar\nu\nu\bar\ell,xyz}/2
\mp
C^{\tV\tL\tS\tR\tV\tL2}_{\bar\nu\nu\bar\ell,xyz}).
\end{align}

\noindent$\bullet~n(p)\to\nu_x(p_1)\bar\nu_y(p_2)\ell_z^-(p_3)K^+(p_4)$ with $x,y=e,\mu,\tau$ and $z=e,\mu$:
\begin{align}
\mathcal{M} =\,&
\frac{ic_1}{F_0}C_{n\Sigma^- K^+}
(\overline{u_1} \gamma_\mu P_\tL v_2)
\nn\\
&\times \Big\{\overline{u_3}
[\gamma^\mu 
(\calK_{\bar\nu\nu\bar\ell,xyz}^{\tV\tL\tV\tL\tS} P_\tL
-\calK_{\bar\nu\nu\bar\ell,xyz}^{\tV\tL\tV\tR\tS} P_\tR)
\nn\\
&-\tilde p^\mu_{\Sigma^-}
(\calK_{\bar\nu\nu\bar\ell,xyz}^{\tV\tL\tS\tL\tV\tR} P_\tL
-\calK_{\bar\nu\nu\bar\ell,xyz}^{\tV\tL\tS\tR\tV\tL} P_\tR)] U_{\Sigma^-}\Big\}
\\
&+(\overline{u_1} \gamma_\mu P_\tL v_2)
\Big\{\overline{u_3}
[\gamma^\mu
(\calC_{\bar\nu\nu\bar\ell,xyz}^{\tV\tL\tV\tL\tS} P_\tL 
+\calC_{\bar\nu\nu\bar\ell,xyz}^{\tV\tL\tV\tR\tS} P_\tR)
\nn\\
&+\tilde p^\mu
(\calC_{\bar\nu\nu\bar\ell,xyz}^{\tV\tL\tS\tL\tV\tR} P_\tL
+\calC_{\bar\nu\nu\bar\ell,xyz}^{\tV\tL\tS\tR\tV\tL} P_\tR )] u_n\Big\} \frac{ic_1}{\sqrt{2}F_0}.
\nn
\end{align}

\subsection{$\Delta L = +3$: 
$p\to\bar\nu_x\bar\nu_y\bar\nu_zK^+,\, n\to\bar\nu_x\bar\nu_y\bar\nu_zK^0$ (e)}

For the five types of dim-9 operators $\tO_{\nu\nu\nu,xyz}^{\tS\tL\tS\tL\tS\tL1,2}$
and $\tO_{\nu\nu\nu,xyz}^{\tS\tL\tS\tL\tS\tR,1,2,3}$ contributing to the $\Delta L=+3$ processes $\tN\to\bar\nu_x\bar\nu_y\bar\nu_z K$, each type contains eight independent flavor-specific operators, as indicated in \cref{tab:dim9ope_L3}. 
For the process $\tN\to\bar\nu_e\bar\nu_\mu\bar\nu_\tau K$, two flavor-specific operators in each type contribute, while for $\tN\to\bar\nu_x\bar\nu_x\bar\nu_{y\neq x} K$, only one flavor-specific operator contributes. 
Therefore, we choose the following eight independent flavor combinations:   
$xyz=e\mu\tau,\mu\tau e, ee\mu,ee\tau,\mu\mu e,\mu\mu\tau,\tau\tau e,\tau\tau\mu$. 
A similar convention applies to the $\Delta L=-3$ processes  $\tN\to \nu_x \nu_y \nu_z K$. 

The relevant WC combinations are:
\begin{align}
\calC^{\tS\tL\tS\tL\tS ,p}_{\nu\nu\nu,xyz}&\equiv\frac{1}{\sqrt{2}}
(C_{\nu\nu\nu,xyz}^{\tS\tL\tS\tL\tS\tR1}
+C_{\nu\nu\nu,xyz}^{\tS\tL\tS\tL\tS\tR3}
+\kappa_2C_{\nu\nu\nu,xyz}^{\tS\tL\tS\tL\tS\tL2}),
\nn\\
\calC^{\tS\tL\tS\tL\tS ,n}_{\nu\nu\nu,xyz}&\equiv
\frac{1}{\sqrt{2}}\big[C_{\nu\nu\nu,xyz}^{\tS\tL\tS\tL\tS\tR2}
+C_{\nu\nu\nu,xyz}^{\tS\tL\tS\tL\tS\tR3}
\nn\\
&\qquad-\kappa_2(C_{\nu\nu\nu,xyz}^{\tS\tL\tS\tL\tS\tL1}-C_{\nu\nu\nu,xyz}^{\tS\tL\tS\tL\tS\tL2})\big],
\nn\\
\calK^{\tS\tL\tS\tL\tS,\Lambda^0}_{\nu\nu\nu,xyz}&\equiv
\frac{1}{\sqrt{6}}\big[
C_{\nu\nu\nu,xyz}^{\tS\tL\tS\tL\tS\tR1}
+C_{\nu\nu\nu,xyz}^{\tS\tL\tS\tL\tS\tR2}
-2C_{\nu\nu\nu,xyz}^{\tS\tL\tS\tL\tS\tR3}
\nn\\
&\qquad+
\kappa_2(C_{\nu\nu\nu,xyz}^{\tS\tL\tS\tL\tS\tL1}-2C_{\nu\nu\nu,xyz}^{\tS\tL\tS\tL\tS\tL2})\big],
\\
\calK^{\tS\tL\tS\tL\tS,\Sigma^0}_{\nu\nu\nu,xyz}&\equiv
\frac{1}{\sqrt{2}}(C_{\nu\nu\nu,xyz}^{\tS\tL\tS\tL\tS\tR1}-C_{\nu\nu\nu,xyz}^{\tS\tL\tS\tL\tS\tR2}
-\kappa_2C_{\nu\nu\nu,xyz}^{\tS\tL\tS\tL\tS\tL1}).
\nn
\end{align}

\noindent$\bullet~p(p)\to \bar\nu_e(p_1) \bar\nu_\mu(p_2) \bar\nu_\tau(p_3)K^+(p_4)$:
\begin{align}
\mathcal{M} =\, &\,
\sum_{B=\Lambda^0,\Sigma^0}\frac{ic_1}{2F_0}C_{pB K^+}
\nn\\
&\times\Big\{\calK_{\nu\nu\nu,e\mu\tau}^{\tS\tL\tS\tL\tS,B}
[(\overline{u_1} P_\tL v_2) 
(\overline{u_3} P_\tL U_B)
+(p_2\leftrightarrow p_3)]
\nn\\
&+ \calK_{\nu\nu\nu,\mu\tau e}^{\tS\tL\tS\tL\tS,B}
[(\overline{u_2} P_\tL v_3)
(\overline{u_1} P_\tL U_B)
+(p_1\leftrightarrow p_3)]\Big\}
\nn\\
&+\Big\{\calC_{\nu\nu\nu,e\mu\tau}^{\tS\tL\tS\tL\tS,p}
[(\overline{u_1} P_\tL v_2) 
(\overline{u_3} P_\tL u_p)
+(p_2\leftrightarrow p_3)]
\nn\\
&+ \calC_{\nu\nu\nu,\mu\tau e}^{\tS\tL\tS\tL\tS,p}
[(\overline{u_2} P_\tL v_3)
(\overline{u_1} P_\tL u_p)+(p_1\leftrightarrow p_3)]
\nn\\
&-\frac{\sqrt{2}}{9}\kappa_{4\chi}C^{\tT\tL\tS\tL\tT\tL}_{\nu\nu\nu,e\mu\tau}
[ (\overline{u_1}\sigma_{\mu\nu} P_\tL v_2) 
(\overline{u_3}\hat{\Gamma}^{\tL}_{\mu\nu\alpha\beta} u_p)
\nn\\
&+ (\overline{u_2}\sigma_{\mu\nu} P_\tL v_3) 
(\overline{u_1}\hat{\Gamma}^{\tL}_{\mu\nu\alpha\beta} u_p)
\nn\\
&+ (\overline{u_3}\sigma_{\mu\nu} P_\tL v_1)
(\overline{u_2}\hat{\Gamma}^{\tL}_{\mu\nu\alpha\beta} u_p) ] p^\alpha p_4^\beta
\Big\}\frac{ic_1}{2F_0},
\end{align}
where $u_{1,2,3}=v_{1,2,3}^\C$.

\noindent$\bullet~p(p)\to \bar\nu_x(p_1) \bar\nu_x(p_2) \bar\nu_{y\neq x}(p_3)K^+(p_4)$ with $x,y=e,\mu,\tau$:
\begin{align}
\mathcal{M} =\, &\,
\frac{ic_1}{2F_0} \Big\{ \sum_{B=\Lambda^0,\Sigma^0}C_{pB K^+} \calK_{\nu\nu\nu,xxy}^{\tS\tL\tS\tL\tS,B}
\nn\\
&\times [(\overline{u_1} P_\tL v_2) 
(\overline{u_3} P_\tL U_B)
+(\overline{u_1} P_\tL v_3)
(\overline{u_2} P_\tL U_B)]
\nn\\
&+\calC_{\nu\nu\nu,xxy}^{\tS\tL\tS\tL\tS,p} 
[(\overline{u_1} P_\tL v_2) 
(\overline{u_3} P_\tL u_p)
\nn\\
&+(\overline{u_1}P_\tL v_3)(\overline{u_2}P_\tL u_p)]\Big\}
-p_1\leftrightarrow p_2,
\end{align}
where $u_{1,2,3}=v_{1,2,3}^\C$.

\noindent$\bullet~n(p)\to \bar\nu_e(p_1) \bar\nu_\mu(p_2) \bar\nu_\tau(p_3)K^0(p_4)$:
\begin{align}
\mathcal{M} =\, &\,\sum_{B=\Lambda^0,\Sigma^0}\frac{ic_1}{2F_0}C_{nB K^0}
\nn\\
&\times \Big\{
\calK_{\nu\nu\nu,e\mu\tau}^{\tS\tL\tS\tL\tS,B}
[(\overline{u_1} P_\tL v_2)
(\overline{u_3} P_\tL U_B)
+(p_2\leftrightarrow p_3) ]
\nn\\
&+ \calK_{\nu\nu\nu,\mu\tau e}^{\tS\tL\tS\tL\tS,B}
[ (\overline{u_2} P_\tL v_3)
(\overline{u_1} P_\tL U_B)
+(p_1\leftrightarrow p_3) ]\Big\}
\nn\\
&+\Big\{ \calC_{\nu\nu\nu,e\mu\tau}^{\tS\tL\tS\tL\tS,n}
[ (\overline{u_1} P_\tL v_2)
(\overline{u_3} P_\tL u_n)
+(p_2\leftrightarrow p_3)]
\nn\\
&+ \calC_{\nu\nu\nu,\mu\tau e}^{\tS\tL\tS\tL\tS,n}
[ (\overline{u_2} P_\tL v_3)
(\overline{u_1} P_\tL u_n)+(p_1\leftrightarrow p_3) ]
\nn\\
&+\frac{\sqrt{2}}{9}\kappa_{4\chi}C^{\tT\tL\tS\tL\tT\tL}_{\nu\nu\nu,e\mu\tau}
[ (\overline{u_1}\sigma_{\mu\nu} P_\tL v_2) 
(\overline{u_3}\hat{\Gamma}^{\tL}_{\mu\nu\alpha\beta} u_n)
\nn\\
&+(\overline{u_2}\sigma_{\mu\nu} P_\tL v_3)
(\overline{u_1}\hat{\Gamma}^{\tL}_{\mu\nu\alpha\beta} u_n)
\nn\\
&+ (\overline{u_3}\sigma_{\mu\nu} P_\tL v_1) 
(\overline{u_2}\hat{\Gamma}^{\tL}_{\mu\nu\alpha\beta} u_n) ] p^\alpha p_4^\beta
\bigg\}\frac{ic_1}{2F_0},
\end{align}
where $u_{1,2,3}=v_{1,2,3}^\C$.

\noindent$\bullet~n(p)\to \bar\nu_x(p_1) \bar\nu_x(p_2) \bar\nu_{y\neq x}(p_3)K^0(p_4)$ with $x,y=e,\mu,\tau$:
\begin{align}
\mathcal{M} =\, &\,
\frac{ic_1}{2F_0}\Big\{
\sum_{B=\Lambda^0,\Sigma^0}C_{nBK^0} 
\calK_{\nu\nu\nu,xxy}^{\tS\tL\tS\tL\tS,B}
\nn\\
&
 \times [ (\overline{u_1} P_\tL v_2) 
(\overline{u_3} P_\tL U_B)
+(\overline{u_1} P_\tL v_3)
(\overline{u_2} P_\tL U_B) ]
\nn\\
&
+\calC_{\nu\nu\nu,xxy}^{\tS\tL\tS\tL\tS,n}
[ (\overline{u_1} P_\tL v_2)
(\overline{u_3} P_\tL u_n)
\nn\\
&+ (\overline{u_1} P_\tL v_3)
(\overline{u_2} P_\tL u_n)]\Big\} 
-(p_1\leftrightarrow p_2),
\end{align}
where $u_{1,2,3}=v_{1,2,3}^\C$. 

\subsection{Remaining cases}

For the remaining primed processes listed in \cref{tab:n2lllK}, including
$n\to\nu_x\nu_y\ell_z^-K^+$,
$\tN\to\ell_x^-\ell_y^+\nu_z K$, 
$\tN\to\nu_x\nu_y\bar\nu_z K$,
$p\to\bar\nu_x\bar\nu_y\ell_z^+K^0$, and $\tN\to\nu_x\nu_y\nu_zK$, 
their WC combinations and decay amplitudes can be directly obtained from their corresponding unprimed processes (those without primes in \cref{tab:n2lllK}) through simultaneous interchange of $\tL\leftrightarrow\tR$ and $\nu\leftrightarrow\bar\nu$. 

\section{Decay width expressions in terms of LEFT WCs}
\label{app:decay_width_WCs}

With the amplitudes given in \cref{app:decay_width_amp}, the squared matrix elements are calculated using the \texttt{FeynCalc} package~\cite{Shtabovenko:2016sxi}. The resulting expressions are integrated over the final-state phase space by means of the Monte Carlo simulation in~\cref{app:4body_PS}, yielding the decay widths in terms of combinations of the relevant WCs. As a cross-check, we have verified our numerical results against the analytical four-body phase-space formula of \cite{Asatrian:2012tp} for the decay modes $p\to \hat{\nu}\hat{\nu}\hat{\nu}K^+$ and $n\to \hat{\nu}\hat{\nu}\hat{\nu}K^0$, where three final-state particles are massless.

The chiral symmetry–breaking scale is fixed to $\Lambda_\chi \simeq 1.2~\text{GeV}$. We adopt the central values of the LECs $c_1$ and $c_2$, while $c_3$ and $c_4$ are parameterized through the ratios $\kappa_3 = c_3/c_1$ and $\kappa_4 = c_4/c_1$. In the following, we present only the squared contributions to the decay widths, while the complete expressions including interference terms, which are represented as ellipses below, are provided in the auxiliary \texttt{Mathematica} notebooks.
\begin{widetext}

\subsection{\tf{$\Delta L=+1$:
$p\to \ell_x^+\ell_y^+\ell_z^-K^0$}{p2lplplmK0}}

\begin{subequations}
\begin{align}
{\Gamma_{p\to e^+e^+e^-K^0} \over 10^{-13}{ \text{GeV}}^{11}}  
=\,& 
5|C_{\ell\ell\bar\ell,eee}^{\tV\tL\tV\tR\tS\tL}|^2
+3|C_{\ell\ell\bar\ell,eee}^{\tV\tL\tV\tR\tS\tR}|^2
+2.5\big(|C_{\ell\ell\bar\ell,eee}^{\tS\tL\tS\tR\tS\tL}|^2+|C_{\ell\ell\bar\ell,eee}^{\tS\tL\tS\tL\tS\tR}|^2\big)
+1.5\big(|C_{\ell\ell\bar\ell,eee}^{\tS\tL\tS\tL\tS\tL}|^2+|C_{\ell\ell\bar\ell,eee}^{\tS\tL\tS\tR\tS\tR}|^2\big)
\nn\\
&+0.3\kappa_3^2|C_{\ell\ell\bar\ell,eee}^{\tV\tL\tS\tL\tV\tR2}|^2
+0.07\kappa_3^2|C_{\ell\ell\bar\ell,eee}^{\tV\tL\tS\tL\tV\tR1}|^2
+\tL\leftrightarrow\tR+\cdots,
\\
{\Gamma_{p\to \mu^+\mu^+\mu^-K^0} \over 10^{-13}{\text{GeV}}^{11}}  
=\,& 
0.2|C_{\ell\ell\bar\ell,\mu\mu\mu}^{\tV\tL\tV\tR\tS\tL}|^2
+0.1|C_{\ell\ell\bar\ell,\mu\mu\mu}^{\tV\tL\tV\tR\tS\tR}|^2
+0.1\big(|C_{\ell\ell\bar\ell,\mu\mu\mu}^{\tS\tL\tS\tR\tS\tL}|^2+|C_{\ell\ell\bar\ell,\mu\mu\mu}^{\tS\tL\tS\tL\tS\tR}|^2\big)
+0.09\big(|C_{\ell\ell\bar\ell,\mu\mu\mu}^{\tS\tL\tS\tL\tS\tL}|^2+|C_{\ell\ell\bar\ell,\mu\mu\mu}^{\tS\tL\tS\tR\tS\tR}|^2\big)
\nn\\
&+0.03\kappa_3^2|C_{\ell\ell\bar\ell,\mu\mu\mu}^{\tV\tL\tS\tL\tV\tR2}|^2
+0.006\kappa_3^2|C_{\ell\ell\bar\ell,\mu\mu\mu}^{\tV\tL\tS\tL\tV\tR1}|^2
+\tL\leftrightarrow\tR+\cdots,
\\
{\Gamma_{p\to e^+e^+\mu^-K^0} \over 10^{-13}{\text{GeV}}^{11}}  
=\,& 
2.5|C_{\ell\ell\bar\ell,ee\mu}^{\tV\tL\tV\tR\tS\tL}|^2
+1.6|C_{\ell\ell\bar\ell,ee\mu}^{\tV\tL\tV\tR\tS\tR}|^2
+1.3\big(|C_{\ell\ell\bar\ell,ee\mu}^{\tS\tL\tS\tR\tS\tL}|^2+|C_{\ell\ell\bar\ell,ee\mu}^{\tS\tL\tS\tL\tS\tR}|^2\big)
+0.8\big(|C_{\ell\ell\bar\ell,ee\mu}^{\tS\tL\tS\tL\tS\tL}|^2+|C_{\ell\ell\bar\ell,ee\mu}^{\tS\tL\tS\tR\tS\tR}|^2\big)
\nn\\
&+0.2\kappa_3^2|C_{\ell\ell\bar\ell,ee\mu}^{\tV\tL\tS\tL\tV\tR2}|^2
+0.04\kappa_3^2|C_{\ell\ell\bar\ell,ee\mu}^{\tV\tL\tS\tL\tV\tR1}|^2
+\tL\leftrightarrow\tR+\cdots,
\\
{\Gamma_{p\to \mu^+\mu^+e^-K^0} \over 10^{-13}{\text{GeV}}^{11}}  
=\,& 
0.8|C_{\ell\ell\bar\ell,\mu\mu e}^{\tV\tL\tV\tR\tS\tL}|^2
+0.5|C_{\ell\ell\bar\ell,\mu\mu e}^{\tV\tL\tV\tR\tS\tR}|^2
+0.5\big(|C_{\ell\ell\bar\ell,\mu\mu e}^{\tS\tL\tS\tR\tS\tL}|^2+|C_{\ell\ell\bar\ell,\mu\mu e}^{\tS\tL\tS\tL\tS\tR}|^2\big)
+0.3\big(|C_{\ell\ell\bar\ell,\mu\mu e}^{\tS\tL\tS\tL\tS\tL}|^2+|C_{\ell\ell\bar\ell,\mu\mu e}^{\tS\tL\tS\tR\tS\tR}|^2\big)
\nn\\
&
+0.1\kappa_3^2|C_{\ell\ell\bar\ell,\mu\mu e}^{\tV\tL\tS\tL\tV\tR2}|^2
+0.02\kappa_3^2|C_{\ell\ell\bar\ell,\mu\mu e}^{\tV\tL\tS\tL\tV\tR1}|^2
+\tL\leftrightarrow\tR+\cdots,
\\
{\Gamma_{p\to e^+\mu^+e^-K^0} \over 10^{-13}{\text{GeV}}^{11}}  
=\,& 30|C_{\ell\ell\bar\ell,e\mu e}^{\tT\tL\tT\tL\tS\tR}|^2
+19|C_{\ell\ell\bar\ell,e\mu e}^{\tT\tL\tT\tL\tS\tL}|^2
+2.5\big(|C_{\ell\ell\bar\ell,e\mu e}^{\tV\tL\tV\tR\tS\tL}|^2+|C_{\ell\ell\bar\ell,\mu ee}^{\tV\tL\tV\tR\tS\tL}|^2\big)
+1.6\big(|C_{\ell\ell\bar\ell,e\mu e}^{\tV\tL\tV\tR\tS\tR}|^2+|C_{\ell\ell\bar\ell,\mu ee}^{\tV\tL\tV\tR\tS\tR}|^2\big)
\nn\\
&+0.6\kappa_3^2|C_{\ell\ell\bar\ell,e\mu e}^{\tT\tL\tV\tL\tV\tR2}|^2
+0.6\big(|C_{\ell\ell\bar\ell,e\mu e}^{\tS\tL\tS\tR\tS\tL}|^2+|C_{\ell\ell\bar\ell,e\mu e}^{\tS\tL\tS\tL\tS\tR}|^2\big)
+0.4\big(|C_{\ell\ell\bar\ell,e\mu e}^{\tS\tL\tS\tL\tS\tL}|^2+|C_{\ell\ell\bar\ell,e\mu e}^{\tS\tL\tS\tR\tS\tR}|^2\big)
\nn\\
&+0.16\kappa_3^2\big(|C_{\ell\ell\bar\ell,e\mu e}^{\tV\tL\tS\tL\tV\tR2}|^2+|C_{\ell\ell\bar\ell,\mu e e}^{\tV\tL\tS\tL\tV\tR2}|^2\big)
+0.15\kappa_3^2|C_{\ell\ell\bar\ell,e\mu e}^{\tT\tL\tV\tL\tV\tR1}|^2
+0.04\kappa_3^2\big(|C_{\ell\ell\bar\ell,e\mu e}^{\tV\tL\tS\tL\tV\tR1}|^2+|C_{\ell\ell\bar\ell,\mu e e}^{\tV\tL\tS\tL\tV\tR1}|^2\big)
\nn\\
&+9\cdot10^{-4}\kappa_4^2|C_{\ell\ell\bar\ell,e\mu e}^{\tT\tL\tS\tL\tT\tL}|^2
+\tL\leftrightarrow\tR+\cdots,
\\
{\Gamma_{p\to e^+\mu^+\mu^-K^0} \over 10^{-13}{\text{GeV}}^{11}}  
=\,& 12|C_{\ell\ell\bar\ell,e\mu \mu}^{\tT\tL\tT\tL\tS\tR}|^2
+8|C_{\ell\ell\bar\ell,e\mu \mu}^{\tT\tL\tT\tL\tS\tL}|^2
+1.0\big(|C_{\ell\ell\bar\ell,e\mu \mu}^{\tV\tL\tV\tR\tS\tL}|^2+|C_{\ell\ell\bar\ell,\mu e\mu}^{\tV\tL\tV\tR\tS\tL}|^2\big)
+0.7\big(|C_{\ell\ell\bar\ell,e\mu \mu}^{\tV\tL\tV\tR\tS\tR}|^2+|C_{\ell\ell\bar\ell,\mu e\mu}^{\tV\tL\tV\tR\tS\tR}|^2\big)
\nn\\
&+0.3\kappa_3^2|C_{\ell\ell\bar\ell,e\mu \mu}^{\tT\tL\tV\tL\tV\tR2}|^2
+0.3\big(|C_{\ell\ell\bar\ell,e\mu \mu}^{\tS\tL\tS\tR\tS\tL}|^2+|C_{\ell\ell\bar\ell,e\mu \mu}^{\tS\tL\tS\tL\tS\tR}|^2\big)
+0.2\big(|C_{\ell\ell\bar\ell,e\mu \mu}^{\tS\tL\tS\tL\tS\tL}|^2+|C_{\ell\ell\bar\ell,e\mu \mu}^{\tS\tL\tS\tR\tS\tR}|^2\big)
\nn\\
&+0.07\kappa_3^2\big(|C_{\ell\ell\bar\ell,e\mu \mu}^{\tV\tL\tS\tL\tV\tR2}|^2+|C_{\ell\ell\bar\ell,\mu e \mu}^{\tV\tL\tS\tL\tV\tR2}|^2\big)
+0.07\kappa_3^2|C_{\ell\ell\bar\ell,e\mu \mu}^{\tT\tL\tV\tL\tV\tR1}|^2
+0.02\kappa_3^2\big(|C_{\ell\ell\bar\ell,e\mu \mu}^{\tV\tL\tS\tL\tV\tR1}|^2+|C_{\ell\ell\bar\ell,\mu e \mu}^{\tV\tL\tS\tL\tV\tR1}|^2\big)
\nn\\
&+4\cdot10^{-4}\kappa_4^2|C_{\ell\ell\bar\ell,e\mu \mu}^{\tT\tL\tS\tL\tT\tL}|^2
+\tL\leftrightarrow\tR+\cdots.
\end{align}
\end{subequations}

\subsection{\tf{$\Delta L = +1$: $p\to\nu_x\bar\nu_y\ell_z^+K^0$}{p2vvblpK0}}

\begin{subequations}
\begin{align}
{\Gamma_{p\to \nu_x\bar{\nu}_y e^+K^0} \over 10^{-13}{\text{GeV}}^{11}} =\,&   
5\big(|C_{\bar\nu \nu \ell,xye}^{\tV\tL\tV\tL\tS\tR}|^2
+|C_{\bar\nu \nu \ell,xye}^{\tV\tL\tV\tR\tS\tL}|^2\big)
+3\big(|C_{\bar\nu \nu \ell,xye}^{\tV\tL\tV\tL\tS\tL}|^2+|C_{\bar\nu \nu \ell,xye}^{\tV\tL\tV\tR\tS\tR}|^2\big)
+ 0.3\kappa_3^2\big(|C_{\bar\nu \nu \ell,xye}^{\tV\tL\tS\tL\tV\tR2}|^2+
|C_{\bar\nu \nu \ell,xye}^{\tV\tL\tS\tR\tV\tL2}|^2
\big)
\nn\\
&+ 0.07\kappa_3^2\big(
|C_{\bar\nu \nu \ell,xye}^{\tV\tL\tS\tL\tV\tR1}|^2
+|C_{\bar\nu \nu \ell,xye}^{\tV\tL\tS\tR\tV\tL1}|^2
\big)
+\cdots,
\\
{\Gamma_{p\to \nu_x\bar{\nu}_y \mu^+K^0} \over 10^{-13}{\text{GeV}}^{11}} =\,&   
2.5\big(|C_{\bar\nu \nu \ell,xy\mu}^{\tV\tL\tV\tL\tS\tR}|^2
+|C_{\bar\nu \nu \ell,xy\mu}^{\tV\tL\tV\tR\tS\tL}|^2\big)
+1.6\big(|C_{\bar\nu \nu \ell,xy\mu}^{\tV\tL\tV\tL\tS\tL}|^2+|C_{\bar\nu \nu \ell,xy\mu}^{\tV\tL\tV\tR\tS\tR}|^2\big)
+ 0.16\kappa_3^2\big(|C_{\bar\nu \nu \ell,xy\mu}^{\tV\tL\tS\tL\tV\tR2}|^2+
|C_{\bar\nu \nu \ell,xy\mu}^{\tV\tL\tS\tR\tV\tL2}|^2
\big)
\nn\\
&+ 0.04\kappa_3^2\big(
|C_{\bar\nu \nu \ell,xy\mu}^{\tV\tL\tS\tL\tV\tR1}|^2
+|C_{\bar\nu \nu \ell,xy\mu}^{\tV\tL\tS\tR\tV\tL1}|^2
\big)
+\cdots.
\end{align}
\end{subequations}

\subsection{\tf{$\Delta L = +1$: $n\to\bar\nu_x\bar\nu_y\ell_z^-K^+$ (a)}{n2vbvblmKp}}

\begin{subequations}
\begin{align}
{\Gamma_{n\to \bar\nu_x\bar\nu_{y\neq x} e^-K^+} \over 10^{-13}{\text{GeV}}^{11}} =\,& 
64|C_{\nu \nu \bar\ell,xye}^{\tT\tL\tT\tL\tS\tR}|^2
+40|C_{\nu \nu \bar\ell,xye}^{\tT\tL\tT\tL\tS\tL}|^2
+1.3(|C_{\nu \nu \bar\ell,xye}^{\tS\tL\tS\tL\tS\tR}|^2
+|C_{\nu \nu \bar\ell,xye}^{\tS\tL\tS\tR\tS\tL}|^2)
+1.2\kappa_3^2|C_{\nu \nu \bar\ell,xye}^{\tT\tL\tV\tL\tV\tR2}|^2
\nn\\
&
+0.8(|C_{\nu \nu \bar\ell,xye}^{\tS\tL\tS\tL\tS\tL}|^2
+|C_{\nu \nu \bar\ell,xye}^{\tS\tL\tS\tR\tS\tR}|^2)
+0.3\kappa_3^2|C_{\nu \nu \bar\ell,xye}^{\tT\tL\tV\tL\tV\tR1}|^2
+2\cdot10^{-3}\kappa_4^2|C_{\nu \nu \bar\ell,xye}^{\tT\tL\tS\tL\tT\tL}|^2+\cdots,
\\
{\Gamma_{n\to \bar\nu_x\bar\nu_x e^-K^+} \over 10^{-13}{\text{GeV}}^{11}} =\,&  
2.7\big(|C_{\nu \nu \bar\ell,xxe}^{\tS\tL\tS\tR\tS\tL}|^2
+|C_{\nu \nu \bar\ell,xxe}^{\tS\tL\tS\tL\tS\tR}|^2\big)
+1.7\big(|C_{\nu \nu \bar\ell,xxe}^{\tS\tL\tS\tL\tS\tL}|^2
+|C_{\nu \nu \bar\ell,xxe}^{\tS\tL\tS\tR\tS\tR}|^2\big)
+\cdots,
\\
{\Gamma_{n\to \bar\nu_x\bar\nu_{y\neq x} \mu^-K^+} \over 10^{-13}{\text{GeV}}^{11}} =\,& 
33|C_{\nu \nu \bar\ell,xy\mu}^{\tT\tL\tT\tL\tS\tR}|^2
+21|C_{\nu \nu \bar\ell,xy\mu}^{\tT\tL\tT\tL\tS\tL}|^2
+0.7\kappa_3^2|C_{\nu \nu \bar\ell,xy\mu}^{\tT\tL\tV\tL\tV\tR2}|^2
+0.7(|C_{\nu \nu \bar\ell,xy\mu}^{\tS\tL\tS\tL\tS\tR}|^2
+|C_{\nu \nu \bar\ell,xy\mu}^{\tS\tL\tS\tR\tS\tL}|^2)
\nn\\
&
+0.4(|C_{\nu \nu \bar\ell,xy\mu}^{\tS\tL\tS\tL\tS\tL}|^2
+|C_{\nu \nu \bar\ell,xy\mu}^{\tS\tL\tS\tR\tS\tR}|^2)
+0.2\kappa_3^2|C_{\nu \nu \bar\ell,xy\mu}^{\tT\tL\tV\tL\tV\tR1}|^2
+10^{-3}\kappa_4^2|C_{\nu \nu \bar\ell,xy\mu}^{\tT\tL\tS\tL\tT\tL}|^2+\cdots,
\\
{\Gamma_{n\to \bar\nu_x\bar\nu_x \mu^-K^+} \over 10^{-13}{\text{GeV}}^{11}} =\,&  
1.4\big(|C_{\nu \nu \bar\ell,xx\mu}^{\tS\tL\tS\tR\tS\tL}|^2
+|C_{\nu \nu \bar\ell,xx\mu}^{\tS\tL\tS\tL\tS\tR}|^2\big)
+
0.9\big(|C_{\nu \nu \bar\ell,xx\mu}^{\tS\tL\tS\tL\tS\tL}|^2
+|C_{\nu \nu \bar\ell,xx\mu}^{\tS\tL\tS\tR\tS\tR}|^2\big)
+\cdots.
\end{align}
\end{subequations}

\subsection{\tf{$\Delta L = +1$: 
$p\to\ell_x^-\ell_y^+\bar\nu_zK^+,\,
n\to\ell_x^-\ell_y^+\bar\nu_z K^0$ (b)}{N2lmlpvbKp}}

$\bullet$ Proton case.
\begin{subequations}	
\begin{align}
{\Gamma_{p\to e^-e^+ \bar{\nu}_xK^+} \over 10^{-13}{\text{GeV}}^{11}} =\,& 
90(|C_{\bar\ell \ell \nu,e e x}^{\tT\tL\tT\tL\tS\tL2}|^2
+|C_{\bar\ell \ell \nu,e e x}^{\tT\tL\tT\tL\tS\tR3}|^2)
+40|C_{\bar\ell \ell \nu,e e x}^{\tT\tL\tT\tL\tS\tR1}|^2
+7\big(
|C_{\bar\ell \ell \nu,e e x}^{\tV\tL\tV\tR\tS\tL3}|^2
+|C_{\bar\ell \ell \nu,e e x}^{\tV\tR\tV\tR\tS\tL3}|^2
+|C_{\bar\ell \ell \nu,e e x}^{\tV\tL\tV\tR\tS\tR2}|^2
+|C_{\bar\ell \ell \nu,e e x}^{\tV\tR\tV\tR\tS\tR2}|^2
\big)
\nn\\
&+3\big(|C_{\bar\ell \ell \nu,e e x}^{\tT\tL\tT\tL\tS\tL1}|^2
+|C_{\bar\ell \ell \nu,e e x}^{\tT\tL\tT\tL\tS\tR2}|^2\big)
+3\big(|C_{\bar\ell \ell \nu,e e x}^{\tV\tL\tV\tR\tS\tL1}|^2
+|C_{\bar\ell \ell \nu,e e x}^{\tV\tR\tV\tR\tS\tL1}|^2\big)
+2\big(|C_{\bar\ell \ell \nu,e e x}^{\tS\tL\tS\tL\tS\tL2}|^2
+|C_{\bar\ell \ell \nu,e e x}^{\tS\tR\tS\tL\tS\tL2}|^2
\nn\\
&+|C_{\bar\ell \ell \nu,e e x}^{\tS\tL\tS\tL\tS\tR3}|^2
+|C_{\bar\ell \ell \nu,e e x}^{\tS\tR\tS\tL\tS\tR3}|^2
\big)
+1.3\kappa_3^2|C_{\bar\ell \ell \nu,e e x}^{\tT\tR\tV\tR\tV\tL2}|^2
+0.9\big(|C_{\bar\ell \ell \nu,e e x}^{\tS\tR\tS\tL\tS\tR1}|^2
+|C_{\bar\ell \ell \nu,e e x}^{\tS\tL\tS\tL\tS\tR1}|^2
\big)
+0.3\kappa_3^2|C_{\bar\ell \ell \nu,e e x}^{\tT\tR\tV\tR\tV\tL3}|^2
\nn\\
&+0.3\kappa_3^2\big(
|C_{\bar\ell \ell \nu,e e x}^{\tV\tL\tS\tL\tV\tR2}|^2
+|C_{\bar\ell \ell \nu,e e x}^{\tV\tR\tS\tL\tV\tR2}|^2
\big)
+0.08\kappa_3^2\big(|C_{\bar\ell \ell \nu,e e x}^{\tV\tL\tS\tL\tV\tR3}|^2
+|C_{\bar\ell \ell \nu,e e x}^{\tV\tR\tS\tL\tV\tR3}|^2
\big)
+0.3\big(|C_{\bar\ell \ell \nu,e e x}^{\tV\tR\tV\tR\tS\tR1}|^2
+|C_{\bar\ell \ell \nu,e e x}^{\tV\tL\tV\tR\tS\tR1}|^2
\nn\\
&+|C_{\bar\ell \ell \nu,e e x}^{\tV\tR\tV\tR\tS\tL2}|^2
+|C_{\bar\ell \ell \nu,e e x}^{\tV\tL\tV\tR\tS\tL2}|^2
\big)
+0.3\kappa_3^2|C_{\bar\ell \ell \nu,e e x}^{\tT\tR\tV\tR\tV\tL1}|^2
+0.08\kappa_3^2\big(|C_{\bar\ell \ell \nu,e e x}^{\tV\tL\tS\tL\tV\tR1}|^2
+|C_{\bar\ell \ell \nu,e e x}^{\tV\tR\tS\tL\tV\tR1}|^2
\big)
\nn\\
&+0.07\big(|C_{\bar\ell \ell \nu,e e x}^{\tS\tL\tS\tL\tS\tL1}|^2
+|C_{\bar\ell \ell \nu,e e x}^{\tS\tR\tS\tL\tS\tL1}|^2
+|C_{\bar\ell \ell \nu,e e x}^{\tS\tL\tS\tL\tS\tR2}|^2
+|C_{\bar\ell \ell \nu,e e x}^{\tS\tR\tS\tL\tS\tR2}|^2\big)
+5\cdot10^{-4}\kappa_4^2|C_{\bar\ell \ell \nu,e e x}^{\tT\tL\tS\tL\tT\tL}|^2
+\cdots,
\\
{\Gamma_{p\to \mu^-\mu^+ \bar{\nu}_xK^+} \over 10^{-13}{\text{GeV}}^{11}}=\,& 
19(|C_{\bar\ell \ell \nu,\mu\mu x}^{\tT\tL\tT\tL\tS\tL2}|^2
+|C_{\bar\ell \ell \nu,\mu\mu x}^{\tT\tL\tT\tL\tS\tR3}|^2)
+9|C_{\bar\ell \ell \nu,\mu\mu x}^{\tT\tL\tT\tL\tS\tR1}|^2
+1.6\big(|C_{\bar\ell \ell \nu,\mu\mu x}^{\tV\tL\tV\tR\tS\tR2}|^2
+|C_{\bar\ell \ell \nu,\mu\mu x}^{\tV\tR\tV\tR\tS\tL3}|^2
+|C_{\bar\ell \ell \nu,\mu\mu x}^{\tV\tL\tV\tR\tS\tL3}|^2
+|C_{\bar\ell \ell \nu,\mu\mu x}^{\tV\tR\tV\tR\tS\tR2}|^2
\big)
\nn\\
&+0.7\big(|C_{\bar\ell \ell \nu,\mu\mu x}^{\tV\tL\tV\tR\tS\tL1}|^2
+|C_{\bar\ell \ell \nu,\mu\mu x}^{\tV\tR\tV\tR\tS\tL1}|^2\big)
+0.6\big(|C_{\bar\ell \ell \nu,\mu\mu x}^{\tT\tL\tT\tL\tS\tL1}|^2
+|C_{\bar\ell \ell \nu,\mu\mu x}^{\tT\tL\tT\tL\tS\tR2}|^2\big)
+0.4\big(|C_{\bar\ell \ell \nu,\mu\mu x}^{\tS\tL\tS\tL\tS\tL2}|^2
+|C_{\bar\ell \ell \nu,\mu\mu x}^{\tS\tR\tS\tL\tS\tL2}|^2
\nn\\
&+|C_{\bar\ell \ell \nu,\mu\mu x}^{\tS\tL\tS\tL\tS\tR3}|^2
+|C_{\bar\ell \ell \nu,\mu\mu x}^{\tS\tR\tS\tL\tS\tR3}|^2
\big)+0.3\kappa_3^2|C_{\bar\ell \ell \nu,\mu\mu x}^{\tT\tR\tV\tR\tV\tL2}|^2
+0.2\big(|C_{\bar\ell \ell \nu,\mu\mu x}^{\tS\tR\tS\tL\tS\tR1}|^2
+|C_{\bar\ell \ell \nu,\mu\mu x}^{\tS\tL\tS\tL\tS\tR1}|^2
\big)
+0.1\kappa_3^2|C_{\bar\ell \ell \nu,\mu\mu x}^{\tT\tR\tV\tR\tV\tL1}|^2
\nn\\
&+0.09\kappa_3^2(|C_{\bar\ell \ell \nu,\mu\mu x}^{\tV\tR\tS\tL\tV\tR2}|^2+|C_{\bar\ell \ell \nu,\mu\mu x}^{\tV\tL\tS\tL\tV\tR2}|^2)
+0.08\kappa_3^2|C_{\bar\ell \ell \nu,\mu\mu x}^{\tT\tR\tV\tR\tV\tL3}|^2
+0.05\big(|C_{\bar\ell \ell \nu,\mu\mu x}^{\tV\tR\tV\tR\tS\tR1}|^2
+|C_{\bar\ell \ell \nu,\mu\mu x}^{\tV\tL\tV\tR\tS\tR1}|^2
\nn\\
&+|C_{\bar\ell \ell \nu,\mu\mu x}^{\tV\tR\tV\tR\tS\tL2}|^2
+|C_{\bar\ell \ell \nu,\mu\mu x}^{\tV\tL\tV\tR\tS\tL2}|^2
\big)
+0.03\kappa_3^2(|C_{\bar\ell \ell \nu,\mu\mu x}^{\tV\tR\tS\tL\tV\tR1}|^2+|C_{\bar\ell \ell \nu,\mu\mu x}^{\tV\tL\tS\tL\tV\tR1}|^2)
+0.02\kappa_3^2(|C_{\bar\ell \ell \nu,\mu\mu x}^{\tV\tR\tS\tL\tV\tR3}|^2+|C_{\bar\ell \ell \nu,\mu\mu x}^{\tV\tL\tS\tL\tV\tR3}|^2)
\nn\\
&+0.01\big(|C_{\bar\ell \ell \nu,\mu\mu x}^{\tS\tL\tS\tL\tS\tL1}|^2
+|C_{\bar\ell \ell \nu,\mu\mu x}^{\tS\tR\tS\tL\tS\tL1}|^2
+|C_{\bar\ell \ell \nu,\mu\mu x}^{\tS\tL\tS\tL\tS\tR2}|^2
+|C_{\bar\ell \ell \nu,\mu\mu x}^{\tS\tR\tS\tL\tS\tR2}|^2\big)
+10^{-4}\kappa_4^2|C_{\bar\ell \ell \nu,\mu\mu x}^{\tT\tL\tS\tL\tT\tL}|^2
+\cdots,
\\
{\Gamma_{p\to e^-\mu^+\bar{\nu}_xK^+} \over 10^{-13}{\text{GeV}}^{11}} =\,&
45\big(
|C_{\bar\ell \ell \nu,e \mu  x}^{\tT\tL\tT\tL\tS\tL2}|^2
+|C_{\bar\ell \ell \nu,e \mu  x}^{\tT\tL\tT\tL\tS\tR3}|^2
\big)
+21|C_{\bar\ell \ell \nu,e \mu  x}^{\tT\tL\tT\tL\tS\tR1}|^2
+4\big(
|C_{\bar\ell \ell \nu,e \mu  x}^{\tV\tL\tV\tR\tS\tR2}|^2
+|C_{\bar\ell \ell \nu,e \mu  x}^{\tV\tR\tV\tR\tS\tR2}|^2
+|C_{\bar\ell \ell \nu,e \mu  x}^{\tV\tL\tV\tR\tS\tL3}|^2
+|C_{\bar\ell \ell \nu,e \mu  x}^{\tV\tR\tV\tR\tS\tL3}|^2
\big)
\nn\\
&+1.8\big(
|C_{\bar\ell \ell \nu,e \mu  x}^{\tV\tL\tV\tR\tS\tL1}|^2
+|C_{\bar\ell \ell \nu,e \mu  x}^{\tV\tR\tV\tR\tS\tL1}|^2
\big)
+1.7\big(
|C_{\bar\ell \ell \nu,e \mu  x}^{\tT\tL\tT\tL\tS\tL1}|^2
+|C_{\bar\ell \ell \nu,e \mu  x}^{\tT\tL\tT\tL\tS\tR2}|^2
\big)
+1.0\big(
|C_{\bar\ell \ell \nu,e \mu  x}^{\tS\tR\tS\tL\tS\tL2}|^2
+|C_{\bar\ell \ell \nu,e \mu  x}^{\tS\tL\tS\tL\tS\tL2}|^2
\nn\\
&+|C_{\bar\ell \ell \nu,e \mu  x}^{\tS\tR\tS\tL\tS\tR3}|^2
+|C_{\bar\ell \ell \nu,e \mu  x}^{\tS\tL\tS\tL\tS\tR3}|^2
\big)
+0.7\kappa_3^2
|C_{\bar\ell \ell \nu,e \mu  x}^{\tT\tR\tV\tR\tV\tL2}|^2
+0.4\big(
|C_{\bar\ell \ell \nu,e \mu  x}^{\tS\tL\tS\tL\tS\tR1}|^2
+|C_{\bar\ell \ell \nu,e \mu  x}^{\tS\tR\tS\tL\tS\tR1}|^2
\big)
\nn\\
&+0.2\kappa_3^2
(|C_{\bar\ell \ell \nu,e \mu  x}^{\tT\tR\tV\tR\tV\tL1}|^2
+|C_{\bar\ell \ell \nu,e \mu  x}^{\tT\tR\tV\tR\tV\tL3}|^2)
+0.2\kappa_3^2
(|C_{\bar\ell \ell \nu,e \mu  x}^{\tV\tR\tS\tL\tV\tR2}|^2
+|C_{\bar\ell \ell \nu,e \mu  x}^{\tV\tL\tS\tL\tV\tR2}|^2)
+0.14\big(
|C_{\bar\ell \ell \nu,e \mu  x}^{\tV\tL\tV\tR\tS\tR1}|^2
+|C_{\bar\ell \ell \nu,e \mu  x}^{\tV\tR\tV\tR\tS\tR1}|^2
\nn\\
&
+|C_{\bar\ell \ell \nu,e \mu  x}^{\tV\tL\tV\tR\tS\tL2}|^2
+|C_{\bar\ell \ell \nu,e \mu  x}^{\tV\tR\tV\tR\tS\tL2}|^2
\big)
+0.05\kappa_3^2\big(
|C_{\bar\ell \ell \nu,e \mu  x}^{\tV\tL\tS\tL\tV\tR1}|^2
+|C_{\bar\ell \ell \nu,e \mu  x}^{\tV\tR\tS\tL\tV\tR1}|^2
\big)
+0.04\kappa_3^2
(|C_{\bar\ell \ell \nu,e \mu  x}^{\tV\tL\tS\tL\tV\tR3}|^2
+|C_{\bar\ell \ell \nu,e \mu  x}^{\tV\tR\tS\tL\tV\tR3}|^2)
\nn\\
&+0.03\big(
|C_{\bar\ell \ell \nu,e \mu  x}^{\tS\tR\tS\tL\tS\tL1}|^2
+|C_{\bar\ell \ell \nu,e \mu  x}^{\tS\tL\tS\tL\tS\tL1}|^2
+|C_{\bar\ell \ell \nu,e \mu  x}^{\tS\tR\tS\tL\tS\tR2}|^2
+|C_{\bar\ell \ell \nu,e \mu  x}^{\tS\tL\tS\tL\tS\tR2}|^2
\big)
+3\cdot10^{-4}\kappa_4^2
|C_{\bar\ell \ell \nu,e \mu  x}^{\tT\tL\tS\tL\tT\tL}|^2+\cdots\;,
\\
{\Gamma_{p\to \mu^-e^+\bar{\nu}_xK^+} \over 10^{-13}{\text{GeV}}^{11}} =\,&
45\big(
|C_{\bar\ell \ell \nu,\mu e  x}^{\tT\tL\tT\tL\tS\tL2}|^2
+|C_{\bar\ell \ell \nu,\mu e  x}^{\tT\tL\tT\tL\tS\tR3}|^2
\big)
+21|C_{\bar\ell \ell \nu,\mu e  x}^{\tT\tL\tT\tL\tS\tR1}|^2
+4\big(
|C_{\bar\ell \ell \nu,\mu e  x}^{\tV\tL\tV\tR\tS\tR2}|^2
+|C_{\bar\ell \ell \nu,\mu e  x}^{\tV\tR\tV\tR\tS\tR2}|^2
+|C_{\bar\ell \ell \nu,\mu e  x}^{\tV\tL\tV\tR\tS\tL3}|^2
+|C_{\bar\ell \ell \nu,\mu e  x}^{\tV\tR\tV\tR\tS\tL3}|^2
\big)
\nn\\
&+1.8\big(
|C_{\bar\ell \ell \nu,\mu e  x}^{\tV\tL\tV\tR\tS\tL1}|^2
+|C_{\bar\ell \ell \nu,\mu e  x}^{\tV\tR\tV\tR\tS\tL1}|^2
\big)
+1.7\big(
|C_{\bar\ell \ell \nu,\mu e  x}^{\tT\tL\tT\tL\tS\tL1}|^2
+|C_{\bar\ell \ell \nu,\mu e  x}^{\tT\tL\tT\tL\tS\tR2}|^2
\big)
+1.0\big(
|C_{\bar\ell \ell \nu,\mu e  x}^{\tS\tR\tS\tL\tS\tL2}|^2
+|C_{\bar\ell \ell \nu,\mu e  x}^{\tS\tL\tS\tL\tS\tL2}|^2
\nn\\
&+|C_{\bar\ell \ell \nu,\mu e  x}^{\tS\tR\tS\tL\tS\tR3}|^2
+|C_{\bar\ell \ell \nu,\mu e  x}^{\tS\tL\tS\tL\tS\tR3}|^2
\big)
+0.7\kappa_3^2
|C_{\bar\ell \ell \nu,\mu e  x}^{\tT\tR\tV\tR\tV\tL2}|^2
+0.4\big(
|C_{\bar\ell \ell \nu,\mu e  x}^{\tS\tL\tS\tL\tS\tR1}|^2
+|C_{\bar\ell \ell \nu,\mu e  x}^{\tS\tR\tS\tL\tS\tR1}|^2
\big)
\nn\\
&+0.2\kappa_3^2
(|C_{\bar\ell \ell \nu,\mu e  x}^{\tT\tR\tV\tR\tV\tL1}|^2
+|C_{\bar\ell \ell \nu,\mu e  x}^{\tT\tR\tV\tR\tV\tL3}|^2)
+0.2\kappa_3^2
(|C_{\bar\ell \ell \nu,\mu e  x}^{\tV\tR\tS\tL\tV\tR2}|^2
+|C_{\bar\ell \ell \nu,\mu e  x}^{\tV\tL\tS\tL\tV\tR2}|^2)
+0.14\big(
|C_{\bar\ell \ell \nu,\mu e  x}^{\tV\tL\tV\tR\tS\tR1}|^2
+|C_{\bar\ell \ell \nu,\mu e  x}^{\tV\tR\tV\tR\tS\tR1}|^2
\nn\\
&+|C_{\bar\ell \ell \nu,\mu e  x}^{\tV\tL\tV\tR\tS\tL2}|^2
+|C_{\bar\ell \ell \nu,\mu e  x}^{\tV\tR\tV\tR\tS\tL2}|^2
\big)
+0.05\kappa_3^2\big(
|C_{\bar\ell \ell \nu,\mu e  x}^{\tV\tL\tS\tL\tV\tR1}|^2
+|C_{\bar\ell \ell \nu,\mu e  x}^{\tV\tR\tS\tL\tV\tR1}|^2
\big)
+0.04\kappa_3^2
(|C_{\bar\ell \ell \nu,\mu e  x}^{\tV\tL\tS\tL\tV\tR3}|^2
+|C_{\bar\ell \ell \nu,\mu e  x}^{\tV\tR\tS\tL\tV\tR3}|^2)
\nn\\
&+0.03\big(
|C_{\bar\ell \ell \nu,\mu e  x}^{\tS\tR\tS\tL\tS\tL1}|^2
+|C_{\bar\ell \ell \nu,\mu e  x}^{\tS\tL\tS\tL\tS\tL1}|^2
+|C_{\bar\ell \ell \nu,\mu e  x}^{\tS\tR\tS\tL\tS\tR2}|^2
+|C_{\bar\ell \ell \nu,\mu e  x}^{\tS\tL\tS\tL\tS\tR2}|^2
\big)
+3\cdot10^{-4}\kappa_4^2
|C_{\bar\ell \ell \nu,\mu e  x}^{\tT\tL\tS\tL\tT\tL}|^2+\cdots\;.
\end{align}
\end{subequations}

$\bullet$ Neutron case.
\begin{subequations}	
\begin{align}
{\Gamma_{n\to e^-e^+ \bar{\nu}_xK^0} \over 10^{-13}{\text{GeV}}^{11}} =\,&
90(|C_{\bar\ell \ell \nu,eex}^{\tT\tL\tT\tL\tS\tL2}|^2
+|C_{\bar\ell \ell \nu,eex}^{\tT\tL\tT\tL\tS\tR3}|^2)
+60|C_{\bar\ell \ell \nu,eex}^{\tT\tL\tT\tL\tS\tL1}|^2
+40|C_{\bar\ell \ell \nu,eex}^{\tT\tL\tT\tL\tS\tR2}|^2
+7\big(|C_{\bar\ell \ell \nu,eex}^{\tV\tL\tV\tR\tS\tL3}|^2
+|C_{\bar\ell \ell \nu,eex}^{\tV\tR\tV\tR\tS\tL3}|^2
\nn\\
&+|C_{\bar\ell \ell \nu,eex}^{\tV\tL\tV\tR\tS\tR2}|^2
+|C_{\bar\ell \ell \nu,eex}^{\tV\tR\tV\tR\tS\tR2}|^2
\big)
+5\big(|C_{\bar\ell \ell \nu,eex}^{\tV\tL\tV\tR\tS\tR1}|^2
+|C_{\bar\ell \ell \nu,eex}^{\tV\tR\tV\tR\tS\tR1}|^2\big)
+3|C_{\bar\ell \ell \nu,eex}^{\tT\tL\tT\tL\tS\tR1}|^2
+3\big(|C_{\bar\ell \ell \nu,eex}^{\tV\tL\tV\tR\tS\tL2}|^2
\nn\\
&+|C_{\bar\ell \ell \nu,eex}^{\tV\tR\tV\tR\tS\tL2}|^2\big)
+2\big(|C_{\bar\ell \ell \nu,eex}^{\tS\tR\tS\tL\tS\tL2}|^2
+|C_{\bar\ell \ell \nu,eex}^{\tS\tL\tS\tL\tS\tL2}|^2
+|C_{\bar\ell \ell \nu,eex}^{\tS\tR\tS\tL\tS\tR3}|^2
+|C_{\bar\ell \ell \nu,eex}^{\tS\tL\tS\tL\tS\tR3}|^2\big)
+1.2\big(|C_{\bar\ell \ell \nu,eex}^{\tS\tL\tS\tL\tS\tL1}|^2
\nn\\
&+|C_{\bar\ell \ell \nu,eex}^{\tS\tR\tS\tL\tS\tL1}|^2\big)
+1.2\kappa_3^2|C_{\bar\ell \ell \nu,eex}^{\tT\tR\tV\tR\tV\tL1}|^2
+0.8\big(|C_{\bar\ell \ell \nu,eex}^{\tS\tL\tS\tL\tS\tR2}|^2
+|C_{\bar\ell \ell \nu,eex}^{\tS\tR\tS\tL\tS\tR2}|^2\big)
+0.3\kappa_3^2\big(|C_{\bar\ell \ell \nu,eex}^{\tT\tR\tV\tR\tV\tL3}|^2
\nn\\
&+|C_{\bar\ell \ell \nu,eex}^{\tT\tR\tV\tR\tV\tL2}|^2\big)
+0.3\big(|C_{\bar\ell \ell \nu,eex}^{\tV\tL\tV\tR\tS\tL1}|^2
+|C_{\bar\ell \ell \nu,eex}^{\tV\tR\tV\tR\tS\tL1}|^2
\big)
+0.3\kappa_3^2\big(|C_{\bar\ell \ell \nu,eex}^{\tV\tL\tS\tL\tV\tR1}|^2
+|C_{\bar\ell \ell \nu,eex}^{\tV\tR\tS\tL\tV\tR1}|^2
\big)
\nn\\
&+0.09\kappa_3^2\big(|C_{\bar\ell \ell \nu,eex}^{\tV\tL\tS\tL\tV\tR2}|^2
+|C_{\bar\ell \ell \nu,eex}^{\tV\tR\tS\tL\tV\tR2}|^2
\big)
+0.07\kappa_3^2\big(|C_{\bar\ell \ell \nu,eex}^{\tV\tL\tS\tL\tV\tR3}|^2
+|C_{\bar\ell \ell \nu,eex}^{\tV\tR\tS\tL\tV\tR3}|^2
\big)
\nn\\
&+0.07\big(|C_{\bar\ell \ell \nu,eex}^{\tS\tL\tS\tL\tS\tR1}|^2
+|C_{\bar\ell \ell \nu,eex}^{\tS\tR\tS\tL\tS\tR1}|^2
\big)
+4\cdot10^{-4}\kappa_4^2|C_{\bar\ell \ell \nu,eex}^{\tT\tL\tS\tL\tT\tL}|^2
+\cdots,
\\
{\Gamma_{n\to \mu^-\mu^+ \bar{\nu}_xK^0} \over 10^{-13}{\text{GeV}}^{11}} =\,&
18\big(
|C_{\bar\ell \ell \nu,\mu\mu x}^{\tT\tL\tT\tL\tS\tL2}|^2
+|C_{\bar\ell \ell \nu,\mu\mu x}^{\tT\tL\tT\tL\tS\tR3}|^2
\big)
+13|C_{\bar\ell \ell \nu,\mu\mu x}^{\tT\tL\tT\tL\tS\tL1}|^2
+9|C_{\bar\ell \ell \nu,\mu\mu x}^{\tT\tL\tT\tL\tS\tR2}|^2
+1.5\big(
|C_{\bar\ell \ell \nu,\mu\mu x}^{\tV\tL\tV\tR\tS\tL3}|^2
+|C_{\bar\ell \ell \nu,\mu\mu x}^{\tV\tL\tV\tR\tS\tR2}|^2
\nn\\
&+|C_{\bar\ell \ell \nu,\mu\mu x}^{\tV\tR\tV\tR\tS\tR2}|^2
+|C_{\bar\ell \ell \nu,\mu\mu x}^{\tV\tR\tV\tR\tS\tL3}|^2
\big)
+1.0\big(
|C_{\bar\ell \ell \nu,\mu\mu x}^{\tV\tL\tV\tR\tS\tR1}|^2
+|C_{\bar\ell \ell \nu,\mu\mu x}^{\tV\tR\tV\tR\tS\tR1}|^2
\big)
+0.7\big(
|C_{\bar\ell \ell \nu,\mu\mu x}^{\tV\tL\tV\tR\tS\tL2}|^2
+|C_{\bar\ell \ell \nu,\mu\mu x}^{\tV\tR\tV\tR\tS\tL2}|^2
\big)
\nn\\
&+0.6|C_{\bar\ell \ell \nu,\mu\mu x}^{\tT\tL\tT\tL\tS\tR1}|^2
+0.4\big(
|C_{\bar\ell \ell \nu,\mu\mu x}^{\tS\tR\tS\tL\tS\tL2}|^2
+|C_{\bar\ell \ell \nu,\mu\mu x}^{\tS\tR\tS\tL\tS\tR3}|^2
+|C_{\bar\ell \ell \nu,\mu\mu x}^{\tS\tL\tS\tL\tS\tL2}|^2
+|C_{\bar\ell \ell \nu,\mu\mu x}^{\tS\tL\tS\tL\tS\tR3}|^2
\big)
+0.3\kappa_3^2|C_{\bar\ell \ell \nu,\mu\mu x}^{\tT\tR\tV\tR\tV\tL1}|^2
\nn\\
&+0.08\kappa_3^2\big(
|C_{\bar\ell \ell \nu,\mu\mu x}^{\tV\tL\tS\tL\tV\tR1}|^2
+|C_{\bar\ell \ell \nu,\mu\mu x}^{\tV\tR\tS\tL\tV\tR1}|^2
\big)
+0.3\big(
|C_{\bar\ell \ell \nu,\mu\mu x}^{\tS\tL\tS\tL\tS\tL1}|^2
+|C_{\bar\ell \ell \nu,\mu\mu x}^{\tS\tR\tS\tL\tS\tL1}|^2
\big)
+0.2\big(
|C_{\bar\ell \ell \nu,\mu\mu x}^{\tS\tL\tS\tL\tS\tR2}|^2
+|C_{\bar\ell \ell \nu,\mu\mu x}^{\tS\tR\tS\tL\tS\tR2}|^2
\big)
\nn\\
&+0.1\kappa_3^2|C_{\bar\ell \ell \nu,\mu\mu x}^{\tT\tR\tV\tR\tV\tL2}|^2
+0.02\kappa_3^2\big(
|C_{\bar\ell \ell \nu,\mu\mu x}^{\tV\tL\tS\tL\tV\tR2}|^2
+|C_{\bar\ell \ell \nu,\mu\mu x}^{\tV\tR\tS\tL\tV\tR2}|^2
\big)
+0.08\kappa_3^2|C_{\bar\ell \ell \nu,\mu\mu x}^{\tT\tR\tV\tR\tV\tL3}|^2
\nn\\
&+0.02\kappa_3^2\big(
|C_{\bar\ell \ell \nu,\mu\mu x}^{\tV\tL\tS\tL\tV\tR3}|^2
+|C_{\bar\ell \ell \nu,\mu\mu x}^{\tV\tR\tS\tL\tV\tR3}|^2
\big)
+0.05\big(
|C_{\bar\ell \ell \nu,\mu\mu x}^{\tV\tL\tV\tR\tS\tL1}|^2
+|C_{\bar\ell \ell \nu,\mu\mu x}^{\tV\tR\tV\tR\tS\tL1}|^2
\big)
\nn\\
&+0.01\big(
|C_{\bar\ell \ell \nu,\mu\mu x}^{\tS\tL\tS\tL\tS\tR1}|^2
+|C_{\bar\ell \ell \nu,\mu\mu x}^{\tS\tR\tS\tL\tS\tR1}|^2
\big)
+10^{-4}\kappa_4^2|C_{\bar\ell \ell \nu,\mu\mu x}^{\tT\tL\tS\tL\tT\tL}|^2
+\cdots,
\\
{\Gamma_{n\to e^-\mu^+ \bar{\nu}_xK^0} \over 10^{-13}{\text{GeV}}^{11}} =\,& 
44(|C_{\bar\ell \ell \nu,e\mu x}^{\tT\tL\tT\tL\tS\tL2}|^2
+|C_{\bar\ell \ell \nu,e\mu x}^{\tT\tL\tT\tL\tS\tR3}|^2)
+30|C_{\bar\ell \ell \nu,e\mu x}^{\tT\tL\tT\tL\tS\tL1}|^2
+20|C_{\bar\ell \ell \nu,e\mu x}^{\tT\tL\tT\tL\tS\tR2}|^2
+4\big(|C_{\bar\ell \ell \nu,e\mu x}^{\tV\tL\tV\tR\tS\tL3}|^2+|C_{\bar\ell \ell \nu,e\mu x}^{\tV\tL\tV\tR\tS\tR2}|^2
\nn\\
&+|C_{\bar\ell \ell \nu,e\mu x}^{\tV\tR\tV\tR\tS\tR2}|^2
+|C_{\bar\ell \ell \nu,e\mu x}^{\tV\tR\tV\tR\tS\tL3}|^2\big)
+3\big(|C_{\bar\ell \ell \nu,e\mu x}^{\tV\tL\tV\tR\tS\tR1}|^2
+|C_{\bar\ell \ell \nu,e\mu x}^{\tV\tR\tV\tR\tS\tR1}|^2\big)
+2\big(|C_{\bar\ell \ell \nu,e\mu x}^{\tV\tL\tV\tR\tS\tL2}|^2
+|C_{\bar\ell \ell \nu,e\mu x}^{\tV\tR\tV\tR\tS\tL2}|^2\big)
\nn\\
&+1.5|C_{\bar\ell \ell \nu,e\mu x}^{\tT\tL\tT\tL\tS\tR1}|^2
+1.0\big(|C_{\bar\ell \ell \nu,e\mu x }^{\tS\tR\tS\tL\tS\tL2}|^2
+|C_{\bar\ell \ell \nu,e\mu x }^{\tS\tR\tS\tL\tS\tR3}|^2
+|C_{\bar\ell \ell \nu,e\mu x }^{\tS\tL\tS\tL\tS\tL2}|^2
+|C_{\bar\ell \ell \nu,e\mu x }^{\tS\tL\tS\tL\tS\tR3}|^2\big)
+0.7\kappa_3^2|C_{\bar\ell \ell \nu,e\mu x}^{\tT\tR\tV\tR\tV\tL1}|^2
\nn\\
&+0.6\big(|C_{\bar\ell \ell \nu,e\mu x}^{\tS\tL\tS\tL\tS\tL1}|^2
+|C_{\bar\ell \ell \nu,e\mu x}^{\tS\tR\tS\tL\tS\tL1}|^2\big)
+0.4\big(|C_{\bar\ell \ell \nu,e\mu x}^{\tS\tL\tS\tL\tS\tR2}|^2
+|C_{\bar\ell \ell \nu,e\mu x}^{\tS\tR\tS\tL\tS\tR2}|^2\big)
+0.2\kappa_3^2\big(|C_{\bar\ell \ell \nu,e\mu x}^{\tV\tL\tS\tL\tV\tR1}|^2
+|C_{\bar\ell \ell \nu,e\mu x}^{\tV\tR\tS\tL\tV\tR1}|^2\big)
\nn\\
&+0.2\kappa_3^2|C_{\bar\ell \ell \nu,e\mu x}^{\tT\tR\tV\tR\tV\tL2}|^2
+0.2\kappa_3^2|C_{\bar\ell \ell \nu,e\mu x}^{\tT\tR\tV\tR\tV\tL3}|^2
+0.05\kappa_3^2\big(|C_{\bar\ell \ell \nu,e\mu x}^{\tV\tL\tS\tL\tV\tR2}|^2
+|C_{\bar\ell \ell \nu,e\mu x}^{\tV\tR\tS\tL\tV\tR2}|^2
\big)
\nn\\
&+0.04\kappa_3^2\big(|C_{\bar\ell \ell \nu,e\mu x}^{\tV\tL\tS\tL\tV\tR3}|^2
+|C_{\bar\ell \ell \nu,e\mu x}^{\tV\tR\tS\tL\tV\tR3}|^2
\big)
+0.1\big(|C_{\bar\ell \ell \nu,e\mu x}^{\tV\tL\tV\tR\tS\tL1}|^2
+|C_{\bar\ell \ell \nu,e\mu x}^{\tV\tR\tV\tR\tS\tL1}|^2
\big)
\nn\\
&
+0.03\big(|C_{\bar\ell \ell \nu,e\mu x}^{\tS\tL\tS\tL\tS\tR1}|^2
+|C_{\bar\ell \ell \nu,e\mu x}^{\tS\tR\tS\tL\tS\tR1}|^2
\big)
+2\cdot10^{-4}\kappa_4^2|C_{\bar\ell \ell \nu,e\mu x}^{\tT\tL\tS\tL\tT\tL}|^2
+\cdots,
\\
{\Gamma_{n\to \mu^-e^+ \bar{\nu}_xK^0} \over 10^{-13}{\text{GeV}}^{11}} =\,& 
44(|C_{\bar\ell \ell \nu,\mu e x}^{\tT\tL\tT\tL\tS\tL2}|^2
+|C_{\bar\ell \ell \nu,\mu e x}^{\tT\tL\tT\tL\tS\tR3}|^2)
+30|C_{\bar\ell \ell \nu,\mu e x}^{\tT\tL\tT\tL\tS\tL1}|^2
+20|C_{\bar\ell \ell \nu,\mu e x}^{\tT\tL\tT\tL\tS\tR2}|^2
+4\big(|C_{\bar\ell \ell \nu,\mu e x}^{\tV\tL\tV\tR\tS\tL3}|^2+|C_{\bar\ell \ell \nu,\mu e x}^{\tV\tL\tV\tR\tS\tR2}|^2
\nn\\
&+|C_{\bar\ell \ell \nu,\mu e x}^{\tV\tR\tV\tR\tS\tR2}|^2
+|C_{\bar\ell \ell \nu,\mu e x}^{\tV\tR\tV\tR\tS\tL3}|^2\big)
+3\big(|C_{\bar\ell \ell \nu,\mu e x}^{\tV\tL\tV\tR\tS\tR1}|^2
+|C_{\bar\ell \ell \nu,\mu e x}^{\tV\tR\tV\tR\tS\tR1}|^2\big)
+2\big(|C_{\bar\ell \ell \nu,\mu e x}^{\tV\tL\tV\tR\tS\tL2}|^2
+|C_{\bar\ell \ell \nu,\mu e x}^{\tV\tR\tV\tR\tS\tL2}|^2\big)
\nn\\
&+1.5|C_{\bar\ell \ell \nu,\mu e x}^{\tT\tL\tT\tL\tS\tR1}|^2
+1.0\big(|C_{\bar\ell \ell \nu,\mu e x}^{\tS\tR\tS\tL\tS\tL2}|^2
+|C_{\bar\ell \ell \nu,\mu e x}^{\tS\tR\tS\tL\tS\tR3}|^2
+|C_{\bar\ell \ell \nu,\mu e x}^{\tS\tL\tS\tL\tS\tL2}|^2
+|C_{\bar\ell \ell \nu,\mu e x}^{\tS\tL\tS\tL\tS\tR3}|^2\big)
+0.7\kappa_3^2|C_{\bar\ell \ell \nu,\mu e x}^{\tT\tR\tV\tR\tV\tL1}|^2
\nn\\
&+0.6\big(|C_{\bar\ell \ell \nu,\mu e x}^{\tS\tL\tS\tL\tS\tL1}|^2
+|C_{\bar\ell \ell \nu,\mu e x}^{\tS\tR\tS\tL\tS\tL1}|^2\big)
+0.4\big(|C_{\bar\ell \ell \nu,\mu e x}^{\tS\tL\tS\tL\tS\tR2}|^2
+|C_{\bar\ell \ell \nu,\mu e x}^{\tS\tR\tS\tL\tS\tR2}|^2\big)
+0.2\kappa_3^2\big(|C_{\bar\ell \ell \nu,\mu e x }^{\tV\tL\tS\tL\tV\tR1}|^2
+|C_{\bar\ell \ell \nu,\mu e x }^{\tV\tR\tS\tL\tV\tR1}|^2\big)
\nn\\
&+0.2\kappa_3^2|C_{\bar\ell \ell \nu,\mu e x }^{\tT\tR\tV\tR\tV\tL3}|^2
+0.2\kappa_3^2|C_{\bar\ell \ell \nu,\mu e x }^{\tT\tR\tV\tR\tV\tL2}|^2
+0.05\kappa_3^2\big(|C_{\bar\ell \ell \nu,\mu e x}^{\tV\tL\tS\tL\tV\tR2}|^2
+|C_{\bar\ell \ell \nu,\mu e x}^{\tV\tR\tS\tL\tV\tR2}|^2
\big)
\nn\\
&+0.04\kappa_3^2\big(|C_{\bar\ell \ell \nu,\mu e x}^{\tV\tL\tS\tL\tV\tR3}|^2
+|C_{\bar\ell \ell \nu,\mu e x}^{\tV\tR\tS\tL\tV\tR3}|^2
\big)
+0.1\big(|C_{\bar\ell \ell \nu,\mu ex }^{\tV\tL\tV\tR\tS\tL1}|^2
+|C_{\bar\ell \ell \nu,\mu ex }^{\tV\tR\tV\tR\tS\tL1}|^2
\big)
\nn\\
&+0.03\big(|C_{\bar\ell \ell \nu,\mu ex }^{\tS\tL\tS\tL\tS\tR1}|^2
+|C_{\bar\ell \ell \nu,\mu ex }^{\tS\tR\tS\tL\tS\tR1}|^2
\big)
+2\cdot10^{-4}\kappa_4^2|C_{\bar\ell \ell \nu,\mu e x}^{\tT\tL\tS\tL\tT\tL}|^2
+\cdots.
\end{align}
\end{subequations}

\subsection{\tf{$\Delta L = +1$: $p\to\bar\nu_x\bar\nu_y\nu_zK^+,\,
n\to\bar\nu_x\bar\nu_y\nu_z K^0$}{N2vbvbvK}}

$\bullet$ Proton case.
\begin{subequations}
\begin{align}
{\Gamma_{p\to \bar\nu_x\bar{\nu}_{y\neq x} \nu_zK^+} \over 10^{-13}{\text{GeV}}^{11}} =\,&
1.9(|C_{\nu \nu \bar\nu,xyz}^{\tS\tL\tS\tR\tS\tR2}|^2
+|C_{\nu \nu \bar\nu,xyz}^{\tS\tL\tS\tR\tS\tL3}|^2)
+
1.3\kappa_3^2|C_{\nu \nu \bar\nu,xyz}^{\tT\tL\tV\tL\tV\tR2}|^2
+0.9|C_{\nu \nu \bar\nu,xyz}^{\tS\tL\tS\tR\tS\tL1}|^2
+0.4\kappa_3^2|C_{\nu \nu \bar\nu,xyz}^{\tT\tL\tV\tL\tV\tR1}|^2
\nn\\
&
+0.3\kappa_3^2|C_{\nu \nu \bar\nu,xyz}^{\tT\tL\tV\tL\tV\tR3}|^2
+0.07(|C_{\nu \nu \bar\nu,xyz}^{\tS\tL\tS\tR\tS\tR1}|^2
+|C_{\nu \nu \bar\nu,xyz}^{\tS\tL\tS\tR\tS\tL2}|^2)
+\cdots,
\\
{\Gamma_{p\to \bar\nu_x\bar{\nu}_x \nu_y K^+} \over 10^{-13}{\text{GeV}}^{11}} =\,&   
3.8(|C_{\nu \nu \bar\nu,xxy}^{\tS\tL\tS\tR\tS\tR2}|^2
+|C_{\nu \nu \bar\nu,xxy}^{\tS\tL\tS\tR\tS\tL3}|^2)
+1.7|C_{\nu \nu \bar\nu,xxy}^{\tS\tL\tS\tR\tS\tL1}|^2
+0.14(|C_{\nu \nu \bar\nu,xxy}^{\tS\tL\tS\tR\tS\tR1}|^2
+|C_{\nu \nu \bar\nu,xxy}^{\tS\tL\tS\tR\tS\tL2}|^2)
+\cdots.
\end{align}
\end{subequations}

$\bullet$ Neutron case.
\begin{subequations}
\begin{align}
{\Gamma_{n\to \bar\nu_x\bar{\nu}_{y\neq x} \nu_zK^0} \over 10^{-13}{\text{GeV}}^{11}} =\,&   
1.8(|C_{\nu \nu \bar\nu,xyz}^{\tS\tL\tS\tR\tS\tR2}|^2
+|C_{\nu \nu \bar\nu,xyz}^{\tS\tL\tS\tR\tS\tL3}|^2)
+1.2|C_{\nu \nu \bar\nu,xyz}^{\tS\tL\tS\tR\tS\tR1}|^2
+1.2\kappa_3^2|C_{\nu \nu \bar\nu,xyz}^{\tT\tL\tV\tL\tV\tR1}|^2
+0.8|C_{\nu \nu \bar\nu,xyz}^{\tS\tL\tS\tR\tS\tL2}|^2
\nn\\
&+0.3\kappa_3^2(|C_{\nu \nu \bar\nu,xyz}^{\tT\tL\tV\tL\tV\tR2}|^2
+|C_{\nu \nu \bar\nu,xyz}^{\tT\tL\tV\tL\tV\tR3}|^2)
+0.07|C_{\nu \nu \bar\nu,xyz}^{\tS\tL\tS\tR\tS\tL1}|^2
+\cdots,
\\
{\Gamma_{n\to \bar\nu_x\bar{\nu}_x \nu_y K^0} \over 10^{-13}{\text{GeV}}^{11}} =\,&   
3.6(|C_{\nu \nu \bar\nu,xxy}^{\tS\tL\tS\tR\tS\tR2}|^2
+|C_{\nu \nu \bar\nu,xxy}^{\tS\tL\tS\tR\tS\tL3}|^2)
+2.5|C_{\nu \nu \bar\nu,xxy}^{\tS\tL\tS\tR\tS\tR1}|^2
+1.6|C_{\nu \nu \bar\nu,xxy}^{\tS\tL\tS\tR\tS\tL2}|^2
+0.14|C_{\nu \nu \bar\nu,xxy}^{\tS\tL\tS\tR\tS\tL1}|^2
+\cdots.
\end{align}
\end{subequations}

\subsection{\tf{$\Delta L = -1$: 
$n\to\ell_x^-\ell_y^-\ell_z^+K^+$}{n2lmlmlpKp}}

\begin{subequations}
\begin{align}
{\Gamma_{n\to e^-e^-e^+K^+} \over 10^{-13}{\text{GeV}}^{11}}  
=\,& 
5.3|C_{\bar\ell\bar\ell\ell,eee}^{\tV\tL\tV\tR\tS\tL}|^2
+3.3|C_{\bar\ell\bar\ell\ell,eee}^{\tV\tL\tV\tR\tS\tR}|^2
+2.7\big(|C_{\bar\ell\bar\ell\ell,eee}^{\tS\tL\tS\tR\tS\tL}|^2+|C_{\bar\ell\bar\ell\ell,eee}^{\tS\tL\tS\tL\tS\tR}|^2\big)
+1.7\big(|C_{\bar\ell\bar\ell\ell,eee}^{\tS\tL\tS\tL\tS\tL}|^2+|C_{\bar\ell\bar\ell\ell,eee}^{\tS\tL\tS\tR\tS\tR}|^2\big)
\nn\\
&+0.3\kappa_3^2|C_{\bar\ell\bar\ell\ell,eee}^{\tV\tL\tS\tL\tV\tR2}|^2
+0.07\kappa_3^2|C_{\bar\ell\bar\ell\ell,eee}^{\tV\tL\tS\tL\tV\tR1}|^2
+\tL\leftrightarrow\tR+\cdots,
\\
{\Gamma_{n\to \mu^-\mu^-\mu^+K^+} \over 10^{-13}{\text{GeV}}^{11}}  
=\,& 
0.2|C_{\bar\ell\bar\ell\ell,\mu\mu\mu}^{\tV\tL\tV\tR\tS\tL}|^2
+0.16|C_{\bar\ell\bar\ell\ell,\mu\mu\mu}^{\tV\tL\tV\tR\tS\tR}|^2
+0.16\big(|C_{\bar\ell\bar\ell\ell,\mu\mu\mu}^{\tS\tL\tS\tR\tS\tL}|^2+|C_{\bar\ell\bar\ell\ell,\mu\mu\mu}^{\tS\tL\tS\tL\tS\tR}|^2\big)
+0.1\big(|C_{\bar\ell\bar\ell\ell,\mu\mu\mu}^{\tS\tL\tS\tL\tS\tL}|^2+|C_{\bar\ell\bar\ell\ell,\mu\mu\mu}^{\tS\tL\tS\tR\tS\tR}|^2\big)
\nn\\
&+0.04\kappa_3^2|C_{\bar\ell\bar\ell\ell,\mu\mu\mu}^{\tV\tL\tS\tL\tV\tR2}|^2
+0.007\kappa_3^2|C_{\bar\ell\bar\ell\ell,\mu\mu\mu}^{\tV\tL\tS\tL\tV\tR1}|^2
+\tL\leftrightarrow\tR+\cdots,
\\
{\Gamma_{n\to e^-e^-\mu^+K^+} \over 10^{-13}{\text{GeV}}^{11}}  
=\,& 
2.8|C_{\bar\ell\bar\ell\ell,ee\mu}^{\tV\tL\tV\tR\tS\tL}|^2
+1.8|C_{\bar\ell\bar\ell\ell,ee\mu}^{\tV\tL\tV\tR\tS\tR}|^2
+1.4\big(|C_{\bar\ell\bar\ell\ell,ee\mu}^{\tS\tL\tS\tR\tS\tL}|^2+|C_{\bar\ell\bar\ell\ell,ee\mu}^{\tS\tL\tS\tL\tS\tR}|^2\big)
+0.9\big(|C_{\bar\ell\bar\ell\ell,ee\mu}^{\tS\tL\tS\tL\tS\tL}|^2+|C_{\bar\ell\bar\ell\ell,ee\mu}^{\tS\tL\tS\tR\tS\tR}|^2\big)
\nn\\
&+0.2\kappa_3^2|C_{\bar\ell\bar\ell\ell,ee\mu}^{\tV\tL\tS\tL\tV\tR2}|^2
+0.04\kappa_3^2|C_{\bar\ell\bar\ell\ell,ee\mu}^{\tV\tL\tS\tL\tV\tR1}|^2
+\tL\leftrightarrow\tR+\cdots,
\\
{\Gamma_{n\to \mu^-\mu^-e^+K^+} \over 10^{-13}{\text{GeV}}^{11}}  
=\,& 
0.9|C_{\bar\ell\bar\ell\ell,\mu\mu e}^{\tV\tL\tV\tR\tS\tL}|^2
+0.6|C_{\bar\ell\bar\ell\ell,\mu\mu e}^{\tV\tL\tV\tR\tS\tR}|^2
+0.6\big(|C_{\bar\ell\bar\ell\ell,\mu\mu e}^{\tS\tL\tS\tR\tS\tL}|^2+|C_{\bar\ell\bar\ell\ell,\mu\mu e}^{\tS\tL\tS\tL\tS\tR}|^2\big)
+0.4\big(|C_{\bar\ell\bar\ell\ell,\mu\mu e}^{\tS\tL\tS\tL\tS\tL}|^2+|C_{\bar\ell\bar\ell\ell,\mu\mu e}^{\tS\tL\tS\tR\tS\tR}|^2\big)
\nn\\
&+0.1\kappa_3^2|C_{\bar\ell\bar\ell\ell,\mu\mu e}^{\tV\tL\tS\tL\tV\tR2}|^2
+0.02\kappa_3^2|C_{\bar\ell\bar\ell\ell,\mu\mu e}^{\tV\tL\tS\tL\tV\tR1}|^2
+\tL\leftrightarrow\tR+\cdots,
\\
{\Gamma_{n\to e^-\mu^-e^+K^+} \over 10^{-13}{\text{GeV}}^{11}}  
=\,& 33|C_{\bar\ell\bar\ell\ell,e\mu e}^{\tT\tL\tT\tL\tS\tR}|^2
+21|C_{\bar\ell\bar\ell\ell,e\mu e}^{\tT\tL\tT\tL\tS\tL}|^2
+3\big(|C_{\bar\ell\bar\ell\ell,e\mu e}^{\tV\tL\tV\tR\tS\tL}|^2+|C_{\bar\ell\bar\ell\ell,\mu ee}^{\tV\tL\tV\tR\tS\tL}|^2\big)
+2\big(|C_{\bar\ell\bar\ell\ell,e\mu e}^{\tV\tL\tV\tR\tS\tR}|^2+|C_{\bar\ell\bar\ell\ell,\mu ee}^{\tV\tL\tV\tR\tS\tR}|^2\big)
\nn\\
&+0.7\kappa_3^2|C_{\bar\ell\bar\ell\ell,e\mu e}^{\tT\tL\tV\tL\tV\tR2}|^2
+0.7\big(|C_{\bar\ell\bar\ell\ell,e\mu e}^{\tS\tL\tS\tR\tS\tL}|^2+|C_{\bar\ell\bar\ell\ell,e\mu e}^{\tS\tL\tS\tL\tS\tR}|^2\big)
+0.4\big(|C_{\bar\ell\bar\ell\ell,e\mu e}^{\tS\tL\tS\tL\tS\tL}|^2+|C_{\bar\ell\bar\ell\ell,e\mu e}^{\tS\tL\tS\tR\tS\tR}|^2\big)
\nn\\
&+0.2\kappa_3^2\big(|C_{\bar\ell\bar\ell\ell,e\mu e}^{\tV\tL\tS\tL\tV\tR2}|^2+|C_{\bar\ell\bar\ell\ell,\mu e e}^{\tV\tL\tS\tL\tV\tR2}|^2\big)
+0.2\kappa_3^2|C_{\bar\ell\bar\ell\ell,e\mu e}^{\tT\tL\tV\tL\tV\tR1}|^2
+0.04\kappa_3^2\big(|C_{\bar\ell\bar\ell\ell,e\mu e}^{\tV\tL\tS\tL\tV\tR1}|^2+|C_{\bar\ell\bar\ell\ell,\mu e e}^{\tV\tL\tS\tL\tV\tR1}|^2\
\big)
\nn\\
&+1\cdot10^{-3}\kappa_4^2|C_{\bar\ell\bar\ell\ell, e\mu e}^{\tT\tL\tS\tL\tT\tL}|^2
+\tL\leftrightarrow\tR+\cdots,
\\
{\Gamma_{n\to e^-\mu^-\mu^+K^+} \over 10^{-13}{\text{GeV}}^{11}}  
=\,& 14|C_{\bar\ell\bar\ell\ell,e\mu \mu}^{\tT\tL\tT\tL\tS\tR}|^2
+9|C_{\bar\ell\bar\ell\ell,e\mu \mu}^{\tT\tL\tT\tL\tS\tL}|^2
+1.0\big(|C_{\bar\ell\bar\ell\ell,e\mu \mu}^{\tV\tL\tV\tR\tS\tL}|^2+|C_{\bar\ell\bar\ell\ell,\mu e\mu}^{\tV\tL\tV\tR\tS\tL}|^2\big)
+0.8\big(|C_{\bar\ell\bar\ell\ell,e\mu \mu}^{\tV\tL\tV\tR\tS\tR}|^2+|C_{\bar\ell\bar\ell\ell,\mu e\mu}^{\tV\tL\tV\tR\tS\tR}|^2\big)
\nn\\
&+0.3\kappa_3^2|C_{\bar\ell\bar\ell\ell,e\mu \mu}^{\tT\tL\tV\tL\tV\tR2}|^2
+0.3\big(|C_{\bar\ell\bar\ell\ell,e\mu \mu}^{\tS\tL\tS\tR\tS\tL}|^2+|C_{\bar\ell\bar\ell\ell,e\mu \mu}^{\tS\tL\tS\tL\tS\tR}|^2\big)
+0.2\big(|C_{\bar\ell\bar\ell\ell,e\mu \mu}^{\tS\tL\tS\tL\tS\tL}|^2+|C_{\bar\ell\bar\ell\ell,e\mu \mu}^{\tS\tL\tS\tR\tS\tR}|^2\big)
\nn\\
&+0.1\kappa_3^2\big(|C_{\bar\ell\bar\ell\ell,e\mu \mu}^{\tV\tL\tS\tL\tV\tR2}|^2+|C_{\bar\ell\bar\ell\ell,\mu e \mu}^{\tV\tL\tS\tL\tV\tR2}|^2\big)
+0.07\kappa_3^2|C_{\bar\ell\bar\ell\ell,e\mu \mu}^{\tT\tL\tV\tL\tV\tR1}|^2
+0.02\kappa_3^2\big(|C_{\bar\ell\bar\ell\ell,e\mu \mu}^{\tV\tL\tS\tL\tV\tR1}|^2+|C_{\bar\ell\bar\ell\ell,\mu e \mu}^{\tV\tL\tS\tL\tV\tR1}|^2\big)
\nn\\
&+4\cdot10^{-4}\kappa_4^2|C_{\bar\ell\bar\ell\ell, e\mu \mu}^{\tT\tL\tS\tL\tT\tL}|^2
+\tL\leftrightarrow\tR+\cdots.
\end{align}
\end{subequations}

\subsection{\tf{$\Delta L = -1$: 
$p\to\nu_x\nu_y\ell_z^+K^0$ (d)}{p2vvlpK0}}

\begin{subequations}
\begin{align}
{\Gamma_{p\to \nu_x\nu_{y\neq x} e^+K^0} \over 10^{-13}{\text{GeV}}^{11}} =\,& 
60|C_{\bar\nu \bar\nu \ell,xye}^{\tT\tR\tT\tR\tS\tL}|^2
+36|C_{\bar\nu \bar\nu \ell,xye}^{\tT\tR\tT\tR\tS\tR}|^2
+1.2(|C_{\bar\nu \bar\nu \ell,xye}^{\tS\tR\tS\tR\tS\tL}|^2
+|C_{\bar\nu \bar\nu \ell,xye}^{\tS\tR\tS\tL\tS\tR}|^2)
+1.1\kappa_3^2|C_{\bar\nu \bar\nu \ell,xye}^{\tT\tR\tV\tR\tV\tL2}|^2
\nn\\
&
+0.8(|C_{\bar\nu \bar\nu \ell,xye}^{\tS\tR\tS\tR\tS\tR}|^2
+|C_{\bar\nu \bar\nu \ell,xye}^{\tS\tR\tS\tL\tS\tL}|^2)
+0.3\kappa_3^2|C_{\bar\nu \bar\nu \ell,xye}^{\tT\tR\tV\tR\tV\tL1}|^2
+2\cdot10^{-3}\kappa_4^2|C_{\bar\nu \bar\nu \ell,xye}^{\tT\tR\tS\tR\tT\tR}|^2
+\cdots,
\\
{\Gamma_{p\to \nu_x\nu_x e^+K^0} \over 10^{-13}{\text{GeV}}^{11}} =\,&  
2.5\big(|C_{\bar\nu \bar\nu \ell,xxe}^{\tS\tR\tS\tL\tS\tR}|^2
+|C_{\bar\nu \bar\nu \ell,xxe}^{\tS\tR\tS\tR\tS\tL}|^2\big)
+
1.5\big(|C_{\bar\nu \bar\nu \ell,xxe}^{\tS\tR\tS\tR\tS\tR}|^2
+|C_{\bar\nu \bar\nu \ell,xxe}^{\tS\tR\tS\tL\tS\tL}|^2\big)
+\cdots,
\\
{\Gamma_{p\to \nu_x\nu_{y\neq x} \mu^+K^0} \over 10^{-13}{\text{GeV}}^{11}} =\,& 
30|C_{\bar\nu \bar\nu \ell,xy\mu}^{\tT\tR\tT\tR\tS\tL}|^2
+19|C_{\bar\nu \bar\nu \ell,xy\mu}^{\tT\tR\tT\tR\tS\tR}|^2
+0.7\kappa_3^2|C_{\bar\nu \bar\nu \ell,xy\mu}^{\tT\tR\tV\tR\tV\tL2}|^2
+0.6(|C_{\bar\nu \bar\nu \ell,xy\mu}^{\tS\tR\tS\tR\tS\tL}|^2
+|C_{\bar\nu \bar\nu \ell,xy\mu}^{\tS\tR\tS\tL\tS\tR}|^2)
\nn\\
&
+0.4(|C_{\bar\nu \bar\nu \ell,xy\mu}^{\tS\tR\tS\tR\tS\tR}|^2
+|C_{\bar\nu \bar\nu \ell,xy\mu}^{\tS\tR\tS\tL\tS\tL}|^2)
+0.1\kappa_3^2|C_{\bar\nu \bar\nu \ell,xy\mu}^{\tT\tR\tV\tR\tV\tL1}|^2
+9\cdot10^{-4}\kappa_4^2|C_{\bar\nu \bar\nu \ell,xy\mu}^{\tT\tR\tS\tR\tT\tR}|^2
+\cdots,
\\
{\Gamma_{p\to \nu_x\nu_x \mu^+K^0} \over 10^{-13}{\text{GeV}}^{11}} =\,&  
1.3\big(|C_{\bar\nu \bar\nu \ell,xx \mu}^{\tS\tR\tS\tL\tS\tR}|^2
+|C_{\bar\nu \bar\nu \ell,xx \mu}^{\tS\tR\tS\tR\tS\tL}|^2\big)
+0.8\big(|C_{\bar\nu \bar\nu \ell,xx \mu}^{\tS\tR\tS\tR\tS\tR}|^2
+|C_{\bar\nu \bar\nu \ell,xx \mu}^{\tS\tR\tS\tL\tS\tL}|^2\big)
+\cdots.
\end{align}
\end{subequations}

\subsection{\tf{$\Delta L = -1$: 
$n\to\nu_x\bar\nu_y\ell_z^-K^+$}{n2vvblmKp}}

\begin{subequations}
\begin{align}
{\Gamma_{n\to \nu_x\bar{\nu}_y e^-K^+} \over 10^{-13}{\text{GeV}}^{11}} =\,&   
5\big(|C_{\bar\nu \nu \bar\ell,xye}^{\tV\tL\tV\tL\tS\tR}|^2
+|C_{\bar\nu \nu \bar\ell,xye}^{\tV\tL\tV\tR\tS\tL}|^2\big)
+3\big(|C_{\bar\nu \nu \bar\ell,xye}^{\tV\tL\tV\tL\tS\tL}|^2+|C_{\bar\nu \nu \bar\ell,xye}^{\tV\tL\tV\tR\tS\tR}|^2\big)
\nn\\
&+ 0.3\kappa_3^2\big(|C_{\bar\nu \nu \bar\ell,xye}^{\tV\tL\tS\tL\tV\tR2}|^2
+|C_{\bar\nu \nu \bar\ell,xye}^{\tV\tL\tS\tR\tV\tL2}|^2
\big)
+ 0.07\kappa_3^2\big(|C_{\bar\nu \nu \bar\ell,xye}^{\tV\tL\tS\tL\tV\tR1}|^2+
|C_{\bar\nu \nu \bar\ell,xye}^{\tV\tL\tS\tR\tV\tL1}|^2
\big)
+\cdots,
\\
{\Gamma_{n\to \nu_x\bar{\nu}_y \mu^-K^+} \over 10^{-13}{\text{GeV}}^{11}} =\,&   
2.8\big(|C_{\bar\nu \nu \bar\ell,xy\mu}^{\tV\tL\tV\tL\tS\tR}|^2
+|C_{\bar\nu \nu \bar\ell,xy\mu}^{\tV\tL\tV\tR\tS\tL}|^2\big)
+1.8\big(|C_{\bar\nu \nu \bar\ell,xy\mu}^{\tV\tL\tV\tL\tS\tL}|^2+|C_{\bar\nu \nu \bar\ell,xy\mu}^{\tV\tL\tV\tR\tS\tR}|^2\big)
\nn\\
&+ 0.2\kappa_3^2\big(
|C_{\bar\nu \nu \bar\ell,xy\mu}^{\tV\tL\tS\tR\tV\tL2}|^2
+|C_{\bar\nu \nu \bar\ell,xy\mu}^{\tV\tL\tS\tL\tV\tR2}|^2
\big)
+ 0.04\kappa_3^2\big(|C_{\bar\nu \nu \bar\ell,xy\mu}^{\tV\tL\tS\tL\tV\tR1}|^2+
|C_{\bar\nu \nu \bar\ell,xy\mu}^{\tV\tL\tS\tR\tV\tL1}|^2
\big)
+\cdots.
\end{align}
\end{subequations}

\subsection{\tf{$\Delta L = +3$: 
$p\to\bar\nu_x\bar\nu_y\bar\nu_zK^+,\, n\to\bar\nu_x\bar\nu_y\bar\nu_zK^0$ (e)}{N2vbvbvbK}}

$\bullet$ Proton case.
\begin{subequations}
\begin{align}
{\Gamma_{p\to \bar\nu_e\bar\nu_\mu \bar\nu_\tau K^+} \over 10^{-13}{\text{GeV}}^{11}} =\,&    
1.4\big(|C_{\nu \nu \nu,e\mu\tau}^{\tS\tL\tS\tL\tS\tL2}|^2
+|C_{\nu \nu \nu,\mu\tau e}^{\tS\tL\tS\tL\tS\tL2}|^2
+|C_{\nu \nu \nu,e\mu\tau}^{\tS\tL\tS\tL\tS\tR3}|^2
+|C_{\nu \nu \nu,\mu\tau e}^{\tS\tL\tS\tL\tS\tR3}|^2\big)
+0.6\big(|C_{\nu \nu \nu,e\mu\tau}^{\tS\tL\tS\tL\tS\tR1}|^2
+|C_{\nu \nu \nu,\mu\tau e}^{\tS\tL\tS\tL\tS\tR1}|^2
\big)
\nn\\
&
+0.06\big(|C_{\nu \nu \nu,e\mu\tau}^{\tS\tL\tS\tL\tS\tL1}|^2
+|C_{\nu \nu \nu,\mu\tau e}^{\tS\tL\tS\tL\tS\tL1}|^2
+|C_{\nu \nu \nu,e\mu\tau}^{\tS\tL\tS\tL\tS\tR2}|^2
+|C_{\nu \nu \nu,\mu\tau e}^{\tS\tL\tS\tL\tS\tR2}|^2
\big)
\nn\\
&+5\cdot10^{-4}\kappa_4^2|C_{\nu \nu \nu,e\mu\tau}^{\tT\tL\tS\tL\tT\tL}|^2
+\cdots,
\\
{\Gamma_{p\to \bar\nu_x\bar\nu_x \bar\nu_{y\neq x}K^+} \over 10^{-13}{\text{GeV}}^{11}}=\,&    
2.1(|C_{\nu \nu \nu,xxy}^{\tS\tL\tS\tL\tS\tL2}|^2+|C_{\nu \nu \nu,xxy}^{\tS\tL\tS\tL\tS\tR3}|^2)
+1.0|C_{\nu \nu \nu,xxy}^{\tS\tL\tS\tL\tS\tR1}|^2
+0.08(|C_{\nu \nu \nu,xxy}^{\tS\tL\tS\tL\tS\tL1}|^2+|C_{\nu \nu \nu,xxy}^{\tS\tL\tS\tL\tS\tR2}|^2)+\cdots.
\end{align}
\end{subequations}

$\bullet$ Neutron case.
\begin{subequations}
\begin{align}
{\Gamma_{n\to \bar\nu_e\bar\nu_\mu \bar\nu_\tau K^0} \over 10^{-13}{\text{GeV}}^{11}} =\,&   
1.3\big(|C_{\nu \nu \nu,e\mu\tau}^{\tS\tL\tS\tL\tS\tL2}|^2
+|C_{\nu \nu \nu,\mu\tau e}^{\tS\tL\tS\tL\tS\tL2}|^2
+|C_{\nu \nu \nu,e\mu\tau}^{\tS\tL\tS\tL\tS\tR3}|^2
+|C_{\nu \nu \nu,\mu\tau e}^{\tS\tL\tS\tL\tS\tR3}|^2\big)
+0.9\big(|C_{\nu \nu \nu,e\mu\tau}^{\tS\tL\tS\tL\tS\tL1}|^2
+|C_{\nu \nu \nu,\mu\tau e}^{\tS\tL\tS\tL\tS\tL1}|^2
\big)
\nn\\
&
+0.6\big(|C_{\nu \nu \nu,e\mu\tau}^{\tS\tL\tS\tL\tS\tR2}|^2
+|C_{\nu \nu \nu,\mu\tau e}^{\tS\tL\tS\tL\tS\tR2}|^2
\big)
+0.05\big(|C_{\nu \nu \nu,e\mu\tau}^{\tS\tL\tS\tL\tS\tR1}|^2
+|C_{\nu \nu \nu,\mu\tau e}^{\tS\tL\tS\tL\tS\tR1}|^2
\big)
\nn\\
&+4\cdot10^{-4}\kappa_4^2|C_{\nu \nu \nu,e\mu\tau}^{\tT\tL\tS\tL\tT\tL}|^2
+\cdots,
\\
{\Gamma_{n\to \bar\nu_x\bar\nu_x \bar\nu_{y\neq x}K^0} \over 10^{-13}{\text{GeV}}^{11}}=\,&    
2.0(|C_{\nu \nu \nu,xxy}^{\tS\tL\tS\tL\tS\tL2}|^2+|C_{\nu \nu \nu,xxy}^{\tS\tL\tS\tL\tS\tR3}|^2)
+1.4|C_{\nu \nu \nu,xxy}^{\tS\tL\tS\tL\tS\tL1}|^2
+0.9|C_{\nu \nu \nu,xxy}^{\tS\tL\tS\tL\tS\tR2}|^2
+0.08|C_{\nu \nu \nu,xxy}^{\tS\tL\tS\tL\tS\tR1}|^2+\cdots.
\end{align}
\end{subequations}
In the above, the flavor indices $x,y,z$ represent any of $e,\mu,\tau$ unless otherwise specified.  
Similarly, for the remaining four classes of processes listed in \cref{tab:n2lllK}, including 
$n\to\ell_x^-\ell_y^+\nu_z K^0$,
$n\to\nu_x\nu_y\bar\nu_z K^0$,
$p\to\bar\nu_x\bar\nu_y\ell_z^+ K^0$,
and $n\to\nu_x\nu_y\nu_z K^0$, their decay widths can be obtained from the corresponding modes above by $\tL\leftrightarrow\tR$ and $\nu\leftrightarrow\bar\nu$ in the WCs.

\end{widetext}


\bibliography{references.bib}{}
\bibliographystyle{apsrev4-1}

\end{document}